\documentclass[12pt,a4paper,twoside,nofootinbib,spanish]{book}
\usepackage[utf8]{inputenc}
\usepackage{babel}
\usepackage{latexsym,amsmath,amssymb}
\usepackage{graphicx,overpic}
\usepackage{enumerate}
\usepackage{color}

\usepackage{geometry}
\usepackage[T1]{fontenc}
\usepackage{textcomp}
\usepackage{cite}
\usepackage{hyperref}

\allowdisplaybreaks

\makeatletter
\def\cleardoublepage{\clearpage\if@twoside \ifodd\c@page\else
    \hbox{}
    \thispagestyle{empty}
    \newpage
    \if@twocolumn\hbox{}\newpage\fi\fi\fi}
\makeatother \clearpage{\pagestyle{empty}\cleardoublepage}

\numberwithin{equation}{section}

\graphicspath{{figures/}}

\newtheorem{theorem}{Teorema}[section]

\newtheorem{proposition}[theorem]{Proposición}

\newenvironment{proof}[1][Proof]{\begin{trivlist}
\item[\hskip \labelsep {\bfseries #1}]}{\end{trivlist}}
\newenvironment{definition}[1][Definición]{\begin{trivlist}
\item[\hskip \labelsep {\bfseries #1}]}{\end{trivlist}}

\newcommand{\qed}{\nobreak \ifvmode \relax \else
      \ifdim\lastskip<1.5em \hskip-\lastskip
      \hskip1.5em plus0em minus0.5em \fi \nobreak
      \vrule height0.75em width0.5em depth0.25em\fi}

\begin{document}
\thispagestyle{empty}

\begin{titlepage}

\begin{center}
 {\large Departamento de F\'{i}sica Te\'{o}rica II} \\[15pt]
 {\large {Facultad de Ciencias F\'{i}sicas}} \\[15pt]
 {\Large \textbf{Universidad Complutense de Madrid}} \\[20pt]
\begin{figure}[htp]
\centering
 \includegraphics[width=3cm]{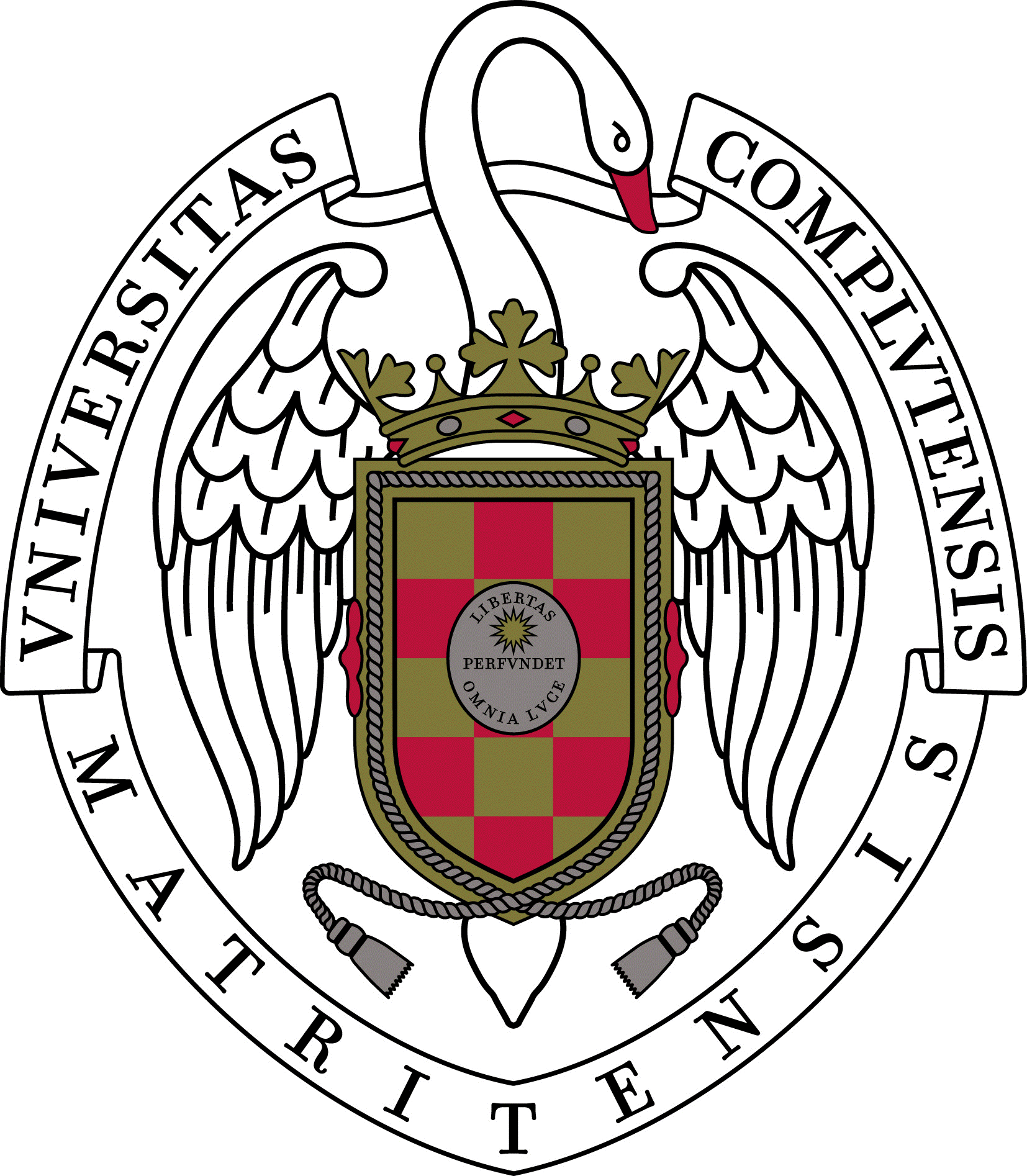}
\end{figure}
\vspace{0.5cm}
{\Large\bf{Tesis Doctoral}}\\[35pt]
{\LARGE \textbf{Cosmología Cuántica de Lazos:\\[10pt]Anisotropías e Inhomogeneidades}}
\bigskip
\vspace*{10mm}\par
{\LARGE \textbf{Mercedes Martín Benito}}\\[40pt]
 {\large {Instituto de Estructura de la Materia}} \\[15pt]
 {\Large \textbf{Consejo Superior de Investigaciones Científicas}}
\vspace*{-4mm}\par
\begin{figure}[htp]
\centering
 \includegraphics[width=8cm]{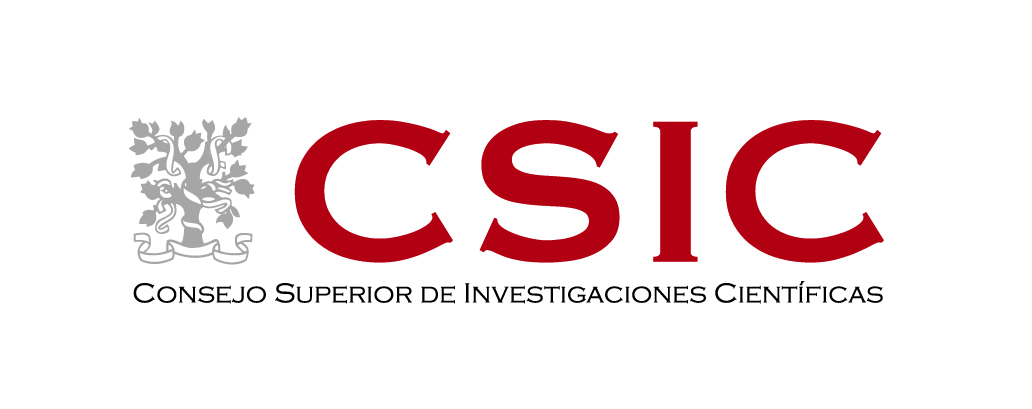}
\end{figure}
\end{center}

\vspace*{-5mm}\par
\hspace*{3mm}
\parbox{5in}{
Directores de tesis:\\
Dr.\ Guillermo A. Mena Marugán\\ Dr.\ Luis J. Garay Elizondo}

\vspace*{-14mm}\par
 \hspace*{8cm} 
\parbox{2.2in}{\flushright
Madrid, 2010}

\setcounter{page}{-5}
\end{titlepage}

\cleardoublepage

\thispagestyle{empty}
\begin{flushright}
\ \\ \par
\vspace{7cm}
{\it A Mª Mercedes Benito Gutiérrez},\\
\vskip 0.2cm
mi madre y maestra,\\
que me dio vida y me educó,\\
que me hizo ser yo.
\end{flushright}

\chapter*{Agradecimientos}
\thispagestyle{empty}

No existen palabras que puedan expresar todo lo agradecida que les estoy a mis
directores de tesis, Guillermo A. Mena Marugán y Luis J. Garay Elizondo, a los que les debo
mi carrera científica. Guillermo, gracias por estar siempre ahí y a veces haber ejercido
casi hasta de padre. Luis, gracias por todo tu apoyo inestimable.

\medskip

A mi madre, a Diego, a Susana, a Javi, y al resto de mi familia les agradezco toda su confianza y
el orgullo que sienten por mí.

\medskip

A David, a Iñaki, a Gil y a Pablo les agradezco que guiaran mis primeros pasos en este bonita época
de doctorado, y a Javi, Dani y Tomek su apoyo desde que están aquí. Gracias chicos por los
buenos ratos disfrutados en el Instituto y fuera de él.
Gracias también al resto de mis compañeros del CSIC y otros compañeros de profesión con los que he
pasado momentos de comida, café, charlas bajo el sol, cañas, fiestas, etc., en especial a Carolina,
a Prado, a Ana y a Natalia.

\medskip

Muchas gracias también a mis amigos de la Universidad, en especial a Jennifer, a Catherine, a Pilar
y a Ángela, por estar cerca. Y también a mis amigos de Condemios de Arriba y alrededores, por los
que merece la pena desconectar del trabajo bajo los pinos de allí. Gracias especiales a mis chicas
del Suanzes, todas ellas como mis hermanas, y a mis chicos ``Los Machos'' también. A Jorge le
agradezco que siempre se haya interesado por mi trabajo (¡uno de los pocos que quiere intentar
leer esta tesis!).

\medskip

A Etera le debo unas gracias enormes (infinitas), porque ha sido un pilar para mí estos últimos
meses.
Y para ti Bea no tengo palabras, muchísimas gracias por estar siempre ahí y facilitarme la vida,
tanto por tu ayuda en casa como por tu apoyo moral. Sin vosotros, la escritura de esta tesis
hubiera sido mucho más difícil.

\medskip

La financiación para la realización de esta tesis doctoral proviene de una beca-contrato del
programa I3P del CSIC junto con el Fondo Social Europeo. También agradezco los fondos recibidos, a
través de proyectos, del MEC/MICINN.

\cleardoublepage

\tableofcontents

\newpage

{\renewcommand{\thechapter}{}\renewcommand{\chaptername}{}
\addtocounter{chapter}{0}
\chapter*{Summary}\markboth{\sl SUMMARY}{\sl SUMMARY}}
\addcontentsline{toc}{chapter}{Summary}

In spite of the impressive progress that cosmology has experienced in recent years, we are still
missing a consistent explanation of the origin of the Universe and the formation of structures,
which should be deduced entirely from a fundamental theory. General Relativity breaks down in the
very initial instants of the history of the Universe, leading to a cosmological singularity of the
big bang type. In this regime General Relativity cannot be trusted, and the very own predictability
of the laws of physics is lost. One expects instead that the physics of the Early Universe
belongs to the realm of Quantum Gravity, namely, a theory of the gravitational field which
incorporates the quantum behavior of nature. One of the most promising candidates for such a theory
is Loop Quantum Gravity. At present, important efforts are being made in order to adapt the
techniques of Loop Quantum Gravity to much simpler settings than those of the complete theory, which
on the other hand remains to be concluded. This is the case of a series of homogeneous cosmological
models obtained from General Relativity by symmetry reduction. The resulting field of research is
known under the general name of Loop Quantum Cosmology (LQC).

As a necessary step towards the extraction of realistic results from LQC, we should consider the
inclusion of inhomogeneities, which play a central role in current cosmology. The main goal of this
thesis is to progress in this direction. With this aim we have studied one of the simplest
inhomogeneous cosmological models, namely the linearly polarized Gowdy $T^3$ model. This model is a
natural test bed to incorporate inhomogeneities in LQC. On the one hand, its quantization by means
of standard techniques has been discussed in detail, and a successful Fock quantization has
already been achieved. On the other hand the subset of its homogeneous solutions describes the
Bianchi I model in vacuo with three-torus topology, and the Bianchi I model has been already
considered within LQC.

We have attained a thorough quantization of this Gowdy model, in which the cosmological singularity
is resolved, by means of a hybrid quantization. This combines the polymeric quantization
characteristic of LQC applied to the homogeneous sector of the (partially reduced) phase space,
which is formed by the set of degrees of freedom that describe the homogeneous Bianchi I
solutions, with a Fock quantization for the inhomogeneities. This approach investigates the effects
of quantum geometry underlying LQC only on the homogeneous sector, while disregards the discreteness
of the geometry encoded by the inhomogeneities. A most natural treatment for the inhomogeneities is
then the Fock quantization. Indeed, one would expect that a quantum field theory for the
inhomogeneities, which can be regarded as a field living on a homogeneous (Bianchi~I) background, be
approximately valid on the polymerically quantized background. 

In order to carry out the loop quantization of the homogeneous sector in the Gowdy model as
rigorously
as possible, we have needed to revisit the very foundations of LQC. We have first reviewed the flat,
homogeneous, and isotropic Friedmann-Robertson-Walker (FRW) universe with a massless scalar field,
which is
a paradigmatic model in LQC. In spite of the prominent role that the model has played in the
development of LQC, there still remained some aspects of its quantization which deserved a more
detailed discussion. These aspects included the kinematical resolution of the cosmological
singularity, the precise relation between the solutions of the densitized and non-densitized
versions of the quantum Hamiltonian constraint, the possibility of identifying superselection
sectors which are as simple as possible, and a clear comprehension of the Wheeler-DeWitt (WDW) limit
associated with
the theory in those sectors. We have revised its quantization by proposing an alternative
prescription when representing the Hamiltonian constraint. Our proposal leads to a
symmetric Hamiltonian constraint operator, which is specially suitable to deal
with all these issues in a detailed and satisfactory way. In particular, with our constraint
operator, the singularity decouples in the kinematical Hilbert space and can be removed already at
this level. Thanks to this fact, we can densitize the quantum Hamiltonian constraint in a
well-controlled manner. Besides, together with the physical observables, this constraint
superselects simple sectors for the universe volume, with a discrete support contained in a single
semiaxis of the real line and for which the basic functions that encode the information about the
geometry possess optimal physical properties: They provide a no-boundary description around
the cosmological singularity and admit a well-defined WDW limit in terms of standing waves.
Both properties explain the presence of a generic quantum bounce replacing the classical singularity
at a fundamental level, in contrast with previous studies where the bounce was proved in concrete
regimes.

Following our program to quantize the Gowdy $T^3$ model, and since the homogeneous sector of this
system coincides with the Bianchi~I model in vacuo with spatial sections of three-torus topology, we
have also faced the quantization of these anisotropic cosmologies in LQC. The implementation of the
so-called \emph{improved dynamics} prescription, which was successfully established for isotropic
situations, gives rise to two different schemes in the presence of anisotropies. The original and
simplest one is a naive adaptation of the isotropic scheme employed for FRW. In turn, the most
recently proposed scheme leads to a more complicated quantum theory. We have investigated the loop
quantization of the Bianchi I model within both
schemes. When representing the Hamiltonian constraint operator, we have followed the same kind of
prescription as in the isotropic FRW model. As a consequence, once again the Hamiltonian constraint
operator leaves suitable subspaces of the kinematical Hilbert space invariant, whose
supports are each contained in a single octant of $\mathbb{R}^3$. This has allowed us to
complete the quantization within both schemes, providing the corresponding physical Hilbert space
and a complete set of physical observables. In both cases, the physical Hilbert space is
superselected in separable sectors. We have analyzed the structure of the resulting superselection
sectors, being those of the second scheme richer and more interesting.

On the other hand, in the vacuum Bianchi I model, and for the simplest scheme of the improved
dynamics, we have explicitly determined the form of the physical solutions to the Hamiltonian
constraint in
terms of elements of the kinematical Hilbert space used to carry out the quantization. This
knowledge makes this model a most appropriate arena to investigate the concept of physical
evolution in LQC in the absence of the massless scalar field which has been used so far in the
literature as an internal time. In order to retrieve the system dynamics when no such a suitable
clock field is present, we have explored different constructions of families of unitarily related
partial observables. These observables are essentially parameterized, respectively, by: $(i)$ one of
the components of the densitized triad, and $(ii)$ its conjugate momentum; each of them
playing the role of an evolution parameter. In order to describe the construction in a simpler
setting, and also for comparison, we have first carried out the analysis in the WDW
quantization of the model, and shown that singularities persist in this quantum approach. In the
loop quantized model the construction is more involved owing to the polymeric nature of the
geometry. Exploiting the properties of the considered example, we have investigated in detail the
domains of applicability of each construction. In both cases the observables possess a neat physical
interpretation only in an approximate sense. However, whereas in case $(i)$ such interpretation is
reasonably  accurate only for a portion of the evolution of the universe, in case $(ii)$ it remains
so during all the evolution (at least in the physically interesting cases). The constructed families
of observables have been next used to describe the evolution of the Bianchi I universe. The
performed
analysis confirms the robustness of the bounces, also in absence of matter fields, as well as the
preservation of the semiclassicality through them. The concept of evolution studied here and the
presented construction of observables are applicable to a wide class of models in LQC, including
the quantization of the Bianchi I model obtained with the other scheme for the improved
dynamics.

Finally, we have implemented the loop quantization of the Bianchi I model corresponding to
each of the two schemes for the improved dynamics in the representation of the homogeneous sector
of the hybrid Gowdy model. For both schemes, the resulting Hamiltonian constraint preserves the
sectors superselected in the Bianchi I model. Thanks to the nice features of these superselection
sectors, we have been able to complete the quantization also in this inhomogeneous system, even
though the homogeneous and inhomogeneous sectors are coupled in a non-trivial way, and now we are
dealing with an infinite number of degrees of freedom. The polymeric quantization of the homogeneous
sector suffices to cure quantum-mechanically the cosmological singularity.
The resulting physical Hilbert space has the
tensor product structure of the physical Hilbert space of the Bianchi I model times a Fock space
which is
unitarily equivalent to the physical space of the conventional Fock quantization of the
deparametrized system. Therefore, we indeed recover the standard quantum field theory for the
inhomogeneities.

In conclusion, we have studied a series of models in the framework of LQC with increasing
complexity, provided by the inclusion of anisotropies and inhomogeneities. We have been able
to improve the LQC techniques that were already developed in homogeneous settings and
complete the quantization of inhomogeneous cosmologies for the first time in the context of LQC.
\cleardoublepage


{\renewcommand{\thechapter}{}\renewcommand{\chaptername}{}
\addtocounter{chapter}{0}
\chapter*{Introducción}\markboth{\sl INTRODUCCIÓN}{\sl INTRODUCCIÓN}}
\addcontentsline{toc}{chapter}{Introducción}

\section*{Planteamiento y objetivos}
\addcontentsline{toc}{section}{Planteamiento y objetivos}

La teoría de la relatividad general es la teoría que explica el comportamiento del campo
gravitacional en la física moderna. Aborda el estudio de la gravedad desde un punto de vista
geométrico: los campos materiales se propagan en un espacio-tiempo dinámico y la teoría, mediante
las ecuaciones de Einstein, describe precisamente la naturaleza dinámica de la geometría de dicho
espacio-tiempo y su interacción con la materia \cite{Eins1,Eins2}. La relatividad general confirma
su propia incompletitud en tanto en cuanto conduce a la aparición de \emph{singularidades}
\cite{HawEll}, o regiones en las que algún observable físico diverge y, por tanto, la teoría pierde
en ellas su capacidad de predicción. Dichas singularidades aparecen justamente en circunstancias en
las que los efectos cuánticos deberían ser importantes. La relatividad general, que es una teoría
clásica, no los contempla. Es natural pensar que las fluctuaciones cuánticas de la geometría puedan
resolver estas singularidades clásicas, en analogía a lo que ocurre por ejemplo con el campo
eléctrico%
\footnote{Mientras que en electrodinámica clásica el modelo del electrón en el átomo de hidrógeno
es singular en tanto en cuanto no explica la estabilidad de la órbita electrónica, la teoría
cuántica de este sistema da lugar a un modelo bien definido.}. 
Así pues parece necesario desarrollar una teoría cuántica de la gravedad para entender la física
gravitacional que la relatividad general no puede explorar \cite{carlip}.

Durante las últimas décadas la gravedad cuántica de lazos \cite{lqg1,lqg2,lqg3} se ha
convertido en uno de los candidatos más prometedores a tal teoría%
\footnote{En la Ref. \cite{lqg4} se expone un resumen no muy técnico del estado actual de la
gravedad cuántica de lazos.}. 
Se trata de una cuantización no perturbativa de la relatividad general que intenta preservar al
máximo sus principios básicos y que, por tanto, mantiene el punto de vista geométrico. En
consecuencia, la teoría es independiente de las estructuras métricas de fondo e incorpora la
covariancia general de la relatividad general. Más concretamente, en su versión canónica, la
gravedad cuántica de lazos sigue el programa de Dirac de cuantización de sistemas con ligaduras
\cite{dirac}. En la formulación canónica de la relatividad general
\cite{adm1,wald,kuc-ish1,kuc-ish2}, la
covariancia
general se traduce en que la teoría está constreñida por cuatro ligaduras: tres ligaduras de
momentos, que generan difeomorfismos espaciales, y una ligadura hamiltoniana o escalar, que
genera reparametrizaciones temporales salvo difeomorfismos. De acuerdo con el programa de Dirac el
espacio de fases de la teoría se representa cuánticamente sobre un espacio de Hilbert, y las
ligaduras clásicas, que son funciones del espacio de fases, se promueven a operadores sobre dicho
espacio. El espacio de Hilbert físico está formado por los estados aniquilados por los operadores de
ligadura o, equivalentemente, por los estados que son invariantes bajo las transformaciones
generadas por dichos operadores. En gravedad cuántica de lazos se usa una formulación en términos de
holonomías de una conexión gauge y flujos de una tríada densitizada, que resulta especialmente
conveniente para aplicar la cuantización de Dirac a la relatividad general.

No obstante, aunque se han hecho muchos avances en el desarrolo de la gravedad cuántica de lazos, a
día de hoy dicha cuantización no se ha completado. Con la intención de progresar en su construcción,
resulta instructivo analizar modelos gravitacionales con alto grado de simetría, cuyo número
reducido de grados de libertad los hacen más manejables. El estudio de este tipo de modelos también
permite extraer predicciones en situaciones concretas de interés físico como, por ejemplo, en
cosmología. Así, en los últimos años, se ha estado desarrollando un programa de cuantización,
aplicado a cosmologías homogéneas, mediante la adaptación de las técnicas de gravedad cuántica de
lazos, que se conoce genéricamente con el nombre de cosmología cuántica de lazos
\cite{lqc1,lqc2a,lqc2b,lqc3}. Haciendo uso de la homogeneidad espacial presente en estos modelos,
las
ligaduras de momento quedan trivialmente satisfechas, de modo que sólo se impone cuánticamente una
ligadura hamiltoniana global (homogénea). Aunque la cosmología cuántica de lazos no es la reducción
de la gravedad cuántica de lazos a situaciones simétricas, se espera que capture los efectos
cuánticos principales de la teoría completa%
\footnote{De nuevo podemos hacer la analogía con el átomo de hidrógeno. El modelo de Bohr del átomo
de hidrógeneo no se obtuvo reduciendo la teoría cuántica del campo electromagnético, la
electrodinámica cuántica, a dicho sistema sencillo. No obstante, dicho modelo captura los efectos
cuánticos más importantes presentes en el sistema.}. En particular, al igual que ésta última, la
cosmología cuántica de lazos también explora la naturaleza discreta de la geometría, con la
finalidad de que en dicha teoría las singularidades cosmológicas clásicas queden resueltas, de modo
que la dinámica cuántica esté perfectamente definida. Por supuesto, antes del desarrollo de la
cosmología cuántica de lazos ya existían tratamientos cuánticos de sistemas cosmológicos (ver
p.~ej.~\cite{witt,mis-a,mis-b,mis-c,hartle,hall}), aunque en el marco de la geometrodinámica
cuántica, también
conocida como
teoría de Wheeler-DeWitt (WDW) \cite{witt,whe}, en la que los operadores tienen una representación
estándar de
tipo Schr\"odinger, no equivalente a la de la cosmología cuántica de lazos y en la que la
singularidad cosmológica no queda resuelta satisfactoriamente.

La cosmología cuántica de lazos surgió con los trabajos pioneros de Bojowald
\cite{boj1a,boj1b,boj1c,boj1d,boj2}, que
muestran los primeros intentos de adaptar las técnicas de gravedad cuántica de lazos en la
cuantización del modelo cosmológico más simple: el modelo plano de Friedmann-Robertson Walker (FRW),
es decir, el modelo homogéneo e isótropo con secciones espaciales planas. Posteriormente, la
correspondiente estructura cinemática fue revisada y establecida más rigurosamente en la
Ref.~\cite{abl}, lo que hizo posible completar la cuantización de dicho modelo en presencia de
un campo homogéneo, escalar y sin masa mínimamente acoplado \cite{aps1a,aps1b}, así como estudiar
en profundidad la evolución cuántica resultante, tras adoptar una prescripción denominada
\emph{dinámica mejorada} \cite{aps3} a la hora de representar el operador ligadura escalar. Aun
tratándose del modelo cosmológico más sencillo, su cuantización de
lazos,
también denominada polimérica%
\footnote{El calificativo \emph{polimérico}, en el contexto de gravedad cuántica de
lazos, hace referencia a que se consideran las excitaciones del campo gravitacional a lo largo
de aristas unidimensionales.}, da lugar
a resultados muy relevantes, tales como la concordancia de la dinámica clásica con la dinámica
cuántica para estados semiclásicos lejos de la singularidad y, más importante, la resolución de la
singularidad cosmológica o \emph{big bang}, que en la teoría cuántica se ve reemplazada por un
\emph{rebote cuántico}.

A partir de la cuantización del modelo isótropo plano, el campo de la cosmología cuántica de lazos
ha experimentado un gran desarrollo. Por ejemplo, todavía en el marco de la cosmología isótropa, se
ha analizado la cuantización de otros modelos, como algunas cosmologías no planas, tanto
con curvatura espacial positiva \cite{apsv,skl} como negativa \cite{vand,luc}, o el modelo plano
con constante cosmológica negativa \cite{tom} y con constante cosmológica positiva
\cite{tom2}. Asimismo, existe una
gran cantidad de trabajos basados en la dinámica clásica efectiva asociada a la cosmología cuántica
de lazos \cite{victor,sv-eff} que proporcionan análisis interesantes acerca de las huellas que
dejarían en el régimen clásico los efectos cuánticos de la geometría, por ejemplo, sobre las
perturbaciones escalares y tensoriales en escenarios inflacionarios
\cite{cop1,cop2,grain1,grain2}.

En vista de los importantes resultados obtenidos en cosmología cuántica de lazos homogénea, sería
interesante implementar sus técnicas poliméricas en la cuantización de los denominados
\emph{midisuperspacios} cosmológicos (El nombre fue acuñado por K.~Kucha\v{r}. Véase p. ej.
\cite{kuchar2}), que son sistemas también reducidos por simetría pero
inhomogéneos. Éste es uno de los principales objetivos afrontados en esta tesis. Muchas
razones motivan este
 análisis. Por una parte, los midisuperespacios preservan una de las
principales características de la teoría completa: el hecho de que ya no se trata de sistemas
mecánicos con un número finito de grados de libertad, sino que se describen mediante campos. Por
tanto, se espera que su cuantización pueda servir para entender mejor algunas de las cuestiones no
resueltas de la teoría completa que están relacionadas con la presencia de un número infinito de
grados de libertad. Por otra parte, para alcanzar predicciones realistas, contrastables con las
observaciones realizadas en cosmología moderna, es necesario comprender el papel
desempeñado por las inhomogeneidades en la física del universo temprano, de cara a progresar más en
el desarrollo de las leyes que explican el origen y la evolución del universo. Además, el análisis
de las inhomogeneidades nos permitiría comprobar la robustez de los resultados obtenidos en
cosmología cuántica de lazos homogénea y, en particular, los que conciernen a la resolución cuántica
de las singularidades cosmológicas clásicas.

Con el propósito de permitir inhomogeneidades en cosmología cuántica de lazos, hemos considerado
uno de los modelos de Gowdy \cite{gowdy1,gowdy2}. Estos modelos cosmológicos representan
espacio-tiempos con
dos vectores de Killing espaciales y secciones espaciales compactas. En particular, hemos escogido
el modelo con la topología del tres-toro y para el caso de un contenido de ondas gravitacionales con
polarización lineal, ya que por su alto grado de simetría resulta ser uno de los modelos
cosmológicos más sencillos y más estudiados. De hecho, no sólo se conocen muy bien sus soluciones
clásicas \cite{mon1,mon2,ise}, sino que además también se ha conseguido una cuantización de Fock
satisfactoria del sistema \cite{men1a,men1b}, y se ha demostrado que es esencialmente única
\cite{men2,men3}. En dicho análisis se resuelve el sistema clásicamente a excepción de una libertad
gauge de traslaciones en el círculo, que se impone cuánticamente \cite{men1a,men1b,men2,men3}. En
cambio, en
nuestro tratamiento, no deparametrizamos completamente el sistema, de modo que también dejamos sin
fijar clásicamente una ligadura hamiltoniana global, con el objetivo de imponerla cuánticamente como
se hace en cosmología cuántica de lazos. La cuantización que implementamos se caracteriza por ser
una cuantización híbrida que combina las técnicas de la cosmología cuántica de lazos con las
técnicas de la cuantización de Fock. Más precisamente, cuantizamos de acuerdo con la cosmología
cuántica de lazos sólo los grados de libertad que parametrizan soluciones homogéneas del sistema,
mientras que las inhomogeneidades son cuantizadas a la Fock. Este tratamiento es minimalista en el
sentido de que explora los efectos de la geometría cuántica subyacentes en cosmología cuántica de
lazos sólo en el sector gravitacional homogéneo. No obstante, aunque no se deriva completamente de
la gravedad cuántica de lazos, retiene propiedades fundamentaless asociadas a la naturaleza discreta
de la geometría. 

Las soluciones homogéneas del modelo de Gowdy representan espacio-tiempos de tipo Bianchi I, es
decir, geometrías con secciones espaciales planas y homogéneas, que además son anisótropas. Por
tanto, otro de los objetivos de esta tesis ha sido investigar la cuantización del modelo de
Bianchi I en cosmología cuántica de lazos, en principio con el propósito de aplicarla luego al
sector homogéneo de Gowdy. No obstante, el estudio del modelo de Bianchi I en el marco de la
cosmología cuántica de lazos ya resulta interesante por sí mismo, en tanto en cuanto significa un
avance más de la teoría, al dar paso a situaciones anisótropas. De hecho, ya existían en la
literatura algunos análisis preliminares sobre su cuantización polimérica \cite{boj,chio}. En esta
tesis, se ha llevado a cabo una cuantización del sistema rigurosa y bien definida a partir de la
revisión del análisis realizado en la Ref. \cite{chio}, que es el primero en el que se intentó
adaptar la cuantización del caso isótropo \cite{aps1a,aps1b,aps3} a este caso anisótropo más
general. A
este respecto hay que comentar que la extensión al caso anisótropo de la prescripción de dinámica
mejorada introducida en la Ref. \cite{aps3}, dio lugar a la propuesta
de dos esquemas de cuantización concurrentes. Inicialmente, seguimos la propuesta más sencilla
adoptada en la Ref. \cite{chio}. Más recientemente se ha argumentado que la otra prescripción 
tiene un comportamiento físico más adecuado bajo cambios de escala, y se ha representado el
operador ligadura escalar de acuerdo con este otro esquema~\cite{awe}. En consecuencia, en esta
tesis también hemos revisado ese estudio alternativo del modelo de Bianchi I, introduciendo algunas
mejoras con el fin de poder completar la nueva cuantización. Asimismo, en vista de los dos esquemas
diferentes de cuantización polimérica para el modelo de Bianchi I, hemos analizado la cuantización
híbrida del modelo de Gowdy aplicando ambos en la representación del sector homogéneo.

Otros modelos homogéneos anisótropos no considerados en esta tesis y que también han sido
estudiados en el marco de la cosmología cuántica de lazos son por ejemplo los modelos espacialmente
curvos de Bianchi II y Bianchi IX \cite{boj,bdv,awe2,bdh,we}. Esos estudios muestran de nuevo que
las singularidades clásicas desaparecen en la teoría cuántica. También se han realizado estudios
sobre la resolución de singularidades mediante las técnicas de la cuantización de lazos en
espacio-tiempos con simetría esférica, y en particular en el interior del agujero
negro de Schwarzschild
\cite{boj-bh1,boj-bh2,boj-bh3,boj-bh4,boj-bh5,boj-bh6,pullin1,pullin2}.

En el proceso de cuantización siempre existe una cierta ambigüedad a la hora de adoptar un orden de
factores concreto cuando las expresiones clásicas de las ligaduras se promueven a operadores. Este
hecho es una consecuencia de la no conmutatividad de los operadores básicos que representan las
variables del espacio de fases. Es deseable que el operador ligadura escalar sea esencialmente
autoadjunto, pues
esta propiedad facilita la determinación del espacio de Hilbert físico. Para ello, primero es
natural requerir que el orden de factores elegido sea simétrico, para al menos garantizar que el
operador resultante lo es. En nuestro primer estudio de la cuantización de Bianchi~I, en
el que adoptamos el esquema más sencillo para la dinámica mejorada, hemos introducido un orden
simétrico concreto que resulta muy apropiado para al final poder caracterizar completamente el
espacio de Hilbert físico, así como conocer de forma explícita la relación que existe entre la
estructura cinemática introducida antes de la imposición de las ligaduras y dicho espacio físico.
Nuestra prescripción, aunque inicialmente aplicada en ese contexto específico (Bianchi I), es
generalizable a cualquier modelo tratado en cosmología cuántica de lazos, y permite comprender mejor
aspectos intrínsecos a la cosmología cuántica de lazos que no quedaban clarificados en estudios
previos. En particular, es especialmente significativo el análisis del modelo isótropo, completado
satisfactoriamente en la Ref.~\cite{aps3}, que supuso un gran avance en el campo de la cosmología
cuántica de lazos. No obstante, el estudio de la Ref.~\cite{aps3} dejaba aún interrogantes sin
responder, algunos tan importantes como el hecho de si ``el rebote cuántico'' es un resultado
genérico de la cosmología cuántica de lazos o si solo se obtiene en regímenes concretos. En esta
tesis, motivados por la experiencia obtenida en los trabajos realizados sobre los modelos de Bianchi
I y Gowdy, también hemos llevado a cabo la cuantización del modelo plano de FRW implementando en
ella nuestra prescripción de simetrización del operador ligadura escalar. Dicha cuantización tiene
como objetivo el esclarecimiento de la cuestión planteada anteriormente, así como de otros detalles
que necesitaban una discusión más profunda, usando para ello el modelo más sencillo de todos, pero
que también es el que ha desempeñado un papel fundamental en el desarrollo de la cosmología cuántica
de lazos.

En los análisis de la dinámica cuántica realizados en cosmología cuántica de lazos, usualmente se
acopla mínimamente a la geometría un campo escalar sin masa que desempeña la función de tiempo. Así,
la ligadura escalar se puede interpretar como una ecuación de evolución temporal con respecto a ese
tiempo interno. Como dicho campo se cuantiza adoptando una representación
estándar de tipo Schr\"odinger, la evolución resultante es unitaria, propiedad esencial a la hora de
interpretar los resultados físicos de modo probabilístico, como se hace en general en cualquier
teoría cuántica. Sin embargo, en muchos casos, tales como el interior de agujeros negros o
universos vacíos de materia, no existe tal campo. Por tanto, sería útil saber describir la evolución
usando como tiempo interno alguno de los grados de libertad de la geometría, pues éstos están
siempre presentes. Éste es otro de los objetivos afrontados en esta tesis. Nuestro propósito es
desarrollar un método genérico que permita definir el concepto de evolución con respecto a alguno de
los grados de libertad geométricos. Para ello hemos elegido como ejemplo el modelo de Bianchi I
en vacío, cuya dinámica no es trivial (a diferencia del caso isótropo en vacío). Hemos considerado
la cuantización del modelo según el esquema original para la dinámica mejorada pues, como ya se
ha señalado, en ese caso conocemos explícitamente la relación entre los espacios de Hilbert
cinemático y físico, que es necasaria para definir la imagen de evolución. El hecho de elegir como
tiempo interno una variable cuantizada poliméricamente dificulta bastante la tarea de garantizar la
unitariedad de la evolución. Para hacer más patente la no trivialidad de nuestra construcción,
primero hemos realizado el estudio en el marco de la cuantización estándar del modelo, o
cuantización de WDW, en la que la evolución unitaria es fácilmente implementada. A su
vez, la relación entre la cuantización de WDW y la cuantización de lazos nos ha
ayudado a determinar cómo construir el correspondiente concepto de evolución unitaria en cosmología
cuántica de lazos. En particular hemos desarrollado dos construcciones, correspondientes a dos
elecciones distintas de tiempo interno.

Aquí concluye la visión global de los objetivos planteados y afrontados en esta tesis. Para que
su motivación se vea más clara, los hemos resumido tal como han surgido cronológicamente, de acuerdo
con los avances que han acontecido en el campo de la cosmología cuántica de lazos durante el período
en el que hemos realizado los trabajos aquí expuestos. No obstante, con el fin de que la exposición
sea
más clara y ordenada, detallaremos la investigación realizada siguiendo un hilo argumental
distinto, que básicamente atiende a un orden creciente en la complejidad de los modelos analizados
en cada momento.

\section*{Estructura de la tesis}
\addcontentsline{toc}{section}{Estructura de la tesis} 

\begin{list}{\labelitemi}{\leftmargin=1em}
 \item La Parte 1, Preliminares, contiene una revisión del trabajo que ya había sido realizado,
previo a nuestra investigación, y que constituyó el punto de partida. Está estructurada en dos
capítulos, que están en consonancia con el posterior cuerpo de la tesis. El Capítulo
\ref{chap:introLQC} repasa el formalismo de cosmología cuántica de lazos para el modelo plano e
isótropo, desarrollado principalmente en las Refs.
\cite{abl,aps1a,aps1b,aps3}, estableciendo así los fundamentos matemáticos de la cuantización
polimérica.
Además, el capítulo resume el análisis de la dinámica cuántica proporcionada por la interpretación
del campo como tiempo interno, que es la usual en cosmología cuántica de lazos, y describe el
concepto de \emph{rebote cuántico}. En el Capítulo \ref{chap:introGowdy} se presenta el modelo
de Gowdy, y se resumen los elementos de su cuantización de Fock \cite{men1a,men1b} que hemos
empleado al
representar su sector inhomogéneo en nuestra cuantización híbrida.
\end{list}
\noindent La investigación que conforma esta tesis está estructura en tres grandes partes:

\begin{list}{\labelitemi}{\leftmargin=1em}
\item  La Parte \ref{part1} recoge la cuantización que hemos llevado a cabo en los modelos
homogéneos investigados:
\begin{list}{\labelitemii}{\leftmargin=1em}
\item El Capítulo \ref{chap:revFRW} está dedicado a la revisión del modelo isótropo.
Hemos partido de una nueva prescripción simétrica para el operador ligadura hamiltoniana. Tras
densitizar la ligadura, hemos llevado a cabo un análisis espectral completo del operador resultante,
lo que nos ha permitido obtener el espacio de Hilbert físico, junto con la estructura de sus
estados, y demostrar que existe un rebote cuántico fundamental que evita la singularidad clásica.
Finalmente, hemos comparado nuestra propuesta de cuantización con los estudios previos de este
modelo.
\item El Capítulo \ref{chap:bianchi} contiene la cuantización de lazos del modelo de Bianchi I en
los dos esquemas de la dinámica mejorada que han sido propuestos en situaciones anisótropas. En
ambos casos, hemos obtenido un operador ligadura hamiltoniana densitizada bien definido en un
espacio de Hilbert cinemático separable. En la cuantización correspondiente al primer esquema de la
dinámica mejorada, hemos determinado explícitamente las soluciones físicas de la ligadura resultante
y la estructura de espacio de Hilbert para los estados físicos. En la cuantización correspondiente
al segundo esquema, hemos argumentado que la ligadura resultante conduce a un problema de
valores iniciales bien definido y hemos determinado el producto interno que dota de estructura de
Hilbert al espacio de los datos iniciales.
\end{list}
\item La Parte \ref{part2} expone nuestro análisis de la evolución temporal física en
cosmología cuántica, en el que hemos empleado como tiempo interno uno de los grados de libertad de
la geometría. Dicho análisis se ha realizado en el ejemplo concreto del modelo de Bianchi I en
vacío.
\begin{list}{\labelitemii}{\leftmargin=1em}
\item El Capítulo \ref{chap:5-evol} contiene la construcción de un concepto de evolución en
cosmología cuántica estándar, esto es, en la cuantización de WDW del modelo. Hemos
llevado a cabo dos construcciones de observables relacionales, en las que la variable que desempeña
la función de tiempo interno es respectivamente: (i)~un parámetro afín asociado a uno de los
coeficientes de la tríada densitizada; (ii) su variable canónicamente conjugada. Los observables
construidos están unitariamente relacionados y su interpretación física
es exacta. Analizando el valor esperado de estos observables físicos hemos demostrado que la
cuantización de WDW no resuelve la singularidad cosmológica clásica. Por otra parte, este
capítulo también contiene la relación exacta que existe entre los estados físicos de la
cuantización de lazos del modelo y los estados físicos de su cuantización de WDW.
\item El capítulo \ref{chap:evolutionLQC} está dedicado a la construcción de la imagen de evolución
en la cuantización polimérica del modelo. La construcción (i) debe ser modificada, con respecto a
la llevada a cabo en la cuantización estándar, para conseguir la unitariedad de la evolución. Los
observables construidos tienen una interpretación física aproximada y no son fiables en las
primeras épocas de la evolución. La construcción (ii) también da lugar a observables cuya
interpretación es aproximada, pero muy precisa durante toda la evolución. Las predicciones de ambas
construcciones muestran que, en la cuantización de lazos, la singularidad clásica se ve reemplazada
por rebotes cuánticos que la evitan. Además, hemos demostrado la conservación del comportamiento
semiclásico a través de estos rebotes.
\end{list}
\item La parte \ref{part3} recoge nuestra cuantización híbrida del modelo de Gowdy, contenida en el
capítulo \ref{chap:gowdy-hyb}. Hemos cuantizado poliméricamente (en los esquemas existentes) el
sector homogéneo, que coincide con el espacio de fases del modelo de Bianchi I, y hemos cuantizado
las inhomogeneidades en un espacio de Fock. En ambos esquemas hemos obtenido un operador ligadura
hamiltoniana bien definido en un dominio denso de un espacio de Hilbert cinemático separable. Este
operador acopla de forma no trivial ambos sectores. También, en ambos esquemas, la ligadura
hamiltoniana densitizada conduce a un problema de valores iniciales bien definido: las soluciones
formales a la ligadura están unívocamente determinadas por un conjunto de datos iniciales. Hemos
identificado un conjunto completo de observables que actúan sobre estos datos iniciales, así como el
producto interno con el que satisfacen las condiciones de realidad requeridas. Éste determina el
espacio de Hilbert físico.
\end{list}

Las Conclusiones recogen un resumen de los principales resultados obtenidos en esta tesis.

Por último, el Apéndice \ref{appA} contiene un resumen de conceptos y teoremas de la teoría de
operadores y el Apéndice \ref{appB} contiene detalles acerca de la evolución analizada en el
Capítulo \ref{chap:evolutionLQC}.

\cleardoublepage

\addcontentsline{toc}{part}{Preliminares}
\part*{Preliminares}

\chapter[Introducción a cosmología cuántica de lazos]{Introducción a la
cosmología cuántica de lazos a través del modelo isótropo}
\label{chap:introLQC}

La cosmología cuántica de lazos adapta las técnicas de la gravedad cuántica de lazos a la
cuantización de modelos gravitacionales con gran
simetría, típicamente homogéneos, con el fin de explorar los efectos de la discretización de la
geometría en situaciones sencillas. En particular se persigue la obtención de una teoría cuántica
bien definida en la que no existan análogos a las singularidades clásicas.

El modelo más sencillo es el espacio-tiempo isótropo y plano, pues su geometría se describe con
un solo grado de libertad, el factor de escala. Este sistema, aun siendo muy simplificado, es de
gran interés físico pues, a grandes escalas, nuestro universo es aproximadamente homogéneo e
isótropo. Además las observaciones cosmológicas son compatibles con una geometría espacial
plana. Con el propósito de que el modelo tenga dinámica no trivial, se introduce el contenido
material más simple, descrito por un campo escalar, homogéneo y sin masa, acoplado mínimamente a la
geometría. Aunque no es el contenido material real de nuestro universo, sirve como punto de partida
en su estudio. Las soluciones clásicas de las ecuaciones de Einstein de dicho modelo representan
universos en expansión con una singularidad inicial o \emph{big bang}
(gran explosión). En ella ciertos observables físicos, como la densidad de materia, divergen. Por
tanto con la teoría clásica no es posible hacer predicciones acerca de los instantes más
tempranos del universo y sobre cómo se originó. Veremos cómo la cuantización de lazos, en cambio,
consigue dar una descripción del modelo en la que los observables físicos están bien definidos a lo
largo de toda la evolución. Pasemos a ver los fundamentos en los que se basa.

\section{Descripción clásica}

\subsection{Formalismo de Ashtekar-Barbero}

En cosmología cuántica de lazos el espacio de fases clásico del modelo se describe usando
la formulación empleada en gravedad cuántica de lazos, es decir, usando las variables de
Ashtekar-Barbero. Empecemos pues haciendo un repaso de este formalismo.

La gravedad cuántica de lazos es una cuantización canónica de la relatividad general. Por tanto, 
se parte de una descripción de la teoría clásica en la formulación hamiltoniana. Para ello, se
consideran espacio-tiempos globalmente hiperbólicos, que admiten una función global de tiempo, y se
lleva a cabo su descomposición 3+1 \cite{adm1,adm2}. De esta manera, la métrica espacio-temporal se
describe mediante una tres-métrica $q_{ab}$ inducida en las secciones espaciales $\Sigma$ en las que
queda foliada la variedad espacio-temporal, la función lapso $N$ y el vector
desplazamiento $N^{a}$. Los índices latinos del inicio del alfabeto, $a,b,...$,
denotan índices espaciales. Las dos últimas cantidades, $N$ y $N^{a}$, no son físicas, ya que son
multiplicadores de Lagrange que acompañan a cuatro ligaduras presentes en la acción que codifican la
covariancia general de la relatividad general: la ligadura escalar o hamiltoniana y las ligaduras de
momentos. Así pues, la información físicamente relevante está contenida en la
tres-métrica espacial y en su momento canónicamente conjugado o, equivalentemente, en la curvatura
extrínseca
$K_{ab}=\mathcal L_n q_{ab}/2$, donde $n$ es la normal unitaria a $\Sigma$ y $\mathcal L_n$ es la
derivada de Lie a lo largo de ella \cite{wald}. Estas variables $q_{ab}$ y $K_{ab}$ son justamente
las empleadas
en la geometrodinámica cuántica \cite{whe}, pero a partir de ellas la teoría cuántica resultante no
está bien definida, en parte porque no se pueden regularizar los operadores que representan las
ligaduras en esas variables \cite{lqg1}.

Motivados por la experiencia ya adquirida en teorías cuánticas de campo no perturbativas, como
teorías gauge de tipo Yang-Mills, se buscó una formulación de la relatividad general en términos de
una
conexión gauge. La conexión es, de forma natural, el objeto asociado a las simetrías gauge y, por
tanto,
cabía esperar que en dicha formulación las ligaduras adquirieran una forma más sencilla y fuera más
abordable su cuantización. Esta idea supuso el punto de partida de la gravedad cuántica de lazos
\cite{lqg0aa,lqg0ab,lqg0ba,lqg0bb,lqg0c}. En dicha teoría las variables empleadas para describir el
espacio de fases clásico son, entonces, una conexión gauge y su momento canónicamente conjugado, que
es la
tríada densitizada, y que desempeña el papel de ``campo eléctrico''. Veamos cómo se construyen a
partir del par $(q,K)$.

En primer lugar se define la co-tríada $e^i_a$ en términos de la tres-métrica $q_{ab}$, a través de
la métrica de Cartan-Killing del grupo $SO(3)$, que es la métrica euclídea, dada por la
delta de Kronecker en tres dimensiones, $\delta_{ij}$, es decir%
\footnote{A lo largo de esta sección usaremos el convenio de sumación de Einstein de índices
repetidos.},
\begin{align}
 q_{ab}=e^i_ae^j_b\delta_{ij}.
\end{align}
Los índices latinos de la mitad del alfabeto, $i,j,...$ son índices $SO(3)$ y
etiquetan nuevos grados de libertad que hemos introducido al pasar a esta formulación triádica.
Estos grados de libertad son gauge ya que la métrica interna introducida es invariante bajo
rotaciones $SO(3)$. Por ello, en esta formulación de la relatividad general, además de las ligaduras
de difeomorfismos espaciales y la escalar, se deben satisfacer también tres ligaduras gauge que
fijan la libertad de rotaciones internas que acabamos de introducir. Una vez definida así la
co-tríada, se introduce la tríada $e^a_i$ como su inversa
\begin{align}
 e^a_ie^j_a=\delta_i^j,\quad e^a_ie^i_b=\delta^a_b,
\end{align}
y la tríada densitizada
\begin{align}
E^a_i=\sqrt{q}e^a_i,
\end{align}
donde $q$ designa el determinante de la tres-métrica espacial. 

El momento conjugado a la tríada densitizada es la curvatura extrínseca en forma triádica,
\begin{align}
K^i_a=K_{ab}e^b_j\delta^{ij}.
\end{align}
Este objeto no es una conexión, sino un vector respecto a su dependencia en el espacio interno.
No obstante, como hemos dicho, se busca una formulación de la teoría en términos de una conexión
gauge. Para ello puede emplearse la denominada conexión de espín, que llamaremos $\Gamma^j_b$.
Recordemos que la conexión determina el transporte paralelo, es decir, recoge la información de cómo
se ve afectada la materia por el campo gravitacional. Veamos cómo se obtiene. La conexión de espín
es
compatible con la tríada densitizada $E^a_i$, en el sentido de
que la derivada covariante de la tríada densitizada (incluida la actuación sobre el espacio interno
introducido) se anula. Es decir, se verifica
\begin{align}
\nabla_b E^a_i+\epsilon_{ijk}\Gamma^j_b E^{ak}=0,
\end{align}
donde $\epsilon_{ijk}$ es el símbolo de Levi-Civit\`a totalmente antisimétrico y $\nabla_b$ denota
la
derivada covariante espacial usual \cite{wald}. De esta relación se deduce que la conexión de espín
viene dada por \cite{spin}
\begin{align}
\Gamma^i_a=-\frac1{2}\epsilon^{ijk}E_{jb}(\partial_a E^b_k+\Gamma^b_{ca}E^c_k).
\end{align}
En esta expresión, $E_{jb}$ es el inverso de la tríada densitizada y $\Gamma^b_{ca}$ denota los
símbolos de Christoffel asociados a la métrica $q_{ab}$ \cite{wald}.
En lugar de considerar el grupo $SO(3)$, se toma su recubridor universal, el grupo $SU(2)$, para así
permitir el acoplo con materia fermiónica. Por tanto, la conexión toma valores en el álgebra
$su(2)$. 

Por último, sumando la conexión de espín y el vector interno $K^i_a$, se obtiene
la conexión de Ashtekar-Barbero \cite{bar}, que por construcción es una conexión
real canónicamente conjugada a la tríada densitizada, es decir
\begin{align}\label{var}
 A^i_a=\Gamma^i_a+\gamma K^i_a,
\end{align}
donde $\gamma$ es un parámetro real arbitrario y no nulo, denominado el parámetro de
Immirzi~\cite{gior1,gior2}.
El par canónico $(A,E)$ tiene el siguiente corchete de Poisson
\begin{align}\label{eq:poisson}
 \{A^a_i(x),E^j_b(y)\}=8\pi G\gamma\delta^a_b\delta^j_i\delta(x-y),
\end{align}
donde $G$ es la constante de Newton, y $\delta(x-y)$ denota la función delta de Dirac
tridimensional sobre la hipersuperficie $\Sigma$.

En estas variables, las ligaduras de gauge $SU(2)$, de difeomorfismos espaciales y hamiltoniana
tienen respectivamente la siguiente expresión (en vacío)%
\footnote{En presencia de materia acoplada a la geometría, un término
material contribuye a cada ligadura.}  \cite{lqg1},
\begin{subequations}\begin{align}
 \mathcal{G}_i&=\partial_aE^a_i+\epsilon_{ijk}\Gamma^j_a E^{ak}=0,\\
\mathcal{C}_a&=F_{ab}^iE^b_i=0,\\
\label{eq:lig-escalar-lqg}
\mathcal{C}&=\frac1{\sqrt{|det(E)|}}\epsilon_{ijk}\left[F^i_{ab}-(1+\gamma^2)\epsilon^i_{mn}
K^m_aK^n_b\right]E^{aj} E^{bk}=0,
\end{align}\end{subequations}
donde $F^i_{ab}$ es el tensor de curvatura de la conexión de Ashtekar-Barbero,
\begin{align}
 F^i_{ab}=\partial_a A^i_b-\partial_b A^i_a+\epsilon_{ijk}A^j_aA^k_b.
\end{align}

En cosmología homogénea, las ligaduras de momento y las de gauge se satisfacen de forma
inmediata haciendo uso de la simetría de homogeneidad, de modo que la ligadura hamiltoniana será la
única remanente en el modelo clásico. Además, las cosmologías homogéneas que consideraremos
en esta tesis tendrán curvatura de espín nula. En estos casos, la expresión de la contribución de
la geometría a la ligadura escalar en forma integral es%
\footnote{La función lapso $N$ sale de la integral debido a la homogeneidad.} $C_\text{grav}(N)=
NC_\text{grav}$, con
\begin{align}\label{eq:lig-escalar-lqg-hom}
C_\text{grav}=\int_{\Sigma} d^3x \mathcal C=-\frac1{\gamma^2}\int_{\Sigma} d^3x 
\frac{\epsilon_{ijk}F^i_{ab}E^{aj}E^{bk}}{\sqrt{|det(E)|}}.
\end{align}

Particularicemos este formalismo al modelo de FRW plano con un campo escalar sin masa y homogéneo,
acoplado mínimamente a la geometría. En este modelo, como las secciones
espaciales $\Sigma$ no son compactas, y las variables que lo describen son espacialmente
homogéneas, varias integrales divergen, como pasa por ejemplo con la integral
\eqref{eq:lig-escalar-lqg-hom}. Este problema se evita restringiendo el análisis a una celda finita
$\mathcal V$. Debido a la homogeneidad, el estudio de esta celda reproducirá lo que ocurre en el
universo completo, y de hecho se puede comprobar que los resultados no dependen de la elección de
celda. Al imponer la homogeneidad y la isotropía, se pueden describir la conexión y la tríada
densitizada (con una elección conveniente de gauge y de la libertad bajo difeomorfismos)
mediante un solo parámetro espacialmente constante, $c$ y $p$ respectivamente, del modo siguiente
\cite{abl}
\begin{align}\label{eq:variables-abl}
 A^i_a=c \;V_o^{-1/3} \;{}^oe^i_a, \qquad E^a_i=p\; V_o^{-2/3}\;\sqrt{{}^oq}\;{}^oe^a_i.
\end{align}
En estas expresiones se ha introducido una co-tríada ${}^oe^i_a$ fiducial y el determinante de la
métrica fiducial correspondiente. Los resultados no dependen de la elección de estas estructuras
fiduciales. En lo sucesivo, elegimos una co-tríada diagonal, tal que ${}^oe^i_a=\delta^i_a$.
En la expresión de $A^i_a$ y $E^a_i$ se introduce además el volumen fiducial (medido con la métrica
fiducial) $V_o$ de la celda $\mathcal V$ para garantizar que el corchete de Poisson%
\footnote{A lo largo de esta tesis consideraremos $\gamma>0$.},
\begin{align}\label{eq:poisson-frw}
 \{c,p\}=\frac{8\pi G\gamma}{3},
\end{align}
no dependa de dicho volumen.
La variable $p$ está relacionada con el factor de escala $a$, usado usualmente en geometrodinámica,
mediante la expresión $a(t)=\sqrt{|p(t)|} V_o^{-1/3}$. Nótese que $p$ es positivo si la tríada
física y la fiducial tienen la misma orientación, y negativo si la tienen opuesta. 

Por otra parte, el campo (homogéneo) escalar sin masa, $\phi$, junto con su momento $P_\phi$, nos
proporcionan otro par canónico de variables con corchete de Poisson
\begin{align}
 \{\phi,P_\phi\}=1.
\end{align}

Así, en este modelo isótropo la ligadura hamiltoniana total tiene contribución material además de
la geométrica, dada en la ec. \eqref{eq:lig-escalar-lqg-hom}, y tiene la forma
\begin{align}\label{eq:ligFRW}
 C=C_\text{grav}+C_\text{mat}=-\frac6{\gamma^2}c^2\sqrt{|p|}+8\pi G\frac{P_\phi^2}{V}=0,
\end{align}
donde $V=|p|^{3/2}$ es el volumen físico de la celda $\mathcal V$.

\subsection{Álgebra de holonomías y flujos}

En gravedad cuántica de lazos, una vez introducida la conexión, se pasa a describir el espacio de
configuración con sus holonomías, que resultan más convenientes por sus propiedades de
transformación bajo transformaciones gauge.
La holonomía de la conexión $A$ a lo largo de la arista $e$ se define como
\begin{align}
 h_e(A)=\mathcal P e^{\int_e dx^a A^i_a(x)\tau_i},
\end{align}
donde $\mathcal P$ es el operador ordenación de camino y $\tau_i$ son los generadores del grupo
$SU(2)$, tal que $[\tau_i,\tau_j]=\epsilon_{ijk}\tau^k$. La variable conjugada a la holonomía
viene dada por el flujo del ``campo eléctrico'' $E^a_i$ sobre superficies $S$ y ``suavizado''
con una función con valores en $su(2)$, $f^i$:
\begin{align}
 E(S,f)=\int_Sf^i E^a_i\epsilon_{abc}dx^bdx^c.
\end{align}

La descripción del espacio de fases en términos de las holonomías y los flujos no solo es apropiada
por sus propiedades de transformación, sino que además estos objetos son invariantes bajo
difeomorfismos y su definición no requiere la introdución de estructuras métricas de fondo. En
efecto, nótese que la conexión $A_a=A^i_a(x)\tau_i$ es una uno-forma, y por tanto se integra de modo
natural a lo largo de un elemento de línea. A su vez $E^a_i$ es una densidad vectorial, y entonces
su dual de Hodge $E^a_i\epsilon_{abc}$ es una dos-forma, integrable de modo natural sobre
dos-superficies. Entonces, con las holonomías y los flujos ya introducidos, hemos suavizado las
variables
del espacio de fases sobre las tres dimensiones espaciales, respetando la invariancia bajo
difeomorfismos. Como resultado, en particular, se obtiene un corchete de Poisson entre holonomías y
flujos libre de divergencias:
\begin{align}
 \{E(S,f),h_e(A)\}=2\pi G\gamma \epsilon(e,S)f^i\tau_i h_e(A),
\end{align}
donde $\epsilon(e,S)$ representa la regularización de la delta de Dirac: se anula si la arista $e$
y la superficie $S$ no intersectan o si $e\subset S$, y $|\epsilon(e,S)|=1$ si $e$ y $S$ intersectan
en un punto. El signo de $\epsilon(e,S)$ depende de la orientación relativa entre $e$ y $S$
\cite{lqg3}.

En cosmología cuántica de lazos y en el caso particular del modelo plano de FRW, debido
a la homogeneidad, es suficiente considerar aristas rectas orientadas en las direcciones
fiduciales, y con longitud orientada fiducial igual a
$\mu V_o^{1/3}$, donde $\mu$ es un número real arbitrario. Por tanto, la holonomía a lo
largo de una de tales aristas, en la dirección $i$-ésima, está dada por
\begin{align}
 h_i^\mu(c)=e^{\mu c \tau_i}=\cos\left(\frac{\mu c}{2}\right)\mathbb{I}+2\sin\left(\frac{\mu
c}{2}\right)\tau_i.
\end{align}
Entonces, la parte gravitacional del álgebra de configuración es el álgebra generada por las
sumas de productos de elementos de matriz de las holonomías, que son las exponenciales complejas
\begin{align}
 \mathcal N_\mu(c)=e^{\frac{i}{2}\mu c}.
\end{align}
Como $\mu$ es cualquier real, éste es el álgebra de funciones cuasi-periódicas de $c$. En analogía
con la terminología usada en gravedad cuántica de lazos \cite{lqg1,lqg3}, al espacio
vectorial de estas exponenciales complejas lo denominaremos espacio de las funciones cilíndricas 
definidas sobre las conexiones simétricas, y será designado con $\text{Cil}_\text{S}$.

A su vez, el flujo viene dado por
\begin{align}
 E(S,f)=p V_o^{-2/3}A_{S,f},
\end{align}
donde $A_{S,f}$ es el área fiducial de $S$ multiplicada por un factor de orientación (que depende de
$f^i$). Entonces, el flujo esencialmente está descrito por $p$.

En resumen, en cosmología cuántica de lazos el espacio de fases se describe mediante las variables
$\mathcal N_\mu(c)$ y $p$, cuyo corchete de Poisson es
\begin{align}
 \{\mathcal N_\mu(c),p\}=i\frac{4\pi G\gamma}{3}\mu\mathcal N_\mu(c).
\end{align}

\section{Estructura cinemática}
\label{1sec:kin}

\subsection{Representación polimérica de la geometría}

En gravedad cuántica de lazos, se representa el álgebra de holonomías y flujos sobre un espacio de
Hilbert cinemático, que es la compleción del espacio de funciones cilíndricas (definidas sobre el
espacio de conexiones generalizadas) con respecto a la medida de
Ashtekar-Lewandowski \cite{ALmeasure1,ALmeasure2,baez,ALmeasure3}. Importantes resultados
\cite{fleis,lost1,lost2} demuestran que ésta es
la única representación cíclica (salvo equivalencia unitaria) que admite un estado (identificado
con el vacío de la representación) invariante bajo los difeomorfismos espaciales generados
por las ligaduras de momento. Dicha representación resulta ser altamente discontinua, de modo que la
conexión no admite un operador que la represente.

Imitando la cuantización implementada en gravedad cuántica de lazos, en cosmología cuántica de lazos
se obtiene una representación del álgebra generada por las variables del espacio de fases $\mathcal
N_\mu(c)$ y $p$, que no es continua en la conexión, y en la que, por tanto, no existe un operador
que represente a $c$ \cite{abl}. Más concretamente, el espacio de configuración cuántico resulta ser
la compactificación de Bohr de la recta real, $\mathbb{R}_\text{Bohr}$, y la correspondiente medida
de Haar que caracteriza el espacio de Hilbert cinemático es la llamada medida de Bohr \cite{Vel}.
Es más sencillo trabajar en la representación de momentos. En efecto, se demuestra que dicho espacio
de Hilbert es isomorfo al espacio de funciones de $\mu\in\mathbb{R}$ de cuadrado sumable con
respecto a la medida discreta \cite{Vel}, que recibe el nombre de espacio \emph{polimérico}. En
otras palabras, si empleamos los \emph{kets} $|\mu\rangle$ para denotar a los estados cuánticos
$\mathcal N_\mu(c)$, cuya envolvente lineal es el espacio $\text{Cil}_\text{S}$, denso en
$\mathbb{R}_\text{Bohr}$, entonces el espacio de Hilbert cinemático es la compleción de
$\text{Cil}_\text{S}$ con respecto al producto interno
\begin{align}\label{inner}
 \langle \mu|\mu'\rangle=\delta_{\mu\mu'}.
\end{align}
Designaremos dicho espacio de Hilbert con el símbolo $\mathcal H_{\text{grav}}$. Nótese que
$\mathcal H_{\text{grav}}$ no es separable, ya que los estados $|\mu\rangle$ forman una base
ortonormal no numerable.

Obviamente, la acción sobre los estados de la base de los operadores $\hat{\mathcal N}_{\mu}$
asociados con las holonomías es
\begin{align}
 \hat{\mathcal N}_{\mu'} |\mu\rangle= |\mu+\mu'\rangle.
\end{align}
Por otra parte, la verificación de la regla de Dirac
$[\hat{\mathcal N}_\mu,\hat p]=i\hbar \widehat{\{\mathcal N_\mu(c),p\}}$
implica que
\begin{align}\label{eq:action-p-frw}
 \hat{p}|\mu\rangle=p(\mu)|\mu\rangle, \qquad p(\mu)=\frac{4\pi l_\text{Pl}^2\gamma}{3} \mu,
\end{align}
donde $l_\text{Pl}=\sqrt{G\hbar}$ es la longitud de Planck. Como vemos, el espectro de este operador
es discreto, como consecuencia de la falta de continuidad en $\mu$ de la representación.

Esta falta de continuidad hace que, en este contexto, no sea aplicable el teorema de
Stone-von Neumann \cite{SvN1,SvN2} de unicidad de la representación para mecánica cuántica, y por
tanto la
cuantización de lazos de este modelo es inequivalente a la cuantización convencional de la teoría a
lo WDW, con resultados físicos muy diferentes. 

\subsection{Representación del contenido material}

Para el campo material, se implementa la representación estándar de tipo Schr\"odinger, en la que
$\hat{\phi}$ actúa por multiplicación y $\hat{P}_\phi=-i\hbar\partial_\phi$ por derivación, ambos
operadores definidos en el espacio de Hilbert $L^2(\mathbb{R},d\phi)$. Como dominio suyo, tomamos
el espacio de
Schwartz $\mathcal{S}(\mathbb{R})$ de las funciones de decrecimiento rápido, que es denso en
$L^2(\mathbb{R},d\phi)$. Llamaremos $|\phi\rangle$
a los autoestados generalizados de $\hat{\phi}$, con autovalor generalizado $\phi\in\mathbb{R}$,
que proporcionan una base ortonormal generalizada (normalizable a la delta de Dirac) de
$L^2(\mathbb{R},d\phi)$.

El espacio de Hilbert cinemático total es, entonces, $\mathcal H_{\text{cin}}=\mathcal
H_{\text{grav}}\otimes L^2(\mathbb{R},d\phi)$. Notemos que los operadores básicos introducidos están
definidos en el producto tensorial de ambos sectores, actuando como la identidad en el sector del
cual no dependen. Es decir que, por ejemplo, el operador $\hat{p}$ definido sobre
$\text{Cil}_\text{S}\otimes
\mathcal{S}(\mathbb{R})$ realmente designa el operador
$\hat{p}\otimes\mathbb{I}$, o que $\partial_\phi$ es realmente $\mathbb{I}\otimes\partial_\phi$. No
obstante, para no complicar la notación, ignoraremos el producto tensorial trivial por la
identidad. 

\section{Operador ligadura hamiltoniana}
\label{1sec:lig-ham-aps}

\subsection{Operador curvatura y dinámica mejorada}

Como la conexión no está bien definida en la teoría cuántica, la expresión clásica de la ligadura
hamiltoniana, dada en la ec. \eqref{eq:ligFRW}, no se puede promover directamente a un operador.
Para obtener el análogo cuántico de la contribución gravitacional, regresamos a la expresión más
general \eqref{eq:lig-escalar-lqg-hom} y expresamos el tensor de curvatura en función de las
holonomías, que sí pueden representarse como operadores.

En cosmología cuántica de lazos, en analogía a lo que se hace en gravedad cuántica de lazos
\cite{lqg1}, se toma un circuito cerrado y cuadrado con
holonomía 
\begin{align}
 h^\mu_{\square_{ij}}=h_i^\mu h_j^\mu (h_i^\mu)^{-1} (h_j^\mu)^{-1},
\end{align}
que encierra un área fiducial $A_{\square}=\mu^2 V_o^{2/3}$.
El tensor de curvatura se expresa, entonces, como~\cite{abl}
\begin{align}\label{eq:curvatura-exacta}
 F^i_{ab}=-2 \;\lim_{A_{\square}\rightarrow 0}
\text{tr}\left(\frac{h^\mu_{\square_{jk}}-\delta_{jk}}{A_{\square}}\tau^i\right){}^oe^j_a{}
^oe^k_b,
\end{align}
donde $\text{tr}$ denota la traza.
Este límite está bien definido clásicamente. Sin embargo, en la teoría cuántica no es posible
contraer el área hasta cero porque ese límite no converge, como consecuencia de la no existencia
del operador conexión. Mientras que en la teoría completa de gravedad cuántica de lazos la
invariancia bajo difeoforfismos permite finalmente obtener un operador bien definido en el espacio
de Hilbert cinemático de estados invariantes bajo difeomorfismos espaciales, en cosmología cuántica
de lazos los difeomorfismos han sido fijados clásicamente y tal solución ya no es posible.

Para superar este problema se apela a la discretización de la geometría. En gravedad cuántica de
lazos el operador área geométrica tiene un espectro discreto con un autovalor mínimo no nulo, que
llamaremos $\Delta$ \cite{area1,area2}. Este carácter discreto, en cierto sentido, indica que no
deberíamos
tomar el límite de área nula, sino considerar solo áreas hasta $\Delta$. Entonces, contraemos el
área del lazo hasta un valor mínimo $A_{\square_\text{min}}=\bar\mu^2 V_o^{2/3}$, tal que el área
geométrica correspondiente a esta área fiducial, medida por el flujo $E(\square_\text{min},f=1)$,
sea igual a $\Delta$. En
definitiva, se define la
curvatura como
\begin{align}\label{eq:curvatura}
{F}^i_{ab}=-2\;
\text{tr}\left( \frac{h^{\bar\mu}_{\square_{jk}}-\delta_{jk}}{\bar\mu^2
V_o^{2/3}}\tau^i\right){}^oe^j_a{}^oe^k_b,
\end{align}
donde $\bar\mu$ verifica la igualdad
\begin{align}\label{mu}
 \frac1{\bar\mu}=\sqrt{\frac{|p|}{\Delta}}.
\end{align}
Seguidamente, se promueve la expresión \eqref{eq:curvatura} a un operador. Este paso de la
cuantización es lo que usualmente se denomina implementación de la {dinámica mejorada} en la
literatura de la cosmología cuántica de lazos \cite{aps3}.

Nótese que, al lazo $h^{\bar\mu}_{\square_{ij}}$, contribuyen términos del tipo ${\mathcal
N}_{\bar\mu}={e^{i\frac{\bar\mu c}{2}}}$. Por tanto, necesitamos definir el operador
\begin{align}
 \hat{\mathcal N}_{\bar\mu}=\widehat{e^{i\frac{\bar\mu c}{2}}},
\end{align}
para poder obtener el operador curvatura y, así, el operador ligadura escalar. Este punto
es sutil ya que, como $\bar\mu$ es función de $p$ y no existe el operador que representa a $c$,
$\hat{\mathcal N}_{\bar\mu}$ no puede expresarse directamente en función de los operadores
elementales $\hat{p}$ y $\widehat{e^{ikc/2}}$, con $k$ constante.
Para definir el operador $\hat{\mathcal N}_{\bar\mu}$ se asume que éste produce traslaciones
de una unidad sobre el parámetro afín asociado al campo vectorial
$\bar\mu[p(\mu)]\partial_\mu$ \cite{aps3}.
En otras palabras, se introduce una transformación canónica en el espacio de fases de la geometría,
tal que éste pasa a ser descrito por la variable $\beta=\hbar\bar\mu c/2$ y su canónica
conjugada $v(p)=(2\pi\gamma l_\text{Pl}^2\sqrt{\Delta})^{-1}\text{sgn}(p)|p|^{3/2}$ ($\text{sgn}$
denota la
función signo), con $\{\beta,v\}=1$. La variable $v(\mu)=v[p(\mu)]$ precisamente
cumple que $\partial_v=\bar\mu(\mu)\partial_\mu$. Entonces, se reetiquetan los estados de la
base de $\mathcal H_{\text{grav}}$ con este nuevo parámetro $v$ que, a diferencia de $\mu$, está
adaptado a la actuación de $\hat{\mathcal N}_{\bar\mu}$. De hecho, introduciendo el operador
$\hat{v}$
con acción $\hat{v}|v\rangle=v|v\rangle$, es inmediato encontrar la acción del operador
$\hat{\mathcal N}_{\bar\mu}=\widehat{e^{i\beta/\hbar}}$ en los estados $|v\rangle$. En efecto,
la regla de
Dirac \linebreak $[\widehat{e^{i\beta/\hbar}},\hat{v}]|v\rangle=i\hbar
\widehat{\{e^{i\beta/\hbar},v\}}|v\rangle$
implica que
\begin{equation}\label{eq:action-N}
 \hat{\mathcal N}_{\bar\mu}|v\rangle=|v+1\rangle.
\end{equation}
Por otra parte, invirtiendo la relación entre $v$ y $p$ tenemos
\begin{align}\label{eq:action-v}
 \hat{p}|v\rangle=(2\pi \gamma l_\text{Pl}^2\sqrt{\Delta})^{2/3}\text{sgn}(v)|v|^{2/3} |v\rangle.
\end{align}

Cabe señalar que el parámetro $v$ tiene una interpretación geométrica: su valor absoluto es
proporcional al volumen físico de la celda $\mathcal V$, dado por
\begin{align}\label{eq:volumen}
 \hat{V}=\widehat{|p|}^{3/2}, \qquad \hat{V}|v\rangle= 2\pi \gamma
l_\text{Pl}^2\sqrt{\Delta}|v||v\rangle.
\end{align}
Por tanto, la dinámica mejorada conlleva el paso de trabajar con los
autoestados del área, esencialmente dada por $\hat{p}$, a trabajar con los autoestados del
volumen. 

La compleción de la cuantización siguiendo este esquema de dinámica mejorada supuso un éxito rotundo
para la cosmología cuántica de lazos \cite{aps3}. Antes de su
implementación, se asumía que la longitud mínima fiducial era simplemente una constante,
relacionada con $\Delta$ \cite{abl}. No obstante, la teoría resultante no era satisfactoria, en
tanto en cuanto los efectos cuánticos de la geometría podían ser muy importantes a escalas donde la
densidad material no necesariamente debía ser grande, 
de modo que en regímenes semiclásicos los resultados físicos se desviaban apreciablemente de las
predicciones de la relatividad general \cite{aps1b}. La dinámica mejorada solventó este problema.
Además, se ha comprobado que es la única cuantización polimérica del sistema (dentro de una cierta
familia de posibilidades) que da lugar a un
modelo físicamente admisible~\cite{cs}, entendiendo como tal que el modelo resultante debe ser
independiente de las estructuras de fondo fiduciales introducidas en la construcción, que debe tener
un límite clásico bien definido que coincida con la relatividad general y que da lugar a una escala
del orden de la de Planck donde se manifiestan los efectos cuánticos importantes.

\subsection{Contribución gravitacional a la ligadura hamiltoniana}

Para poder promover  la expresión \eqref{eq:lig-escalar-lqg-hom} a un operador
encontramos otra dificultad adicional que concierne al inverso del volumen,
\begin{align}
 \frac1{V}=\frac{\sqrt{{}^o q}}{\sqrt{|det(E)|}V_o}.
\end{align}
De acuerdo con la ec. \eqref{eq:volumen}, el operador volumen tiene un espectro discreto y
el autovalor nulo está incluído en él. Esta característica imposibilita definir su inverso en
términos del teoremal espectral (enunciado en el Apéndice \ref{appA}), porque estaría mal definido
en cero.

No obstante, siguiendo el procedimiento llevado a cabo en gravedad cuántica de lazos
\cite{inv-vol-lqg1,inv-vol-lqg2}, podemos partir de
la identidad clásica
\begin{align}\label{eq:identidad-lig}
\frac{\epsilon_{ijk}E^{aj}E^{bk}}{\sqrt{|det(E)|}}&=\sum_{k=1}^3\frac{\text{sgn}(p)}{2\pi\gamma
GV_o^{1/3}}\frac1{l}\; {}^o e^k_c \;{}^o \epsilon^{abc}\,
\text{tr}\left(h_k^{l}(c)\big\{[h_k^{l}(c)]^{-1},
V\big\}\tau_i\right),
\end{align}
y representar con un operador el miembro izquierdo de esta igualdad promoviendo las funciones del
miembro derecho a los operadores correspondientes y haciendo la sustitución
\begin{align}
 \{\widehat{\;\;,\;\;}\}\rightarrow -\frac{i}{\hbar}[\hat{\;\;},\hat{\;\;}].
\end{align}
Nótese que el parámetro $l$ etiqueta una ambigüedad en la cuantización. Para no introducir nuevas
escalas en la teoría, que la harían más complicada, se toma como elección natural para $l$ el valor
$\bar\mu=\sqrt{{\Delta}/{|p|}}$
\cite{aps3}.

Insertando este resultado en la ligadura hamiltoniana \eqref{eq:lig-escalar-lqg-hom},
así como la curvatura dada en la ec. \eqref{eq:curvatura}, se llega, tras algunas operaciones,
a que la contribución de la geometría al operador de ligadura escalar viene dada por 
\begin{align}
 \widehat{C}_\text{grav}=i\frac{\widehat{\text{sgn}(p)}}{2\pi\gamma^3
l_\text{Pl}^2\Delta^{3/2}}\hat{V}\sum_{ijk}\epsilon^{ijk}\text{tr}
\left(\widehat{h}_i^{\bar\mu}\; \widehat{h}_j^{\bar\mu}\; (\widehat{h}_i^{\bar\mu})^{-1}\;
(\widehat{h}_j^{\bar\mu})^{-1}\; \widehat{h}_k^{\bar\mu}\;\big[ (\widehat{h}_k^{\bar\mu})^{-1},
\hat{V}\big]\right),
\end{align}
donde 
\begin{align}
 (\widehat{h}_i^{\bar\mu})^{\pm1}=\widehat{\cos\left(\frac{\bar\mu
c}{2}\right)}\mathbb{I}\pm2\widehat{\sin\left(\frac{\bar\mu
c}{2}\right)}\tau_i,
\end{align}
con
\begin{align}
 \widehat{\cos\left(\frac{\bar\mu c}{2}\right)}=\frac{\hat{\mathcal N}_{\bar\mu}+\hat{\mathcal
N}_{-\bar\mu}}{2},\qquad \widehat{\sin\left(\frac{\bar\mu c}{2}\right)}=\frac{\hat{\mathcal
N}_{\bar\mu}-\hat{\mathcal N}_{-\bar\mu}}{2i}.
\end{align}
Desarrollando la traza y adoptando un orden adecuado, se obtiene \cite{aps3}:
\begin{align}\label{eq:operador-lig-aps}
 \widehat{C}_\text{grav}=\widehat{\sin\left(\bar\mu c\right)}\left[
i\frac{3\widehat{\text{sgn}(p)}}{2\pi\gamma^3
l_\text{Pl}^2\Delta^{3/2}}\hat{V}\left(\hat{\mathcal
N}_{\bar{\mu}}\hat{V}\hat{\mathcal N}_{-\bar{\mu}}- \hat{\mathcal
N}_{-\bar{\mu}}\hat{V}\hat{\mathcal N}_{\bar{\mu}}\right)\right]\widehat{\sin\left(\bar\mu
c\right)}.
\end{align}
El operador \eqref{eq:operador-lig-aps} resulta ser simétrico porque $(\hat{\mathcal
N}_{\bar{\mu}}\hat{V}\hat{\mathcal N}_{-\bar{\mu}}- \hat{\mathcal
N}_{-\bar{\mu}}\hat{V}\hat{\mathcal N}_{\bar{\mu}})$ y $\hat{V}$ conmutan.
Finalmente, la actuación del operador gravitacional es
\cite{aps3}
\begin{align}
\widehat{C}_\text{grav}|v\rangle=\tilde f_+(v)|v+4\rangle+ \tilde f_o(v)|v\rangle+\tilde
f_-(v)|v-4\rangle,
\end{align}
con
\begin{align}
 \tilde f_+(v)&=\frac{3\pi l_\text{Pl}^2}{2\gamma\sqrt{\Delta}}
|v+2|\big||v+1|-|v+3|\big|,\nonumber\\
\tilde f_-(v)&=\tilde f_+(v+4),\nonumber\\
\tilde f_o(v)&=-\tilde f_+(v)-\tilde f_-(v).
\end{align}

Llegados a este punto, es importante señalar que la construcción del operador
$\widehat{C}_\text{grav}$ contiene varias ambigüedades relevantes. Por un lado, presenta la
ambigüedad, ya comentada, de escoger la longitud del camino a lo largo del cual se definen las
holonomías, que aquí se ha fijado a $\bar\mu$. Por otro lado, esta construcción 
también está sujeta a otra ambigüedad, menos evidente, en cómo se tratan y simetrizan los productos
que contienen al operador $\widehat{\text{sgn}(p)}$. Esto tiene consecuencias físicas
importantes, como veremos en el Capítulo~\ref{chap:revFRW}, donde se
investiga otra prescripción para el operador $\widehat{C}_\text{grav}$, que resulta ser más
apropiada que la descrita aquí. Finalmente, otro tipo de ambigüedad concierne al orden
que se elige para los factores que involucran potencias de $\hat{p}$
cuando se representa el inverso del volumen, aunque esta ambigüedad apenas influye en los
resultados físicos.

\subsection{Contribución material a la ligadura hamiltoniana}
\label{1subsec:mat}

Para completar la representación de la ligadura escalar en la teoría cuántica falta promover a
operador la contribución material, dada en el segundo sumando de la ec.~\eqref{eq:ligFRW}.
De nuevo, nos vemos en la necesidad de representar como operador el inverso del volumen.
Seguimos la misma estrategia que en la sección anterior, esta vez partiendo de la identidad clásica
\begin{align}\label{eq:identidad-inv-vol}
\frac{\text{sgn}(p)}{|p|^{1-a}}&=\frac{1}{a4\pi\gamma
G}\frac1{l}\text{tr}\left(\sum_i\tau^ih_i^{l}(c)\big\{[h_i^{l}(c)]^{-1},
|p|^a\big\}\right),
\end{align}
y de nuevo eligiendo en la teoría cuántica $l$ igual a $\bar\mu$.
Para fijar la ambigüedad en la constante $a>0$, se elije por simplicidad $a=1/2$. En
definitiva, se obtiene
\begin{align}
\widehat{\left[\frac{1}{\sqrt{|p|}}\right]}&=-\frac{i}{2\pi\gamma
l_\text{Pl}^2\sqrt{\Delta}}\widehat{\text{sgn}(p)}\widehat{\sqrt{|p|}}\text{tr}
\left(\sum_i\tau^i\widehat{
h_i^ { \bar\mu } } \big [ (\widehat{h_i^{
\bar\mu } )^ { -1 } }, \widehat{\sqrt { |p|}}\big]\right)\nonumber\\
&=\frac{3}{4\pi\gamma
l_\text{Pl}^2\sqrt{\Delta}}\widehat{\text{sgn}(p)}\widehat{\sqrt{|p|}}\left(\hat{\mathcal
N}_{-\bar{\mu}}\widehat{\sqrt{|p|}}\hat{\mathcal
N}_{\bar{\mu}}- \hat{\mathcal
N}_{\bar{\mu}}\widehat{\sqrt{|p|}}\hat{\mathcal
N}_{-\bar{\mu}}\right).
\end{align}
De las ecs. \eqref{eq:action-N} y \eqref{eq:action-v}, se deduce que su acción sobre los estados de
la base es diagonal, y está dada por
\begin{align}\label{eq:triad-operator}
 \widehat{\left[\frac1{\sqrt{|p|}}\right]}|v\rangle=b(v)|v\rangle,\qquad
b(v)=\frac{3}{2}\frac1{(2\pi\gamma l_\text{Pl}^2\sqrt{\Delta})^{1/3}}|v|^{1/3}
\big||v+1|^{1/3}-|v-1|^{1/3}\big|.
\end{align}
Mientras que, para valores grandes de $v$, $b(v)$ aproxima muy bien el valor clásico de
$1/\sqrt{|p|}$, para valores pequeños de $v$ ambos difieren apreciablemente. En
particular, el operador obtenido es acotado superiormente y aniquila a los estados de volumen
cero.

Finalmente la contribución material a la ligadura escalar viene dada por el operador
\begin{align}
\widehat{C}_\text{mat}=-8\pi l_\text{Pl}^2\hbar
\widehat{\left[\frac1{\sqrt{|p|}}\right]}^3\partial_\phi^2.
\end{align}

\section{Estructura física}
\label{1sec:phys}

\subsection{Imposición de la ligadura hamiltoniana cuántica}

En la teoría clásica las ligaduras son funciones del espacio de fases que deben anularse sobre
soluciones. El análogo cuántico de esta condición es que los operadores que representan las
ligaduras deben aniquilar a los estados físicos \cite{dirac}. 

En general, los estados no triviales aniquilados por el operador ligadura escalar,
$\widehat{C}=\widehat{C}_\text{grav}+\widehat{C}_\text{mat}$, no son normalizables en el espacio
de Hilbert cinemático $\mathcal H_{\text{cin}}$ en el cual se ha representado dicho operador. 
Por otra parte, el operador $\widehat{C}$ deja invariante su dominio $\text{Cil}_\text{S}\otimes
\mathcal{S}(\mathbb{R})$. Esta propiedad permite definir (p. ej. mediante el procedimiento de
promedio sobre grupos \cite{gave1a,gave1b,gave1c,gave1d,gave2}) las soluciones a la ligadura escalar
en el dual algebraico de dicho dominio, $\left(\text{Cil}_\text{S}\otimes
\mathcal{S}(\mathbb{R})\right)^*$, que es un espacio bastante más grande que el espacio de
Hilbert cinemático. En concreto se tiene la relación
\begin{align}
\text{Cil}_\text{S}\otimes\mathcal{S}(\mathbb{R})\subset\mathcal H_{\text{cin}}\subset
 \left(\text{Cil}_\text{S}\otimes\mathcal{S}(\mathbb{R})\right)^*.
\end{align}
Por tanto, impondremos la ligadura en dicho espacio dual. Sus elementos son de la forma
\begin{align}\label{eq:estado-dual}
 (\psi|=\int_\mathbb{R}d\phi\sum_v \psi(v,\phi)\langle\phi|\otimes\langle v|,
\end{align}
con $\psi(v,\phi)<\infty$ para todo $v,\phi\in\mathbb{R}$.
Los distinguiremos con los paréntesis $(\,\cdot\,|$ para diferenciarlos de los estados
$\langle\,\cdot\,|$ pertenecientes al dual de $\mathcal H_{\text{cin}}$ (e identificable con él).

Las soluciones a la ligadura son entonces aquéllas que verifican 
\begin{align}
 (\psi|\widehat{C}^\dagger=0
\;\longleftrightarrow\;(\psi|\widehat{C}_\text{mat}^\dagger=-(\psi|\widehat{C}_\text{grav}^\dagger.
\end{align}
Sustituyendo la expansión \eqref{eq:estado-dual} y teniendo en cuenta la acción de
$\widehat{C}_\text{grav}$ y $\widehat{C}_\text{mat}$ ,
la ligadura escalar se reescribe como
\begin{align}
 8\pi l_\text{Pl}^2\hbar [b(v)]^3\partial^2_\phi\psi(v,\phi)=-\left[\tilde
f_+(v)\psi(v+4,\phi)+\tilde f_o(v)\psi(v,\phi)+\tilde
f_-(v)\psi(v-4,\phi)\right].
\end{align}
A su vez, en la Ref. \cite{aps3}, esta ecuación fue reescrita en la forma
\begin{align}\label{eq:Klein-Gordon}
 \partial_\phi^2\psi(v,\phi)=-\widehat{\Theta}\psi(v,\phi),
\end{align}
introduciendo el operador
\begin{align}\label{eq:theta-aps}
 \widehat{\Theta}=[B(v)]^{-1}\widehat{C}_\text{grav},\qquad B(v)=8\pi l_\text{Pl}^2\hbar[b(v)]^{3},
\end{align}
es decir, se ``densitizó'' la ligadura. Como veremos en el Capítulo~\ref{chap:revFRW}, nosotros la
hemos densitizado siguiendo un procedimiento ligeramente distinto. 
Además, por conveniencia, en la Ref. \cite{aps3} se reemplazó el espacio de Hilbert cinemático
$\mathcal H_{\text{grav}}$,
caracterizado por el producto interno $\langle v|v'\rangle=\delta_{vv'}$, por un nuevo espacio de
Hilbert $\mathcal H'_{\text{grav}}$ con producto interno
\begin{align}\label{eq:inner-prod-aps}
 \langle v|v'\rangle=B(v)\delta_{vv'},
\end{align}
en el que el operador $\widehat{\Theta}$ sí es simétrico.

\subsection{Propiedades del operador $\widehat{\Theta}$}

El operador $\widehat{\Theta}$, al igual que el operador $\widehat{C}_\text{grav}$, es un operador
en diferencias de paso cuatro. Por tanto, los estados con soporte perteneciente a
diferentes redes de paso cuatro, definidas como
\begin{align}
\mathcal
L_{\pm|\epsilon|}:=\{v=\pm|\epsilon|+4n,\,n\in\mathbb{Z}\}, \quad|\epsilon|\in [0,2],
\end{align}
no se mezclan bajo la acción de $\widehat{\Theta}$.
En otras palabras, los subespacios de Hilbert $\mathcal H_{\text{grav}}^{\pm|\epsilon|}$, cuyos
estados tienen soporte en las redes $\mathcal L_{\pm|\epsilon|}$, proveen sectores de
superselección. Nótese que estos subespacios de $\mathcal H'_{\text{grav}}$ son separables, a
diferencia de él. Además conviene señalar que $\mathcal
L_{+2}=\mathcal L_{-2}$ y $\mathcal L_{+0}=\mathcal L_{-0}$.

Puede demostrarse rigurosamente \cite{kale} que $\widehat{\Theta}$, definido en dichos
sectores de superselección $\mathcal H_{\text{grav}}^{\pm|\epsilon|}$, es esencialmente
autoadjunto%
\footnote{Los conceptos de teoría de operadores aquí empleados están definidos en el Apéndice
\ref{appA}.},
y que las partes absolutamente continua y esencial del espectro son positivas. Por otra parte, por
evidencia numérica, se puede confirmar que los espectros discreto y singular son vacíos \cite{aps3}.
Por tanto, su espectro coincide con el absolutamente continuo.
Además, cada autovalor generalizado está doblemente
degenerado. Se eligió una base generalizada de autofunciones ortonormales
específica \cite{aps3}. Sus elementos fueron denominados%
\footnote{El superíndice $\pm$ hace referencia al sector de superselección considerado, 
$\mathcal H_{\text{grav}}^{+|\epsilon|}$ o $\mathcal H_{\text{grav}}^{-|\epsilon|}$.}
$e^\pm_{k}(v)$ con
$k\in\mathbb{R}$, de modo que $e^\pm_{|k|}(v)$ y $e^\pm_{-|k|}(v)$ tienen el mismo autovalor, igual
a $12\pi l_\text{Pl}^2 k^2$, y tales que $\langle e^\pm_{k}|e^\pm_{k'}\rangle=\delta(k-k')$. Esta
base está
definida de tal forma que $e^\pm_{-|k|}(v)$ tiende a $e^{-ik\ln|v|}/\sqrt{2\pi}$ para valores de $v$
grandes y positivos. Este límite es una autofunción del operador análogo a $\widehat{\Theta}$ que se
obtiene al hacer la cuantización estándar o de WDW del modelo. Realizando un análisis
numérico, se observa que el comportamiento asintótico de $e^\pm_{-|k|}(v)$ resulta entonces venir
dado por
\begin{align}\label{eq:wdw-limit-aps}
e^\pm_{-|k|}(v)
\xrightarrow{v\gg 1}\frac1{\sqrt{2\pi}}e^{-ik\ln|v|},\qquad
e^\pm_{-|k|}(v) \xrightarrow{v\ll -1}
\frac{A}{\sqrt{2\pi}}e^{-ik\ln|v|}+ \frac{B}{\sqrt{2\pi}}e^{ik\ln|v|},
\end{align}
donde se comprueba que, para grandes valores de $|k|$, $A$ y $B$ satisfacen $|A|^2-|B|^2=1$
y $|A|\sim |B|\gg1$. Como resultado, las autofunciones $e^\pm_{-|k|}(v)$ sufren una amplificación
en el semieje negativo de $v$. Además, esta amplificación es mayor cuanto mayor es $k$. 

Adicionalmente, la transformación de paridad $v\rightarrow -v$ es una simetría del
sistema, es decir, si $e(v)$ es autofunción de $\widehat{\Theta}$, entonces $e(-v)$ también lo es.
Apelando a esta simetría, se restringió la teoría al sector simétrico de
funciones invariantes bajo paridad, para el que se construyó la correspondiente base de
autofunciones de $\widehat{\Theta}$, denominadas $e^{(s)}_{k}(v)$. Éstas tienen soporte en la
unión $\mathcal L_{+|\epsilon|}\cup\mathcal L_{-|\epsilon|}$. Gracias a la introducción de esta
simetría, el estudio puede restringirse al semieje positivo de $v$. Veremos en el
Capítulo~\ref{chap:revFRW} que nuestro procedimiento alternativo nos permite restingir el estudio a
dicho semieje sin la necesidad de recurrir a la simetría bajo paridad.

\subsection{Espacio de Hilbert físico}

Una vez realizado el análisis espectral del operador $\widehat{\Theta}$ y determinada una base de
autofunciones del mismo, podemos expandir las soluciones de la ligadura en dicha base. Asimismo,
recordemos de la mecánica cuántica elemental que el operador $\partial_\phi^2$ es esencialmente
autoadjunto en su dominio $\mathcal{S}(\mathbb{R})$, con espectro absolutamente continuo y
doblemente degenerado, siendo sus autofunciones generalizadas, de autovalor $-\nu^2$, las ondas
planas $e^{\pm i|\nu|\phi}$. 
En esta base, las soluciones simétricas a la ligadura son de la forma
\begin{align}\label{eq:sol-lig-aps}
 \psi(v,\phi)=\int_{-\infty}^\infty dk \int_{-\infty}^{\infty} d\nu\;
e^{(s)}_k(v)\big[\tilde\psi_+(k)e^{i\nu(k)\phi}+\tilde\psi_-(k)e^{-i\nu(k)\phi}
\big],
\end{align}
con $\nu(k)^2=12\pi l_\text{Pl}^2k^2$, donde los perfiles espectrales $\tilde\psi_\pm(k)$ de los
estados
físicos pertenecen al espacio de Hilbert
\begin{align}
 \mathcal H^\epsilon_\text{fis}=L^2(\mathbb{R},dk).
\end{align}
Equivalentemente, en representación de $v$, el producto escalar físico que caracteriza a $\mathcal
H^\epsilon_\text{fis}$ es
\begin{align}
 \langle\psi_1|\psi_2\rangle_\epsilon=\sum_{v\in\mathcal L_{+|\epsilon|}\cup\mathcal
L_{-|\epsilon|}}B(v)\psi^*_1(v,\phi)\psi(_2v,\phi),
\end{align}
para cualquier valor de $\phi$, donde el asterisco denota conjugación compleja.

\subsection{Imagen en evolución y observables físicos}
\label{1subsec:evol}

En cualquier sistema gravitacional, como el considerado aquí, el hamiltoniano es una combinación
lineal de las ligaduras y, por tanto, es nulo. En otras palabras, la coordenada temporal de la
métrica no es un tiempo físico, y proporciona una noción de evolución congelada, a diferencia
de lo que ocurre en teorías en las que la métrica es una estructura de fondo estática, como son las
teorías de campos usuales. Con el propósito de interpretar los resultados físicos en un
escenario de evolución temporal, necesitamos definir pues cuál es este concepto de evolución. Para
ello, se elije una variable o función del espacio de fases que se interpreta como tiempo
\cite{kuchar2}.

En el modelo que estamos describiendo, vemos que la ec. \eqref{eq:Klein-Gordon} tiene la forma de la
ecuación de Klein-Gordon para un campo escalar sin masa, donde $\phi$ desempeña el papel de
tiempo y el operador $\widehat{\Theta}$ desempeña el papel de laplaciano. Por tanto, es
natural elegir $\phi$ como el tiempo. Así, la ligadura escalar se puede
interpretar como una ecuación de evolución en este tiempo interno. A su vez, $\nu$ hace las
veces de frecuencia asociada a ese tiempo.

Como vemos en la ec. \eqref{eq:sol-lig-aps}, las soluciones de la ligadura se pueden descomponer en
componentes de frecuencia positiva y negativa
\begin{align}
\psi_\pm(v,\phi)=\int_{-\infty}^\infty dk\;\tilde\psi_\pm(k)
e^{(s)}_k(v)e^{\pm i\nu(k)\phi},
\end{align}
que además están determinadas por los datos iniciales $\psi_\pm(v,\phi_0)$, a través de la evolución
unitaria
\begin{align}\label{eq:evol-aps}
 \psi_\pm(v,\phi)=U_\pm(\phi-\phi_0)\psi_\pm(v,\phi_0), \qquad U_\pm(\phi-\phi_0)=e^{\pm
i\sqrt{\widehat{\Theta}}(\phi-\phi_0)}.
\end{align}

Este concepto de evolución nos permite definir observables de Dirac ``en evolución'', con los que
interpretar los resultados físicos. Señalemos primero que, en la teoría clásica, aunque $v$ no es
una constante de movimiento, $v(\phi)$ resulta ser una función monovaluada de $\phi$ en cada
trayectoria dinámica \cite{aps3} y, por tanto, $|v|_{\phi=\phi_0}$ es un observable bien definido
para cada valor fijo $\phi_0$. El análogo cuántico a dicho observable es el operador
\begin{align}
 \widehat{|v|}_{\phi_0} \psi(v,\phi)= U_+(\phi-\phi_0)|v| \psi_+(v,\phi_0)+
U_-(\phi-\phi_0)|v|\psi_-(v,\phi_0).
\end{align}
Vemos que, dada una solución física $\psi(v,\phi)$, la actuación de este operador consiste en:
\begin{enumerate}
\item [i)]descomponer la solución en sus componentes de frecuencia positiva y negativa,
\item [ii)] congelarlas en el tiempo inicial $\phi=\phi_0$,
\item [iii)] multiplicar su dato inicial por $|v|$, y
\item [iv)] evolucionar mediante la ec. \eqref{eq:evol-aps}.
\end{enumerate}
El resultado es otra vez una solución física y el operador $\widehat{|v|}_{\phi_0}$ así
definido es de hecho un observable de Dirac. Entonces los operadores $\widehat{|v|}_{\phi_0}$ y
$P_\phi=-i\hbar\partial_\phi$ forman un conjunto completo de observables de Dirac o físicos.

Notemos que tanto los observables físicos como el producto interno físico preservan los subespacios
de frecuencia positiva y negativa. Por consiguiente, hay más superselección: cualquiera de estos
subespacios proporciona una representación irreducible del álgebra de observables y se puede
restringir el estudio al sector de frecuencias positivas, por ejemplo.

\subsection{Resultados físicos: rebote cuántico}

En la Ref. \cite{aps3} se hizo un análisis detallado de la dinámica cuántica resultante. El análisis
fue limitado a los estados más interesantes físicamente: aquéllos que son semiclásicos a tiempos
tardíos. En concreto, se estudiaron perfiles gaussianos picados a tiempo grande $\phi^*$ en un valor
grande del momento del campo o, equivalentemente, picados en un valor de $k$ muy negativo, denotado
por $k^*$. Es decir,
\begin{align}\label{eq:perfil-aps}
 \tilde\psi_+(k)=e^{-\frac{(k-k^*)^2}{2\sigma^2}}e^{-i\nu(k)\phi^*}.
\end{align}
Como el soporte de este perfil en valores positivos de $k$ es prácticamente nulo, en este régimen
de estados semiclásicos la contribución de $e^{(s)}_{|k|}(v)$ es despreciable, y se puede escribir
la solución como 
\begin{align}
\psi_+(v,\phi)=\int_{-\infty}^0 dk\;e^{-\frac{(k-k^*)^2}{2\sigma^2}}
e^{(s)}_{-|k|}(v)e^{ i\nu(k)(\phi-\phi^*)}.
\end{align}
Se determinaron numéricamente las autofunciones $e^{(s)}_{-|k|}(v)$, y se computó esta integral para
un rango amplio de valores de $\phi$ y de $v$, obteniéndose la gráfica de la norma compleja de la
función de ondas. Ésta se mantiene picada a lo largo de toda la evolución. Por tanto, se comprobó
que la evolución conserva el comportamiento semiclásico. Asimismo, se calcularon numéricamente los
valores
esperados y dispersiones del observable $\widehat{|v|}_{\phi_o}$, que recordemos representa
esencialmente el volumen físico. Mientras que clásicamente, para un universo en expansión,
este observable se anula a tiempo inicial $\phi=-\infty$, los resultados cuánticos muestran que el
volumen no llega hasta cero, sino que ocurre un rebote, que se da cuando la densidad material
alcanza un valor crítico que es aproximadamente 0.41 veces la densidad de Planck. Este fenómeno se
conoce con el nombre de \emph{rebote cuántico} o \emph{gran rebote} (big bounce) y reemplaza
cuánticamente la
\emph{gran explosión} clásica o big bang. En definitiva, la cosmología cuántica de lazos predice,
para este
modelo, y restringiéndose a la teoría efectiva correspondiente a la clase de estados estudiada
\cite{victor}, que un universo clásico en expansión proviene de un universo, también clásico en un
pasado lejano, que se contrajo y rebotó a escalas del orden de Planck hasta expandirse a su estado
actual. 

\cleardoublepage

\chapter[Introducción al modelo de Gowdy]{Introducción al modelo de Gowdy $T^3$ con polarización
lineal}
\label{chap:introGowdy}

En el capítulo anterior, hemos visto con un ejemplo concreto cómo los modelos gravitacionales
reducidos por simetría ofrecen un marco sencillo donde poner a prueba las técnicas y estrategias
de cuantización de la gravedad. En este sentido, son especialmente interesantes los modelos
inhomogéneos, pues son teorías de campos con infinitos grados de libertad, al igual que la teoría
completa. Por otra parte, en situaciones de interés físico tales como la cosmología, los modelos
homogéneos son en primera aproximación buenos candidatos para obtener predicciones preliminares.
Sin embargo son insuficientes a la hora de obtener resultados verdaderamente realistas. La
cosmología moderna indica que las fluctuaciones cuánticas de las inhomogeneidades desempeñaron una
función clave en la formación de estructuras presentes en el universo y su posterior evolución.

Las cosmologías inhomogéneas mejor conocidas corresponden a los modelos de
Gowdy~\cite{gowdy1,gowdy2}. Éstos
representan espacio-tiempos vacíos, globlalmente hiperbólicos, con secciones de tiempo constante
compactas y dos campos vectoriales de Killing espaciales (que resultan conmutar). Las secciones
espaciales admiten tres posibles topologías: la de la tres-asa ($S^1\times S^2$), la de la
tres-esfera ($S^3$) y la del tres-toro ($T^3$), o la de alguna variedad que esté cubierta por una de
éstas. El modelo de Gowdy más sencillo de
todos es el modelo de Gowdy $T^3$ polarizado linealmente, por lo que ha sido ampliamente estudiado
en la literatura. De hecho, sus soluciones clásicas son conocidas exactamente \cite{mon1, mon2,ise}
y representan ondas gravitacionales que se propagan en un universo cerrado en expansión. Además, su
cuantización fue considerada ya desde los años 70, con los trabajos pioneros de Misner y Berger
\cite{misner,berger1,berger2,berger3}. Posteriormente, gracias al hecho de que el modelo puede ser
tratado como un
campo escalar sin masa acoplado a gravedad en 2+1 dimensiones y que, por tanto, el problema se
reduce al de cuantizar un campo en un espacio-tiempo curvo, se llevó a cabo la cuantización de Fock
correspondiente \cite{pierri}. No obstante, se demostró que la dinámica resultante no puede
implementarse de forma unitaria \cite{ccq,torre}. La unitariedad en la evolución es necesaria para
entender los resultados físicos mediante la interpretación probabilística estándar de la física
cuántica. De ahí que se investigaran las razones de por qué falla dicha unitariedad \cite{men0}, con
el propósito de poder entender cómo se puede restablecer. La
introducción de una transformación canónica dependiente
del tiempo apropiada con la que el modelo de la Ref. \cite{pierri} pasa a poder ser tratado (salvo
por una ligadura global
remanente) como un campo escalar con masa dependiente del tiempo en un espacio-tiempo estático, de
tipo Minkowski en 2+1, pero con topología $T^2\times\mathbb{R}^+$, proporciona la solución al
problema \cite{men1a,men1b}. En efecto, cuantizando el campo resultante en un espacio de Fock
fiducial, en el que luego se impone cuánticamente la ligadura remanente para obtener el Fock físico,
se obtiene una
cuantización de Fock con dinámica unitaria, tanto a nivel fiducial como físico \cite{men1a,men1b}.
Mas aún, se ha demostrado que esta cuantización de Fock es esencialmente única \cite{men2,men3}.
Además, se han estudiado otros aspectos de su cuantización, como la obtención del operador de
evolución correspondiente \cite{dani1} y la formulación equivalente en la representación de
Schrödinger \cite{men4}. 

En la literatura también se ha estudiado la cuantización de otros midisuperespacios
correspondientes a cosmologías de Gowdy con otras topologías, para las que también se ha
demostrado la unicidad de la cuantización de Fock \cite{men5,dani2}. Asimismo, también se han
obtenido resultados similares de unicidad para la cuantización de Fock de un campo
escalar con masa dependiente del tiempo sobre ciertos espacio-tiempos no estacionarios
\cite{men6,men7,men8}.

A continuación, se resumen tanto los aspectos clásicos del modelo de Gowdy como los de su
cuantización de Fock según la Ref. \cite{men1b}, que emplearemos posteriormente en el tratamiento
híbrido desarrollado en esta tesis.

\section{Descripción clásica del modelo}

Como se ha señalado anteriormente, el modelo de Gowdy $T^3$ representa las soluciones de vacío de
las ecuaciones de Einstein, con dos campos vectoriales de Killing espaciales que conmutan y con
hipersuperficies espaciales homeomorfas a una variedad tres-toroidal. Consideraremos el modelo con
polarización lineal, que posee una simetría adicional: los vectores de Killing son ortogonales a
hipersuperficies y, por tanto, son mutuamente ortogonales en todo punto. Otra de las propiedades de
este modelo, que contribuye a su sencillez, es que la raíz cuadrada del determinante de la
dos-métrica inducida en el grupo de órbitas generadas por los vectores de Killing es una función
global de tiempo. Sean $\partial_\sigma$ y $\partial_\delta$ los
campos de Killing, entonces, es posible escoger coordenadas globales $\{t,\theta,\sigma,\delta\}$
adaptadas a las simetrías para describir la métrica espacio-temporal, con $\theta,\sigma,\delta\in
S^1$.

En la descomposición 3+1, podemos describir la métrica espacio-temporal mediante la
tres-métrica $q_{ab}$ inducida en las secciones espaciales que folian la variedad
cuatro-dimensional, el lapso densitizado ${N_{_{_{\!\!\!\!\!\!\sim}}\;}}=N/\sqrt{q}$ y el vector
desplazamiento $N^{a}$, con \linebreak$a,b\in\{\theta, \sigma,\delta\}$. Debido a la propiedad de
polarización lineal, la métrica satisface la condición $q_{\sigma\delta}=0$. Ademaś, debido a
la simetría en las coordenadas $\sigma$ y $\delta$, las
componentes de la métrica dependen solo de $t$ y de $\theta$, y son periódicas en ésta última, ya
que
es una coordenada circular. Por tanto, podemos expandir las componentes de la métrica en serie de
Fourier. Adoptaremos el siguiente convenio para definir los modos de Fourier $\phi_{m}$ de un campo
genérico $\phi(\theta)$:
\begin{align}\label{eq:Fourier-modes}
\phi(\theta)=\sum_{m\in\mathbb{Z}}\frac1{\sqrt{2\pi}}\phi_{m}e^{im\theta},
\qquad \phi_{m}=\frac1{\sqrt{2\pi}}\oint d\theta
\phi(\theta)e^{-im\theta}.
\end{align}

Como es general para espacio-tiempos con un grupo de isometría bidimensional compacto, la anulación
de las componentes cruzadas de la métrica correspondientes a las direcciones definidas por los
vectores de Killing, es decir $q_{\theta\sigma}=0=q_{\theta\delta}$, fija la libertad gauge
asociada a las ligaduras de momentos en esas direcciones \cite{man}, e implica que
$N^\sigma=0=N^\delta$. Imponiendo, por tanto, estas fijaciones de gauge, la tres-métrica toma una
forma diagonal y queda descrita por tres campos $(\tau,\xi,\bar\gamma)$, los cuales, en esencia,
caracterizan el área de las órbitas de isometrías, la norma de uno de los vectores de Killing y
el factor conforme de la métrica inducida en el conjunto de las órbitas de isometría. En
términos de
estos campos, $(\tau,\xi,\bar\gamma)$ la métrica adquiere la forma \cite{men1b}
\begin{align}\label{eq:gowdy-metric-ccm}
ds^2&=q_{\theta\theta}\big[-\tau^2{N_{_{_{\!\!\!\!\!\!\sim}}\;}}^2dt^2\!+\!(d\theta+N^\theta
dt)^2\big]\!+\!q_{\sigma\sigma}d\sigma^2\!+\!q_{\delta\delta}d\delta^2,\nonumber\\
q_{\theta\theta}&=\frac{4G}{\pi}e^{\bar\gamma-\frac{\xi}{\sqrt{\tau}}-\frac{\xi^2}{4\tau}},
\quad
q_{\sigma\sigma}=\frac{\pi}{4G}\tau^2e^{-\frac{\xi}{\sqrt{\tau}}},\quad
q_{\delta\delta}=\frac{4G}{\pi}e^{\frac{\xi}{\sqrt{\tau}}}.
\end{align}
Como ya hemos señalado anteriormente, la raíz cuadrada del área de las órbitas, $\tau>0$, es una
función global de tiempo.

Entonces, el espacio de fases está constituido por los tres campos $(\tau,\xi,\bar\gamma)$ y sus
respectivos momentos conjugados $(P_\tau,P_\xi,P_{\bar\gamma})$. A su vez, la acción que gobierna
este sistema es \cite{men1b}
\begin{subequations}\label{gowdy-class}\begin{align}\label{eq:action-gowdy}
S&=\int_{t_i}^{t_f} dt\oint
d\theta\big[P_\tau\dot{\tau}+P_{\bar\gamma}\dot{\bar\gamma}+P_\xi\dot{\xi}-({N_{_{_{\!\!\!\!\!\!\sim
} } \; }}\tilde{\mathcal C}+N^\theta\mathcal C_\theta)\big],\\
\label{eq:Ctheta-gowdy-class}
\mathcal C_\theta&=P_\tau\tau^\prime+P_{\bar\gamma}
\bar\gamma^\prime+P_\xi\xi^\prime-2P_{\bar\gamma}^\prime=0,
\\
\label{eq:Cscalar-gowdy-class}
\tilde{\mathcal C}&=\frac{4G}{\pi}\left[\frac{\tau}{2}P_\xi^2+\frac{\xi^2}{8\tau} P_{\bar\gamma}^2
-\tau P_\tau P_{\bar\gamma}\right]+\frac{\pi}{4G}\frac{\tau}{2}\Bigg[4\tau^{\prime\prime}
-2\bar\gamma^\prime\tau^\prime-\bigg(\frac{\xi\tau^\prime}{2\tau}\bigg)^2+(\xi^\prime)^2\Bigg]=0,
\end{align}\end{subequations}
donde el punto denota derivada con respecto al tiempo $t$, y la prima denota derivada con
respecto a $\theta$. Aquí, $\mathcal C_\theta$ es la ligadura de difeomorfismos en la coordenada
$\theta$ y $\tilde{\mathcal C}$ es la ligadura escalar densitizada.

\subsection{Fijación parcial de gauge y deparametrización}

Con el objetivo de deparametrizar el modelo y fijar (casi toda) la libertad gauge remanente se
imponen condiciones de gauge adicionales.

En concreto, se imponen las condiciones
\begin{subequations}\label{gauge}\begin{align}\label{eq:g1-ccm}
&g_1\equiv P_{\bar\gamma}-\frac{P_{\bar\gamma_{_0}}}{\sqrt{2\pi}}=0,\\
\label{eq:g2-ccm} &g_2\equiv \tau+t\frac{P_{\bar\gamma_{_0}}}{\sqrt{2\pi}}=0,
\end{align}\end{subequations}
donde recordamos que
\begin{align}
 P_{\bar\gamma_{_0}}=\frac1{\sqrt{2\pi}}\oint d\theta P_{\bar\gamma}(\theta).
\end{align}
Estas condiciones $g_1$ y $g_2$, junto con las ligaduras \eqref{eq:Ctheta-gowdy-class} y
\eqref{eq:Cscalar-gowdy-class}, deben formar un sistema de ligaduras de segunda clase (en la
terminología de Dirac \cite{dirac}) para que la fijación de gauge esté bien planteada.
Esto significa que los corchetes de Poisson
\begin{subequations}\begin{align}
\label{eq:gauge-property1}
\Big\{g_1(\theta),\oint d\bar\theta[G(\bar\theta)&\mathcal
C_\theta(\bar\theta)+F_{_{_{\!\!\!\!\!\!\sim}}\;}(\bar\theta)\tilde{\mathcal
C}(\bar\theta)]\Big\}=\frac{P_{\bar\gamma_{_0}}}{\sqrt{2\pi}}G^\prime(\theta),\\
\label{eq:gauge-property2}
\Big\{g_2(\theta), \oint d\bar\theta[G(\bar\theta)&\mathcal
C_\theta(\bar\theta)+F_{_{_{\!\!\!\!\!\!\sim}}\;}(\bar\theta)\tilde{\mathcal
C}(\bar\theta)]\Big\}=t\left(\frac{P_{\bar\gamma_{_0}}}{\sqrt{2\pi}}\right)^2
F_{_{_{\!\!\!\!\!\!\sim}}\;}(\theta),
\end{align}\end{subequations}
deben ser diferentes de cero para funciones periódicas $G(\theta)$ y densidades periódicas 
$F_{_{_{\!\!\!\!\!\!\sim}}\;}(\theta)$ arbitrarias. Por tanto, $P_{\bar\gamma_{_0}}$
no puede ser nulo. En la Ref. \cite{men1b}, el estudio fue restringido al sector $t>0,
P_{\bar\gamma_{_0}}<0$.
Además, la fijación de gauge debe ser compatible con la dinámica para estar bien definida. Esto
significa que se debe cumplir
\begin{align}
\dot g_i= \partial_tg_i+\Big\{g_i, \oint
d\theta[N^\theta(\theta)\mathcal
C_\theta(\theta)+N_{_{_{\!\!\!\!\!\!\sim}}\;}(\theta)\tilde{\mathcal
C}(\theta)]\Big\}=0,
\end{align}
para $i=1,2$. De la ecuación para $g_1$, obtenemos que la componente $\theta$ del vector
desplazamiento debe ser una función homogénea $N^\theta=N^\theta(t)$ y de la ecuación para $g_2$
obtenemos que el lapso densitizado queda totalmente fijado al valor
\begin{align}
 {N_{_{_{\!\!\!\!\!\!\sim}}\;}}=-\frac{\sqrt{2\pi}}{tP_{\bar\gamma_{_0}}}.
\end{align}
En consecuencia, el sistema queda totalmente deparametrizado (se resuelve la ligadura escalar) y
se fija la libertad gauge asociada con los modos de Fourier no cero de la ligadura de momentos en la
dirección inhomogénea $\theta$.
Nótese además que la ligadura~\eqref{eq:Ctheta-gowdy-class} y las condiciones de fijación de gauge 
\eqref{gauge} implican la igualdad
\begin{align}
-\frac{P_{\bar\gamma_{_0}}}{\sqrt{2\pi}}\bar\gamma^\prime=P_\xi\xi^\prime,
\end{align}
de modo que $\xi$, $P_\xi$ y $P_{\bar\gamma_{_0}}\neq0$ determinan todos los modos de Fourier no
cero de $\bar\gamma$:
\begin{align}\label{eq:gamma}
\bar\gamma(\theta)=\frac{\bar\gamma_{_0}}{\sqrt{2\pi}}+\sum_{m\neq
0}\frac{i}{\sqrt{2\pi}mP_{\bar\gamma_{_0}}}\oint d\bar\theta
e^{im(\theta-\bar\theta)}P_\xi(\bar\theta)\xi^\prime(\bar\theta).
\end{align}

En resumen, el par de grados de libertad tipo ``partícula puntual''
$(\bar\gamma_{_0},P_{\bar\gamma_{_0}})$ y los campos $(\xi,P_\xi)$ a los que no les afecta la
fijación de gauge constituyen el espacio de fases reducido. La acción reducida del sistema
resulta ser \cite{men1b}
\begin{subequations}\label{action-gow}\begin{align}
S&=\int_{t_i}^{t_f}
dt\left[P_{\bar\gamma_{_0}}\dot{\bar\gamma}_{_0}+\oint d\theta
P_\xi\dot{\xi}-(N^\theta C_\theta+\bar{H}_r)\right],\\
\label{eq:Ctheta-gow-red}
C_\theta&=\oint d\theta P_\xi\xi^\prime=0,\\
\bar{H}_r&=\frac1{2}\oint d\theta \left[P_\xi^2+(\xi')^2+\frac{\xi^2}{4t^2}\right].
\end{align}\end{subequations}
La ligadura $C_\theta$, que se impone cuánticamente, genera traslaciones en el
círculo y permanece en el sistema porque el modo cero de $N^\theta$ no se ha fijado.
El hamiltoniano reducido $\bar{H}_r$ corresponde al hamiltoniano de un campo escalar, libre,
axialmente simétrico y con masa dependiente del tiempo, que se propaga en un fondo estático
ficticio, con topología $T^2\times\mathbb{R}^+$ y métrica plana
$h_{ab}=-(dt)_a(dt)_b+(d\theta)_a(d\theta)_b+(d\sigma)_a(d\sigma)_b$.

\section{Fundamentos de la cuantización de Fock}
\label{2sec:fock}

Las ecuaciones de movimiento correspondientes a la acción \eqref{action-gow} implican que los
grados de libertad $(\bar\gamma_{_0},P_{\bar\gamma_{_0}})$ son constantes del movimiento, mientras
que el campo $\xi$ satisface la ecuación de Klein-Gordon
\begin{align}
 \ddot{\xi}-\xi''+\frac{\xi}{4t^2}=0.
\end{align}

El problema de cuantizar el sistema se reduce al de llevar a cabo la cuantización
de este campo $\xi$, por ejemplo a lo Fock. Como es habitual en teoría de campos, existe una
ambigüedad a la hora de descomponer $\xi$ en modos de frecuencia positiva
y negativa, lo que, a priori, puede dar lugar a teorías cuánticas inequivalentes.
No obstante, nótese que en el límite de tiempos grandes, la ecuación de movimiento se simplifica a
la de un campo escalar sin masa, para cuya descripción existe una elección natural de modos,
apelando a las simetrías del fondo plano. En vista
de esta propiedad, en la Ref.~\cite{men1b}, se descompuso el campo $\xi$ en los modos de frecuencia
positiva y negativa que se corresponderían con los de un campo escalar sin masa. Equivalentemente,
el campo se describió con las variables de tipo aniquilación y destrucción que se asocian usualmente
a un campo libre escalar sin masa, esto es
\begin{subequations}\begin{align}
a_0&=\sqrt{\frac{\pi}{8G}}\left(\xi_{_{0}}+i\frac{4G}{\pi}P_{\xi_{0}}\right),\qquad\qquad\qquad\,
a^*_0=\sqrt{\frac{\pi}{8G}}\left(\xi_{_{0}}-i\frac{4G}{\pi}P_{\xi_{0}}\right),\\
a_m&=\sqrt{\frac{\pi}{8G|m|}}\left(|m|\xi_{_{m}}+i\frac{4G}{\pi}P_{\xi_{m}}\right),\qquad
a^*_{-m}=\sqrt{\frac{\pi}{8G|m|}}\left(|m|\xi_{_{m}}-i\frac{4G}{\pi}P_{\xi_{m}}\right),
\label{eq:variables-fock}
\end{align}\end{subequations}
tales que $\{a_m,a_{\tilde m}^*\}=-i\delta_{m\tilde m}$.
Con este cambio de variables en la ec. \eqref{eq:Ctheta-gow-red}, el generador de
traslaciones en el círculo se reescribe de la forma
\begin{align}
 C_\theta&=\sum_{m=1}^\infty m(a_m^*a_m-a_{-m}^*a_{-m})=0.
\end{align}

En la cuantización de Fock, las variables $a_m$ y $a_m^*$ se promueven a operadores de aniquilación
y destrucción, $\hat a_m$ y $\hat a^\dagger_m$ respectivamente, tales
que $[\hat a_m,\hat a^\dagger_{\tilde m}]=\delta_{m\tilde m}$. A partir del estado de
vacío $|0\rangle$, caracterizado por las ecuaciones
\begin{equation}
\hat a_m|0\rangle=0,\qquad \forall m\in\mathbb{Z},
\end{equation}
se construye el espacio de Hilbert de una partícula y el espacio simétrico de Fock asociado
$\mathcal{F}$ \cite{wald2}. Sobre este espacio de Hilbert, se representa la ligadura, que implementa
cuánticamente el hecho de que el momento angular total del campo $\xi$ es nulo. El correspondiente
operador es
\begin{equation}\label{eq:Ctheta-operator}
\widehat C_\theta=\hbar\sum_{m>0}^\infty m(\hat
a^\dagger_m \hat a_m-\hat a^\dagger_{-m} \hat a_{-m}),
\end{equation}
donde se ha tomado orden normal \cite{men1b}. Este operador es autoadjunto en el
espacio de Fock $\mathcal F$. Entonces, el espacio de Hilbert físico $\mathcal F_f$, formado por
los estados aniquilados por $\widehat C_\theta$, es un subespacio propio del espacio de Fock
fiducial. 

En la Ref. \cite{men2} se demostró que ésta es la única cuantización de Fock del campo
$\xi$ (salvo equivalencia unitaria) en la cual la dinámica se implementa de forma unitaria
y en la que el vacío de la teoría es invariante bajo las transformaciones de simetría del grupo
$S^1$. Por tanto, la ambigüedad existente a la hora de definir los modos de frecuencia positiva y
negativa no da lugar a teorías inequivalentes con estas características. Posteriormente se demostró
también que ninguna otra
elección de las variables del espacio de fases, correspondiente a otra parametrización de los campos
de la métrica, da lugar a una cuantización de Fock con estas propiedades (salvo por la posibilidad
de hacer una transformación canónica dependiente del tiempo que suma un término lineal en $\xi$ a
$P_\xi$).

\cleardoublepage

\part{Cosmología cuántica de lazos homogénea}
\label{part1}

\chapter[Revisión del modelo plano de FRW]{Revisión del modelo plano de
Friedmann-Robertson-Walker}
\label{chap:revFRW}

Como ya hemos señalado anteriormente, el modelo plano e isótropo con un campo escalar sin masa ha
sido el modelo paradigmático en cosmología cuántica de lazos. La construcción rigurosa de su
estructura cuántica cinemática,
establecida en la Ref.~\cite{abl} y mejorada en la Ref. \cite{aps3}, junto con algunos análisis
adicionales sobre propiedades específicas del espacio de Hilbert cinemático \cite{Vel} y de los
operadores sobre él definidos \cite{kale,Pol}, han servido para fijar meticulosamente los
fundamentos matemáticos de la cosmología cuántica de lazos. Este rigor matemático permitió completar
la cuantización del modelo \cite{aps1a,aps1b} y llevar a cabo el análisis de su dinámica
\cite{aps3}, lo que reveló resultados tan sorprendentes como el del rebote cuántico ya comentado.

No obstante, el estudio del modelo de FRW plano con un campo escalar sin masa de la Ref. \cite{aps3}
no aclaraba completamente ciertos aspectos de la cuantización que, por tanto, necesitaban un
análisis más
cuidadoso con el fin de comprender mejor la teoría. Este capítulo está dedicado precisamente a la
investigación de algunos de esos aspectos y refleja principalmente el contenido de nuestra Ref.
\cite{mmo}, cuyos puntos esenciales también han sido recogidos en la Ref. \cite{lqc3}.
En particular, los aspectos que hemos revisado incluyen, por una parte, la densitización de la
ligadura
escalar cuántica con respecto al volumen del universo, paso que se realiza para
obtener una ligadura escalar \emph{densitizada} que es más fácil de imponer cuánticamente que la
ligadura \emph{no densitizada}%
\footnote{Denominamos ligadura escalar no densitizada a la que tiene la misma densitización que
la
ligadura hamiltoniana considerada en gravedad cuántica de lazos.},
y, por otra parte, la superselección del espacio de Hilbert cinemático en sectores.

Pretendemos que la densitización cuántica de la ligadura esté rigurosamente definida. Asimismo,
queremos que los sectores de superselección sean lo más simple posible y que, a la vez,
sus estados posean propiedades físicas óptimas. Entre tales propiedades, estamos especialmente
interesados en la existencia de un régimen en el que se obtenga la cuantización estándar o de
WDW como límite de la cuantización de lazos y en el que las características genéricas de
los estados expliquen la aparición del \emph{rebote cuántico}, no solo en un contexto limitado
como el analizado en previos estudios, sino de modo general.

Nuestro análisis parte de adoptar, en la cuantización de la Ref. \cite{aps3}, una prescripción de
operadores (con un orden de factores) diferente del escogido en la ec. \eqref{eq:operador-lig-aps}
al representar la ligadura hamiltoniana. A continuación, veremos que, efectivamente, nuestra
prescripción presenta ventajas con respecto a las anteriores propuestas y analizaremos las
consecuencias de implementar la cuantización con esta nueva ligadura hamiltoniana, tanto desde un
punto de vista cinemático como físico, de modo que los puntos expuestos previamente quedarán
aclarados.

\section{Operador ligadura escalar: nueva prescripción}
\label{3sec:sym}

A la vista de la ligadura hamiltoniana de la Ref. \cite{aps3}, dada en las ecs.
\eqref{eq:Klein-Gordon} y \eqref{eq:theta-aps}, una primera razón muy sencilla nos ha llevado a
cuestionar la prescripción allí adoptada al representar la ligadura como un operador simétrico: el
operador $\widehat{\Theta}$ introducido no está bien definido en $v=0$. En efecto, nótese que la
función $B(v)$ se anula en $v=0$ y, por tanto, su inverso diverge en dicho punto. Más aún, el
rescalado del producto interno introducido en la ec. \eqref{eq:inner-prod-aps} no se corresponde con
un cambio de representación mediante una transformación unitaria, ya que, mientras el estado
$|0\rangle$ tiene norma unidad en el espacio
de Hilbert cinemático original, $\mathcal H_{\text{grav}}$, dicho estado tiene norma nula en
$\mathcal H'_{\text{grav}}$, cuyo producto interno es el dado en la ec. \eqref{eq:inner-prod-aps}. 

Para solucionar este problema, hemos aprovechado la libertad que se tiene en la construcción del
operador para elegir una prescripción simétrica que desacople precisamente el estado de volumen
nulo $|0\rangle$,
de modo que dicho estado se pueda eliminar de la cuantización. Así, conseguimos que el cero ya no
esté en el espectro discreto del operador inverso del volumen y que, por tanto, este operador se
pueda invertir sin problema.

Por otra parte, en el modelo de Bianchi I, que es una generalización del modelo plano de FRW a
situaciones anisótropas, el signo del coeficiente de la tríada desempeña un papel importante. En
efecto, como existen tres direcciones diferentes en lugar de una, los productos de dos signos no son
necesariamente iguales a la identidad, como ocurre en FRW. Más concretamente, en la identidad
clásica análoga a la ec.\eqref{eq:identidad-lig} aparece, un producto de tres signos
diferentes en lugar de uno \cite{chio}. En consecuencia, se debe tener en cuenta la presencia
de los tres signos en la simetrización del operador que representa la ligadura. Precisamente, como
veremos en detalle en el próximo capítulo dedicado al modelo de Bianchi I, no solo
hemos desacoplado los estados de volumen nulo en nuestra cuantización del modelo, sino que también
hemos tenido especial cuidado en el tratamiento aplicado al operador signo.

Motivados por estos estudios del modelo de Bianchi I%
\footnote{Recordamos de la Introducción que cronológicamente hablando este modelo fue el
primero
que
cuantizamos.}, en el modelo isótropo simetrizamos el operador ligadura hamiltoniana de la siguiente
manera,
\begin{equation}\label{eq:operador-lig-mmo} 
\widehat{C}:=\widehat{\left[
\frac{1}{V}\right]}^{1/2}\left(-\frac{6}{\gamma^{2}}
\widehat{\Omega}^2+8\pi
G\hat{P}_{\phi}^2\right)\widehat{\left[
\frac{1}{V}\right]}^{1/2}.
\end{equation}
En el término que da cuenta de la contribución de la geometría, hemos representado el operador
inverso del volumen siguiendo la misma prescripción que en el término material%
\footnote{Siempre que aparezca alguna función de un operador (como la potencia $1/2$ del
operador inverso del volumen) la supondremos definida a partir de la descomposición espectral del
mismo, en virtud del teorema espectral (Sección \ref{teorema-espectral}).}. La actuación de este
operador sobre los
estados de la base $|v\rangle$ se obtiene de la ec.
\eqref{eq:triad-operator} mediante la relación
\begin{align}\label{inv-volum}
 \widehat{\left[\frac{1}{V}\right]}&=
\widehat{\left[\frac{1}{\sqrt{|p|}}\right]}^{3}.
\end{align}
Por otra parte, el operador $\widehat\Omega$ se define como
\begin{align}\label{eq:operador-grav-mmo}
\widehat\Omega&=\frac1{4i\sqrt{\Delta}}
\widehat{\sqrt{|p|}} 
\widehat{\left[\frac1{\sqrt{|p|}}\right]}^{-1/2}
\left[\left(\hat{\mathcal
N}_{2\bar\mu}-\hat{\mathcal
N}_{-2\bar\mu}\right)\widehat{\text{sgn}(p)}
+\widehat{\text{sgn}(p)}\left(\hat{\mathcal
N}_{2\bar\mu}-\hat{\mathcal
N}_{-2\bar\mu}\right)\right]\nonumber\\
&\times\widehat{\left[\frac1{\sqrt{|p|}}\right]}^{-1/2}
\widehat{\sqrt{|p|}},
\end{align}
con las definiciones de los operadores involucrados dadas en la Sección \ref{1sec:lig-ham-aps}.
Tanto la acción de $\widehat{\text{sgn}(p)}$ como la de
$\widehat{\left[1/{\sqrt{|p|}}\right]}^{-1/2}$ en el estado $|v=0\rangle$ se pueden definir de
modo arbitrario, ya que la acción final de $\widehat\Omega$ es independiente de esa elección, pues
tenemos que $\widehat\Omega|0\rangle=0$.

El orden de factores adoptado en estas expresiones garantiza que este operador ligadura escalar
coincide con el análogo del modelo de Bianchi I si identificamos en él las tres direcciones.
Además, si comparamos el término gravitacional o geométrico del operador
\eqref{eq:operador-lig-mmo} con el operador
\eqref{eq:operador-lig-aps} vemos que efectivamente difieren en la prescripción de simetrización.

Dos características principales distinguen nuestra prescripción:
\begin{enumerate}
 \item [i)] repartimos las potencias de los operadores $\hat{p}$ y $\widehat{[1/\sqrt{|p|}]}$ a la
derecha y a la izquierda de cada factor en el que están presentes y
\item[ii)] simetrizamos el término que involucra el signo de $p$ del siguiente modo:
\begin{equation}\label{eq:sym-sgn}
{\sin(\bar\mu c)}{\text{sgn}(p)}\rightarrow\frac{1}{2}\left[\widehat{\sin(\bar\mu
c)}
\widehat{\text{sgn}(p)}+\widehat{\text{sgn}(p)}\widehat{\sin(\bar\mu c)}\right].
\end{equation}
\end{enumerate}

En la expresión \eqref{eq:operador-grav-mmo} del operador $\widehat\Omega$, los operadores que
representan las potencias de $p$ aparecen a la derecha y a la izquierda de una manera muy
particular, no trivial, ya que en la representación polimérica el producto de $\widehat{|p|}$ por
$\widehat{[1/|p|]}$ no es la identidad. No obstante, cabe comentar que,
si bien este orden contreto se toma así por analogía con el modelo de Bianchi I, como después
veremos, 
los resultados físicos no dependen apreciablemente de él, y se podría tomar un orden más
sencillo. De hecho, en la literatura ya se había propuesto una simplificación del
modelo de la Ref.~\cite{aps3}, que se denominó cosmología cuántica de lazos simplificada
\cite{acs}, en la que precisamente se ignoran los efectos cuánticos debidos a la regularización
implementada para representar el inverso de $p$ y se comprueba que no son cualitativamente
importantes.
Esta simplificación también eliminaría posibles dependencias en la elección de celda
fiducial que pueden aparecer a órdenes subdominantes en la dinámica clásica efectiva asociada a la
cosmología cuántica de lazos \cite{chinos}.

\subsection{Resolución cinemática de la singularidad clásica}

Es inmediato comprobar que nuestro operador $\widehat{C}$, definido en
${\mathcal{H}}_{\textrm{grav}}$ con dominio $\text{Cil}_\text{S}$, aniquila al estado
$|v=0\rangle$ y deja invariante su complemento ortogonal, designado por
$\widetilde{\mathcal{H}}_{\textrm{grav}}$. Es decir, los estados con volumen nulo se desacoplan y
podemos restringir el estudio al subespacio $\widetilde{\mathcal{H}}_{\textrm{grav}}$ del espacio de
Hilbert cinemático. Este hecho resuelve
el problema existente en la Ref. \cite{aps3}, señalado al inicio de esta sección. A este respecto,
cabe añadir que posteriormente apareció en la literatura una alternativa para corregir dicho
problema \cite{Pol}, en la que se desacopla el estado $|0\rangle$ redefiniendo la actuación del
operador \eqref{eq:operador-lig-aps} sobre él y en sus estados aledaños, esto es, sobre $|-4\rangle$
y $|4\rangle$.

La singularidad clásica se alcanza cuando el factor de escala o, equivalentemente, el coeficiente
$p$ de la tríada densitizada, se anulan. En nuestra teoría cuántica, una vez que hemos restringido
el estudio a $\widetilde{\mathcal{H}}_{\textrm{grav}}$, el operador $\hat{p}$ ya no
tiene autoestados de autovalor nulo, pues el cero ha sido eliminado de su espectro.
Entonces
podemos decir que resolvemos el big bang cinemáticamente%
\footnote{En cierto sentido este concepto de resolución de la singularidad implementa ideas de
los
trabajos originales de Bojowald \cite{boj2}.}. Veremos además que esto implica también su
resolución en el espacio de Hilbert físico.

\section{Densitización del operador ligadura escalar}
\label{3sec:dens}

A la hora de resolver la ligadura hamiltoniana cuántica es conveniente reescribirla en una forma
densitizada, pues esto permite separar las variables geométricas de las variables materiales. 

Recordemos primero que imponemos la ligadura cuántica en el dual del dominio denso del
operador
ligadura $\widehat{C}$. Definamos el espacio
\begin{align}
 \widetilde{\textrm{Cil}}_\text{S}=\text{lin}\{|v\rangle;\;v\in\mathbb{R}-\{0\}\},
\end{align}
como la envolvente lineal de los vectores $|v\rangle$ de volumen no nulo,
cuya compleción de Cauchy con respecto a la medida discreta es el espacio de Hilbert
$\widetilde{\mathcal{H}}_{\textrm{grav}}$ 
introducido antes. Entonces, podemos buscar las soluciones a la
ligadura escalar en el espacio dual $\left(\widetilde{\text{Cil}}_\text{S}\otimes
\mathcal{S}(\mathbb{R})\right)^*$, donde recordemos que $\mathcal{S}(\mathbb{R})$ es el dominio de
los operadores del sector material. 

Nuestro procedimiento de densitización consiste en introducir la siguiente aplicación
\begin{equation}\label{eq:mapeo-dens}
(\tilde\psi|\longrightarrow(\psi|=(\tilde\psi|
\widehat{\left[\frac{1}{V}\right]}^{1/{2}},
\end{equation}
para cualquier elemento $(\tilde\psi|$ perteneciente a
$\left(\widetilde{\text{Cil}}_\text{S}\otimes\mathcal{S}(\mathbb{R})\right)^*$. Nótese que esta
transformación está bien definida, ya que $\widehat{[1/V]}$ deja invariante su dominio
$\widetilde{\text{Cil}}_\text{S}$. Es más, es sencillo ver que la transformación es una biyección.
Ahora bien, si $(\tilde\psi|$ es solución de la
ligadura hamiltoniana: $(\tilde\psi|\widehat{C}^\dagger=0$, entonces el transformado  $(\psi|$ es
solución de la ligadura hamiltoniana densitizada, definida como
\begin{equation}\label{eq:densitization}
 (\psi|\widehat{\mathcal C}^\dagger=0, \qquad \widehat{{\cal C}}=
\widehat{\left[\frac1{V}\right]}^{-1/{2}}
\widehat{C}\widehat{\left[\frac{1}{V}
\right]}^{-1/{2}}.
\end{equation}
Nótese que, si no hubiéramos desacoplado los estados de volumen nulo, el cero estaría en el espectro
discreto del operador $\widehat{[1/V]}$ y el operador $\widehat{\left[1/{V}\right]}^{-1/2}$
(obtenido aplicando el teorema espectral) estaría mal definido. Sin embargo, en
$\widetilde{\mathcal{H}}_{\textrm{grav}}$ (con dominio $\widetilde{\text{Cil}}_\text{S}$) está bien
definido. De la Ec.
\eqref{eq:operador-lig-mmo} deducimos que el operador (simétrico) correspondiente a la ligadura
densitizada, $\widehat{\mathcal C}$, tiene la expresión explícita
\begin{equation}\label{eq:operador-lig-dens-mmo}
\widehat{{\cal C}}=-\frac{6}{\gamma^{2}}
\widehat{\Omega}^2+8\pi G\hat{P}_{\phi}^2.
\end{equation}

Esta forma equivalente de expresar la ligadura hamiltoniana es más fácil de imponer porque, como
vemos, los operadores $\widehat{\Omega}^2$ y $\hat{P}_{\phi}^2=-\hbar^2\partial_\phi^2$ pasan a ser
observables de Dirac compatibles, pues conmutan entre ellos y con el operador ligadura escalar
densitizada.

Otra forma de obtener la ligadura densitizada cuántica consiste en representar directamente con un
operador la ligadura densitizada clásica $\tilde{\mathcal C}=\sqrt{q}\;\mathcal C$, que en la acción
va multiplicada por la función lapso densitizado ${N_{_{_{\!\!\!\!\!\!\sim}}\;}}=N/\sqrt{q}$. En
cosmología homogénea, equivalentemente, se puede tomar su integral
\begin{align}
 \tilde{C}=\int_\mathcal V d^3x \;\mathcal{\tilde C}=VC.
\end{align}
Este procedimiento se está empleando últimamente en la literatura de cosmología cuántica de lazos
(véase por ejemplo la
Ref. \cite{awe}) y equivale a ignorar los
efectos cuánticos debidos al operador inverso del volumen, una vez se tiene garantizado que la
densitización es factible. Por tanto, coincide, en ese aspecto, con la denominada
cosmología cuántica de lazos simplificada, introducida en la Ref. \cite{acs}. No obstante, en la
teoría completa se ha demostrado que la ligadura
clásica densitizada $\mathcal{\tilde{C}}$ no da lugar a una teoría cuántica bien definida
\cite{lqg1}. Así pues, aunque en cosmología homogénea la diferencia entre este procedimiento y el
nuestro no sea cuantitativamente relevante desde un punto de vista físico, es preferible
partir del objeto clásico análogo al que se representa en gravedad cuántica de lazos y densitizarlo
cuánticamente de forma rigurosa, por consistencia y fidelidad a la teoría completa. 

\section[Operador ligadura gravitacional]{Caracterización del operador ligadura\\ gravitacional}
\label{3sec:grav}

Como hemos visto, el operador $\widehat\Omega^2$, que proporciona la parte gravitacional de la
ligadura escalar \eqref{eq:operador-lig-dens-mmo}, es un observable de Dirac. Como la parte
material es bien conocida y, de hecho, ya ha sido analizada en el Capítulo \ref{chap:introLQC},
será suficiente analizar las propiedades de $\widehat\Omega^2$ y llevar a cabo su análisis
espectral para resolver la ligadura.

\subsection{Sectores de superselección}

En el Capítulo \ref{chap:introLQC} vimos que el operador de ligadura gravitacional es un
operador en diferencias que produce una traslación de cuatro unidades en el número cuántico $v$
de los vectores $|v\rangle$. Como consecuencia, solo los estados con soporte en redes discretas de
paso 4 están relacionados bajo su acción. Adicionalmente, los espacios de Hilbert cuyos
estados tienen soporte en tales redes están superseleccionados, ya que tanto la ligadura escalar
total como los observables físicos preservan estos espacios. Si bien, estos resultados son
generales en cosmología cuántica de lazos isótropa, veremos en esta subsección que las redes
asociadas a nuestro operador 
$\widehat\Omega^2$, y los correspondientes sectores de superselección, son más sencillos que los
obtenidos en los estudios previos del modelo.

Es fácil comprobar que la acción de $\widehat\Omega^2$ sobre la base de estados $|v\rangle$ de 
$\widetilde{\mathcal{H}}_{\textrm{grav}}$ es
\begin{align}\label{grav-op-action}
\widehat\Omega^2|v\rangle&=-f_+(v)f_+(v+2)|v+4\rangle+
\left[f_+^2(v)+f_-^2(v)\right]|v\rangle-f_-(v)f_-(v-2)
|v-4\rangle,
\end{align}
donde
\begin{align}\label{eq:f}
f_\pm(v)&=\frac{\pi\gamma l_{\textrm{Pl}}^2}{3}
g(v\pm2)s_\pm(v)g(v), \\
s_\pm(v)&=\text{sgn}(v\pm2)+\text{sgn}(v),\\
g(v)&=
\begin{cases}
\left|\left|1+\frac1{v}\right|^{\frac1{3}}
-\left|1-\frac1{v}\right| ^{\frac1{3}}
\right|^{-\frac1{2}} & {\text{si}} \quad v\neq 0,
\vspace*{.2cm}\\
0 & {\text{si}} \quad v=0.\\
\end{cases}
\end{align}

Nótese que la combinación de signos en las funciones $f_\pm(v)$, la cual se debe a nuestra 
prescripción para simetrizar el signo de $p$ en la ec. \eqref{eq:sym-sgn}, da lugar a
una propiedad muy destacable: 
\begin{itemize}
 \item $f_-(v)f_-(v-2)=0,\quad \text{si} \quad v\in(0,4]$,
  \item $f_+(v)f_+(v+2)=0,\quad \text{si} \quad v\in[-4,0)$.
\end{itemize}
Como consecuencia, los semiejes $v>0$ y $v<0$ se desacoplan el uno del otro bajo la acción de
nuestro operador $\widehat\Omega^2$, como se puede ver a partir de la ec.
\eqref{grav-op-action}, al aplicarla en $|v|\leq4$. En conclusión, este operador solo
relaciona los estados $|v\rangle$ cuyo número cuántico $v$ pertenece a una de las semirredes de paso
cuatro dadas por
\begin{align}\label{eq:redes-4}
{\mathcal L}_{\tilde\varepsilon}^\pm=\{v=\pm(\tilde\varepsilon+4n),
\,n\in\mathbb{N}\},\qquad \tilde\varepsilon\in(0,4].
\end{align}
En otras palabras, el operador $\widehat\Omega^2$ está bien definido en cualquiera de los espacios
de Hilbert $\mathcal H^\pm_{\tilde\varepsilon}$ obtenidos como cierre de los respectivos dominios
\begin{align}
 \text{Cil}_{\tilde\varepsilon}^{\pm}=\text{lin}
\{|v\rangle,\,v\in{\mathcal L}_{\tilde\varepsilon}^\pm\}
\end{align}
con respecto al producto interno discreto. El espacio de Hilbert cinemático y no separable
$\widetilde{\mathcal H}_{\text{grav}}$ se puede entonces escribir como la suma directa de
subespacios separables:
\begin{align}
 \widetilde{\mathcal
H}_{\text{grav}}=\oplus_{\tilde\varepsilon}
(\mathcal H^+_{\tilde\varepsilon}\oplus\mathcal
H^-_{\tilde\varepsilon}).
\end{align}

La acción de la ligadura hamiltoniana (y la de los observables físicos) preserva los espacios
$\mathcal H^\pm_{\tilde\varepsilon}\otimes L^2(\mathbb{R},d\phi)$, como ocurría en el Capítulo
\ref{chap:introLQC}. Por tanto, estos espacios de
Hilbert proporcionan sectores de superselección en nuestra teoría y podemos restringir el estudio a
cualquiera de ellos. Nosotros hemos restringido nuestra formulación a $\mathcal
H^+_{\tilde\varepsilon}\otimes~L^2(\mathbb{R},d\phi)$, para un valor arbitrario de
$\tilde\varepsilon\in(0,4]$.

La diferencia con respecto a los trabajos previos en la cuantización polimérica de este modelo
\cite{abl,aps1a,aps1b,aps3,kale,acs}, parte de los cuales se ha resumido en el
Capítulo~\ref{chap:introLQC}, es que el soporte de nuestros sectores está contenido en un solo
semieje de la recta real, mientras que en esos estudios los sectores tienen contribuciones tanto
del semieje negativo como del positivo, pues no se desacoplan. Veremos que la
simplicidad de nuestros sectores aporta ventajas significativas.

Conviene enfatizar que el estado $|v=0\rangle$ no está incluido en ninguno de los sectores de
superselección de nuestro modelo. En principio, se podría pensar que los sectores etiquetados con
$\tilde\varepsilon=4$ están conectados con ese estado bajo la acción del operador
$\widehat\Omega^2$. Sin embargo, se puede comprobar explícitamente que esto no es cierto. 
Por tanto, nada especial ocurre en estos sectores de
superselección en comparación con los otros con $\tilde\varepsilon\neq4$.

\subsection{Autoadjunción}

Hemos podido demostrar rigurosamente que el operador simétrico $\widehat\Omega^2$ (con dominio
$\text{Cil}_{\tilde\varepsilon}^{+}$) es esencialmente autoadjunto. Para ello, nos hemos basado en
los resultados de autoadjunción de la Ref. \cite{kale} y en la teoría de Kato de perturbaciones de
operadores resumida en la Sección \ref{perturbaciones}. En la Ref. \cite{kale} se introduce un
operador en diferencias de
paso 4, denotado%
\footnote{Frecuentemente, en la literatura de cosmología cuántica de lazos, se emplea el
acrónimo
``APS'' para referirse al modelo de la Ref. \cite{aps3} de Ashtekar, Pawlowski y Singh.} por
$\widehat{H}'_\text{APS}$,
cuya expresión explícita es
\begin{align}
 \widehat{H}'_\text{APS}=-\frac{4\pi G}{3}\big[\hat{\mathcal
N}_{2\bar\mu}(\hat{v}^2-2-a)\hat{\mathcal
N}_{2\bar\mu}+\hat{\mathcal N}_{-2\bar\mu}(\hat{v}^2-2-a)\hat{\mathcal
N}_{-2\bar\mu}-2\hat{v}^2+2a\big],
\end{align}
donde $a$ es cualquier constante real. Este operador está definido sobre los siguientes espacios de
Hilbert:
\begin{itemize}
 \item $\mathcal H^+_{\tilde\varepsilon}\oplus\mathcal H^-_{4-\tilde\varepsilon}$, con dominio
natural $\text{Cil}_{\tilde\varepsilon}^{+}\cup
\text{Cil}_{4-\tilde\varepsilon}^{-},\;\text{ si }\tilde\varepsilon\neq4$;
\item $\mathcal H^+_{4}\oplus\mathcal H^-_{4}\oplus\mathcal H_0$, (siendo $\mathcal H_0$ el espacio
de Hilbert unidimensional generado por el estado de norma unidad $|v=0\rangle$), con dominio natural
$\text{Cil}_{4}^{+}\cup\text{Cil}_{-4}^{-}\cup\text{lin}\{|0\rangle\},\;
\text{ si}$ $\tilde\varepsilon=4$.
\end{itemize}

En la Ref. \cite{kale} se demuestra que este operador, definido en cualquiera de los espacios de
Hilbert antes especificados, está unitariamente relacionado mediante una tranformación de Fourier
con el hamiltoniano de una partícula puntual en la recta real sometida a un potencial de
P\"osch-Teller, operador diferencial bien conocido, que es esencialmente
autoadjunto. Por tanto, el operador $H'_\text{APS}$ también lo es. 

Comparemos este operador con nuestro operador $\widehat\Omega^2$, definidos ambos en el mismo
dominio (en el caso $\tilde\varepsilon=4$ definimos $\widehat\Omega^2|0\rangle=0$). Para ello,
primero introducimos los operadores $f_\pm(\hat{v})$, obtenidos a partir de la
resolución espectral del operador de multiplicación $\hat{v}$. A continuación, reescribimos el
operador $\widehat\Omega^2$ de la forma
\begin{align}
 \widehat{\Omega}^2=-\hat{\mathcal N}_{2\bar\mu}f_-(\hat{v})f_+(\hat{v})\hat{\mathcal N}_{2\bar\mu}-
\hat{\mathcal N}_{-2\bar\mu}f_+(\hat{v})f_-(\hat{v})\hat{\mathcal N}_{-2\bar\mu}+
\left\{[f_+(\hat{v})]^2+[f_-(\hat{v})]^2\right\}.
\end{align}
Por último, definimos el operador diferencia $\widehat\Omega^2-\frac{4}{3 \pi
G}\widehat{H}'_\text{APS}$:
\begin{align}
\widehat{\Omega}^2-\frac{4}{3 \pi
G}\widehat{H}'_\text{APS}=-\hat{\mathcal N}_{2\bar\mu}h_1(\hat{v})\hat{\mathcal
N}_{2\bar\mu}-
\hat{\mathcal N}_{-2\bar\mu}h_1(\hat{v})\hat{\mathcal N}_{-2\bar\mu}+
h_2(\hat{v}),
\end{align}
con
\begin{align}
 h_1(\hat{v})&=f_-(\hat{v})f_+(\hat{v})-(\hat{v}^2-2-a),\\
h_2(\hat{v})&=\left\{[f_+(\hat{v})]^2+[f_-(\hat{v})]^2\right\}-2\hat{v}^2+2a.
\end{align}
Tomando $a=10/27$ no es difícil demostrar que $h_1({v})\sim\mathcal{O}(v^{-2})$ y
$h_2({v})\sim\mathcal{O}(v^{-2})$ en el límite
asintótico $v\rightarrow\pm\infty$. Como consecuencia, para este valor de $a$, los
operadores $\widehat\Omega^2$ y $\frac{4}{3 \pi
G}\widehat{H}'_\text{APS}$ difieren en un operador simétrico de la clase de traza y, por tanto,
dicho operador diferencia es $\frac{4}{3 \pi
G}\widehat{H}'_\text{APS}$-pequeño (veáse la definición en la Sección \ref{perturbaciones}).
Entonces, el teorema \ref{kato-rellich} garantiza que $\widehat\Omega^2$, definido en el mismo
espacio de Hilbert que $\widehat{H}'_\text{APS}$, es esencialmente autoadjunto.

Para demostrar que la restricción al espacio $\mathcal H^+_{\tilde\varepsilon}$, por ejemplo, es
también esencialmente autoadjunta podemos hacer uso del teorema \ref{th:esa}.
Como hemos visto, $\widehat\Omega^2$ es esencialmente autoadjunto en $\mathcal
H^+_{\tilde\varepsilon}\oplus\mathcal H^-_{4-\tilde\varepsilon}$. Supongamos que
su restricción en $\mathcal H^+_{\tilde\varepsilon}$ no fuera esencialmente
autoadjunta. Esto implicaría que existe alguna solución no trivial a su ecuación de índices de
defecto normalizable en $\mathcal H^+_{\tilde\varepsilon}$. A partir de esta solución, podríamos
construir una solución normalizable de la ecuación de índices de defecto cuando el operador se
define en el espacio de Hilbert mayor $\mathcal H^+_{\tilde\varepsilon}\oplus\mathcal
H^-_{4-\tilde\varepsilon}$, igual a la primera en $\mathcal H^+_{\tilde\varepsilon}$ e
idénticamente nula en $\mathcal
H^-_{4-\tilde\varepsilon}$. Esto significaría que $\widehat\Omega^2$ no es esencialmente autoadjunto
en dicho espacio de Hilbert mayor, en contradicción con nuestro resultado de que sí lo es.
Por tanto, el operador $\widehat\Omega^2$ es esencialmente autoadjunto también en $\mathcal
H^+_{\tilde\varepsilon}$, como queríamos demostrar.

\subsection{Análisis espectral}

En la Ref. \cite{kale} se demostró que los espectros esencial y absolutamente
continuos del operador $H'_\text{APS}$ son ambos $[0,\infty)$. Los teoremas \ref{weyl} y
\ref{kato-birman}
nos permiten extender estos resultados a nuestro operador $\widehat\Omega^2$,
debido al hecho de que (salvo una constante multiplicativa global) ambos operadores difieren en un
operador de la
clase de traza. Además, teniendo en cuenta la simetría del operador $\widehat\Omega^2$ bajo
un cambio de signo en $v$, $f_\pm(-v)=-f_\mp(v)$, y asumiendo la independencia del
espectro en la etiqueta $\tilde\varepsilon$, concluimos que el operador $\widehat\Omega^2$ definido
en $\mathcal H^+_{\tilde\varepsilon}$ tiene también espectros
absolutamente continuo y esencial iguales a $[0,\infty)$. 

Adicionalmente, como veremos en la sección \ref{3subsec:wdw}, las autofunciones generalizadas de
$\widehat\Omega^2$ convergen para valores grandes de $v$ a las autofunciones de su operador análogo
en la cuantización estándar geometrodinámica de WDW. Este hecho, junto con la
continuidad del espectro en geometrodinámica cuántica, es suficiente para concluir que los espectros
discreto y singular son vacíos. 

En definitiva, el operador $\widehat\Omega^2$ definido en $\mathcal
H^+_{\tilde\varepsilon}$ es un operador positivo y esencialmente autoadjunto con espectro
absolutamente continuo dado por $\mathbb{R}^+$.

\subsection{Autofunciones generalizadas}

Denominemos los autoestados generalizados de $\widehat\Omega^2$, correspondientes al autovalor
(en sentido generalizado) $\lambda\in[0,\infty)$, como $
 |e^{\tilde\varepsilon}_{\lambda}\rangle=
\sum_{v\in{\mathcal L}^+_{\tilde\varepsilon}}
e^{\tilde\varepsilon}_{\lambda}(v)|v\rangle.
$
Mediante el análisis de la correspondiente ecuación de autovalores
$\widehat\Omega^2|e^{\tilde\varepsilon}_{\lambda}\rangle=\lambda|e^{\tilde\varepsilon}_{\lambda}
\rangle$, hemos demostrado que todos los coeficientes
$e^{\tilde\varepsilon}_{\lambda}(\tilde\varepsilon+4n)$ 
($n\in \mathbb{N}^+$) de la correspondiente autofunción generalizada están unívocamente
determinados por el dato inicial $e^{\tilde\varepsilon}_{\lambda}(\tilde\varepsilon)$ y hemos
determinado su expresión explícita. Veámoslo:

Si tenemos en cuenta la acción \eqref{grav-op-action} del operador $\widehat\Omega^2$ y las
propiedades de las funciones $f_\pm(v)$, en particular que $f_+(v)=f_-(v+2)$, la ecuación de
autovalores da lugar a la ecuación recursiva
\begin{align}
e^{\tilde\varepsilon}_{\lambda}(v+4)&=\left[\frac{f_-(v+2)}{f_+(v+2)}+\frac{f_-(v)}{f_+(v)}
\frac{f_+(v-2)}{f_+(v+2)}-\frac{\lambda}{f_+(v)f_+(v+2)}\right]e^{\tilde\varepsilon}_{\lambda}
(v)\nonumber\\
&-\frac{f_-(v)}{f_+(v)}\frac{f_-(v-2)}{f_+(v+2)}e^{\tilde\varepsilon}_{\lambda}(v-4).
\end{align}
Gracias a que $f_-(\tilde\varepsilon)f_-(\tilde\varepsilon-2)=0$, esta ecuación, que en principio
es de segundo orden para un valor de $v$ arbitrario, da lugar a una relación de consistencia entre
los dos primeros datos iniciales $e^{\tilde\varepsilon}_{\lambda}(\tilde\varepsilon+4)$ y
$e^{\tilde\varepsilon}_{\lambda}(\tilde\varepsilon)$. Haciendo iteraciones se comprueba entonces
que los distintos coeficientes de la autofunción generalizada verifican la siguiente
expresión:
\begin{align}\label{eq:eigenstates-omega2}
e^{\tilde\varepsilon}_{\lambda}(\tilde\varepsilon+4n)&=
\left[\mathcal S_{\tilde\varepsilon}(0,2n)
+\frac{F(\tilde\varepsilon)}{G_\lambda(\tilde
\varepsilon-2)}\mathcal
S_{\tilde\varepsilon}(1,2n)\right]
e^{\tilde\varepsilon}_{\lambda}(\tilde\varepsilon),\quad n\in\mathbb{N}^+,
\end{align}
donde
\begin{subequations}\begin{align}\label{FG}
F(v)&=\frac{f_-(v)}{f_+(v)}, \qquad G_\lambda(v)=
-\frac{i\sqrt{\lambda}}{f_+(v)},\\
\label{S}
\mathcal
S_{\tilde\varepsilon}(a,b)&=\sum_{O(a\rightarrow b)}
\left[\prod_{\{r_p\}}
F(\tilde\varepsilon+2r_p+2)\prod_{\{s_q\}}
G_\lambda(\tilde\varepsilon+2s_q) \right],
\end{align}\end{subequations}
y $O(a\rightarrow b)$ denota el conjunto de todas las posibles maneras de
ir desde $a$ hasta $b$ a saltos de una o dos unidades. Para cada elemento en el conjunto
$O(a\rightarrow b)$, $\{r_p\}$ es el subconjunto de enteros seguidos por un salto de dos
unidades mientras que $\{s_q\}$ es el subconjunto de enteros seguidos de un salto de una unidad.
Ilustremos estos subconjuntos con un ejemplo: si $a=1$ y $b=4$, entonces se tienen tres elementos en
el conjunto
$O(1\rightarrow4)$, dados por
\begin{align*}
 1\rightarrow2\rightarrow3\rightarrow4&:\quad\{r_p\}=\emptyset, \quad\{s_q\}=\{1,2,3\};\\
1\rightarrow2\rightarrow4&:\quad \{r_p\}=\{2\}, \quad\{s_q\}=\{1\};\\
1\rightarrow3\rightarrow4&: \quad\{r_p\}=\{1\}, \quad\{s_q\}=\{3\}.
\end{align*}
Nótese que $F(\tilde\varepsilon)=0$ para todo $\tilde\varepsilon\leq2$. Por tanto, en estos casos
el segundo término en la ec. \eqref{eq:eigenstates-omega2} no contribuye.

Como he señalado anteriormente, el espectro de $\widehat\Omega^2$ es positivo y absolutamente
continuo. Vemos que, además, es no degenerado, en tanto en cuanto solo un dato inicial determina
completamente cada autofunción. Elegimos una base de autoestados
$|e^{\tilde\varepsilon}_{\lambda}\rangle$ normalizados a la delta de Dirac tal que 
\begin{align}
 \langle e^{\tilde\varepsilon}_{\lambda}
|e^{\tilde\varepsilon}_{\lambda'}\rangle=
\delta(\lambda-\lambda').
\end{align}
Esta condición fija la norma compleja de $e^{\tilde\varepsilon}_{\lambda}(\tilde\varepsilon)$
en la ec. \eqref{eq:eigenstates-omega2}. La única libertad remanente en la elección de este dato
inicial está en la elección de su fase, que fijamos de tal modo que
$e^{\tilde\varepsilon}_{\lambda}(\tilde\varepsilon)$ sea positivo. Entonces, las autofunciones de la
base así construidas son reales, como consecuencia de que el operador en diferencias
$\widehat{\Omega}^2$ [ec. \eqref{grav-op-action}] tiene coeficientes reales. En definitiva, la
resolución espectral de la identidad en el espacio de Hilbert cinemático $\mathcal
H^+_{\tilde\varepsilon}$ asociada a $\widehat{\Omega}^2$ puede expresarse como
\begin{equation}\label{identity}
\mathbb{I}=\int_{\mathbb{R^+}} d\lambda
|e_{\lambda}^{\tilde\varepsilon}\rangle \langle
e_{\lambda}^{\tilde\varepsilon}|.
\end{equation}

Cabe destacar que hemos sido capaces de resolver la ecuación general de autovalores de nuestro
operador gravitacional $\widehat\Omega^2$, determinando explícitamente la forma de sus
autofunciones generalizadas. Este hecho contrasta con el nivel de resolución alcanzado, por ejemplo,
en la Ref.
\cite{aps3}, donde las autofunciones del operador análogo $\widehat{\Theta}$ se proporcionaron solo
a través de una relación de recurrencia y se generaron numéricamente.

\section{Espacio de Hilbert físico}
\label{3sec:phys}

Estamos ya en disposición de completar la cuantización. Para ello, podemos seguir cualquiera de las
dos estrategias siguientes:
\begin{itemize}
\item Aplicar el procedimiento de promedio sobre grupos \cite{gave1a,gave1b,gave1c,gave1d,gave2}.
Los estados físicos
son los estados invariantes bajo la acción del grupo generado por la extensión autoadjunta del
operador ligadura y los podemos obtener promediando sobre dicho grupo. Además, este
promedio determina (salvo equivalencia unitaria) un produnto interno natural que dota a los
estados físicos de una estructura de espacio de Hilbert.
 \item Resolver la ligadura buscando los elementos
$(\psi|\in\left(\widetilde{\text{Cil}}_\text{S}\otimes\mathcal{S}(\mathbb{R})\right)^*$ que
verifican $(\psi|\widehat{\mathcal C}^\dagger=0$ y dotarlos de estructura de espacio de Hilbert. Si
imponemos que un conjunto completo de observables sean autoadjuntos, determinamos (salvo
equivalencia unitaria) el producto interno físico \cite{red1,red2}.
\end{itemize}
Como en la Sección \ref{1sec:phys} del Capítulo \ref{chap:introLQC} ya hemos descrito cómo se
obtiene la estructura física siguiendo la segunda estrategia, por completitud expositiva, ahora,
aplicaremos el procedimiento de promedio sobre grupos.

Sea $\widehat{\bar{\mathcal C}}$ la extensión autoadjunta del operador ligadura hamiltoniana
$\widehat{\mathcal C}$ en el espacio de Hilbert cinemático $\mathcal
H^\pm_{\tilde\varepsilon}\otimes
L^2(\mathbb{R},d\phi)$ y sea $\mathcal D$ su dominio denso de definición. Dado un elemento 
$|\psi\rangle\in\mathcal D$, con función de onda $\psi(v,\phi)$, podemos actuar sobre él con el
promedio del grupo uniparamétrico
generado por $\widehat{\bar{\mathcal C}}$. Entonces, el elemento resultante,
\begin{align}\label{ga}
 \Psi(v,\phi)=[\mathcal P \psi](v,\phi)=\int_\mathbb{R} dt e^{it\widehat{\bar{\mathcal C}}}
\psi(v,\phi),
\end{align}
es invariante bajo la acción de este grupo. 

El promedio $\mathcal P$ ``proyecta'' el espacio de Hilbert cinemático sobre el
espacio de Hilbert físico. En términos de la base de autofunciones
$e^{\tilde\varepsilon}_\lambda(v)$ de la parte geométrica y de la base de autofunciones de la parte
material $e^{\pm i\nu\phi}$ (recordemos que $\nu\in\mathbb{R}$), los elementos de $\mathcal D$
se pueden expandir de la forma
\begin{align}\label{eq:expansion-kin}
 \psi(v,\phi)=\int_0^\infty d\lambda \int_{-\infty}^{\infty} d\nu\;
e^{\tilde\varepsilon}_\lambda(v)\big[\tilde\psi_+(\lambda)e^{i\nu\phi}+\tilde\psi_-(\lambda)e^{
-i\nu\phi}
\big].
\end{align}
Insertando esta expresión en la ec. \eqref{ga} y operando obtenemos (salvo por una constante de
normalización)
\begin{align}\label{ga3}
 \Psi(v,\phi)&=\int_0^\infty \frac{d\lambda}{\nu(\lambda)}
e^{\tilde\varepsilon}_\lambda(v)\big[\tilde\psi_+(\lambda)e^{i\nu(\lambda)\phi}
+\tilde\psi_-(\lambda)e^ {
-i\nu(\lambda)\phi }
\big], 
\end{align}
donde
\begin{align}
 \nu(\lambda)&=\sqrt{\frac{3\lambda}{4\pi l_\text{Pl}^2\hbar\gamma^2}}.
\end{align}

El promedio $\mathcal P$ también proporciona un producto interno natural. En efecto, dados dos
estados físicos $|\Psi_1 \rangle$ y $|\Psi_2 \rangle$ se tiene
\begin{align}
 \langle \Psi_1|\Psi_2 \rangle_{\text{fis}}=\langle \mathcal P
\psi_1|\psi_2\rangle_{\text{cin}}= \int_0^\infty \frac{d\lambda}{\nu(\lambda)}\big[
\tilde\psi_{1+}^*(\lambda)\tilde\psi_{2+}(\lambda)+\tilde\psi_{1-}^*
(\lambda)\tilde\psi_{2-}(\lambda)\big ].
\end{align}
Nótese que este producto es independiente de cuál de los dos elementos cinemáticos se promedie.
Por tanto, el espacio de Hilbert físico, al cual pertenecen los perfiles espectrales
$\tilde\psi_\pm(\lambda)$, es
\begin{align}
 \mathcal
H^{\tilde\varepsilon}_{\text{fis}}=L^2\left(\mathbb{R}^+,\frac{d\lambda}{\nu(\lambda)} \right).
\end{align}

De nuevo, como en el Capítulo \ref{chap:introLQC}, si interpretamos $\phi$
como un tiempo interno,
vemos que las soluciones se descomponen en componentes de frecuencia positiva ($+$) y negativa
($-$), $\Psi_\pm(v,\phi)$, que están determinadas por el dato inicial $\Psi_\pm(v,\phi_0)$ mediante
una evolución unitaria:
\begin{align}
 \Psi_\pm(v,\phi)=U_\pm(\phi-\phi_0)\Psi_\pm(v,\phi_0),\qquad U_\pm(\phi-\phi_0)=\exp{\left[\pm
i\sqrt{\frac{3}{4\pi l_\text{Pl}^2\hbar\gamma^2}\,\widehat{\Omega}^2}(\phi-\phi_0)
\right]}.
\end{align}
Una vez más, en completa analogía con lo descrito en la sección \ref{1subsec:evol}, la constante del
movimiento $\hat{P}_{\phi}$ y el observable $\hat{v}|_{\phi_0}$, que mide el valor del
volumen cuando el tiempo vale $\phi_0$, proporcionan un conjunto completo de observables. El
segundo además permite interpretar el sistema en la imagen de evolución. 

\section{Límite de WDW}
\label{3subsec:wdw}
 
Otro aspecto importante que hemos investigado es el comportamiento de nuestra teoría en la
región de volúmenes grandes. Hemos comprobado que, en esa
región, se recupera la cuantización estándar implementada en la geometrodinámica, es decir, la
teoría
de WDW, cuyas predicciones para los valores esperados de los observables sobre estados
semiclásicos coinciden con las predicciones clásicas de la relatividad general. 

Para llevar a cabo tal análisis, solo se necesita saber cómo se comportan las
autofunciones del operador gravitacional $\widehat\Omega^2$ en el límite $v\rightarrow\infty$,
puesto
que ya hemos cuantizado el otro operador involucrado en la ligadura, $\hat{P}_\phi^2$, de acuerdo
con la representación estándar de tipo Schr\"odinger que se adopta en la teoría de
WDW.
En esta sección, veremos cómo son las autofunciones del operador análogo a $\widehat\Omega^2$ en la
teoría de WDW y cuál es su relación con las autofunciones de la cuantización de lazos
para valores grandes de $v$.

\subsection{Cuantización de WDW del modelo de FRW}
\

Como ya hicimos en la cuantización de lazos, trabajaremos en la representación de momentos (es
decir, de $v$). La parte gravitacional del espacio de Hilbert cinemático de la teoría de
WDW es el espacio de Hilbert de las funciones de cuadrado integrable
con respecto a la medida usual de Lebesgue, $L^2(\mathbb{R},dv)$, en lugar de la discreta. En esta
representación, el operador $\hat{p}$ actúa por multiplicación por el factor
\begin{align}
p=\text{sgn}(v) (2\pi\gamma l_{\text{Pl}}^2\sqrt{\Delta}|v|)^{2/3}, 
\end{align}
como en la cuantización
polimérica. Además, como la representación ahora es continua, la conexión tiene un operador
asociado
\begin{align}
 \hat c=i\,2(2\pi\gamma l_{\text{Pl}}^2/\Delta)^{1/3}|v|^{1/6}{\partial}_{v}|v|^{1/6},
\end{align}
tal que se respeta la regla de Dirac: $[\hat c,\hat{p}]=i\hbar\widehat{\{c,p\}}$.

Sea $\widehat{\underline\Omega}^2$ el operador cuántico que representa a la cantidad
clásica $(cp)^2$ en la teoría de WDW%
\footnote{Distinguiremos las cantidades correspondientes a la teoría de WDW de
sus análogas en cosmología cuántica de lazos mediante un subrayado.}, definido en el espacio de
Schwartz
$\mathcal{S}(\mathbb{R})$. 
Como queremos comparar sus propiedades con las de su análogo $\widehat\Omega^2$, escogemos para
$\widehat{\underline\Omega}^2$ un orden simétrico similar, que da lugar a un operador bien definido
en sentido distribucional:
\begin{align}
\widehat{\underline\Omega}^2&=-\frac{\alpha^2}{4}\sqrt{|v|}
\left[\text{sgn}(v)\partial_{v}+\partial_
{v}\text{sgn}(v)\right]|v|\left[\text{sgn}(v)
\partial_{v}+\partial_{v}\text{sgn}(v)\right]
\sqrt{|v|}\nonumber\\
&=-\frac{\alpha^2}{4}\left[1+4v\partial_{v}+
4(v\partial_v)^2\right],
\end{align}
donde $\alpha=4\pi\gamma l_{\text{Pl}}^2$.
Nótese que en la segunda igualdad hemos eliminado el término $|v|\delta(v)$, que no contribuye,
con el fin de simplificar la expresión.

Debido a las propiedades bien conocidas del operador $\widehat{\Omega}=-i\alpha(v\partial_v+1/2)$,
podemos asegurar que $\widehat{\underline\Omega}^2$ no solo es esencialmente autoadjunto en
$L^2(\mathbb{R},dv)$ sino que también lo es en cada uno de los subespacios
$L^2(\mathbb{R}^\pm,dv)$. Por tanto, su acción en el semieje positivo $v>0$ está desacoplada de la
del negativo $v<0$, al igual que ocurre con su análogo $\widehat{\Omega}^2$ en cosmología cuántica
de lazos, y podemos, también en este caso, restringir el estudio a las funciones con soporte en
$v>0$, es decir, a $L^2(\mathbb{R}^+,dv)$. Además, en este último espacio de Hilbert, el espectro de
$\widehat{\underline\Omega}^2$ es positivo y absolutamente continuo. Sus autofunciones
generalizadas, correspondientes al autovalor generalizado $\lambda\in[0,\infty)$, se pueden
etiquetar con $\omega=\pm\sqrt{\lambda}\in\mathbb{R}$ y están dadas por
\begin{equation}\label{eq:wdw-eig-mmo}
\underline{e}_{\omega}(v)= \frac{1}{\sqrt{2\pi\alpha
|v|}}\exp\left({-i\omega\frac{\ln{|v|}}{\alpha}}\right).
\end{equation}
Estas funciones proporcionan una base ortonormal generalizada para $L^2(\mathbb{R}^+,dv)$, con
normalización
\begin{align}
\langle \underline{e}_{\omega} |\underline{e}_{\omega^{\prime}}\rangle
=\delta(\omega-\omega^{\prime}).
\end{align}

\subsection{Límite de WDW de las autofunciones}
\label{3sec:lim-wdw-frw}

Se puede probar que las autofunciones $e^{\tilde{\varepsilon}}_{\lambda}(v)$ convergen en el límite
de $v$ grande a una autofunción del operador $\widehat{\underline\Omega}^2$ de la teoría de
WDW. Para ello hemos aplicado un método
recientemente publicado en la Ref. \cite{kp-posL}, que, resumidamente, consiste en los siguientes
pasos:
\begin{itemize}
\item[i)] Se introduce la notación vectorial
\begin{align}
 {\vec e}^{\;\tilde{\varepsilon}}_{\lambda}(v)=
\left(
\begin{array}{l}
 e^{\tilde{\varepsilon}}_{\lambda}(v)\\
e^{\tilde{\varepsilon}}_{\lambda}(v-4)
\end{array}\right)
\end{align}
y se escribe la ecuación de las autofunciones en forma matricial, mediante una matriz de
transferencia $\boldsymbol{A}(v)$, es decir, 
${\vec
e}^{\;\tilde{\varepsilon}}_{\lambda}(v+4)=\boldsymbol{A}(v){\vec
e}^{\;\tilde{\varepsilon}}_{\lambda}(v)$.
\item[ii)] Por otra parte, se representan los valores de la autofunción en dos puntos consecutivos
$v$ y $v+4$ de la red ${\mathcal L}_{\tilde{\varepsilon}}^+$ por vectores $\vec{\psi}(v)$ formados
por los coeficientes de su descomposición en términos de las autofunciones
$\underline{e}_{\pm|\omega|}$ de la teoría de WDW evaluadas en ese par de puntos, es
decir
${\vec e}^{\;\tilde{\varepsilon}}_{\lambda}(v)=\boldsymbol{B}(v)\vec{\psi}(v)$, con
\begin{align}
 \boldsymbol{B}(v)=
\left(
\begin{array}{ll}
 \underline{e}_{\omega}(v) & \underline{e}_{-\omega}(v)\\
\underline{e}_{\omega}(v-4) & \underline{e}_{-\omega}(v-4)
\end{array}\right).
\end{align}
\item[iii)] Se reescribe la ecuación de autovalores como una ecuación para los vectores
$\vec{\psi}(v)$: $\vec{\psi}(v+4)=\boldsymbol{M}(v)\vec{\psi}(v)$, donde
$\boldsymbol{M}(v)=\boldsymbol{B}^{-1}(v+4)\boldsymbol{A}(v)\boldsymbol{B}(v)$.
\item[iv)] Se calcula la expansión asintótica de la matriz $\boldsymbol{M}(v)$ cuando
$v\to\infty$.
\end{itemize}
El cálculo explícito muestra que la matriz considerada es de la forma \cite{kp-posL}
\begin{align}
\boldsymbol{M}(v) = \mathbb{I} + \mathcal{O}(v^{-3}).
\end{align}
Entonces, existe un límite bien definido
\begin{equation}
\vec{\psi} = \lim_{v\to\infty} \vec{\psi}(v) .
\end{equation}
Es más, las propiedades asintóticas del operador $\widehat{\Omega}^2$, el cual, 
salvo por una constante global y después de un conveniente cambio de representación,
difiere del operador $\widehat{\Theta}$ de la Ref. \cite{aps3} solo en un operador compacto (véase
la definición en la Sección \ref{perturbaciones}), nos permiten aplicar aquí los resultados
numéricos de dicha referencia, que muestran que $\vec{\psi}(v)$ converge a $\vec{\psi}$ de la
siguiente manera:
\begin{equation}\label{eq:conv-rate}
\vec{\psi}(v) = \vec{\psi} +\mathcal{O}(v^{-2}).
\end{equation}

Una vez que se ha demostrado que cada autofunción
$e^{\tilde{\varepsilon}}_{\lambda}(v)$ efectivamente converge a cierta función de la teoría de
WDW, es fácil ver cómo debe ser ese límite. Vemos que el espectro de
$\widehat{\underline\Omega}^2$ tiene degeneración 2, mientras que habíamos demostrado que el de
$\widehat{\Omega}^2$ es no degenerado. Entonces, cada una de las autofunciones de la cosmología
cuántica de lazos debe converger en el límite de $v$ grande a una combinación lineal de dos
autofunciones de la teoría de WDW. Es más, como las autofunciones de la
cosmología cuántica de lazos son reales, las dos componentes WDW deben contribuir con
igual amplitud en dicha combinación lineal. En definitiva, el límite de WDW debe ser de
la forma
\begin{equation}\label{wdw-limit-mmo}
e^{\tilde{\varepsilon}}_{\lambda}(v) \xrightarrow{v\gg 1} r\left\{
\exp\left[{i\phi_{\tilde\varepsilon}(\omega)}\right]
\,\underline{e}_{\omega}(v) +
\exp\left[{-i\phi_{\tilde\varepsilon}(\omega)}\right]
\,\underline{e}_{-\omega}(v) \right\},
\end{equation}
donde $r$ es cierta constante de normalización. A su vez, se ha comprobado numéricamente que la fase
$\phi_{\tilde\varepsilon}(\omega)$ tiene el siguiente comportamiento \cite{kp-posL}
\begin{equation}\label{alpha-form}
\phi_{\tilde\varepsilon}(\omega) = T(|\omega|) +
c_{\tilde{\varepsilon}} +
R_{\tilde{\varepsilon}}(|\omega|),
\end{equation}
donde $T$ es una función solo de $|\omega|$, $c_{\tilde{\varepsilon}}$ es una constante y
$\lim_{\omega\rightarrow0}R_{\tilde{\varepsilon}}(|\omega|)=0$.

\section[Resolución de la singularidad: rebote cuántico]{Resolución de la singularidad
cosmológica:\\ rebote cuántico}

A la vista de los resultados obtenidos, estamos en disposición de revisar el concepto de
\emph{rebote cuántico} y de replantearnos, al menos en situación isótropa, la pregunta formulada
en la introdución de esta tesis acerca de si este resultado es genérico en el contexto de la
cosmología cuántica de lazos.

Nuestra cuantización da respuesta afirmativa a dicho interrogante gracias a dos propiedades
básicas que la distinguen de otras propuestas de cuantización, como veremos en la
siguiente y última sección de discusión. Estas características son:
\begin{itemize}
  \item {\bf{Comportamiento de onda estacionaria exacta}}:\\
Como acabamos de ver, las autofunciones de nuestro operador gravitacional convergen en el límite
de volúmenes grandes a una combinación de dos autofunciones de la teoría de WDW. Estas
autofunciones, dadas en la ec.~\eqref{eq:wdw-eig-mmo}, se pueden interpretar como ondas
correspondientes a universos en
contracción y en expansión (esencialmente en $v$) o, equivalentemente, como ondas entrantes y
salientes, respectivamente. Estas componentes contribuyen con igual amplitud ya que las
autofunciones son reales y, en este sentido, el límite es una onda estacionaria exacta.
 \item {\bf{Ausencia de frontera}}:\\
Por otra parte, nuestras autofunciones tienen soporte en un solo semieje, que además no contiene a
la singularidad potencial. Este comportamiento no se deriva de imponer ninguna condición
particular, como podría ser una condición de frontera, sino que es una característica natural de
nuestro modelo debida solo a las propiedades funcionales de nuestro operador gravitacional. Desde
esta perspectiva, consideramos que nuestro modelo proporciona una descripción intrínseca de tipo
\emph{ausencia de frontera}%
\footnote{En cosmología cuántica se había empleado anteriormente el concepto de ausencia de frontera
o \emph{no-boundary}, en un sentido distinto al aquí presentado \cite{haw1,haw2,haw3}.}.
\end{itemize}

Estos dos resultados implican que la componente saliente debe evolucionar a una componente entrante
y viceversa, ya que el flujo no puede escapar a través de $v=0$. Por tanto, en la solución física
de la ec. \eqref{ga3}, restringida por ejemplo al sector de frecuencias positivas, las componentes
en expansión y en contracción deben representar las dos ramas de un universo que sufre un
\emph{rebote cuántico} para algún valor esperado de $\hat{v}|_{\phi_0}$ positivo,
independientemente del perfil espectral físico $\tilde\psi_+(\lambda)$ que se considere, que puede
ser, por tanto, todo lo cuántico que se quiera. Obviamente, las componentes entrantes y salientes
comentadas estarán picadas sobre trayectorias bien diferenciadas solo para cierto tipo de estados,
aquéllos que son semiclásicos en la región de $v$ grande y, solo para ellos, se obtendrán
trayectorias que muestren un \emph{gran rebote} en sustitución de la \emph{gran explosión} clásica.

En resumen, hemos obtenido un rebote cuántico completamente genérico. 

\section{Discusión y comparación con trabajos previos}

Terminaremos el capítulo discutiendo las similitudes y diferencias con respecto a los trabajos
previos
en cosmología cuántica de lazos isótropa, especialmente el de la Ref.~\cite{aps3}, que fue el
primero donde se determinó
el espacio de Hilbert físico y donde se estudió satisfactoriamente por primera vez el rebote
cuántico de los estados que son semiclásicos a tiempos tardíos.

Como vimos en el Capítulo \ref{chap:introLQC}, para el tratamiento de la Ref. \cite{aps3} los
sectores de
superselección tienen soporte en redes $\mathcal L_{\pm|\epsilon|}$ que se extienden sobre toda la
recta real. Asociados con esos sectores, el espectro del operador $\widehat{\Theta}$ (análogo a
nuestro operador $\widehat{\Omega}^2$) tiene degeneración 2, y las correspondientes autofunciones
generalizadas tienen el comportamiento asintótico dado en la ec. \eqref{eq:wdw-limit-aps}. Además,
haciendo uso de la simetría bajo paridad ($v\to -v$), el estudio se restringe al sector simétrico.
Para ello, es necesario unir dos redes diferentes de tal modo que el soporte de los estados esté
simétricamente distribuido en torno a $v=0$. Gracias a la introducción de esta simetría, el análisis
se puede restringir al semieje de $v$ positivo. Por otro lado, en ese trabajo, el estudio se limita
solo a estados semiclásicos, cuyo soporte en vectores de onda $k$ positivo es despreciable.

Una vez que se introduce la simetría de paridad y en la práctica se restringen los estados
físicos a la región de $k$ negativo, en la Ref. \cite{aps3} se obtiene un escenario similar al
nuestro, en el sentido de que, para valores grandes de $v$, las autofunciones simétricas que
más contribuyen tienen un límite de WDW en \emph{cada} red $\mathcal L_{\pm|\epsilon|}$
que es \emph{aproximadamente} de tipo onda estacionaria. Esto es una consecuencia de la ec.
\eqref{eq:wdw-limit-aps} y de las propiedades comentadas de los coeficientes $A$ y $B$,
junto con la implementación de la simetría de paridad. Sin embargo, es de destacar que, mientras
que este comportamiento de onda estacionaria es aproximado y solo válido para $k\ll -1$ en el caso
de la Ref. \cite{aps3}, en nuestro modelo este comportamiento del límite de WDW se
obtiene de forma \emph{exacta} y \emph{para todos} los autovalores del momento del campo escalar.

Por otra parte, el procedimiento de la Ref. \cite{aps3} de restringir los estados al sector
simétrico bajo paridad, que exige la unión de dos redes diferentes para un valor genérico de
$|\epsilon|$ (es decir cuando $|\epsilon|\neq0,2$), tiene algunas consecuencias que merecen
comentarse. Incluso cuando el análisis se restringe al sector $k\ll -1$, el límite de WDW
de las autofunciones lleva una fase que depende de $\epsilon$. De hecho, se puede ver numéricamente
que la fase relativa de los coeficientes $A$ y $B$ presenta el mismo tipo de
dependencia encontrada para nuestro modelo en la ec. \eqref{alpha-form} (con $\tilde\varepsilon$ y
$|\omega|$ ahora reemplazados por $\epsilon$ y $|k|$). Entonces, incluso en la región de interés,
dos redes diferentes poseen diferentes límites y, por tanto, su unión no admite un límite
global. 
Aunque cabe añadir que para los estados semiclásicos considerados en la Ref. \cite{aps3}, que están
picados para $v\gg 1$ en torno a dos trayectorias clásicas que no se solapan (una la
de un universo en expansión y la otra la de un universo en contracción), la diferencia en el límite
de WDW de las dos redes es simplemente una fase global para cada una de las dos
mencionadas ramas, que no afecta la norma del estado. No obstante, esta propiedad del límite de
WDW de la norma no es válida para estados más generales.

El modelo estudiado en la Ref. \cite{aps3} fue después simplificado en la Ref. \cite{acs}, de modo
que se obtuvo un modelo analíticamente resoluble con el que se demostró que, de hecho, el rebote
cuántico es independiente del estado físico considerado. Pero este resultado no es genérico, pues se
demostró solo para un sector de superselección específico, precisamente el que contiene al estado
$|v=0\rangle$.

En contraste, en nuestra propuesta todos los sectores de superselección tienen el soporte contenido
en un semieje. Esto nos permite restringir el estudio a, por ejemplo, $v\in\mathbb{R}^+$ de un modo
natural, sin necesidad de apelar a un proceso de simetrización bajo paridad como el descrito antes.
En
cualquier caso, nótese que los sectores de nuestro modelo son perfectamente compatibles con la
imposición de la simetría de paridad, la cual puede ser directamente implementada tomando la suma
directa de dos sectores tal que su soporte sea la unión de las semirredes
${\mathcal L}^+_{\tilde\varepsilon}\cup{\mathcal L}^-_{\tilde\varepsilon}$. Debido a la
simplicidad de nuestros sectores, el espectro de $\widehat{\Omega}^2$ es no degenerado, propiedad
que facilita el cálculo exacto y explícito de la base de autofunciones generalizadas [veáse la ec.
\eqref{eq:eigenstates-omega2}]. Una ventaja analítica y numérica del hecho de que no haya
degeneración es que, para fijar cada autofunción, solo hemos necesitado imponer que el dato inicial
$e^{\tilde\varepsilon}_{\lambda}(\tilde\varepsilon)$ sea positivo, porque la condición de
normalización fija completamente su norma. Finalmente, como ya hemos visto en la sección anterior,
las propiedades de nuestro modelo nos llevan a demostrar que el rebote cuántico es realmente
genérico, no como en la Ref. \cite{aps3}, donde solo se considera un cierto régimen. 

Conviene destacar además que nuestros resultados son válidos para cualquier elección de sector de
superselección (es decir para todos los valores de $\tilde{\varepsilon}$) y que el tipo de
descripción de {ausencia de frontera} que hemos obtenido, que desempeña un papel fundamental en
los
argumentos que explican la existencia del rebote cuántico, es una característica de nuestro modelo
que no es compartida por ninguno de los trabajos previos realizados sobre el modelo isótropo plano
\cite{abl,aps1a,aps1b,aps3,kale,Pol,acs}.

En definitiva, nuestro estudio complementa de modo significativo los
resultados obtenidos en cosmología cuántica de lazos. El panorama actual nos permite
afirmar que la cuantización de lazos del modelo plano de FRW acoplado a un campo escalar sin masa
(el modelo paradigmático en cosmología) resuelve satisfactoriamente y sin excepción la
singularidad inicial clásica a la vez que está de acuerdo con la relatividad general en regímenes
semiclásicos. Además, recientemente, en el estudio de la Ref.~\cite{kp-posL}, se
ha analizado un aspecto no tratado aquí%
\footnote{Este punto fue también parcialmente considerado en la Ref. \cite{cs2}.}: el hecho de
si el
rebote cuántico respeta el comportamiento semiclásico. Se ha demostrado que cualquier estado
que es marcadamente
semiclásico para un valor grande de $v$ se mantiene así a lo largo de toda la evolución cuántica. En
la parte II de esta tesis veremos explícitamente un resultado similar en el contexto del modelo de
Bianchi I, cuya cuantización pasamos a estudiar a continuación.

\chapter[Cuantización del modelo de Bianchi I]{Cuantización del modelo de Bianchi
I}
\label{chap:bianchi}

Con el próposito de extender el formalismo de cosmología cuántica de lazos isótropa a situaciones
cada vez más generales, el siguiente paso natural es analizar cosmologías todavía homogéneas pero
anisótropas. El espacio-tiempo anisótropo más sencillo de todos es el que tiene secciones espaciales
planas. Éste es el modelo de Bianchi I, de acuerdo con la clasificación de álgebras de Lie
tridimensionales hecha por Bianchi~\cite{bianchi,kramer}.
Nosotros, además, estamos interesados en este modelo porque representa el
sector homogéneo de nuestra cuantización híbrida del modelo de Gowdy.

El modelo de Bianchi I ha sido ampliamente estudiado debido a su simplicidad y aplicaciones en
cosmología. 
De hecho, antes del desarrollo de la cosmología cuántica de lazos ya existían en la literatura
algunos análisis preliminares sobre su cuantización empleando en su descripción variables de
Ashtekar \cite{aspu,neg2,neg1}. Los primeros intentos de construir un espacio de Hilbert cinemático
e
introducir el operador ligadura hamiltoniana en un formalismo polimérico se deben a Bojowald 
\cite{boj}. La estructura cinemática fue después revisada en la Ref.~\cite{chio} empleando las
técnicas desarrolladas para el caso isótropo en la Ref. \cite{abl}. El trabajo de la
Ref.~\cite{chio} también fue el primero en el que se intentó completar el programa de cuantización
de Dirac para obtener el espacio de Hilbert físico. Posteriormente, también se estudió la
correspondiente dinámica clásica efectiva \cite{chi1,chi2}.
Nosotros, en nuestra motivación de aplicar la representación polimérica del modelo de Bianchi I al
sector homogéneo de Gowdy, hemos revisado la cuantización de la Ref.~\cite{chio}, aprovechando
la oportunidad no solo para corregir los errores allí cometidos, sino también para aplicar nuestra
prescripción de simetrización del operador ligadura hamiltoniana, que, como acabamos de ver en el
contexto del modelo de FRW,
resulta ser muy conveniente para completar el análisis de los estados físicos. Esta investigación
ha dado lugar a nuestra publicación \cite{mmp}. Por otra parte, en la Ref. \cite{luc2} se
revisó también el trabajo de la Ref.~\cite{chio}, simultáneamente a nosotros y de modo
independiente. No obstante, ese estudio se encuadra en el contexto de la cosmología cuántica de
lazos simplificada, siguiendo la línea de la Ref. \cite{acs}, y tampoco explota las ventajas de una
prescripción de simetrización como la que hemos escogido nosotros.

El aspecto más intrincado que se encuentra al intentar adaptar la cuantización del caso isótropo a
este modelo anisótropo reside en la implementación de la llamada {dinámica mejorada},
explicada en la Sección \ref{1sec:lig-ham-aps}. En
presencia de anisotropías, es necesario introducir tres longitudes fiduciales mínimas a la hora de
definir el tensor de curvatura en términos de un lazo de holonomías. Originalmente surgieron dos
esquemas distintos para definir dichas longitudes mínimas, que en la literatura usualmente se
denominan esquema $\bar\mu$ y esquema $\bar\mu'$ \cite{chi2}. Como veremos, estos
esquemas dan lugar, repectivamente, a una teoría factorizable en las distintas direcciones
fiduciales y a una teoría no factorizable. Por simplificar su nomenclatura, a estos esquemas
factorizable y no factorizable los denominaremos esquemas A y B, respectivamente. Debido a la no
factorización, a partir del esquema B es más difícil llevar a cabo la correspondiente teoría
cuántica. Esto, unido al hecho de que a priori no había argumentos físicos claros que seleccionaran
de modo categórico uno de los esquemas
frente al otro, llevó a
que, originalmente, se implementara la cuantización de acuerdo con el esquema A \cite{chio}, que es
también el que nosotros adoptamos en la Ref. \cite{mmp}. No obstante, dicho esquema presenta
inconvenientes en situaciones con topología espacial no compacta \cite{chi2,luc2}. Esto ha
motivado
que recientemente se haya vuelto a investigar el esquema B. En concreto, se ha defendido su
plausibilidad apelando a la hipotética relación entre los grados de libertad de la
cosmología cuántica de lazos y de la gravedad cuántica de lazos y, además, Ashtekar y Wilson-Ewing
han conseguido construir el operador ligadura hamiltoniana correspondiente \cite{awe}. A la vista
de estos resultados, también hemos revisado esta reciente cuantización con el propósito de
completarla. En particular, una vez más, hemos aplicado nuestra prescripción de simetrización, pues
la propiedad asociada de {ausencia de frontera} facilita significativamente la determinación
del espacio de Hilbert físico \cite{mmw}.

Hemos considerado el modelo en vacío y con la topología de un tres-toro, pues, como hemos
comentado, estamos especialmente interesados en analizar el sector de soluciones homogéneas del
modelo de Gowdy $T^3$. A diferencia del modelo de FRW, que en vacío es estático, el modelo de
Bianchi I en vacío tiene dinámica no trivial. Sus soluciones son del llamado tipo
Kasner \cite{kasner}, con dos factores de escala en expansión y el tercero en contracción o
viceversa. A este respecto, el hecho de estudiar el modelo en vacío también tiene como motivo el
análisis (recogido en la siguiente parte de esta tesis) de la dinámica cuántica en un escenario en
el que todos los grados de libertad son geométricos y en el que, entonces, la variable interpretada
como tiempo interno experimenta una cuantización polimérica.

Este capítulo, por tanto, recoge la cuantización que hemos llevado a cabo en el modelo de
Bianchi I, con ambos esquemas de la dinámica mejorada. Realizaremos un análisis paralelo de estas
dos teorías cuánticas
para hacer más patente los aspectos que comparten y sus diferencias.

\section[Formulación clásica]{Formulación clásica en variables de Ashtekar- \\Barbero}
\label{4sec:clas}

A diferencia del modelo plano de FRW analizado en el Capítulo \ref{chap:revFRW}
, en el modelo de
Bianchi I con la topología del tres-toro no
es necesario introducir ninguna celda fiducial pues el modelo ya proporciona una celda natural
finita, la del tres-toro, que tomamos con lados de longitud coordenada igual a $2\pi$.

Al igual que en el caso isótropo, a la hora de describir el sistema en el formalismo de
Ashtekar-Barbero, fijamos el gauge y elegimos una co-tríada plana diagonal, ${}^o
e_a^i=\delta_a^i$.
Por tanto, la presencia de tres direcciones espaciales diferentes
requiere la introducción de tres parámetros para describir las componentes no triviales de la
conexión de Ashtekar-Barbero y otros tres para la tríada densitizada, esto es%
\footnote{En lo que sigue, no usaremos el convenio de sumación de Einstein, a no ser que
se especifique lo contrario.},
\begin{equation}\label{eq:var-ashtekar-bianchi}
A_a^i=\frac{c^{i}}{2\pi}\delta_a^i, \qquad
E_i^a=\frac{p_{i}}{4\pi^2}\delta_i^a\sqrt{{}^oq}.
\end{equation}
De este modo se tiene que los corchetes de Poisson no triviales entre las variables $c^i$ y $p_i$
que parametrizan el espacio de fases son
\begin{align}\label{eq:poisson-bianchi}
 \{c^i,p_j\}=8\pi G\gamma\delta^i_j.
\end{align}
La métrica espacio-temporal escrita en estas variables adquiere la forma
\begin{equation} \label{metric-bianchi}
ds^2= -N^2
dt^2+\frac{|p_\theta p_\sigma p_\delta|}{4\pi^2}\left(
\frac{d\theta^2}{p_\theta^2}+\frac{d\sigma^2}{p_\sigma^2}+\frac{d\delta^2}{p_\delta^2}\right),
\end{equation}
donde se han introducido coordenadas espaciales circulares $\{\theta,\sigma,\delta\}$.
A su vez, el espacio de fases está constreñido por la ligadura hamiltoniana
\begin{align}\label{eq:ligBianchi}
 C_\text{BI}=-\frac2{\gamma^2}\frac{c^\theta p_\theta c^\sigma p_\sigma+c^\theta p_\theta c^\delta
p_\delta+c^\sigma p_\sigma c^\delta p_\delta}{V}=0.
\end{align}
En esta expresión $V=\sqrt{|p_\theta p_\sigma p_\delta|}$ es el volumen físico del universo.

Nótese, comparando con las ecs.~\eqref{eq:variables-abl} y \eqref{eq:ligFRW}, que si identificamos
las tres direcciones, haciendo $p_\theta=p_\sigma=p_\delta=p$ y $c^\theta=c^\sigma=c^\delta=c$,
obtenemos el sector geométrico del modelo de FRW (con $V_0=(2\pi)^3$) como era
de esperar.

\section{Representación cuántica}

Para obtener la representación polimérica de este sistema, se implementa para cada dirección
fiducial el procedimiento descrito en la Sección \ref{1sec:kin} \cite{chio}.

Recordemos que el espacio de configuración se describe con holonomías 
\begin{align}
h_i^{\mu_i}(c^i)=e^{\mu_{i}c^{i}\tau_{i}} 
\end{align}
definidas a lo largo de aristas de longitud fiducial
$2\pi\mu_i\in\mathbb{R}$ y orientadas en las direcciones fiduciales, aquí etiquetadas con
$i=\theta,\sigma,\delta$. Los flujos de la tríada densitizada a través de superficies rectangulares
de área fiducial $A_{\square}^i$ ortogonales a la dirección $i$-ésima, dados por
\begin{align}\label{eq:flujo-bianchi}
E(A_{\square}^i,f=1)=\frac{p_{i}}{4\pi^2}A_{\square}^i, 
\end{align}
completan la descripción del espacio de
fases que se representa en la teoría cuántica. El álgebra de configuración es el producto tensorial
de las álgebras de funciones cuasi periódicas de la conexión para cada dirección fiducial.
Consecuentemente, se tiene que
\begin{align}
\text{Cil}_\text{S}=\otimes_i\text{Cil}_\text{S}^i=\text{lin}\{|\mu_\theta,\mu_\sigma,
\mu_\delta\rangle\},
\end{align}
donde los kets $|\mu_i\rangle$ denotan los estados cuánticos correspondientes a los elementos de
matriz de las holonomías $\mathcal N_{\mu_i}(c^i)=e^{\frac{i}{2}\mu_{i}c^{i}}$ en representación
de momentos.
Por tanto, el espacio de Hilbert cinemático es el producto tensorial de los espacios de Hilbert
poliméricos para cada dirección fiducial:
\begin{align}
\mathcal H_{\text{grav}}=\otimes_i\mathcal H_{\text{grav}}^i,
\end{align}
donde $\mathcal H_{\text{grav}}^i$ es la compleción de $\text{Cil}_\text{S}^i$ con respecto al
producto
interno discreto 
\begin{align}
 \langle\mu_i|\mu_i^\prime\rangle=\delta_{\mu_i\mu_i^\prime}.
\end{align}

De nuevo se asumirá que los operadores básicos actúan como la identidad en el sector del que no
dependen. Estos operadores son los operadores $\hat p_i$ asociados a los flujos y los operadores 
$\hat{\mathcal N}_{\mu_i^\prime}$ asociados a las holonomías. Su acción sobre los estados
$|\mu_i\rangle$ de la base de $\mathcal H_{\text{grav}}^i$ es
\begin{subequations}\begin{align}\label{eq:action-p-bianchi}
\hat p_i|\mu_i\rangle&=p_i(\mu_i)|\mu_i\rangle,\qquad p_i(\mu_i)=4\pi\gamma l_\text{Pl}^2\mu_i,\\
\hat{\mathcal N}_{\mu_i^\prime}|\mu_i\rangle&=|\mu_i+\mu_i^\prime\rangle,
\end{align}\end{subequations}
tal que $[\hat{\mathcal N}_{\mu_i},\hat p_j]=i\hbar \widehat{\{\mathcal N_{\mu_i}(c^i),p_j\}}$.

\section{Dinámica mejorada}
\label{4sec:improved}

Recordemos que la denominada prescripción de {dinámica mejorada} está basada en el hecho de
que, debido a la existencia de un salto mínimo de área, $\Delta$, en el espectro del
operador geométrico de área en gravedad cuántica de lazos, las áreas fiduciales no pueden
ser arbitrariamente pequeñas, sino que existe una área fiducial mínima $A_{\square_\text{min}}$,
cuyo valor es tal que el área geométrica correspondiente, dada por $E(A_{\square_\text{min}},f)$,
sea igual a $\Delta$. Como consecuencia, las longitudes fiduciales, como por ejemplo
la longitud de las aristas a lo largo de las cuales se definen las holonomías, también
presentan un valor mínimo, que llamamos $2\pi\bar\mu_i$.

Este valor mínimo no es constante, porque depende de los coeficientes de la tríada densitizada. Los
aquí denominados esquemas A y B se diferencian en cómo incorporan la relación concreta que
existe entre las longitudes mínimas $\bar\mu_i$ y los coeficientes de la tríada $p_i$.

\begin{itemize}
 \item {\bf{Esquema A}}:\\
Este esquema asume que el área mínima $A_{\square_\text{min}}^i$ contenida en el
plano ortogonal a la dirección $i$-ésima, por ejemplo, debe ser un cuadrado fiducial de lados de
longitud fiducial $2\pi\bar\mu_i$.
Por tanto, de acuerdo con esta prescripción, $A_{\square_\text{min}}^i=4\pi^2\bar\mu_i^2$. El área
geométrica de esta área fiducial es $E(A_{\square_\text{min}}^i,f=1)=\bar\mu_i^2p_i^2$, como se
deduce de la ec. \eqref{eq:flujo-bianchi}. La condición de que sea igual a $\Delta$ implica que
\begin{align}\label{eq:mubarraA}
 {\frac1{\bar\mu_i}}=\frac{\sqrt{|p_i|}}{\sqrt{\Delta}}.
\end{align}

Originalmente, se presentaron argumentos heurísticos en favor de este esquema~\cite{chio}, que
apelaban a que en
gravedad cuántica de lazos, los autovalores del espectro del operador área, correspondiente a un
área fiducial $S$, están determinados por los números cuánticos de las holonomías cuyas aristas
asociadas $e$ intersectan a $S$ \cite{area1,area2}. Entonces, si
$E(A_{\square_\text{min}}^i,f=1)=\Delta$, esto implica que existe una holonomía en la dirección
$i$-ésima cuya arista asociada atraviesa a la superficie $A_{\square_\text{min}}^i$ y es de longitud
mínima $2\pi\bar\mu_i$. Además, el área geométrica $E(A_{\square_\text{min}}^i,f=1)$, dada en la ec.
\eqref{eq:flujo-bianchi}, tiene que estar determinada por el número cuántico $\bar\mu_i$ asociado a
esta holonomía. Así que al área fiducial $A_{\square_\text{min}}^i$ se le hace depender del
parámetro $\bar\mu_i$ de la forma más natural posible, es decir, tomando esa área como el cuadrado
de superficie fiducial $4\pi^2\bar\mu_i^2$.

  \item {\bf{Esquema B:}}\\
De acuerdo con este esquema, lo natural es que el área fiducial mínima $A_{\square_\text{min}}^i$,
contenida en el plano%
\footnote{Aquí, y en lo que sigue, siempre que aparezcan los índices $i,j,k$ en la misma
cantidad
o expresión consideraremos que etiquetan direcciones diferentes, es decir, que $\epsilon_{ijk}\neq
0$.}  $j-k$, por ejemplo, sea un rectángulo de lados mínimos $2\pi\bar\mu_j$ y
$2\pi\bar\mu_k$, y no el
cuadrado anterior, por puras razones geométricas \cite{chi2}. También se han dado
argumentos heurísticos que apoyan esta propuesta apelando a la hipotética relación que pueda existir
entre los estados cuánticos de la cosmología cuántica de lazos y los estados cuánticos de la
gravedad cuántica de lazos \cite{awe}. Entonces, de acuerdo con este esquema,
$A_{\square_\text{min}}^i=4\pi^2\bar\mu_j\bar\mu_k$. El área
geométrica de esta área fiducial es $E(A_{\square_\text{min}}^i,f=1)=\bar\mu_j\bar\mu_k^2p_i^2$ y su
igualdad con $\Delta$ implica
\begin{equation}\label{eq:mubarraB}
{\frac1{\bar\mu_i}}=\frac1{\sqrt{\Delta}}\sqrt{\left|\frac{p_j p_k}{p_i}\right|}.
\end{equation}
\end{itemize}

Obviamente, en ambos casos, si identificamos las tres direcciones obtenemos la prescripción del caso
isótropo \eqref{mu}. De hecho, la prescripción A da lugar exactamente a la misma relación que la del
caso isótropo para cada dirección fiducial por separado, de ahí que hayamos indicado anteriormente
que la teoría resultante es factorizable, mientras que la prescripción B mezcla las
distintas direcciones. 
Es más, como luego se verá, el esquema B está determiando unívocamente por la condición de que, para
todas las direcciones, los exponentes $\bar\mu_i c^i$ de los elementos $\mathcal
N_{\bar\mu_i}(c^i)$ tengan un corchete de Poisson constante y fijo (salvo por un signo) con la
variable
\begin{align}\label{v}
 v=\text{sgn}(p_\theta p_\sigma p_\delta)\frac{\sqrt{|p_\theta p_\sigma p_\delta|}}{2\pi\gamma
l_\text{Pl}^2\sqrt{\Delta}},
\end{align}
que, además, coincide con el parámetro $v(p)$ del caso isótropo si
identificamos las tres direcciones fiduciales.

\section{Operador de curvatura}

Su construcción es enteramente análoga a la del operador de curvatura del caso isótropo, descrita en
la Sección \ref{1sec:lig-ham-aps}, ahora teniendo en cuenta la
diferencia entre las tres direcciones fiduciales. Es decir, el operador de curvatura se obtiene de
promover a operador la expresión clásica del tensor de curvatura expresado en términos de un
circuito de holonomías para aristas de longitud mínima,
\begin{align}\label{eq:curvatura-bianchi}
{F}^i_{ab}=-2\sum_{j,k}\;
\text{tr}\left(
\frac{h^{\bar\mu}_{\square_{jk}}-\delta_{jk}}{4\pi^2\bar\mu_j\bar\mu_k}\tau^i\right)\delta^j_a
\delta^k_b,
\end{align}
con
\begin{align}\label{eq:loop-holonomy-bianchi}
 h^{\bar\mu}_{\square_{jk}}=h_j^{\bar\mu_j} h_k^{\bar\mu_k} (h_j^{\bar\mu_j})^{-1}
(h_k^{\bar\mu_k})^{-1}.
\end{align}

Una vez más, la construcción del operador curvatura requiere la definición de los operadores
$\hat{\mathcal N}_{\bar\mu_i}$, los cuales no se pueden escribir directamente en términos de
operadores básicos debido a la dependencia de $\bar\mu_i$ en los coeficientes de la tríada
densitizada. Para definirlos, se sigue una estrategia análoga a la del caso isótropo. Veámoslo para
cada esquema.

\subsection{Operador $\hat{\mathcal N}_{\bar\mu_i}$ y reparametrización afín}

\subsubsection{Esquema A}

Como en este esquema cada longitud mínima $\bar\mu_i$ depende solo del coeficiente de la
tríada en la dirección $i$-ésima, podemos encontrar un parámetro afín $v_i(\mu_i)$ tal que
$\partial_{v_i}=\bar\mu_i[p_i(\mu_i)]\partial_{\mu_i}$, exactamente igual que en el caso isótropo.
Por tanto, obtenemos el mismo resultado, salvo por un factor
$3$ que se debe al hecho de que ${p_i(\mu_i)}/{\mu_i}=3{p(\mu)}/{\mu}$ (compárense
las ecs. \eqref{eq:action-p-frw} y \eqref{eq:action-p-bianchi}).
Es decir, $v_i(p_i)=3^{-1}(2\pi\gamma
l_{\text{Pl}}^2\sqrt{\Delta})^{-1}\text{sgn}(p_i)|p_i|^{3/2}$. Recordemos que $\hat{\mathcal
N}_{\bar\mu_i}$ se define sobre la base de vectores $|v_i\rangle$ del sector $\mathcal
H_{\text{grav}}^i$ como el operador que produce la misma traslación que el campo vectorial
$\bar\mu_i[p_i(\mu_i)]\partial_{\mu_i}$ sobre la variable $v_i$. Por tanto, se tiene
\begin{subequations}\begin{align}
\hat{\mathcal N}_{\pm\bar\mu_i}|v_i\rangle&=|v_i\pm 1\rangle,\\
\hat p_i|v_i\rangle&=3^{2/3}(2\pi\gamma
l_{\text{Pl}}^2\sqrt{\Delta})^{2/3}\,\text{sgn}(v_i)|v_i|^{2/3}|v_i\rangle \label{repA}.
\end{align}\end{subequations}

La gran ventaja de este esquema es que no mezcla las tres direcciones fiduciales. Gracias a
esta propiedad, el correspondiente operador ligadura hamiltoniana preserva la
factorización del espacio de Hilbert cinemático, pues tampoco mezcla las distintas direcciones. Como
veremos, además, esto permite caracterizar completamente la descomposición espectral de dicho
operador, en una manera muy similar a como se ha hecho para el operador ligadura en el caso
isótropo.

\subsubsection{Esquema B}

En este caso, el campo vectorial $\bar\mu_i\partial_{\mu_i}$ depende de los
tres coeficientes de la tríada, del modo siguiente
\begin{align}
 \bar\mu_i\partial_{\mu_i}=4\pi\gamma
l_\text{Pl}^2\sqrt{\Delta}\sqrt{\left|\frac{p_i}{p_jp_k}\right|}\partial_{p_i}.
\end{align}
De nuevo existe la posibilidad de reparametrizar $p_i(\mu_i)$ con un parámetro afín $\lambda_i(p_i)$
tal que este campo vectorial produzca traslaciones en el nuevo parámetro, no dependientes de él pero
dependientes de los parámetros correspondientes a las otras dos direcciones $j$- y $k$-ésima
\cite{awe}. Es decir, este parámetro verifica%
\footnote{Introducimos el factor 2 para seguir el convenio de la Ref. \cite{awe}.}
\begin{align}
 \partial_{\lambda_i}=2\sqrt{|p_i|}\partial_{p_i},
\end{align}
cuya solución es $\lambda_i(p_i)={\text{sgn}(p_i)\sqrt{|p_i|}}/{(4\pi\gamma
l_\text{Pl}^2\sqrt{\Delta})^{1/3}}$. Haciendo este cambio de variable se tiene%
\begin{align}
 \bar\mu_i\partial_{\mu_i}=\frac{1}{2|\lambda_j\lambda_k|}\partial_{\lambda_i}.
\end{align}
En analogía con los casos anteriores, definimos el operador $\hat{\mathcal N}_{\bar\mu_i}$ tal que
su
acción sobre los estados $|\lambda_i\rangle$ sea la misma que la transformación generada por
$\bar\mu_i\partial_{\mu_i}$ sobre el parámetro $\lambda_i$, es decir
\begin{align}\label{eq:N-operator-bianchi} \hat{\mathcal
N}_{\pm\bar\mu_\theta}|\lambda_\theta,\lambda_\sigma,
\lambda_\delta\rangle=\bigg|\lambda_\theta\pm\frac1 {
2|\lambda_\sigma\lambda_\delta|},\lambda_\sigma,\lambda_\delta\bigg\rangle,
\end{align}
y análogamente para $\hat{\mathcal N}_{\pm\bar\mu_\sigma}$ y $\hat{\mathcal N}_{\pm\bar\mu_\delta}$.
Además, invirtiendo el cambio de variable, obtenemos
\begin{align}\label{p-B}
 \hat{p}_i|\lambda_\theta,\lambda_\sigma,\lambda_\delta\rangle=(4\pi\gamma
l_\text{Pl}^2\sqrt{\Delta})^{2/3}\text{sgn}(\lambda_i)\lambda_i^2|\lambda_\theta,\lambda_\sigma.
\lambda_\delta\rangle.
\end{align}

Al igual que ocurre en el caso isótropo y como luego veremos, gracias a nuestra
prescripción de simetrización podremos eliminar el núcleo del operador volumen, que está generado
por los estados con
$\lambda_\theta\lambda_\sigma\lambda_\delta=0$. Por tanto, los operadores $\hat{\mathcal
N}_{\bar\mu_i}$ están bien definidos.

Vemos que su acción es más complicada que en el esquema A, pues ya no producen traslaciones
constantes. Sin embargo, dicha acción se simplifica si introducimos la variable $v$ \cite{awe},
definida en la ec. \eqref{v}, que en términos de las variables $\lambda_i$ se escribe como
\begin{align}
 v=2\lambda_\theta\lambda_\sigma\lambda_\delta.
\end{align}

En efecto, si hacemos el cambio de las variables
$(\lambda_\theta,\lambda_\sigma,\lambda_\delta)$ a las variables
$(v,\lambda_\sigma,\lambda_\delta)$, por ejemplo, la representación cuántica de los operadores
fundamentales
pasa a ser
\begin{subequations}\label{eq:rep-B}\begin{align}\label{eq:rep-N-B}
\hat{\mathcal
N}_{\pm\bar\mu_\theta}\big|v,\lambda_\sigma,\lambda_\delta\big\rangle&=\big|v\pm\text{sgn}
(\lambda_\sigma\lambda_\delta),\lambda_\sigma,\lambda_\delta\big\rangle,\\
\hat p_\theta\big|v,\lambda_\sigma,\lambda_\delta\big\rangle&=(4\pi\gamma
l_{\text{Pl}}^2\sqrt{\Delta})^{\frac2{3}}\text{sgn}
\left(\frac{v}{\lambda_\sigma\lambda_\delta}\right)\frac{v^2}{4\lambda_\sigma^2\lambda_\delta^2}
\big|v,\lambda_\sigma,\lambda_\delta\big\rangle,\\
\hat{\mathcal N}_{\pm\bar\mu_\sigma}\big|v,\lambda_\sigma,\lambda_\delta\big\rangle&=
\bigg|v\pm\text{sgn}(\lambda_\sigma
v),\left(\frac{v\pm\text{sgn}(v\lambda_\sigma)}{v}\right)\lambda_\sigma,
\lambda_\delta\bigg\rangle,\\
\hat p_\sigma\big|v,\lambda_\sigma,\lambda_\delta\big\rangle&=(4\pi\gamma
l_{\text{Pl}}^2\sqrt{\Delta})^{\frac2{3}}\text{sgn}
(\lambda_\sigma)\lambda_\sigma^2\big|v,\lambda_\sigma,\lambda_\delta\big\rangle,\\
\hat{\mathcal N}_{\pm\bar\mu_\delta}\big|v,\lambda_\sigma,\lambda_\delta\big\rangle&=
\bigg|v\pm\text{sgn}(\lambda_\delta v),\lambda_\sigma,
\left(\frac{v\pm\text{sgn}(v\lambda_\delta)}{v}\right)\lambda_\delta\bigg\rangle,\\
\hat p_\delta|v,\lambda_\sigma,\lambda_\delta\rangle&=(4\pi\gamma
l_{\text{Pl}}^2\sqrt{\Delta})^{\frac2{3}}\text{sgn}
(\lambda_\delta)\lambda_\delta^2|v,\lambda_\sigma,\lambda_\delta\rangle \label{eq:rep-p-B}.
\end{align}\end{subequations}
Vemos que la variable $v$ sufre un desplazamiento constante en cada octante de los
definidos por los signos de los coeficientes de la tríada densitizada y este desplazamiento es igual
a $1$ o a $-1$ dependiendo de la orientación de dichos coeficientes. Además, las variables
$\lambda_\sigma$ y $\lambda_\delta$ sufren una ``dilatación'' o ``contracción'' que sólo depende de
su propio signo y de $v$. Como luego veremos, nosotros hemos simetrizado los
signos presentes en el operador ligadura hamiltoniana siguiendo el mismo procedimiento que para el
caso isótropo, con lo que se desacoplan estados con diferentes orientaciones de
los coeficientes de la tríada densitizada. Esto nos ha permitido, en la práctica, 
restringir el estudio al octante $\{v>0,\lambda_\sigma>0,\lambda_\delta>0\}$, lo que simplifica
mucho el análisis, pues los signos presentes en las expresiones anteriores se reducen a la
unidad.

Nótese que $v$ es proporcional al volumen físico, al igual que en el caso isótropo:
\begin{align}
 \hat{V}=\widehat{\sqrt{|p_\theta p_\sigma p_\delta|}}, \qquad 
\hat{V}|v,\lambda_\sigma,\lambda_\delta\rangle=2\pi\gamma
l_{\text{Pl}}^2\sqrt{\Delta}|v||v,\lambda_\sigma,\lambda_\delta\rangle.
\end{align}
Por tanto, como ocurría en el caso isótropo, este esquema lleva de forma natural a trabajar con
los autoestados del volumen, que es la única variable que sufre desplazamientos constantes en esta
representación. Las dos variables restantes miden el grado de anisotropía del sistema.

A diferencia del esquema A, las diferentes direcciones ahora están mezcladas y los operadores
$\hat{\mathcal N}_{\bar\mu_i}$ definidos para distintas direcciones no conmutan. 

\section{Operador ligadura escalar}

Como ocurría en el caso isótropo, para obtener el operador ligadura hamiltoniana no podemos
representar directamente su expresión clásica \eqref{eq:ligBianchi}, sino su expressión en términos
del tensor de curvatura. Para modelos homogéneos y con conexión de espín nula, como es el
modelo de Bianchi I, esta expresión viene dada en la ec. \eqref{eq:lig-escalar-lqg-hom}. Si
sustituimos la expresión de la tríada densitizada \eqref{eq:var-ashtekar-bianchi} y la de la
curvatura \eqref{eq:curvatura-bianchi}, podemos reescribir esta ligadura de Bianchi I de la forma
\begin{align}\label{eq:ligadura-bianchi-generica}
 C_\text{BI}=\frac{2}{\gamma^2}\frac{1}{V}\sum_{i,j,k}\epsilon^{ijk}p_jp_k\frac{\text{tr}
\left(\tau_i h^{\bar\mu}_{\square_{jk}}\right)}{\bar\mu_j\bar\mu_k}.
\end{align}

Cada esquema de dinámica mejorada da lugar a un operador ligadura diferente. Como ya hemos
comentado,
en nuestro análisis ambas construcciones de dicho operador comparten con la
construcción del operador ligadura del caso isótropo una característica fundamental: la prescripción
adoptada al simetrizar el operador, descrita en la Sección \ref{3sec:sym}.
Esta propiedad es la que diferencia nuestras cuantizaciones de las de las Refs.
\cite{chio,awe,luc2}. Gracias a esta prescripción, en ambos esquemas hemos podido desacoplar los
estados de volumen nulo y eliminarlos de la teoría. Asimismo, hemos
podido aplicar nuestro procedimiento de densitización de la ligadura. Además, hemos podido
restringir el estudio al octante de orientación positiva para los tres coeficientes de la tríada,
obteniendo de nuevo, por tanto, una descripción de tipo {ausencia de frontera} en las tres
direcciones
fiduciales.

\subsection{Esquema A}

En este esquema, $\bar\mu_i$ se define según la ec. \eqref{eq:mubarraA}. Si lo sustituimos en
la ec. \eqref{eq:ligadura-bianchi-generica} obtenemos
\begin{align}\label{eq:ligadura-bianchi-generica-A}
 C_\text{BI}^\text{A}=\frac{2}{\gamma^2\Delta}\sum_{i,j,k}\epsilon^{ijk}\frac1{\sqrt{|p_i|}}|p_j|
|p_k|\text{sgn}(p_j)\text{sgn}(p_k)\text{tr}
\left(\tau_i h^{\bar\mu}_{\square_{jk}}\right).
\end{align}
Desarrollamos la traza, promovemos a operador y simetrizamos siguiendo las pautas del procedimiento
expuesto para el caso isótropo en la Sección \ref{3sec:sym}, de modo que obtemos:
\begin{align}\label{eq:ligadura-bianchi-A}
 \widehat{C}_\text{BI}^\text{A}&=-\frac{2}{\gamma^2}
\left\{\widehat{\Lambda}_\theta\widehat{\Lambda}_\sigma\widehat{\left[\frac1{\sqrt{|p_\delta|}}
\right]}+\widehat{\Lambda}_\theta\widehat{\Lambda}_\delta\widehat{\left[\frac1{\sqrt{|p_\sigma|}}
\right]}+\widehat{\Lambda}_\sigma\widehat{\Lambda}_\delta\widehat{\left[\frac1{\sqrt{|p_\theta|}}
\right]}\right\},
\end{align}
donde
\begin{align}\label{Lambdasym}
\widehat{\Lambda}_i&=
\frac{1}{4i\sqrt{\Delta}}\widehat{\sqrt{|p_i|}}\bigg[(\hat{\mathcal
N}_{2\bar\mu_i}-\hat{\mathcal N}_{-2\bar\mu_i})
\widehat{\text{sgn}(p_i)}+\widehat{\text{sgn}(p_i)}
(\hat{\mathcal N}_{2\bar\mu_i}-\hat{\mathcal N}_{-2\bar\mu_i})\bigg]
\widehat{\sqrt{|p_i|}},\\
\widehat{\left[\frac{1}{\sqrt{|p_i|}}\right]}&=\frac1{4\pi\gamma
l_{\text{Pl}}^2\sqrt{\Delta}}\widehat{\text{sgn}(p_i)}\widehat{\sqrt{|p_i|}}\left(\hat{\mathcal
N}_{-\bar\mu_i}\widehat{\sqrt{|p_i|}}\hat{\mathcal
N}_{\bar\mu_i}-\hat{\mathcal
N}_{\bar\mu_i}\widehat{\sqrt{|p_i|}}\hat{\mathcal
N}_{-\bar\mu_i}\right)\label{invA}.
\end{align}
Este último operador, regularización del inverso de $\sqrt{|p_i|}$, se construye según la
descripción de la Subsección \ref{1subsec:mat}. Su acción en la base de estados
$|v_i\rangle$ es diagonal, con autovalores iguales a $b(v_i)/3$, donde $b(v)$ es la función
definida en la ec.~\eqref{eq:triad-operator}. A partir de este operador, se construye el operador
inverso del volumen:
\begin{align}
\widehat{\left[\frac{1}{V}\right]}&=
\otimes_i \widehat{\left[\frac{1}{\sqrt{|p_i|}}\right]}.
\end{align}

Vemos que el operador \eqref{eq:ligadura-bianchi-A} es una suma de productos de operadores
simétricos direccionales, cada uno de ellos definido en uno de los sectores ${\cal
H}^i_{\text{grav}}$ del
espacio de Hilbert cinemático total. Como entre ellos conmutan (recordemos que actúan como la
identidad en los sectores del espacio de Hilbert correspondiente a otras direcciones), el operador
ligadura es efectivamente simétrico, definido en ${\cal H}_{\text{grav}}$.

\subsection{Esquema B}

En el esquema B, $\bar\mu_i$ se define según la ec. \eqref{eq:mubarraB}. Si lo sustituimos en
la ec. \eqref{eq:ligadura-bianchi-generica} obtenemos
\begin{align}\label{eq:ligadura-bianchi-B}
 C_\text{BI}^\text{B}=\frac{2}{\gamma^2\Delta}\frac{1}{V}V^2\sum_{i,j,k}\epsilon^{ijk}\text{sgn}
(p_j)\text{sgn}(p_k)\text{tr}\left(\tau_i h^{\bar\mu}_{\square_{jk}}\right).
\end{align}
En esta expresión no hemos simplicado $V^2$ con $1/V$ para hacer explícito
que, al representar la ligadura, dichos factores se van a representar por sus operadores
correspondientes. Esta
ambigüedad en el orden de factores es irrelevante en los regímenes de interés físico, como ya
comentamos para el caso isótropo, ya que típicamente el rebote
cuántico ocurre para valores de $v$ lo suficientemente grandes como para que la diferencia entre el
espectro del operador inverso de volumen y su valor clásico sea despreciable. La razón por la que
hemos tomado este orden es que, de este modo, nuestro operador ligadura densitizada tiene, en lo que
a las potencias de $V$ se refiere, el mismo orden de factores que el operador ligadura de la
Ref. \cite{awe} y así es más fácil la comparación entre ambos trabajos. La diferencia
verdaderamente relevante la marcará el tratamiento dado al signo de $p_i$.

A partir de la expresión \eqref{eq:ligadura-bianchi-B}, desarrollamos la traza, promovemos a
operador y simetrizamos siguiendo las mismas pautas que en los casos anteriores, de
modo que obtemos el siguiente operador definido en ${\cal H}_{\text{grav}}$ con dominio denso
$\text{Cil}_\text{S}$:
\begin{subequations}\label{eq:op-lig-ham-bianchiB}\begin{align}\label{eq:op-lig-ham-bianchiB-a}
\widehat{C}_{\text{BI}}^\text{B}&=\widehat{\left[\frac{1}{V}\right]}^{1/2}\left[\widehat{\mathcal
C}^{(\theta)}+\widehat{\mathcal C}^{(\sigma)}+\widehat{\mathcal
C}^{(\delta)}\right]\widehat{\left[\frac{1}{V}\right]}^{1/2},\\
\widehat{\mathcal C}^{(i)}&=-\frac1{4\gamma^2\Delta}\widehat{\sqrt{V}}[\hat F_j\hat{V}\hat F_k+\hat
F_k\hat{V}\hat F_j]\widehat{\sqrt{V}},\label{Ci}\\
\hat F_i&=\frac{\hat{\mathcal N}_{2\bar\mu_i}-\hat{\mathcal N}_{-2\bar\mu_i}}{2i}
\widehat{\text{sgn}(p_i)}+\widehat{\text{sgn}(p_i)}\frac{\hat{\mathcal N}_{2\bar\mu_i}-\hat{\mathcal
N}_{-2\bar\mu_i}}{2i} \label{oper-F}.
\end{align}\end{subequations}

El operador inverso del volumen $\widehat{\left[{1}/{V}\right]}$ es también sensible al esquema
elegido. 
Para definirlo, partimos de nuevo de la identidad clásica \eqref{eq:identidad-inv-vol} con $p$
reemplazado por $3p_i$
(el factor $3$ da cuenta de las diferencias en los corchetes de Poisson)
y tomamos $l$ igual a $\bar\mu_i$, dado en la ec. \eqref{eq:mubarraB}. Con
respecto al valor de la constante $a>0$, esta vez la elección $a=1/2$, hecha en el caso isótropo
y en el esquema A, solo daría una representación no trivial del operador identidad, y no de una
potencia inversa de $|p_i|$, por lo que debemos elegir otro valor. Por simplicidad, hemos tomado
$a=1/4$. En
definitiva, obtenemos
\begin{align}\label{invB-def}
\widehat{\left[\frac1{|p_\theta|^{\frac1{4}}}\right]}&=\frac{\widehat{\text{sgn}(p_\theta)}}
{2\pi\gamma l_{\text{Pl}}^2\sqrt{\Delta}}\widehat{\sqrt{|p_\sigma p_\delta|}}\big[\hat{\mathcal
N}_{-\bar\mu_\theta}\widehat{|p_\theta|}^{\frac1{4}}\hat{\mathcal
N}_{\bar\mu_\theta}-\hat{\mathcal
N}_{\bar\mu_\theta}\widehat{|p_\theta|}^{\frac1{4}}\hat{\mathcal N}_{-\bar\mu_\theta}\big],
\end{align}
y análogamente para $\widehat{[1/|p_\sigma|^{1/4}]}$ y $\widehat{[1/|p_\delta|^{1/4}]}$. A partir
de la representación de los operadores involucrados, dada en las
ecs.~\eqref{eq:rep-B}, se comprueba que su acción sobre los estados de la base 
de ${\cal H}_{\text{grav}}$ es diagonal y dada por
\begin{align}\label{invB}
 \widehat{\left[\frac1{|p_i|^{\frac1{4}}}\right]}&|v,\lambda_\sigma,\lambda_\delta\rangle=
\frac{
b^\star_i(v,\lambda_\sigma,\lambda_\delta)}{(4\pi\gamma
l_{\text{Pl}}^2\sqrt{\Delta})^{\frac1{6}}}|v,\lambda_\sigma, \lambda_\delta\rangle ,
\end{align}
donde
\begin{subequations}\label{b-inv-vol-B}\begin{align}\label{btheta}
b^\star_\theta&(v,\lambda_\sigma,\lambda_\delta)=
\sqrt{2|\lambda_\sigma\lambda_\delta|}\left|\sqrt{|v+1|}-\sqrt{|v-1|}\right|,\\
b^\star_a&(v,\lambda_\sigma,\lambda_\delta)=
\sqrt{\left|\frac{v}{\lambda_a}\right|}\left|\sqrt{|v+1|}-\sqrt{|v-1|}\right|, \quad
a=\sigma,\delta.
\end{align}\end{subequations}
Por tanto, hemos representado el inverso del volumen como el operador regularizado
\begin{align}
\widehat{\left[\frac{1}{V}\right]}&=
\otimes_i \widehat{\left[\frac{1}{|p_i|^{\frac1{4}}}\right]}^2.
\end{align}

Vemos que el operador \eqref{eq:op-lig-ham-bianchiB-a} contiene más términos que su análogo en el
caso A porque ahora los operadores no diagonales correspondientes a diferentes
direcciones ya no conmutan. En particular, no es difícil comprobar que la acción de los operadores 
$\hat{F}_i$ viene dada por:
\begin{subequations}\label{eq:F-oper}\begin{align}\label{eq:Ftheta}
\hat{F}_\theta\big|v,\lambda_\sigma,\lambda_\delta\big\rangle&=\frac{\text{sgn}
(\lambda_\sigma\lambda_\delta)}{2i}\bigg\{\big[\text{sgn}(v-2\text{sgn}
(\lambda_\sigma\lambda_\delta))+\text{sgn}(v)\big]\big|v-2\text{sgn}
(\lambda_\sigma\lambda_\delta),\lambda_\sigma,\lambda_\delta\big\rangle\nonumber\\
&-\big[\text{sgn}(v)+\text{sgn}(v+2\text{sgn}(\lambda_\sigma\lambda_\delta))\big]\big|v+2\text{sgn}
(\lambda_\sigma\lambda_\delta),\lambda_\sigma,\lambda_\delta\big\rangle\bigg\},
\end{align}
\begin{align}\label{eq:Fsigma}
\hat{F}_\sigma|v,\lambda_\sigma,\lambda_\delta\rangle&=\frac{\text{sgn}
(\lambda_\sigma)}{2i}\nonumber\\
&\times\bigg\{\big[1+\text{sgn}(|v|-2\text{sgn}
(\lambda_\sigma))\big]\bigg|v-2\text{sgn}
(v\lambda_\sigma),\frac{v-2\text{sgn}(v\lambda_\sigma)}{v}\lambda_\sigma,
\lambda_\delta\bigg\rangle\nonumber\\
&-\big[1+\text{sgn}(|v|+2\text{sgn}
(\lambda_\sigma))\big]\bigg|v+2\text{sgn}(v\lambda_\sigma),\frac{v+2\text{sgn}
(v\lambda_\sigma)}{v}\lambda_\sigma,\lambda_\delta\bigg\rangle\bigg\}.
\end{align}\end{subequations}
La acción de $\hat{F}_\delta$ es análoga a la de $\hat{F}_\sigma$, intercambiando
$\lambda_\sigma$ por $\lambda_\delta$. Por tanto, los distintos operadores $\hat F_i$ no conmutan
entre sí.

\section{Densitización del operador ligadura escalar}
\label{4sec:dens}

Gracias a la prescripción de simetrización aplicada y, en particular, al reparto de potencias de
$p_i$ a la derecha
y a la izquierda de los operadores involucrados en las expresiones de
$\widehat{C}_{\text{BI}}$, tanto en el caso A como en el B, así como al comportamiento de las
funciones
$b(v_i)$ en el esquema A, y $b^\star(v_i)$, en el caso B, que se anulan en cero, el operador
ligadura escalar obtenido en ambos esquemas aniquila a los estados de volumen nulo,
que son aquéllos que están en el núcleo de alguno de los operadores $\hat{p}_i$, y deja invariante
su complemento ortogonal $\widetilde{\mathcal H}_\text{grav}$, igual que ocurría en nuestro
tratamiento del modelo isótropo. Por tanto, de nuevo, los estados de volumen nulo se desacoplan y
podemos restringir el estudio a $\widetilde{\mathcal H}_\text{grav}=\otimes_i\widetilde{\mathcal
H}_\text{grav}^i$. Como hemos visto, esta restricción equivale a resolver cuánticamente, en el
espacio de Hilbert cinemático, la singularidad clásica.

Sea
\begin{align}\label{dual}
 \widetilde{\text{Cil}}_\text{S}=\otimes_i\widetilde{\text{Cil}}_\text{S}{}^i&=\text{lin}\{|v_\theta
,v_\sigma,v_\delta\rangle;\;v_\theta v_\sigma v_\delta\neq0\}\nonumber\\
&=\text{lin}\{|v,\lambda_\sigma,\lambda_\delta\rangle;\;v\lambda_\sigma\lambda_\delta\neq0\}
\end{align}
el dominio de definición de $\widehat C_\text{BI}$, tanto en el caso A como en el B, cuya compleción
de Cauchy con respecto al producto interno discreto para cada dirección
es el espacio de Hilbert $\widetilde{\mathcal H}_\text{grav}$. En la última línea de la ec.
\eqref{dual}, hemos reflejado de forma explícita que se excluyen los valores nulos de
$\lambda_\sigma$ y $\lambda_\delta$, aunque esto esté implícito en el hecho de que
$v=2\lambda_\theta\lambda_\sigma\lambda_\delta$ debe ser distinto de cero.

Una vez que hemos desacoplado los estados de volumen nulo, que están también en el núcleo del
operador inverso de volumen, hemos densitizado la ligadura $(\tilde\psi|\widehat
C_\text{BI}^\dagger=0$ siguiendo exactamente el mismo procedimiento que el descrito en la
Sección \ref{3sec:dens}, es decir, mediante la biyección
\eqref{eq:mapeo-dens}, donde ahora $(\tilde\psi|$ y $(\psi|$ pertenencen a
$\widetilde{\text{Cil}}_\text{S}{}^{^*}$. Los estados transformados $(\psi|$ son entonces solución
de
la ligadura densitizada $(\psi|\widehat {\mathcal C}_\text{BI}^\dagger=0$, con
\begin{align}\label{eq:dens-bianchi}
\widehat{\mathcal C}_\text{BI}=\widehat{\left[\frac1{V}\right]}^{-1/{2}}
\widehat{C}_\text{BI}\widehat{\left[\frac{1}{V}\right]}^{-1/{2}}.
\end{align}

Pasamos a analizar, en cada esquema, las propiedades del operador resultante.

\section[Operador ligadura densitizada]{Operador ligadura densitizada}

\subsection{Esquema A}
\label{3sec:op-dens-lig-A}

En el esquema A,  la expresión explícita del operador densitizado $\widehat {\mathcal
C}_\text{BI}^\text{A}$ es
\begin{align}\label{eq:lig-dens-bian-A}
\widehat{\mathcal C}_\text{BI}^\text{A}&=-\frac{2}{\gamma^2}\big(\widehat{\Omega}_\theta
\widehat{\Omega}_\sigma+\widehat{\Omega}_\theta\widehat{\Omega}_\delta+
\widehat{\Omega}_\sigma\widehat{\Omega}_\delta\big).
\end{align}
Vemos, por tanto, que solo depende de un operador direccional $\widehat{\Omega}_i$, que es el
operador simétrico
\begin{align}
\widehat{\Omega}_i&= \widehat{\left[\frac{1}{\sqrt{|p_i|}}\right]}^{-\frac1
{2}}\widehat{\Lambda}_i\widehat{\left[\frac{1}{\sqrt{|p_i|}}\right]}^{-\frac1
{2}},
\end{align}
explícitamente dado por
\begin{align}\label{omega-op}
\widehat{\Omega}_i&=\frac{1}{4i\sqrt{\Delta}}\widehat{\left[\frac{1}{\sqrt{|p_i|}}\right]}^{-\frac1
{2}}\widehat{\sqrt{ |p_i|}}\bigg[(\hat{\mathcal N}_{2\bar\mu_i}-\hat{\mathcal N}_{-2\bar\mu_i})
\widehat{\text{sgn}(p_i)}+\widehat{\text{sgn}(p_i)}
(\hat{\mathcal N}_{2\bar\mu_i}-\hat{\mathcal N}_{-2\bar\mu_i})\bigg]\nonumber\\
&\times
\widehat{\sqrt{|p_i|}}\widehat{\left[\frac{1}{\sqrt{|p_i|}}\right]}^{-\frac1{2}}.
\end{align}
Comparando con la ec. \eqref{eq:operador-grav-mmo} comprobamos que, en efecto, como ya habíamos
anticipado, el operador $\widehat{\Omega}$ del modelo de FRW es de la misma forma que el operador
análogo $\widehat{\Omega}_i$ del modelo de Bianchi I, de modo que, si en el modelo de Bianchi I
identificamos las tres direcciones espaciales mediante las igualdades
$\hat{p}_i|v_i\rangle=3\hat{p}|v\rangle$ y $\hat{\mathcal N}_{\bar\mu_i}|v_i\rangle=\hat{\mathcal
N}_{\bar\mu}|v\rangle$, obtenemos el sector geométrico de nuestro modelo de FRW analizado en el
Capítulo \ref{chap:revFRW}.

La gran ventaja de esta cuantización es que las propiedades del operador
ligadura escalar $\widehat{\mathcal C}_\text{BI}^\text{A}$ se derivan directamente de las del
operador direccional
$\widehat{\Omega}_i$, que es observable de Dirac para todo $i=\theta,\sigma,\delta$, y cuyo
cuadrado ya ha sido completamente analizado en el contexto del modelo de FRW. Por ello, las
propiedades ya demostradas para $\widehat{\Omega}^2$ en la Sección \ref{3sec:grav} nos servirán para
caracterizar totalmente el operador $\widehat{\Omega}_i$.
A continuación, analizamos sus propiedades.

\subsubsection{Sectores de superselección}

La acción del operador $\widehat{\Omega}_i$ sobre los vectores $|v_i\rangle$ es
\begin{equation}\label{eq:omega-op}
\widehat{\Omega}_i|v_i\rangle=-i3
\big[f_+(v_i)|v_i+2\rangle-f_-(v_i)|v_i-2\rangle\big],
\end{equation}
donde las funciones $f_\pm(v)$ son las dadas en la ec. \eqref{eq:f}.

En completa analogía con lo que ocurría en el caso isótropo, este operador deja invariantes
subsectores del espacio direccional $\widetilde{\mathcal H}_\text{grav}^i$ cuyos estados tienen
soporte en redes discretas, esta vez de paso 2, pues el operador es un operador en diferencias de
ese paso. Además, de nuevo, gracias a las propiedades de las funciones $f_\pm(v_i)$,
$\widehat{\Omega}_i$ no mezcla estados correspondientes a orientaciones opuestas del coeficiente
$i$-ésimo de la tríada densitizada. Por tanto, dadas las semirredes de paso 2
\begin{align}
\mathcal L_{\varepsilon_i}^\pm=\{\pm(\varepsilon_i+2k),k\in\mathbb{N}\},\qquad
\varepsilon_i\in(0,2],
\end{align}
los subespacios de Hilbert $\mathcal H_{\varepsilon_i}^{\pm}$, definidos como la compleción de
Cauchy de los espacios
\begin{equation}\label{cyl-epsilon}
\text{Cil}_{\varepsilon_i}^{\pm}=\text{lin}\{|v_i\rangle;
v_i\in\mathcal L_{\varepsilon_i}^\pm\}
\end{equation}
con respecto al producto interno discreto,
proporcionan sectores de superselección, a los que se puede restringir el estudio. Nosotros
hemos restringido el estudio al espacio \linebreak $\mathcal H_{\vec\varepsilon}^+=\otimes_i\mathcal
H_{\varepsilon_i}^+$, con $\vec\varepsilon=(\varepsilon_1,\varepsilon_2,
\varepsilon_3)$ arbitrario pero fijo. Equivalentemente, podríamos haberlo restringido a cualquier
otro sector, correspondiente a otro octante (definido por otras orientaciones de las componentes de
la tríada densitizada), ya que el
operador ligadura es simétrico bajo cambio de signo de $v_i$, por la identidad
$f_\pm(-v_i)=-f_\mp(v_i)$, como se deduce de la ec. \eqref{eq:f}.

\subsubsection{Análisis espectral y autofunciones generalizadas}

El operador  $\widehat{\Omega}_i^2$ de nuestro modelo de Bianchi I coincide esencialmente con el
operador $\widehat{\Omega}^2$ de nuestro modelo de FRW y, por tanto, ya conocemos sus propiedades
espectrales. Por una parte, $\widehat\Omega^2_i$, definido en el espacio de Hilbert
${}^{(4)}\mathcal H^+_{\tilde\varepsilon_i}$, cuya base de
vectores $|v_i\rangle$ tiene soporte en la semirred de paso cuatro ${}^{(4)}{\mathcal
L}^+_{\tilde\varepsilon_i}$ introducida en la ec. \eqref{eq:redes-4}%
\footnote{Se ha añadido el superíndice 4 a estos objetos definidos en el Capítulo
\ref{chap:revFRW}
para no confundirlos con los de paso 2 definidos en este capítulo.}, es un operador positivo y
esencialmente autoadjunto con espectro absolutamente continuo y no degenerado dado por
$\mathbb{R}^+$. Por otra parte, los coeficientes
$e^{\tilde\varepsilon_i}_{\lambda_i}(\tilde\varepsilon_i+4n)$ de la función de onda de sus
autoestados generalizados $|{}^{(4)}e^{\tilde\varepsilon_i}_{\lambda_i}\rangle$, correspondientes al
autovalor $\lambda_i\in[0,\infty)$, están explícitamente determinados por el dato inicial
$e^{\tilde\varepsilon_i}_{\lambda_i}(\tilde\varepsilon_i)$, mediante la
relación \eqref{eq:eigenstates-omega2} con $\lambda_i=9\lambda$ (de nuevo, el rescalado numérico
aparece por las diferencias en los corchetes de Poisson).
Escogiendo dicho dato inicial positivo y normalizado tal que $\langle
{}^{(4)}e^{\tilde\varepsilon_i}_{\lambda_i}|{}^{(4)}e^{\tilde\varepsilon_i}_{\lambda'_i}
\rangle=\delta(\lambda_i-\lambda'_i)$, estos autoestados proporcionan la resolución espectral de la
identidad en ${}^{(4)}\mathcal H^+_{\tilde\varepsilon_i}$:
\begin{align}\label{eq:spectral-ident-4}
\mathbb{I}=\int_{\mathbb{R^+}} d\lambda_i
|{^{(4)}e_{\lambda_i}^{\tilde\varepsilon_i}}\rangle
\langle{^{(4)}e_{\lambda_i}^{\tilde\varepsilon_i}}|.
\end{align}
Conocidas estas propiedades del operador $\widehat{\Omega}_i^2$ hemos podido determinar
también las del operador $\widehat{\Omega}_i$. 

En primer lugar, es inmediato percatarse de que el operador 
$\widehat{\Omega}_i$ no deja invariantes los sectores de superselección ${}^{(4)}\mathcal
H^+_{\tilde\varepsilon_i}$ asociados a $\widehat{\Omega}_i^2$, sino la suma directa de dos de
ellos%
\footnote{Nótese que $\tilde\varepsilon_i\in(0,4]$ mientras que $\varepsilon_i\in(0,2]$.}:
\begin{align}\label{eq:hilbert-suma}
 \mathcal
H_{\varepsilon_i}^{+}={}^{(4)}\mathcal
H^+_{\tilde\varepsilon_i=\varepsilon_i}\oplus{}^{(4)}\mathcal
H^+_{\tilde\varepsilon_i=\varepsilon_i+2},
\end{align}
ya que la semirred de paso dos ${\mathcal L}^+_{\varepsilon_i}$ es la unión de las semirredes de
paso cuatro ${}^{(4)}{\mathcal L}^+_{\tilde\varepsilon_i=\varepsilon_i}$ y ${}^{(4)}{\mathcal
L}^+_{\tilde\varepsilon_i=\varepsilon_i+2}$. Definiendo ambos operadores $\widehat{\Omega}_i$ y
$\widehat{\Omega}_i^2$ en el mismo espacio de Hilbert \eqref{eq:hilbert-suma}, hemos
demostrado que $\widehat{\Omega}_i$ es esencialmente autoadjunto en $\mathcal
H_{\varepsilon_i}^{+}$. En efecto, supongamos que no lo fuera. En virtud del teorema
\ref{th:esa}, eso implicaría que existe alguna
solución no trivial a su ecuación de índices de defecto normalizable en $\mathcal
H_{\varepsilon_i}^{+}$. Dicha solución se podría descomponer en suma directa de dos componentes,
una normalizable en ${}^{(4)}\mathcal H^+_{\tilde\varepsilon_i=\varepsilon_i}$ y la otra en
${}^{(4)}\mathcal H^+_{\tilde\varepsilon_i=\varepsilon_i+2}$, y cada una por separado proporcionaría
una solución a la ecuación de índices de defecto de $\widehat{\Omega}_i^2$ en estos últimos espacios
de Hilbert. Esto significaría que $\widehat\Omega_i^2$ no es esencialmente autoadjunto, en
contradicción con lo que ya sabemos. Por tanto, el operador $\widehat\Omega_i$ es
esencialmente autoadjunto en $\mathcal H^+_{\varepsilon_i}$, como hemos anticipado.

Por otra parte, dados dos autoestados de $\widehat{\Omega}_i^2$,
$|{}^{(4)}e^{\varepsilon_i}_{\lambda_i}\rangle$ y $|{}^{(4)}e^{\varepsilon_i+2}_{\lambda_i}\rangle$,
uno normalizable (en sentido generalizado) en ${}^{(4)}\mathcal
H^+_{\tilde\varepsilon_i=\varepsilon_i}$ y el otro en ${}^{(4)}\mathcal
H^+_{\tilde\varepsilon_i=\varepsilon_i+2}$, y correspondientes al autovalor
$\lambda_i\in\mathbb{R}^+$, el estado
\begin{align}\label{eq:descomp-lineal}
 |e^{\varepsilon_i}_{\pm|\omega_i|}\rangle=|X_{\pm|\omega_i|}|\big\{|{^{(4)}e_{\lambda_i}^{
\varepsilon_i}}\rangle\mp i|Y_{\pm|\omega_i|}||{^{(4)}e_{\lambda_i}^{
\varepsilon_i+2}}\rangle\big\}
\end{align}
es por construcción autoestado generalizado de $\widehat{\Omega}_i$ en $\mathcal
H^+_{\varepsilon_i}$ con autovalor $\pm|\omega_i|=\pm\sqrt{\lambda_i}$, como se deduce de la
ecuación de autovalores de $\widehat{\Omega}_i$. Como consecuencia, el espectro de
$\widehat{\Omega}_i$ resulta ser
absolutamente continuo e igual a la recta real. Hemos determinado los coeficientes de normalización
$|X_{\pm|\omega_i|}|$ e $|Y_{\pm|\omega_i|}|$ tales que la resolución espectral de la identidad en
$\mathcal H^+_{\varepsilon_i}$ se escribe
\begin{align}\label{eq:spectral-ident-2}
\mathbb{I}=\int_{\mathbb{R}} d\omega_i|e_{\omega_i}^{\varepsilon_i}\rangle
\langle e_{\omega_i}^{\varepsilon_i}|.
\end{align}
En vista de las ecs. \eqref{eq:spectral-ident-4} y \eqref{eq:hilbert-suma}, ésta también se puede
escribir de la forma
\begin{align}\label{eq:spectral-ident-2bis}
\mathbb{I}
=\int_{\mathbb{R^+}} d\lambda_i
|{^{(4)}e_{\lambda_i}^{\varepsilon_i}}\rangle
\langle{^{(4)}e_{\lambda_i}^{\varepsilon_i}}|+\int_{\mathbb{R^+}} d\lambda_i
|{^{(4)}e_{\lambda_i}^{\varepsilon_i+2}}\rangle
\langle{^{(4)}e_{\lambda_i}^{\varepsilon_i+2}}|.
\end{align}
Para hacer que ambas expresiones sean compatibles se debe verificar que
\begin{align}
 |Y_{-|\omega_i|}|=|Y_{+|\omega_i|}|=1, \qquad
|X_{+|\omega_i|}|^2+|X_{-|\omega_i|}|^2=2|\omega_i|.
\end{align}
Además, la resolución espectral de la identidad en cada espacio de Hilbert involucrado
equivale a que la normalización de los estados sea
\begin{align}
 \langle
e_{\omega_i}^{\varepsilon_i}|e_{\omega'_i}^{\varepsilon_i}\rangle=\delta(\omega_i-\omega'_i),
\qquad \langle
{}^{(4)}e^{\tilde\varepsilon_i}_{\lambda_i}|{}^{(4)}e^{\tilde\varepsilon_i}_{\lambda'_i}
\rangle=\delta(\lambda_i-\lambda'_i).
\end{align}
Teniendo esto en cuenta se obtiene la condición
\begin{align}
 |X_{\omega_i}||X_{\omega'_i}|(1+|Y_{\omega_i}||Y_{\omega'_i}|)=2|\omega_i|.
\end{align}
Concluimos, por tanto, que la relación precisa entre la base ortonormal (en sentido generalizado)
de $\mathcal H^+_{\varepsilon_i}$ y las bases de ${}^{(4)}\mathcal
H^+_{\tilde\varepsilon_i=\varepsilon_i}$ y ${}^{(4)}\mathcal
H^+_{\tilde\varepsilon_i=\varepsilon_i+2}$ es
\begin{align}\label{eq:estados-2}
|e_{\omega_i}^{\varepsilon_i}\rangle&=\sqrt{|\omega_i|}\left[
|{^{(4)}e_{\omega_i^2}^{\varepsilon_i}}\rangle-
i\text{sgn}(\omega_i)|{^{(4)}e_{\omega_i^2}^{2+\varepsilon_i}}\rangle
\right],
\end{align}
para $\omega_i\neq0$. Para  $\omega_i=0$ hemos definido
\begin{align}
|e_{0}^{\varepsilon_i}\rangle=|{^{(4)}e_{0}^{\varepsilon_i}}\rangle,
\end{align}
elección que no es relevante, en tanto en cuanto  $\omega_i=0$ es un punto de medida nula en el
espectro. 

Nótese que, para $\omega_i\neq 0$, las proyecciones de $|e_{\omega_i}^{\varepsilon_i}\rangle$ en los
subespacios de Hilbert ${^{(4)}{\mathcal H}_{\varepsilon_i}^{+}}$ y ${^{(4)}{\mathcal
H}_{\varepsilon_i+2}^{+}}$ tienen un desfase relativo de $\pm\pi/2$. Como consecuencia, la fase de
los coeficientes de la función de onda, $e_{\omega_i}^{\varepsilon_i}(v_i)= \langle
v_i|e^{\varepsilon_i}_{\omega_i}\rangle$, oscila rápidamente cuando $v_i$ varía en la semirred
${\mathcal L}_{\varepsilon_i}^+$, siendo constante en cada red de paso cuatro ${}^{(4)}{\mathcal
L}^+_{\varepsilon_i}$ y ${}^{(4)}{\mathcal
L}^+_{\varepsilon_i+2}$ por separado.

Recordemos que conocemos explícitamente la forma de las autofunciones
${}^{(4)}e^{\tilde\varepsilon_i}_{\lambda_i}(v_i)$ del operador $\widehat{\Omega}_i^2$, dada en la
ec. \eqref{eq:eigenstates-omega2} con $\lambda=\lambda_i/9$. Haciendo uso de esta expresión y de la
relación \eqref{eq:estados-2}, o bien resolviendo directamente la ecuación de autovalores de
$\widehat{\Omega}_i$, comprobamos que cada autoestado generalizado
\begin{align}
|e^{\varepsilon_i}_{\omega_i}\rangle=\sum_{v_i\in\mathcal
L^+_{\varepsilon_i}}e^{\varepsilon_i}_{\omega_i}(\varepsilon_i)|v_i\rangle
\end{align}
está determinado por el dato inicial $e^{\varepsilon_i}_{\omega_i}
(\varepsilon_i)$ de su función de onda de acuerdo con la expresión ($n\in\mathbb{N}^+$)
\begin{subequations}\label{eq:eigenstates-bianchi-A}\begin{align}
e^{\varepsilon_i}_{\omega_i}(\varepsilon_i+2n)&=\sum_{O(0\rightarrow n)}
\left[\prod_{\{r_p\}} F(\varepsilon_i+2r_p+2)\prod_{\{s_q\}}
G_{\omega_i}(\varepsilon_i+2s_q)\right]e^{\varepsilon_i}_{\omega_i}
(\varepsilon_i),
\end{align}
\begin{align}\label{def-F-G}
 F(v_i)=\frac{f_-(v_i)}{f_+(v_i)},\qquad G_{\omega_i}(v_i)=\frac{-i\omega_i}{3f_+(v_i)}.
\end{align}\end{subequations}
La definición del conjunto $O(0\rightarrow n)$ y la de los subconjuntos $\{r_p\}$ y $\{s_q\}$ es la
explicada bajo la ec. \eqref{S}. La norma de $e^{\varepsilon_i}_{\omega_i}
(\varepsilon_i)$ queda fijada por la condición de normalización $\langle
e_{\omega_i}^{\varepsilon_i}|e_{\omega'_i}^{\varepsilon_i}\rangle=\delta(\omega_i-\omega'_i)$ y su
fase se elige de modo que dicho dato inicial sea positivo.

\subsection{Esquema B}

La aplicación de la densitización \eqref{eq:dens-bianchi} al operador
$\widehat{C}_{\text{BI}}^\text{B}$ da como resultado que nuestro operador ligadura densitizada en el
esquema B es
\begin{align}\label{densCB}
\widehat{\mathcal C}_{\text{BI}}^\text{B}&=\widehat{\mathcal C}^{(\theta)}+\widehat{\mathcal
C}^{(\sigma)}+\widehat{\mathcal C}^{(\delta)},
\end{align}
con las definiciones de la ec. \eqref{eq:op-lig-ham-bianchiB}.
Este operador ligadura coincide con el de la Ref.~\cite{awe} salvo por
el tratamiento de los signos de $p_i$ en el orden simétrico adoptado.

Al contrario de lo que pasa en el esquema A, ahora el operador de ligadura densitizada no está
formado por una suma de observables de Dirac, pues los operadores $\widehat{\mathcal C}^{(i)}$
no conmutan entre ellos. Además, los operadores involucrados son tan complicados que no hemos sido
capaces de determinar las propiedades espectrales del operador $\widehat{\mathcal
C}_{\text{BI}}^\text{B}$, a diferencia del esquema A. No obstante, este operador también
superselecciona el espacio de Hilbert cinemático en diferentes sectores separables, que presentan
la misma descripción de tipo {ausencia de frontera}, muy importante a la hora de caracterizar
el espacio de Hilbert físico.

\subsubsection{Sectores de superselección}

Los operadores $\hat{F}_i$, cuya actuación sobre los
vectores de la base está dada en la ec. \eqref{eq:F-oper}, no mezclan estados
que correspondan a orientaciones contrarias de alguno de los coeficientes de la tríada densitizada,
es decir, estados $|v,\lambda_\sigma,\lambda_\delta\rangle$ con signo opuesto de alguno de sus
números cuánticos. Por tanto, el operador ligadura $\widehat{\mathcal{C}}_{\text{BI}}^\text{B}$
deja invariantes todos los octantes en el espacio tridimensional definido por $v$, $\lambda_\sigma$
y $\lambda_\delta$. Así pues, podemos restringir el estudio a cualquiera de ellos. Una vez más,
nosotros nos hemos restringido al subespacio de coeficientes de la tríada densitizada positivos,
dado por
\begin{align}\label{cylpositive}
{\text{Cil}}_\text{S}^+&=\text{lin}\{|v,\lambda_\sigma,\lambda_\delta\rangle;\,
v,\lambda_\sigma,\lambda_\delta>0\}.
\end{align}
Conviene hacer notar que el operador ligadura de la Ref. \cite{awe} no deja invariantes estos
octantes. No obstante, apelando a
la simetría bajo paridad, también allí se restringe el estudio al espacio ${\text{Cil}}_\text{S}^+$.

La acción de nuestro operador $\widehat{\mathcal{C}}_{\text{BI}}^\text{B}$ sobre los estados de
${\text{Cil}}_\text{S}^+$ resulta ser:
\begin{align}\label{eq:actionCB}
\widehat{\mathcal{C}}_{\text{BI}}^\text{B}|v,\lambda_\sigma,\lambda_\delta\rangle&=\frac{(\pi
l_{\text{Pl}}^2)^2}{4}\big[x_-(v)|v-4,\lambda_\sigma,\lambda_\delta\rangle_-
-x^-_0(v)|v,\lambda_\sigma,\lambda_\delta\rangle_{0^-}\nonumber\\&-
x^+_0(v)|v,\lambda_\sigma,\lambda_\delta\rangle_{0^+}+x_+(v)|v+4,\lambda_\sigma,
\lambda_\delta\rangle_+\big],
\end{align}
donde se han introducido los coeficientes
\begin{subequations}\label{coeficientes}\begin{align}
 x_-(v)&=2\sqrt{v}(v-2)\sqrt{v-4}[1+\text{sgn}(v-4)], \label{coefficient1}\qquad
&x_+(v)=x_-(v+4),\\
x^-_0(v)&=2(v-2)v[1+\text{sgn}(v-2)], \label{coefficient3}\qquad &x^+_0(v)=x^-_0(v+2),
\end{align}\end{subequations}
y las siguientes combinaciones lineales de estados
\begin{subequations}\label{eq:combi}\begin{align}\label{eq:combi1}
|v\pm4,\lambda_\sigma,\lambda_\delta\rangle_\pm&=\bigg|v\pm4,\lambda_\sigma,\frac{v\pm4}
{v\pm2}\lambda_\delta\bigg\rangle+\bigg|v\pm4,\frac{v\pm4}
{v\pm2}\lambda_\sigma,\lambda_\delta\bigg\rangle\nonumber\\&+\bigg|v\pm4,\frac{v\pm2}
{v}\lambda_\sigma,\lambda_\delta\bigg\rangle+\bigg|v\pm4,\lambda_\sigma,
\frac{v\pm2} {v}\lambda_\delta\bigg\rangle\nonumber\\&+\bigg|v\pm4,\frac{v\pm2}
{v}\lambda_\sigma,\frac{v\pm4} {v\pm2}\lambda_\delta\bigg\rangle+\bigg|v\pm4,\frac{v\pm4}
{v\pm2}\lambda_\sigma,\frac{v\pm2}{v}\lambda_\delta\bigg\rangle,
\end{align}
\begin{align}\label{eq:combi2}
|v,\lambda_\sigma,\lambda_\delta\rangle_{0^\pm}&=\bigg|v,\lambda_\sigma,\frac{v}
{v\pm2}\lambda_\delta\bigg\rangle+\bigg|v,\frac{v}
{v\pm2}\lambda_\sigma,\lambda_\delta\bigg\rangle+\bigg|v,\lambda_\sigma,\frac{v\pm2}
{v}\lambda_\delta\bigg\rangle\nonumber\\&+\bigg|v,\frac{v\pm2}
{v}\lambda_\sigma,\lambda_\delta\bigg\rangle+\bigg|v,\frac{v\pm2}
{v}\lambda_\sigma,\frac{v}
{v\pm2}\lambda_\delta\bigg\rangle+\bigg|v,\frac{v}
{v\pm2}\lambda_\sigma,\frac{v\pm2}
{v}\lambda_\delta\bigg\rangle.
\end{align}\end{subequations}
Nótese que, en efecto, el operador $\widehat{\mathcal{C}}_{\text{BI}}^\text{B}$ está bien definido
en ${\text{Cil}}_\text{S}^+$, ya que, $x_-(v)=0$ si $v\leq4$, y $x^-_0(v)=0$ si $v\leq2$.

El análisis de la acción del operador $\widehat{\mathcal{C}}_{\text{BI}}^\text{B}$ sobre un
estado genérico $|v,\lambda_\sigma^\star,\lambda_\delta^\star\rangle$ muestra lo siguiente:
\begin{itemize}
 \item[i)] En lo que concierne a la variable $v$, ésta sufre un desplazamiento constante
igual a $4$ ó $-4$, el último solo en caso de que $v>4$. Por tanto,
$\widehat{\mathcal{C}}_{\text{BI}}^\text{B}$ conserva los subespacios de estados cuyo número
cuántico $v$ pertenece a alguna de las semirredes de paso cuatro
\begin{equation}
\mathcal
L_{\tilde\varepsilon}^+=\{\tilde\varepsilon+4k,k=0,1,2...\},\qquad\tilde\varepsilon\in(0,4],
\end{equation}
al igual que ocurre en el caso isótropo.
\item[ii)] En lo concerniente a las variables de anisotropía, $\lambda_\sigma$ y $\lambda_\delta$,
se comprueba que los estados relacionados con
$|v,\lambda_\sigma^\star,\lambda_\delta^\star\rangle$
mediante la acción de $\widehat{\mathcal{C}}_{\text{BI}}^\text{B}$ tienen los siguientes números
cuánticos $(\lambda_a,\lambda_b)$, donde indistintamente se puede hacer la identificación
$(a,b)=(\sigma,\delta)$ o la identificación $(a,b)=(\delta,\sigma)$:
\begin{itemize}
 \item Sección $v-4>0$:
\begin{align*}
 \left(\lambda_a^\star,\frac{v-4}{v-2}\lambda_b^\star\right),\quad
\left(\lambda_a^\star,\frac{v-2}{v}\lambda_b^\star\right),\quad
\left(\frac{v-2}{v}\lambda_a^\star,\frac{v-4}{v-2}\lambda_b^\star\right),
\end{align*}
\item Sección $v$:
\begin{align*}
 \left(\lambda_a^\star,\frac{v}{v+2}\lambda_b^\star\right), \quad
\left(\lambda_a^\star,\frac{v+2}{v}\lambda_b^\star\right), \quad
 \left(\frac{v+2}{v}\lambda_a^\star,\frac{v}{v+2}\lambda_b^\star\right),
\end{align*}
y, si $v>2$, también 
\begin{align*}
\left(\lambda_a^\star,\frac{v-2}{v}\lambda_b^\star\right),\quad
\left(\lambda_a^\star,\frac{v}{v-2}\lambda_b^\star\right),\quad
\left(\frac{v-2}{v}\lambda_a^\star,\frac{v}{v-2}\lambda_b^\star\right),
\end{align*}
 \item Sección $v+4$:
\begin{align*}
 \left(\lambda_a^\star,\frac{v+4}{v+2}\lambda_b^\star\right), \quad
\left(\lambda_a^\star,\frac{v+2}{v}\lambda_b^\star\right), \quad
\left(\frac{v+2}{v}\lambda_a^\star,\frac{v+4}{v+2}\lambda_b^\star\right).
\end{align*}
\end{itemize}
\end{itemize}
Vemos que el efecto causado en las variables de anisotropía no depende de los números cuánticos de
referencia $\lambda_\sigma^\star$ y $\lambda_\delta^\star$, sino que solo depende de
$v=\tilde\varepsilon+4k$. Además, esta dependencia tiene lugar mediante fracciones cuyo denomidador
es dos unidades mayor o menor que el denominador. 

Hemos determinado que, en cada sección de
$v\in\mathcal L_{\tilde\varepsilon}^+$ constante, solo un cierto subespacio de estados están
conectados con $|v,\lambda_\sigma^\star,\lambda_\delta^\star\rangle$ mediante la actuación
reiterada de la
ligadura. Además, hemos demostrado que dichos subespacios son los formados por los estados cuyos
números cuánticos
$\lambda_\sigma$ y $\lambda_\delta$ son de la forma
$\lambda_a=\omega_{\tilde\varepsilon}\lambda_a^\star$, con $\omega_{\tilde\varepsilon}$
perteneciente al
conjunto
\begin{equation}\label{W-set}
\mathcal
W_{\tilde\varepsilon}=\left\{\left(\frac{\tilde\varepsilon-2}{\tilde\varepsilon}\right)^z\prod_{m,
n\in\mathbb{N}}\left(\frac{\tilde\varepsilon+2m}{\tilde\varepsilon+2n}\right)^{k_n^m};\quad
k_n^m\in\mathbb{N},\;\; z\in\mathbb{Z}\text{ si } \tilde\varepsilon>2,\;\;z=0\text{ si }
\tilde\varepsilon<2\right\}.
\end{equation}

Probemos que, en efecto, los valores que toman las variables
$\lambda_a$ son los mismos en todas las secciones de $v$ constante. 
\begin{itemize}
 \item
Para demostrar este resultado,
basta con comprobar que un estado cualquiera $|v,\lambda_\sigma,\lambda_\delta\rangle$ está
conectado mediante la actuación de la ligadura con los estados
$|v\pm4,\lambda_\sigma,\lambda_\delta\rangle$. Tras una actuación de
$\widehat{\mathcal{C}}_{\text{BI}}^\text{B}$ esto no se cumple, como se observa de las relaciones
anteriores dadas en el punto ii). Sin embargo, vemos que el operador relaciona, por ejemplo,
\begin{align*}
\big|v,\lambda_\sigma,\lambda_\delta\big\rangle&\quad\text{con}\quad\bigg|v+4,\lambda_\sigma,
\frac{v+4}{v+2}\lambda_\delta\bigg\rangle,\\
\big|v,\lambda_\sigma,\lambda_\delta\big\rangle&\quad\text{con}\quad\bigg|v,\lambda_\sigma,
\frac{v-2}{v}\lambda_\delta\bigg\rangle, \quad \text{si }v>2.
\end{align*}
Entonces, el operador ligadura relaciona
\begin{align*}
\bigg|v+4,\lambda_\sigma,
\frac{v+4}{v+2}\lambda_\delta\bigg\rangle\quad\text{con}\quad\bigg|v+4,\lambda_\sigma,
\frac{v+2}{v+4}\frac{v+4}{v+2}\lambda_\delta\bigg\rangle=
\big|v+4,\lambda_\sigma,\lambda_\delta\big\rangle,
\end{align*}
en virtud de la segunda conexión entre estados. La conexión con
$|v-4,\lambda_\sigma,\lambda_\delta\rangle$ se demuestra de igual modo. Por tanto, tras dos
actuaciones de $\widehat{\mathcal{C}}_{\text{BI}}^\text{B}$ el estado
$|v,\lambda_\sigma,\lambda_\delta\rangle$ queda conectado con los estados
$|v\pm4,\lambda_\sigma,\lambda_\delta\rangle$, como queríamos demostrar.
\end{itemize}

Probemos también que el recorrido de valores de $\lambda_\sigma/\lambda_\sigma^\star$ es
exactamente el mismo que el recorrido de valores de $\lambda_\delta/\lambda_\delta^\star$ e igual a
todos los elementos del conjunto $\mathcal W_{\tilde\varepsilon}$. 

\begin{itemize}
 \item
Demostrar que, en cada sección de $v$ constante, los valores de una de las variables $\lambda_a$ sea
tal que el cociente $\lambda_a/\lambda_a^\star$ tome el
valor de todos los puntos de $\mathcal W_{\tilde\varepsilon}$ es obvio, en tanto en cuanto las
relaciones del punto ii), especificadas ahí para un valor de referencia $\lambda_a^\star$ son
válidas para cualquier valor que tome $\lambda_a$ mediante la acción reiterada de la ligadura.
Como es de esperar, también se puede demostrar que los valores que toma el cociente
$\lambda_\sigma/\lambda_\sigma^\star$ son los mismos que los que toma
$\lambda_\delta/\lambda_\delta^\star$
aunque, si examinamos los pares de puntos $(\lambda_a,\lambda_b)$ dados anteriormente, observemos
que
tras una actuación del operador ligadura nunca se alcanzan estados que tengan
$\lambda_\sigma/\lambda_\sigma^\star=\lambda_\delta/\lambda_\delta^\star$. En efecto, se puede
comprobar que tales estados están conectados si consideramos de nuevo actuaciones repetidas de
$\widehat{\mathcal{C}}_{\text{G}}^\text{B}$. Lo ilustraremos con un ejemplo concreto: tras
una actuación el operador ligadura relaciona
\begin{align*}
\big|v,\lambda_\sigma^\star,\lambda_\delta^\star\big\rangle&\quad\text{con}\quad\bigg|v,
\lambda_\sigma^\star,\frac{v+2}{v}\lambda_\delta^\star\bigg\rangle,\\
\big|v,\lambda_\sigma^\star,\lambda_\delta^\star\big\rangle&\quad\text{con}\quad\bigg|v,\frac{v+2}
{ v
}\lambda_\sigma^\star,\lambda_\delta^\star\bigg\rangle.
\end{align*}
Entonces, el operador ligadura relaciona
\begin{align*}
\bigg|v,\frac{v+2}{v}
\lambda_\sigma^\star,\lambda_\delta^\star\bigg\rangle\quad\text{con}\quad\bigg|v,\frac{v+2}{v}
\lambda_\sigma^\star,\frac{v+2}{v}\lambda_\delta^\star\bigg\rangle,
\end{align*}
en virtud de la primera conexión entre estados.
Por tanto, $|v,\lambda_\sigma^\star,\lambda_\delta^\star\rangle$ está conectado, por ejemplo,
con el estado $|v,\lambda_\sigma,\lambda_\delta\rangle$, tal que
\begin{align}
 \frac{\lambda_\sigma}{\lambda_\sigma^\star}=\frac{\lambda_\delta}{\lambda_\delta^\star}=
\frac{v+2}{v}.
\end{align}
Análogamente, podemos demostrar lo mismo para cualesquiera valores en el conjunto $\mathcal
W_{\tilde\varepsilon}$ de estos cocientes.
\end{itemize}

Nótese que el conjunto discreto $\mathcal W_{\tilde\varepsilon}$ es infinito y numerable. Además,
este conjunto es denso en la semirrecta real positiva. La demostración es muy simple:
\begin{itemize}
 \item Definamos los conjuntos de números reales
\begin{align}
U_{\tilde\varepsilon}&=\left\{\frac{\tilde\varepsilon+4m}{\tilde\varepsilon+4n};\,m,n\in\mathbb{N}
\right\}
\subset\mathcal W_{\tilde\varepsilon},\qquad
V_{\tilde\varepsilon}=\left\{\frac{\tilde\varepsilon}{4}+n;\, n\in\mathbb{N}\right\}.
\end{align}
Sean $c$ y $d$ dos números reales positivos cualesquiera, tales que $d-c>0$. Entonces existe
$s=(\tilde\varepsilon/4+n)\in V_{\tilde\varepsilon}$ tal que $1<s(d-c)$ o, equivalentemente,
\begin{align}\label{eq:a}
 sc+1<sd.
\end{align}
Llamemos $t=\tilde\varepsilon/4+m$ el mayor número del subconjunto
$V_{\tilde\varepsilon}$ que es más pequeño o igual que $sc+1$, esto es,
\begin{equation}\label{eq:b}
 sc< \frac{\tilde\varepsilon}{4}+m \leq sc+1.
\end{equation}
Las expresiones anteriores \eqref{eq:a} y \eqref{eq:b} implican que $sc<\tilde\varepsilon/4+m<sd$,
que a su vez se puede reescribir como
\begin{align}
 c<\frac{\tilde\varepsilon+4m}{\tilde\varepsilon+4n}<d.
\end{align}
En conclusión, dados dos números positivos $c$ y $d$ con $c<d$, siempre existe un número
$r\in U_{\tilde\varepsilon}$ tal que $c<r<d$. Por tanto, $U_{\tilde\varepsilon}\subset\mathcal
W_{\tilde\varepsilon}$ es denso en los números positivos y, en consecuencia, el subespacio mayor
$\mathcal W_{\tilde\varepsilon}$ también lo es, como queríamos demostrar.
\end{itemize}

Así pues, mientras que la variable $v$ tiene soporte en semirredes sencillas de paso constante, por
el contrario las variables $\lambda_a$ toman valores pertenecientes a conjuntos complicados, pero
también proporcionan sectores de superselección separables, hecho que no fue advertido en la
Ref.~\cite{awe}. Como un ejemplo particular, vemos que, si tanto $\tilde\varepsilon$ como
$\lambda_a^\star$ son enteros, entonces $\lambda_a$ toma valores en los números racionales
positivos.

En conclusión, el operador $\widehat{\mathcal{C}}_{\text{BI}}^\text{B}$ conserva los subespacios de
Hilbert $\mathcal H^+_{\tilde\varepsilon,\lambda_\sigma^\star,\lambda_\delta^\star}$, definidos
como
la compleción de Cauchy de los subespacios 
\begin{align}\label{cyl-lambda}
\text{Cil}^+_{\tilde\varepsilon,\lambda_\sigma^\star,\lambda_\delta^\star}=\text{lin}\{
&|v,\lambda_\sigma,\lambda_\delta\rangle;\;
v\in\mathcal L_{\tilde\varepsilon}^+,\;\lambda_a=\omega_{\tilde\varepsilon}\lambda_a^\star,\;
\omega_{\tilde\varepsilon}\in\mathcal W_{\tilde\varepsilon},\;\lambda_a^\star\in\mathbb{R}^+\},
\end{align}
con respecto al producto interno discreto
\begin{align}
\langle
v,\lambda_\sigma,\lambda_\delta|v',\lambda'_\sigma,\lambda'_\delta\rangle=\delta_{vv'}\delta_{
\lambda_\sigma \lambda'_{\sigma}}\delta_{\lambda_\delta \lambda'_{\delta}}.
\end{align}
Por tanto, los espacios de Hilbert separables
$\mathcal H^+_{\tilde\varepsilon,\lambda_\sigma^\star,\lambda_\delta^\star}$ están
superseleccionados y podemos restringir el estudio a cualquiera de ellos.

\section{Espacio de Hilbert físico}

\subsection{Esquema A}
\label{sec:group-ave}

En el esquema A tenemos un conocimiento completo de las propiedades del operador ligadura, en
particular, de su descomposición espectral. Esto nos ha permitido determinar explícitamente los
estados físicos. A continuación, explicamos cómo se obtienen siguiendo el procedimiento de promedio 
sobre grupos, en completa analogía con lo expuesto en la Seccción \ref{3sec:phys}.

Sea $|\phi\rangle$ un elemento del dominio denso de la extensión autoadjunta 
$\widehat{\bar{\cal{C}}}{}_{\text{BI}}^\text{A}$ del operador ligadura en el sector de
superselección ${\mathcal H}_{\vec\varepsilon}^+$ con función de onda $\phi(\vec{v})$, donde 
$\vec{v}=(v_\theta,v_\sigma,v_\delta)$. Si aplicamos la descomposición espectral asociada a los
operadores $\widehat{\Omega}_i$ para las tres direcciones espaciales, podemos expresar la función de
onda $\phi(\vec{v})$ en términos de las autofunciones generalizadas
$e_{\omega_i}^{\varepsilon_i}(v_i)$ de dichos operadores:
\begin{equation}\label{eq:fdec}
\phi(\vec{v}) = \int_{\mathbb{R}^3} d\vec{\omega} \,
\tilde{\phi}(\vec{\omega})
e^{\varepsilon_\theta}_{\omega_\theta}(v_\theta)
e^{\varepsilon_\sigma}_{\omega_\sigma} (v_\sigma)
e^{\varepsilon_\delta}_{\omega_\delta}(v_\delta) ,
\end{equation}
donde $\vec{\omega}=(\omega_\theta,\omega_\sigma,\omega_\delta)$, $\tilde{\phi}(\vec\omega)\in
L^2(\mathbb{R}^3,d\vec{\omega})$ y $v_i\in\mathcal L_{\varepsilon_i}^+$.

A este elemento cinemático $|\phi\rangle$ le aplicamos el promedio $\mathcal P$ del grupo
uniparamétrico generado por el operador autoadjunto
$\widehat{\bar{\cal{C}}}{}_{\text{BI}}^\text{A}$ para proyectarlo sobre el espacio de
soluciones físicas. El resultado es
\begin{align}\label{eq:gave-bianchi}
\Phi(\vec{v})=[{\cal P} \phi](\vec{v})&=\int_{\mathbb{R}} d t\,
e^{it\frac{\gamma^2}{2}\widehat{\bar{\cal C}}{}_{\text{BI}}^\text{A}}
\phi(\vec{v})\nonumber\\&= \int_{\mathbb{R}^3}d\vec{\omega}
\delta(\omega_\theta\omega_\sigma+\omega_\theta\omega_\delta+
\omega_\sigma\omega_\delta)\tilde{\phi}(\vec{\omega})e^{\varepsilon_\theta}_{\omega_\theta}
(v_\theta)
e^{\varepsilon_\sigma}_{\omega_\sigma}(v_\sigma)
e^{\varepsilon_\delta}_{\omega_\delta}(v_\delta).
\end{align}

Por tanto, solo contribuyen al espacio de Hilbert físico productos de estados
$|e^{\varepsilon_i}_{\omega_i}\rangle$ tales que se verifique
\begin{align}
 \sum_{i}\frac1{\omega_i}=0.
\end{align}
Conviene entonces eliminar una de las variables $\omega_i$ en función de las otras dos. Nosotros
hemos elegido $\omega_\theta$ de modo que
\begin{equation}\label{eq:om-func}
\omega_\theta(\omega_\sigma,\omega_\delta)=
-\frac{\omega_\sigma\omega_\delta}{\omega_\sigma+\omega_\delta}.
\end{equation}
Con esta elección, la función de onda de la proyección del estado cinemático $|\phi\rangle$
en el espacio de Hilbert físico se escribe
\begin{align}\label{sol-fis-A}
\Phi(\vec{v}) &= \int_{\mathbb{R}^2} {d\omega_\sigma
d\omega_\delta}\, \tilde\Phi(\omega_\sigma,\omega_\delta)\,
e^{\varepsilon_\theta}_{\omega_\theta(\omega_\sigma,\omega_\delta)}(v_\theta)
e^{\varepsilon_\sigma}_{\omega_\sigma} (v_\sigma)
e^{\varepsilon_\delta}_{\omega_\delta}(v_\delta),
\end{align}
donde hemos definido
$\tilde\Phi(\omega_\sigma,\omega_\delta)=\tilde{\phi}[\omega_\theta(\omega_\sigma,
\omega_\delta),
\omega_\sigma, \omega_\delta]\cdot{|\omega_\sigma+\omega_\delta|^{-1}}$,
absorbiendo el factor $|\omega_\sigma+\omega_\delta|$ por conveniencia.
Recordamos que las autofunciones $e^{\varepsilon_i}_{\omega_i}(v_i)$ que aparecen en la
expresión anterior están dadas explícitamente en la ec. \eqref{eq:eigenstates-bianchi-A}.

A su vez, podemos calcular el producto interno físico entre dos estados $|\Phi_1\rangle$ y
$|\Phi_2\rangle$, de acuerdo con la expresión
\begin{align}
\langle \Phi_1|\Phi_2\rangle_\text{fis} &= \langle
{\cal P}\phi_1|\phi_2\rangle_\text{cin}
= \int_{\mathbb{R}^2} {d\omega_\sigma
d\omega_\delta}{|\omega_\sigma+\omega_\delta|}\,
\tilde\Phi^{*}_1(\omega_\sigma,\omega_\delta)
\tilde\Phi_2(\omega_\sigma,\omega_\delta).
\end{align}
Por tanto, el espacio de Hilbert físico que se obtiene al cuantizar este modelo de Bianchi~2
I con en
el esquema A de la dinámica mejorada es el espacio de funciones en $\mathbb{R}^2$ de cuadrado
integrable con medida de integración $|\omega_\sigma+\omega_\delta|d\omega_\sigma d\omega_\delta$,
es decir,
\begin{align}\label{eq:phys-A}
\mathcal H_\text{fis}^{\vec\varepsilon}= L^2\left(\mathbb{R}^2,
|\omega_\sigma+\omega_\delta|{d\omega_\sigma d\omega_\delta}\right).
\end{align}

Como se comentó en el Capítulo \ref{chap:revFRW}, una manera alternativa de obtener el espacio de
Hilbert físico consiste en buscar el espacio de soluciones a la ligadura y un conjunto completo de
observables (reales) e imponer la condición de que estos observables sean representados por
operadores autoadjuntos con el fin
de determinar una estructura de Hilbert en el espacio de soluciones. Una vez que conocemos la
descomposición espectral de los operadores $\widehat{\Omega}_i$, podemos resolver la ligadura
mediante una expansión formal (en el dual del dominio del operador ligadura) de los estados en
términos de los autoestados generalizados $|e_{\omega_i}^{\varepsilon_i}\rangle$, ya que
$\widehat{{\cal C}}_\text{BI}^\text{A}$ actúa de forma diagonal sobre ellos. De hecho, si
representamos el elemento al cual le imponemos la ligadura por la función de onda
$\phi'(\omega_\theta,\omega_\sigma,\omega_\delta)$, entonces las soluciones físicas están descritas
por funciones de la forma
$\Phi'(\omega_\sigma,\omega_\delta)=\phi'[\omega_\theta(\omega_\sigma,\omega_\delta),\omega_\sigma,
\omega_\delta]$, donde
$\omega_\theta(\omega_\sigma,\omega_\delta)$ es la dada en la ec.~\eqref{eq:om-func}. Un conjunto
completo de observables lo forman, por ejemplo, los operadores $\widehat{\Omega}_\sigma$ y
$\widehat{\Omega}_\delta$, que multiplican la función de onda física por $\omega_\sigma$ y
$\omega_\delta$, respectivamente, y por los operadores de derivación $-i\partial_{\omega_\sigma}$ y
$-i\partial_{\omega_\delta}$. Imponiendo que sean autoadjuntos, llegamos al espacio de Hilbert
$L^2(\mathbb{R}^2,d\omega_\sigma d\omega_\delta)$. Si se divide la función de onda resultante
por el factor $\sqrt{|\omega_\sigma+\omega_\delta|}$, se obtiene, de hecho, una representación
unitariamente equivalente del álgebra de observables en el espacio de Hilbert dado en la
ec.~\eqref{eq:phys-A}.

\subsection{Esquema B}
\label{phys-b}

En este caso, no hemos sido capaces de encontrar una base del espacio de Hilbert $\mathcal
H^+_{\tilde\varepsilon,\lambda_\sigma^\star,\lambda_\delta^\star}$ que diagonalice al
operador ligadura, $\widehat{\mathcal C}_\text{BI}^\text{B}$. Más aún, desconocemos sus propiedades
espectrales. Por tanto, el procedimiento de promedio sobre grupos no es útil en esta ocasión,
y hemos tenido que imponer directamente la ligadura y hacer un análisis de sus soluciones.

Sea $(\psi|$ un elemento del espacio dual
$(\text{Cil}^{+}_{\tilde\varepsilon,\lambda_\sigma^\star,\lambda_\delta^\star})^*$, cuya expansión
formal en la
base de estados $|v,\lambda_\sigma,\lambda_\delta\rangle$ viene dada por
\begin{align}
(\psi|&=\sum_{v\in\mathcal
L_{\tilde\varepsilon}^+}\sum_{\omega_{\tilde\varepsilon}\in\mathcal
W_{\tilde\varepsilon}}\sum_{\bar\omega_{\tilde\varepsilon}\in\mathcal W_{\tilde\varepsilon}}
\psi(v,\omega_{\tilde\varepsilon}\lambda_\sigma^\star,\bar\omega_{\tilde\varepsilon}
\lambda_\delta^\star)
\big\langle
v,\omega_{\tilde\varepsilon}\lambda_\sigma^\star,\bar\omega_{\tilde\varepsilon}
\lambda_\delta^\star\big|.
\end{align}
A partir de la acción del operador $\widehat{\mathcal C}_\text{BI}^\text{B}$, dada en la ec.
\eqref{eq:actionCB} deducimos que la ligadura $\big(\psi\big|\widehat{\mathcal
C}_\text{BI}^\text{B}{}^\dagger=0$ da lugar a la siguiente relación de recurrencia:
\begin{align}\label{eq:solution-bianchi-B}
\psi_+(v+4,\lambda_\sigma,\lambda_\delta)=\frac1{x_+(v)}&
\bigg[{x^-_0(v)}\psi_{0^-}(v,\lambda_\sigma,\lambda_\delta)+{x^+_0(v)}\psi_{0^+}(v,
\lambda_\sigma,\lambda_\delta)\nonumber\\&-{x_-(v)}\psi_-(v-4,\lambda_\sigma,\lambda_\delta)\bigg].
\end{align}
En esta expresión, a fin de abreviar su escritura, hemos introducido las proyecciones de
$(\psi|$ sobre las combinaciones lineales de estados definidas en la ec.
\eqref{eq:combi}, es decir,
\begin{align}
 \psi_\pm(v\pm4,\lambda_\sigma,\lambda_\delta)=(\psi|v\pm4,\lambda_\sigma,\lambda_\delta\rangle_\pm,
\qquad
\psi_{0^\pm}(v,\lambda_\sigma,\lambda_\delta)=(\psi|v,\lambda_\sigma,\lambda_\delta\rangle_{0^\pm
}.
\end{align}
Recordamos que cada una de estas combinaciones está formada por seis términos.

Debido a que $x_-(\tilde\varepsilon)=0$, la anterior relación de recurrencia, que es de orden 2 en
lo que a la variable $v$ se refiere, da una relación de primer orden entre doce datos en la sección
de $v$ inicial, $v=\tilde\varepsilon$, (seis si $\tilde\varepsilon\leq2$) y seis datos en la sección
de $v=\tilde\varepsilon+4$:
\begin{align}\label{eq:condicion-bianchi-B}
\psi_+(\tilde\varepsilon+4,\lambda_\sigma,\lambda_\delta)=\frac1{x_+(\tilde\varepsilon)}&
\bigg[{x^-_0(\tilde\varepsilon)}\psi_{0^-}(\tilde\varepsilon,\lambda_\sigma,\lambda_\delta)+{
x^+_0(\tilde\varepsilon)}\psi_{0^+}(\tilde\varepsilon,\lambda_\sigma,\lambda_\delta)\bigg].
\end{align}
Por tanto, conocidos todos los datos en la sección inicial $v=\tilde\varepsilon$ obtenemos, en la
sección siguiente, todas las combinaciones de seis términos dadas por
\begin{align}\label{eq:combinations-v4}
\psi_+(\tilde\varepsilon+4,\lambda_\sigma,\lambda_\delta)
&=\psi\bigg(\tilde\varepsilon+4,\lambda_\sigma,\frac{\tilde\varepsilon+4}{\tilde\varepsilon+2}
\lambda_\delta\bigg)
+\psi\bigg(\tilde\varepsilon+4,\frac{\tilde\varepsilon+4}{\tilde\varepsilon+2}\lambda_\sigma,
\lambda_\delta\bigg)
\nonumber\\&
+\psi\bigg(\tilde\varepsilon+4,\lambda_\sigma,
\frac{\tilde\varepsilon+2}{\tilde\varepsilon}\lambda_\delta\bigg)
+\psi\bigg(\tilde\varepsilon+4,\frac{\tilde\varepsilon+2}{\tilde\varepsilon}\lambda_\sigma,
\lambda_\delta\bigg)
\nonumber\\&
+\psi\bigg(\tilde\varepsilon+4,\frac{\tilde\varepsilon+2}{\tilde\varepsilon}\lambda_\sigma,
\frac{\tilde\varepsilon+4}{\tilde\varepsilon+2} \lambda_\delta\bigg)
+\psi\bigg(\tilde\varepsilon+4,\frac{\tilde\varepsilon+4}{\tilde\varepsilon+2}\lambda_\sigma,
\frac{\tilde\varepsilon+2}{\tilde\varepsilon} \lambda_\delta\bigg).
\end{align}

Puede argumentarse que formalmente esta relación se puede invertir, no solo para
$v=\tilde\varepsilon$ sino para todo $v\in\mathcal L_{\tilde\varepsilon}^+$, es decir, que el
conocimiento de todas las combinaciones
\begin{align*}
 \{\psi_+(v+4,\lambda_\sigma,\lambda_\delta);\;\lambda_\sigma=\omega_{\tilde\varepsilon}
\lambda_\sigma^\star,\;\lambda_\delta=\bar\omega_{\tilde\varepsilon}
\lambda_\delta^\star,\;\omega_{\tilde\varepsilon},\bar\omega_{\tilde\varepsilon}\in\mathcal
W_{\tilde\varepsilon}\}
\end{align*}
determina unívocamente todos los términos
\begin{align*}
 \{\psi(v+4,\lambda_\sigma,\lambda_\delta);\;\lambda_\sigma=\omega_{\tilde\varepsilon}
\lambda_\sigma^\star,\;\lambda_\delta=\bar\omega_{\tilde\varepsilon}
\lambda_\delta^\star,\;\omega_{\tilde\varepsilon},\bar\omega_{\tilde\varepsilon}\in\mathcal
W_{\tilde\varepsilon}\},
\end{align*}
por separado, para cualquier $v$. Para verlo, es conveniente reescribir las
combinaciones $\psi_+(v+4,\lambda_\sigma,\lambda_\delta)$ como
\begin{align}
 \psi_+(v+4,\lambda_\sigma,\lambda_\delta)=\widehat{U}_6(v+4)
\psi(v+4,\lambda_\sigma,\lambda_\delta),
\end{align}
donde $\widehat{U}_6(v+4)$ es el operador con dominio
\begin{align}
 \text{Cil}_{\lambda_\sigma^\star,\lambda_\delta^\star}&=\text{Cil}_{
\lambda_\sigma^\star}\otimes\text{Cil}_{
\lambda_\delta^\star}=\text{lin}
\{|\lambda_\sigma,\lambda_\delta\rangle;
\;\lambda_\sigma=\omega_{\tilde\varepsilon}
\lambda_\sigma^\star,\;\lambda_\delta=\bar\omega_{\tilde\varepsilon}
\lambda_\delta^\star,\;\omega_{\tilde\varepsilon},\bar\omega_{\tilde\varepsilon}\in\mathcal
W_{\tilde\varepsilon}\}
\end{align}
cuya actuación sobre los estados $|\lambda_\sigma,\lambda_\delta\rangle$ es
\begin{align}
\widehat{U}_6(v+4)|\lambda_\sigma,\lambda_\delta\rangle&=
\bigg|\frac{v}{v+2}\lambda_\sigma,\lambda_\delta\bigg\rangle+\bigg|\lambda_\sigma,
\frac{v}{v+2}\lambda_\delta\bigg\rangle+\bigg|\frac{v+2}{v+4}\lambda_\sigma,
\lambda_\delta\bigg\rangle+\bigg|\lambda_\sigma,
\frac{v+2}{v+4}\lambda_\delta\bigg\rangle\nonumber\\&
+\bigg|\frac{v}{v+2}\lambda_\sigma,\frac{v+2}{v+4}\lambda_\delta\bigg\rangle+\bigg|\frac{v+2}{v+4}
\lambda_\sigma,\frac{v}{v+2}\lambda_\delta\bigg\rangle.
\end{align}
Nótese que el espacio de Hilbert cinemático se factoriza del modo $\mathcal
H^+_{\tilde\varepsilon,\lambda_\sigma^\star,\lambda_\delta^\star}= \mathcal
H^+_{\tilde\varepsilon}\otimes\mathcal H_{\lambda_\sigma^\star,\lambda_\delta^\star}$, siendo 
$\mathcal H_{\lambda_\sigma^\star,\lambda_\delta^\star}$
la complección de $\text{Cil}_{\lambda_\sigma^\star,\lambda_\delta^\star}$ en el producto interno
discreto. 

Basta pues con ver si $\widehat{U}_6(v)$ es invertible.

\subsubsection{Caracterización del operador $\widehat{U}_6(v)$}

Introduzcamos las variables
$x_a=\ln(\lambda_a)=\ln(\lambda_a^\star)+\varpi_{\tilde\varepsilon}$. Nótese que
$\varpi_{\tilde\varepsilon}$ toma valores en un conjunto denso de la recta real, formado por el
logaritmo neperiano de los puntos del conjunto $\mathcal W_{\tilde\varepsilon}$ dado en la ec.
\eqref{W-set}. Llamaremos a dicho conjunto $\mathcal Z_{\tilde\varepsilon}$.
Introduzcamos también los operadores de traslación
\begin{align}
 \widehat{U}^{(\varpi_\sigma,\varpi_\delta)}\psi(v,x_\sigma,x_\delta)=\psi(v,x_\sigma+\varpi_\sigma,
x_\delta+\varpi_\delta) ,
\end{align}
con $\varpi_\sigma,\varpi_\delta\in\mathcal Z_{\tilde\varepsilon}$ arbitrarios.

Entonces, el operador $\widehat{U}_6(v)$ es una combinación de seis operadores de traslación de este
tipo. Estos operadores son unitarios, ya que la suma de
$\|\widehat{U}^{(\varpi_\sigma,\varpi_\delta)}\psi(v,x_\sigma,
x_\delta)\|^2$ sobre todos los puntos $x_\sigma$ y
$x_\delta$ en el sector de superselección coincide con la suma de
$\|\psi(v,x_\sigma,x_\delta)\|^2$. De esta propiedad y
de la desigualdad de Schwarz concluimos fácilmente que el operador $\widehat{U}_6(v)$ es
acotado y su norma está acotada por 6. Por ser acotado, su espectro es no vacío. Más aún, su dominio
se puede extender a todo el espacio de Hilbert $\mathcal
H^+_{\lambda_\sigma^\star,\lambda_\delta^\star}$ y, ya
extendido, $\widehat{U}_6(v)$ resulta ser un operador normal (véase el Apéndice
\ref{appA}). Como consecuencia, su espectro residual es vacío. Además, 
existe una descomposición espectral natural asociada a $\widehat{U}_6(v)$, por
ser operador normal (como se describe en la Sección \ref{normales}). 

Se tiene, en definitiva, que el
inverso del operador $\widehat{U}_6(v)$
está bien definido si el cero no está en el espectro puntual de $\widehat{U}_6(v)$.
Para ver que ocurre así, empleamos primero la propiedad de que las traslaciones 
$\widehat{U}^{(\varpi_\sigma,\varpi_\delta)}$, para cualquier valor de $\varpi_\sigma$
y $\varpi_\delta$ pertenecientes a $\mathcal Z_{\tilde\varepsilon}$, conmutan entre sí y, como
consecuencia, también con 
$\widehat{U}_6(v)$. Esto implica que existe una base de autofunciones (que pueden ser generalizadas)
comunes a todos estos operadores y, por tanto, que los diagonalizan simultáneamente. Introduzcamos
la factorización
$\widehat{U}^{(\varpi_\sigma,\varpi_\delta)}=\widehat{U}_\sigma^{\varpi_\sigma}\otimes
\widehat{U}_\delta^{\varpi_\delta}$, donde el operador $\widehat{U}_a^{\varpi_a}$ actúa sobre 
$\text{Cil}_{\lambda_a^\star}$ ($a=\sigma,\delta$). Dada una autofunción común a todos estos
operadores de traslación, designamos el correspodiente autovalor de $\widehat{U}_a^{\varpi_a}$ por
$\rho_a(\varpi_a)$. Este autovalor debe ser un número complejo de norma unidad, ya que
$\widehat{U}_a^{\varpi_a}$ es un operador unitario. Además, como
$\widehat{U}_a^{\varpi_a}\widehat{U}_a^{\bar\varpi_a}=\widehat{U}_a^{\varpi_a+\bar\varpi_a}$
se cumple que
\begin{align}\label{compoeigen}
\rho_a(\varpi_a)\rho_a(\bar{\varpi}_a)=\rho_a(\varpi_a +\bar{\varpi}_a).
\end{align}

Recordando que todos los puntos en el sector de superselección de la variable $x_a$ se pueden
alcanzar a partir de $\ln({\lambda_a^\star})$ por una traslación $\widehat{U}_a^{\varpi_a}$, es
fácil darse cuenta de que las autofunciones son proporcionales a
$\rho_\sigma(\varpi_\sigma)\rho_\delta(\varpi_\delta)$. Siempre podemos elegirlas de tal modo que 
$\rho_a(0)=1$. Además, con el fin de determinar la autofunción completa, solo necesitamos conocer el
valor de $\rho_a(\varpi_a)$ en un subconjunto apropiado de $\mathcal Z_{\tilde\varepsilon}$, a
saber, en cualquier colección de puntos no conmensurables que puedan generar todo el conjunto
por
multiplicación por enteros. Es posible ver que la propiedad \eqref{compoeigen} proporciona,
entonces, toda la información sobre $\rho_a$ en el resto de puntos de $\mathcal
Z_{\tilde\varepsilon}$. En particular, $\rho_a(n \varpi_a)=[\rho_a(\varpi_a)]^n$.

Las funciones de onda $\rho_a(\varpi_a)$ son claramente no normalizables con respecto al producto
interno discreto en $\mathcal H_{\lambda_a^\star}$, ya que tienen norma compleja unidad en cada
punto del
sector de superselección (dado por el desplazamiento de $\mathcal Z_{\tilde\varepsilon}$ en
$\ln{\lambda_a^\star}$) y este sector contiene un número infinito de puntos. Por otra parte,
diferentes
funciones de onda $\rho_a(\varpi_a)$ deben ser ortogonales, porque siempre existe un operador de
traslación unitario en $\mathcal H_{\lambda_a^\star}$ cuyo autovalor difiere para las dos
autofunciones.

Llegados a este punto, cabe señalar que, por la construcción original del álgebra de operadores de
la cosmología cuántica de lazos, previa a la introducción de los sectores de superselección, los
operadores $\widehat{U}_a^{\varpi_a}$, que actúan por traslación en la representación de $x_a$, se
pueden identificar en la representación dual de holonomías, en la que actúan por multiplicación,
como elementos de la compactificación de Bohr de la recta real \cite{Vel,lqc3}. Estos elementos se
pueden entender como aplicaciones $\rho_a$ desde la recta real (correspondiente a todos los posibles
valores de $x_a$ o, equivalentemente, de $\varpi_a$) al círculo, tal que satisfacen la condición 
\eqref{compoeigen} y $\rho_a(0)=1$. Sin embargo, debido a la superselección, los valores de 
$\varpi_a$ están ahora restringidos al conjunto $\mathcal Z_{\tilde\varepsilon}$. Entonces,
podemos identificar las funciones de onda $\rho_a(\varpi_a)$ como clases de equivalencia de
elementos en $\mathbb{R}_{\text{Bohr}}$, siendo la relación de equivalencia la identificación de
todos las aplicaciones $\rho_a$ que difieran solo por su actuación en el conjunto complementario a 
$\mathcal Z_{\tilde\varepsilon}$ en la recta real, es decir,
$\mathbb{R}\setminus\mathcal{Z}_{\tilde\varepsilon}$. Los mapas exponenciales $\exp{(ik_a
\varpi_a)}$ desde $\mathcal Z_{\tilde\varepsilon}$ a $S^1$ proporcionan ejemplos de
$\rho_a(\varpi_a)$. Como $\mathcal Z_{\tilde\varepsilon}$ contiene números no conmensurables, estas
exponenciales separan todos los valores reales de $k_a$, esto es, por cada dos valores de $k_a$ se
puede encontrar un valor de $\varpi_a$ para el que las exponenciales $\exp{(ik_a \varpi_a)}$ son
diferentes. En consecuencia, el conjunto de todas las posibles aplicaciones $\rho_a$ distintas
contiene todas las exponenciales con $k_a\in \mathbb{R}$.

Volviendo al operador $\widehat{U}_6(v)$, es inmediato encontrar su autovalor para cada una de las
autofunciones
analizadas. Está dado por
\begin{align} \label{eigen2}
\omega_6(\rho_\sigma,\rho_\delta)&=
\sum_{a=\sigma, \delta}\left\{
\rho_a\left[\ln{\left(\frac{v}{v-2}\right)}\right]
+ \rho_a\left[\ln{\left(\frac{v-2}{v-4}\right)}\right]\right\}\nonumber\\
&+\sum_{a,b=\sigma, \delta; a\neq b}\left\{
\rho_a\left[\ln{\left(\frac{v}{v-2}\right)}\right]
\rho_b\left[\ln{\left(\frac{v-2}{v-4}\right)}\right]\right\}.
\end{align}
Recordamos que aquí $v>4$.
El espectro puntual de $\widehat{U}_6(v)$ no contendrá el cero si no existe una superposición lineal
de las anteriores funciones de onda con $\omega_6(\rho_\sigma,\rho_\delta)=0$ que sea normalizable.
En esta superposición la medida de $\rho_a$ es continua. La restricción al núcleo de
$\widehat{U}_6(v)$ se obtiene, entonces, introduciendo una función delta de Dirac de
$\omega_6(\rho_\sigma,\rho_\delta)$ (centrada en cero). Computando la norma de esta superposición,
la ortogonalidad de las funciones de onda $\rho_a(\varpi_a)$ da lugar a integrales sobre la norma
compleja al cuadrado de cada contribución en $(\rho_\sigma,\rho_\delta)$. Pero ésta contiene una
delta al cuadrado; por tanto, la norma ha de divergir. En consecuencia, el espectro puntual del
operador $\widehat{U}_6(v)$ no puede contener el cero, para cualquier valor de $v$, como queríamos
comprobar. De hecho, se puede aplicar la misma línea de razonamiento para cualquier otro autovalor
de $\widehat{U}_6(v)$, no solo para el cero, viendo así que el espectro puntual de dicho operador ha
de
ser vacío. 

\subsubsection{Resolución formal de la ligadura y espacio de Hilbert físico}

La conclusión que se sigue del anterior análisis es que, conocidas las combinaciones
$\psi_+(v,x_\sigma,x_\delta)$, debe ser posible determinar cada uno de los términos
$\psi(v,x_\sigma,x_\delta)$ que las componen, pues el sistema de ecuaciones que
relacionan las primeras con los segundos resulta invertible formalmente. Entonces, volviendo a la
ec. \eqref{eq:condicion-bianchi-B}, tendremos que, dados los datos iniciales
\begin{align}\label{eq:datos-iniciales}
 \{\psi(\tilde\varepsilon,x_\sigma,
x_\delta)=\psi(\tilde\varepsilon,\ln(\lambda_\sigma^\star)+\varpi_\sigma,
\ln(\lambda_\delta^\star)+\bar\varpi_\delta) \} ,
\end{align}
para todo $\varpi_\sigma,\bar\varpi_\delta\in\mathcal Z_{\tilde\varepsilon}$, podemos determinar
todos los datos en la sección $v=\tilde\varepsilon+4$.
Análogamente, con estos datos en las secciones $v=\tilde\varepsilon$ y $v=\tilde\varepsilon+4$,
podemos determinar todos los datos en todas las secciones siguientes, en virtud de la ec.
\eqref{eq:solution-bianchi-B} y de la invertibilidad del operador $\widehat{U}_6(v)$ aducida para
todo $v$. En definitiva, las soluciones de la ligadura cuántica de Bianchi I quedan totalmente
determinadas por el conjunto de datos iniciales \eqref{eq:datos-iniciales} y podemos identificar las
soluciones con este conjunto. 

Conviene resaltar que, de no ser por la invertibilidad del operador $\widehat{U}_6(v)$, el problema
de valores iniciales (en la variable $v$) estaría mal puesto para el modelo de Bianchi I en vacío
con el esquema B, y con ello la propia evolución.

Así pues, podemos caracterizar el espacio de
Hilbert físico como el espacio de Hilbert de los datos iniciales
$\tilde\psi(x_\sigma,x_\delta):=\psi(\tilde\varepsilon,x_\sigma,
x_\delta)$. Para determinar su producto
interno, tomamos un conjunto (super) completo de observables que formen un álgebra cerrada, e
imponemos sus relaciones de conjugación compleja como relaciones de adjunción entre operadores.
Un conjunto tal, que actúa sobre $\tilde\psi(x_\sigma,x_\delta)$, es el formado por los operadores
$\widehat{e^{ix_a}}$ y $\widehat{U}_a^{\varpi_a}$, con $\varpi_a\in\mathcal
Z_{\tilde\varepsilon}$ y $a=\sigma,\delta$, definidos como
\begin{subequations}\label{conjunto-completo}\begin{align}
 \widehat{e^{ix_\sigma}}\tilde\psi(x_\sigma,x_\delta)&={e^{ix_\sigma}}\tilde\psi(x_\sigma,x_\delta),
\\
\widehat{U}_\sigma^{\varpi_\sigma}\tilde\psi(x_\sigma,x_\delta)&=
\tilde\psi(x_\sigma+\varpi_\sigma,x_\delta),
\end{align}\end{subequations}
y análogamente para $\widehat{e^{ix_\delta}}$ y $\widehat{U}_\delta^{\varpi_\delta}$.
Estos operadores son unitarios en $\mathcal H_{\lambda_\sigma^\star,\lambda_\delta^\star}$, de
acuerdo con las condiciones de realidad existentes sobre ellos.
Por tanto, éste es el espacio de Hilbert físico del modelo de Bianchi I cuantizado según el
esquema B de la dinámica mejorada.

\section{Discusión}
\label{4sec:dis}

En este capítulo, hemos presentado dos cuantizaciones poliméricas distintas de las cosmologías de
Bianchi I en vacío. La diferencia entre ellas surge de una implementación distinta de la
dinámica mejorada. En el modelo concreto de Bianchi I con topología compacta en el que las hemos
aplicado, ambos esquemas dan lugar a teorías cuánticas bien definidas. Sin embargo, 
el estudio del comportamiento de la densidad material, tanto en el contexto de la dinámica clásica
efectiva en el orden dominante \cite{chi2}, como en el análisis cuántico de la Ref. \cite{luc2},
muestran que el esquema A (que respeta la factorización del espacio de Hilbert cinemático en tres
sectores direccionales) aplicado a modelos no compactos provoca dependencias de los
resultados físicos en la celda fiducial que se introduce para reducir el estudio a una región
finita del espacio (a no ser que el esquema se modifique en esos modelos para permitir que
$\bar\mu_i$ dependa de la celda fiducial). Por el contrario, el análisis similar para el esquema
B (en el que los operadores básicos no conservan la factorización del espacio de Hilbert cinemático
en tres sectores direccionales), en el contexto de la dinámica clásica efectiva, muestra que estos
problemas no aparecen \cite{chi2}.

Más recientemente, dentro de este contexto de dinámica efectiva, se han analizado otros observables
geométricos, como el grado de expansión de congruencias de observadores cosmológicos y el escalar de
cizalladura (véase definición en la Ref. \cite{wald}), con el fin de demostrar que única y
exclusivamente con el esquema B de la dinámica mejorada se curan los valores divergentes que estos
observables toman en la singularidad clásica y, por tanto, demostrar la invalidez del esquema A por
no curar todas las singularidades, independientemente de que la topología de las secciones
espaciales del modelo sean compactas o no \cite{cs3}. Sin embargo, la dinámica efectiva que se
analiza en ese estudio no se corresponde en realidad con la
que emergería de nuestra cuantización del modelo en el esquema A. En efecto, el estudio de la
Ref.~\cite{cs3} argumenta que dichos observables divergen justamente en la región donde los
coeficientes
de la tríada se anulan, al igual que en relatividad general. No
obstante, en nuestro modelo cuántico en el esquema A, el espectro de cada uno de los operadores
$\hat{p}_i$ presenta un mínimo \emph{no nulo} en cada sector de superselección.
En consecuencia, los observables geométricos analizados en la Ref.~\cite{cs3} presentarían una
cota superior finita, invalidando parte de la discusión presentada en esa referencia.

Volviendo al modo en el que se deriva el valor de $\bar\mu_i$ en ambos esquemas, explicado en la
Sección \ref{4sec:improved}, a priori, parece natural el procedimiento del esquema B. Si
recordamos la expresión clásica de la curvatura,
dada en la ec. \eqref{eq:curvatura-exacta}, los problemas a la hora de definirla surgen porque el
límite de las áreas fiduciales arbitrariamente pequeñas no existe, así que se procede a evaluar la
curvatura en un área mínima, la delimitada por un circuito rectangular de holonomías evaluadas a lo
largo de aristas de longitud fiducial mínima, dada por $2\pi\bar\mu_i$. De hecho, esta área fiducial
es la única que interviene en la construcción de la teoría. Por tanto, parece lógico igualar el
flujo a través de esta área fiducial  (la encerrada por dicho lazo de
holonomías) con el autovalor mínimo no nulo $\Delta$ del operador geométrico que representa el área
en gravedad cuántica de lazos. Esto es precisamente lo que hace se hace en el esquema B, pero no así
en el A. En efecto, si dicho circuito de holonomías de longitud mínima está contenido, por ejemplo,
en
el plano
$j-k$, su área fiducial es $A^i=4\pi^2\bar\mu_j\bar\mu_k$ y, en el esquema A, a dicha área fiducial
le corresponde un flujo dado por
\begin{align}
 E(A^i=4\pi^2\bar\mu_j\bar\mu_k,f=1)=p_i\bar\mu_j\bar\mu_k=\frac{\Delta p_i}{\sqrt{p_j p_k}},
\end{align}
donde hemos usado las ecs. \eqref{eq:flujo-bianchi} y \eqref{eq:mubarraA}.
Obviamente este flujo no coincide en general con $\Delta$.

Por otra parte, observando más a fondo la estructura de las teorías cuánticas resultantes, también
la del esquema B parece más correcta físicamente, pese a ser más complicada. En efecto, en la
teoría correspondiente al esquema A, hemos podido determinar la estructura física y su relación
precisa con la estructura cinemática explícitamente. Esto es consecuencia de que el problema está
completamente
factorizado y, en cierto sentido, es como si representaran tres copias de FRW, con tres direcciones
distintas e independientes, salvo por la relación algebraica que liga los autovalores
$\omega_i$. Por el contrario, el esquema B implementa de una manera muy adecuada la
interrelación entre las anisotropías y el volumen: el comportamiento de las anisotropías está
estrechamente relacionado con el del volumen, de modo que el cambio de éste afecta a las primeras.
Además, es de destacar que el esquema B se distingue como la única prescripción en la que el 
volumen sufre desplazamientos constantes, al igual que en el caso isótropo. En este sentido el
volumen es privilegiado en el esquema B. En el esquema A, por el contrario, las variables
afines son las variables direccionales $v_i$, que no tienen una interpretación física tan directa.

La desventaja del esquema B, como hemos visto, es que da lugar a operadores tan
complicados que no parece que se pueda determinar, explícitamente y sin recurrir al cálculo
numérico, la relación entre la estructura física y la cinemática, como hemos podido hacer en el
esquema A. No obstante, hemos sido capaces de caracterizar el espacio de Hilbert físico. Esto ha
sido
posible gracias a la descripción de tipo {ausencia de frontera}, obtenida como
consecuencia de nuestra prescripción de simetrización. En efecto, gracias a que la variable $v$
presenta un
valor mínimo no nulo, las soluciones quedan completamente determinadas por los datos en la sección
de dicho valor inicial de $v$, como en el esquema A, y hemos podido identificar las soluciones con
dichos datos iniciales. Es de señalar también la complicada y, a la vez, rica estructura que
presentan las variables
de anisotropía, las cuales, por cierto, no toman un valor mínimo, aunque sí tienen un ínfimo nulo.
Estas variables también dan lugar a sectores de superselección, de modo que la teoría está
perfectamente definida en espacios de Hilbert separables, característica necesaria para obtener una
teoría satisfactoria \cite{Vel2}. No obstante, los valores que toman cada una de las variables de
anisotropía, además de ser numerables, también son densos en la recta real positiva, de modo que
realmente se guarda información sobre un conjunto denso de datos acerca de la variabilidad
de las anisotropías.

\cleardoublepage

\part[Evolución en cosmología cuántica: modelo de Bianchi I]{Evolución en cosmología cuántica:
modelo de Bianchi I como ejemplo}
\label{part2}

\chapter[Evolución en cosmología cuántica estándar]{Evolución física en cosmología cuántica
estándar}
\label{chap:5-evol}

En teorías como la mecánica clásica, la mecánica cuántica o
la teoría cuántica de campos, por ejemplo, existe un marco geométrico de fondo fijo, y es habitual
tomar unas
coordenadas fijas y, en particular, hacer una elección concreta de tiempo
coordenado~$t$. Si, además, la teoría es no relativista, como la mecánica cuántica usual, esta
coordenada mide un tiempo \emph{absoluto}. En dichos contextos, el hamiltoniano de
la teoría es el generador de la evolución temporal con respecto a $t$ y podemos
describir los observables en un escenario de evolución dando su dependencia en dicho tiempo. No
obstante, en relatividad general, este paradigma cambia completamente. Debido a la
invariancia bajo difeomorfismos y reparametrizaciones temporales, la teoría está totalmente
constreñida: el
hamiltoniano es puramente una combinación lineal de ligaduras (salvo quiźas términos de frontera
que no afectan a las ecuaciones de movimiento locales \cite{front}). Por tanto, las transformaciones
de coordenadas que genera pueden entenderse como transformaciones gauge (cuando los términos de
frontera son nulos, lo que puede conllevar condiciones de contorno). Como consecuencia, los
observables físicos deben ser
independientes de la
elección de coordenadas para que sean invariantes gauge. 

No obstante, el hecho de que, en particular, la evolución con respecto al tiempo coordenado no sea
física no implica
que en relatividad general se tenga que renunciar a la imagen de evolución temporal. Los
observables físicos se pueden entender como cantidades relativas que expresan correlaciones entre
variables dinámicas \cite{rovelli-obs1}. Más explícitamene, se puede introducir el concepto de
\emph{observable completo} u \emph{observable relacional}, construido a partir de dos variables
dinámicas del espacio de fases denominadas \emph{observables parciales}. Entonces, el observable
completo mide el valor que toma uno de los observables parciales cuando el otro toma cierto valor
\cite{rovelli-obs2,rovelli-obs3,bianca-obs1,bianca-obs2}. De este modo, si interpretamos el
observable parcial
de referencia como un tiempo interno, el observable completo nos da la evolución con respecto a
este tiempo interno del otro observable parcial. Es en este sentido en el que muchas veces se habla
de \emph{evolución}, tanto en relatividad general como en las teorías cuánticas no
perturbativas de la gravedad \cite{thiemann-obs,thiemann-obs2}. Asimismo, se han empleado ideas
similares para discutir la relación entre diferentes gauges y definir invariantes gauge en teoría
de perturbaciones \cite{david1,david2}.

En el programa de cuantización de Dirac, los
observables parciales son operadores cinemáticos con los que se construye el observable completo que
actúa en el espacio de Hilbert físico, al que también se le denomina observable de Dirac. El
observable parcial que se interprete como tiempo interno debe ser adecuado, en el sentido de que la
evolución temporal que defina debería ser unitaria, con el fin de poder llevar a cabo la
interpretación
probabilística asociada, típica de la mecánica cuántica. En el Capítulo~\ref{chap:revFRW} ya
analizamos este concepto de evolución en el contexto del modelo de FRW. Concretamente, en la Sección
\ref{1subsec:evol} vimos que el operador cinemático $\hat{\phi}$ puede interpretarse como tiempo
interno y, a partir de él y el operador $\widehat{|v|}$, se puede construir el observable completo
dado por $\widehat{|v|}_\phi$, que permite analizar los resultados en un escenario de evolución. 

En la mayoría de los trabajos desarrollados en cosmología cuántica de lazos
\cite{aps1a,aps1b,aps3,apsv,vand,luc,tom,chio,awe,luc2}, el papel del tiempo lo desempeña tal campo
escalar, homogéneo y sin masa,  ${\phi}$, que, a diferencia de los grados de libertad
geométricos, se cuantiza adoptando
la representación estándar de tipo Schr\"odinger. Como ya comentamos en la Introducción, uno de
nuestros objetivos ha sido precisamente desarrollar metodológicamente el concepto de evolución en
cosmología cuántica de lazos cuando tal campo no está presente y cuando, por tanto, el papel de
tiempo interno lo tiene que desempeñar uno de los grados de libertad geométricos. En particular,
hemos construido y estudiado las propiedades de varias familias de observables
en el ejemplo concreto del modelo de Bianchi I en vacío. Este estudio está recogido en nuestra
publicación \cite{mmp2}.

El modelo de Bianchi I en vacío es especialmente idóneo para llevar a cabo este análisis puesto que
su cuantización ha sido ampliamente analizada. Para este estudio, hemos considerado la
cuantización polimérica correspondiente al esquema A, no solo
porque en el momento de realizar este trabajo era la única para la que se había determinado el
espacio de Hilbert físico y un conjunto completo de observables, sino también porque, a
diferencia de la cuantización del esquema B, en el esquema A conocemos explícitamente la
estructura y propiedades de las soluciones físicas. Este detalle es necesario para establecer la
conexión entre intuición física y la implementación matemática exacta de nuestro método. Por
ejemplo, gracias a la comentada descripción de tipo
{ausencia de frontera} que verifican los estados físicos, es de esperar que las singularidades
clásicas
de Bianchi I sean
resueltas, además de cinemáticamente (como hemos visto en el capítulo anterior), también
dinámicamente. Nuestra construcción de la imagen de evolución muestra de forma precisa que, en
efecto, así ocurre.
Además, hemos pretendido que nuestra construcción, aunque aplicada en el ejemplo concreto
del modelo de Bianchi I en el esquema A, sea fácilmente generalizable. En particular, como veremos,
es aplicable casi directamente al modelo de Bianchi I en el esquema~B.

Hemos investigado dos posibles construcciones de familias de observables relacionados
unitariamente que están parametrizados respectivamente por: $(i)$ uno de los parámetros afines
$v_i$ asociados a los coeficientes de la tríada densitizada y $(ii)$ la variable conjugada a
dicho parámetro. Por otra parte, con el propósito de comparar los resultados físicos de
cosmología cuántica de lazos con los de geometrodinámica cuántica estándar, primero hemos llevado a
cabo nuestro análisis en la cuantización de WDW del modelo. Así, también presentamos la
construcción en un contexto más sencillo, en el que la evolución se implementa de forma unitaria
muy fácilmente gracias a las propiedades de la representación de Sch\"odinger empleada en esta
cuantización. Veremos que la adaptación directa a la teoría de lazos de la construcción aplicada
en la teoría estándar no da lugar a una imagen de evolución satisfactoria, como una consecuencia de
la naturaleza polimérica de la variable identificada como tiempo interno.

En este capítulo, analizaremos la imagen de evolución en la cuantización de WDW del
modelo. Pospondremos al siguiente capítulo la construcción en el caso de la cuantización de
lazos, que es más complicada. Además, este capítulo también contiene un amplio análisis
acerca de la relación que existe entre los estados físicos de la cuantización de WDW y
los estados físicos de la cuantización de lazos. Esta relación nos ha ayudado a entender las
causas que explican por qué en la cuantización de lazos la implementación de una evolución
unitaria no es tan sencilla, así como a idear un método apropiado que dé como resultado la
descripción de una evolución unitaria bien definida.

\section[Cuantización de WDW]{Cuantización de WDW}
\label{sec:wdw-analog}

En esta sección describiremos la cuantización de WDW del modelo de Bianchi I en vacío.
Como ya hemos comentado y hemos visto en el ejemplo concreto del modelo de FRW (Sección
\ref{3subsec:wdw}), en la cuantización de WDW se emplea la representación estándar de
tipo Schr\"odinger. Como queremos comparar esta cuantización con la cuantización polimérica del
modelo en el esquema A, emplearemos también en este contexto la
formulación del modelo en términos de las variables de Ashtekar-Barbero, que ya ha sido introducida
en la Sección \ref{4sec:clas}, es decir, en términos de los coeficientes $p_i$ de la tríada
densitizada y de los coeficientes $c^i$ de la conexión o, equivalentemente, de los parámetros afines
$v_i$ y de sus variables canónicamente conjugadas $\beta_i$.

\subsection{Estructura cinemática y ligadura escalar}

Al igual que en la cuantización polimérica, trabajamos en la representación de momentos. El espacio
de Hilbert cinemático es el espacio de funciones de cuadrado integrable con la medida de Lebesgue
usual, es decir%
\footnote{Recordamos que la notación con subrayado hace referencia a las magnitudes
de la teoría de WDW para distinguirlas de las de sus análogas en cosmología cuántica de
lazos.},
\begin{align}
 \underline{\mathcal{H}}_{\text{grav}}=\otimes_i\underline{\mathcal{H}}_{\text{grav}}^i,\qquad
\underline{\mathcal{H}}_{\text{grav}}^i=L^2(\mathbb{R},dv_i).
\end{align}
En esta representación el operador $\hat{p}_i$ actúa por multiplicación por el factor
\begin{align}\label{v-p-rel}
 p_i=(6\pi\gamma
l_{\text{Pl}}^2\sqrt{\Delta})^{2/3}\,\text{sgn}(v_i)|v_i|^{2/3},
\end{align}
como en la cuantización de lazos. Por otra parte, promovemos los coeficientes de la conexión a
operadores de derivación,
\begin{equation}
  \hat c^i=i2(6\pi\gamma
l_{\text{Pl}}^2/\Delta)^{1/3}|v_i|^{1/6}{\partial}_{v_i}|v_i|^{1/6},
\end{equation}
tales que se satisface la regla de Dirac $[\hat c {}^i,\hat{p}_j]=i\hbar\widehat{\{c^i,p_j\}}$.

Denominamos $\widehat{\underline\Omega}_i$ al operador que representa la cantidad clásica
$(-c^ip_i)$ en esta teoría de WDW, que está definido en el espacio de Schwartz
$\mathcal{S}(\mathbb{R})$. Hemos escogido para él un orden de factores similar al
empleado en la simetrización de su análogo $\widehat{\Omega}_i$ en la cuantización de lazos, para
simplificar la comparación entre ambas teorías, de modo que este operador se escribe
\begin{align}
 \widehat{\underline\Omega}_i=-i\bar\alpha\sqrt{|v_i|}[\text{sgn}(v_i)\partial_
{v_i}+\partial_
{v_i}\text{sgn}(v_i)]\sqrt{|v_i|},
\end{align}
donde $\bar\alpha=3\alpha=12\pi\gamma l_{\text{Pl}}^2$. Este operador está bien definido en sentido
distribucional y puede ser reescrito en la forma más sencilla
\begin{align}
 \widehat{\underline\Omega}_i=-i\bar\alpha(2v_i\partial_{v_i}+1),
\end{align}
donde se ha despreciado el término $|v_i|\delta(v_i)$, pues no contribuye.

En vista de la ec. \eqref{eq:lig-dens-bian-A}, el operador
ligadura escalar densitizada en la cuantización de WDW es
\begin{equation}\label{CWDW}
\widehat{{\underline{\cal C}}}_{\text{BI}} =
-\frac{2}{\gamma^2}\big(\widehat{\underline\Omega}_\theta
\widehat{\underline\Omega}_\sigma
+ \widehat{\underline\Omega}_\theta\widehat{\underline\Omega}_\delta
+ \widehat{\underline\Omega}_\sigma\widehat{\underline\Omega}_\delta
\big).
\end{equation}

El operador $\widehat{\underline\Omega}_i$ es esencialmente autoadjunto en
$\underline{\mathcal{H}}_{\text{grav}}^i$. Más aún, su restricción a cada uno de los subespacios
$\underline{\mathcal{H}}_{\text{grav}}^{i,\pm}=L^2(\mathbb{R}^\pm,dv_i)$ es también esencialmente
autoadjunta (nótese que genera una dilatación). En cada uno de estos subespacios, el espectro de
$\widehat{\underline\Omega}_i$ es absolutamente continuo, coincide con la recta real y es no
degenerado, como ocurre con su análogo $\widehat{\Omega}_i$ en la cuantización polimérica. Además,
sus autofunciones generalizadas, correspondientes al autovalor $\omega_i$, tienen la forma
\begin{align}\label{eq:WDW-eig}
\underline{e}_{\omega_i}(v_i)&= \frac{1}{\sqrt{2\pi\bar\alpha
|v_i|}}\exp\left({-i\omega_i\frac{\ln{|v_i|}}{\bar\alpha}}\right)
\end{align}
y proporcionan una base ortonormal para $\underline{\mathcal{H}}_{\text{grav}}^{i,\pm}$, tal que
\begin{align}
 \langle \underline{e}_{\omega_i}|\underline{e}_{\omega'_i}\rangle=\delta(\omega_i-\omega'_i).
\end{align}
En analogía a la cuantización de lazos, hemos restringido el estudio a
$\underline{\mathcal{H}}_{\text{grav}}^{+}=\otimes_i\underline{\mathcal{H}}_{\text{grav}}^{i,+}$.

\subsection{Espacio de Hilbert físico}
\label{sec:WDWphys-obs}

Para obtener el espacio de Hilbert físico aplicamos el procedimiento de promedio sobre grupos, en
completo paralelismo con la Sección \ref{sec:group-ave}, simplemente reemplazando los elementos
básicos ${e}_{\omega_i}(v_i)$ de la teoría de lazos por los análogos $\underline{e}_{\omega_i}(v_i)$
en la teoría de WDW [véanse las ecs. \eqref{eq:fdec}-\eqref{eq:phys-A}]. Obviamente, el
espacio de Hilbert que obtenemos es
\begin{align}\label{physpaWDW}
\underline{\mathcal H}_\text{fis}= L^2\left(\mathbb{R}^2,
|\omega_\sigma+\omega_\delta| d\omega_\sigma d\omega_\delta\right).
\end{align}
A su vez, la función de onda de los estados físicos es de la forma
\begin{align}\label{phystawdw}
{\underline\Phi}(\vec{v}) &= \int_{\mathbb{R}^2}
d\omega_\sigma d\omega_\delta\,
{\underline{\tilde\Phi}}(\omega_\sigma,\omega_\delta)\, \underline
e_{\omega_\theta(\omega_\sigma,\omega_\delta)}(v_\theta)
\underline e_{\omega_\sigma} (v_\sigma)\underline e_{\omega_\delta}(v_\delta),
\end{align}
con $ {\underline{\tilde\Phi}}(\omega_\sigma,\omega_\delta)
\in\underline{\mathcal
H}_\text{fis}$ y donde $\omega_\theta(\omega_\sigma,\omega_\delta)$ es la función dada en la ec.
\eqref{eq:om-func}%
\footnote{En lo que sigue y a no ser que se indique lo contrario, emplearemos
$\omega_\theta$ para referirnos a esta función.}.

Conocido el espacio de Hilbert físico, un conjunto completo de observables que~podemos identificar
fácilmente es el formado por los operadores $\widehat{\underline{\Omega}}_a$ ($a=\sigma,\delta$),
que multiplican la función de onda por $\omega_a$, junto con los operadores
$-i|\omega_\sigma+\omega_\delta|^{-1/2}\partial_{\omega_a}
|\omega_\sigma+\omega_\delta|^{1/2}$. En efecto, son observables puesto que son operadores
esencialmente autoadjuntos definidos en el dominio denso $\mathcal
S(\mathbb{R}^2)\subset\underline{\mathcal{H}}_{\text{fis}}$. Sin embargo, este conjunto de
observables no permite introducir un concepto de evolución no trivial, como el que hemos discutido
al
principio de este capítulo. Para ello, debemos construir algún observable \emph{relacional}. En la
próxima sección, describiremos la construcción de un par de tales observables que, junto con las
constantes de movimiento $\widehat{\underline{\Omega}}_a$, formarán un conjunto completo de
observables con los que, además, se podrá interpretar el sistema en evolución.

\section[Evolución con respecto a $\upsilon_\theta$]{Evolución con respecto a $v_\theta$}
\label{sec:WDWphys-evo}

En la introducción de este capítulo hemos centrado la discusión en cómo alcanzar un concepto de
evolución usando observables relacionales. No obstante, podemos lograr una visión equivalente del
concepto de evolución a partir de una composición adecuada de aplicaciones entre distintos espacios
de Hilbert de datos iniciales. 

\subsection{Evolución a partir de aplicaciones entre espacios de $v_\theta$ constante}

En esta descripción, también se empieza por seleccionar una función o variable del espacio
de fases a la que se le asigna el papel de tiempo. Denominamos $T$ a este tiempo y
$\underline{\mathcal{H}}_{T}$ a los espacios de Hilbert de ``datos
iniciales'', dados por la restricción de la función de onda física a la sección con
$T=\text{constante}$. Si existe una transformación unitaria
\begin{align}\label{mapeo-uni}
 \underline{\hat P}_{T}:\;\underline{\mathcal{H}}_\text{fis}\;\to\;\underline{\mathcal{H}}_{T},
\end{align}
entonces cada sección de $T$ constante contiene toda la información necesaria para determinar el
estado físico, esto es, el sistema es cerrado. Más aún, si, además, la aplicación ``identidad''
entre
los espacios de Hilbert $\underline{\mathcal{H}}_{T}$, dada por la identificación trivial de datos
a diferentes tiempos $T$, también es unitaria, se puede definir entonces una evolución unitaria en 
$\underline{\mathcal{H}}_{T}$. Ésta se obtiene mediante la composición de la transformación inversa
$\underline{P}_{T}^{-1}$ con una transformación en la familia $\underline{P}_{T}$ para un valor
diferente de $T$ y con la anterior identificación de datos a diferentes tiempos.

En nuestro caso del modelo de Bianchi I, clásicamente, los coeficientes $p_i$ de la tríada
densitizada o, equivalentemente, las variables asociadas $v_i$, son funciones monótonas a lo largo
de las trayectorias dinámicas definidas por el tiempo coordenado \cite{chi2}. Por tanto, cualquiera
de ellas representa un tiempo natural interno.
Recordemos que en la descripción de los estados físicos, ec.~\eqref{phystawdw}, ya hemos eliminado
la variable  $\omega_\theta$ en términos de $\omega_\sigma$ y $\omega_\delta$. Por tanto, lo más
natural es escoger como tiempo interno $v_\theta$.

Podemos introducir fácilmente los espacios de datos iniciales $\underline{\mathcal{H}}_{
v_\theta}$ mediante la transformación unitaria
\begin{align}\label{eq:wdw-sl-def}
\underline{\hat
P}_{v_\theta}:\underline{\mathcal{H}}_\text{fis}\;&\to\;\underline{\mathcal{H}}_{
v_\theta}=L^2(\mathbb{R}^2,|\omega_\sigma+\omega_\delta|
d\omega_\sigma d\omega_\delta) \\
\;\underline{\tilde{\Phi}}(\omega_\sigma,\omega_\delta)\;&\mapsto\;
\underline{\tilde{\Phi}}_{v_\theta}(\omega_\sigma,\omega_\delta)=
\underline{\tilde{\Phi}}(\omega_\sigma,\omega_\delta)
\sqrt{2\pi\bar\alpha v_\theta}\underline{e}_{\omega_\theta}(v_\theta),\nonumber
\end{align}
\begin{align}\label{slices-states}
\underline{\Phi}_{v_\theta}(v_\sigma,v_\delta) &= \int_{\mathbb{R}^2}
\frac{d\omega_\sigma d\omega_\delta}{\sqrt{2\pi\bar\alpha v_\theta}}
\underline{\tilde{\Phi}}_{v_\theta}(\omega_\sigma,\omega_\delta)
\underline{e}_{\omega_\sigma} (v_\sigma) \underline{e}_{\omega_\delta}(v_\delta).
\end{align}
De este modo, los espacios de datos iniciales $\underline{\mathcal{H}}_{v_\theta}$ coinciden y, por
tanto, la operación de identificar estados a diferentes valores de $v_\theta$ es ciertamente
unitaria. La composición de los mapeos considerados da lugar, efectivamente, a una evolución
unitaria.

\subsection{Construcción de los observables completos}
\label{5sec:obs-rel}

La otra construcción más elaborada de una familia uniparamétrica de
observables parciales (que en su conjunto forma un observable relacional o completo~$\hat O_T$) es
viable si existe una descomposición del espacio de Hilbert cinemático de la forma
\begin{align*}
\underline{\mathcal H}(T)\otimes\underline{\mathcal H}',
\end{align*}
donde el sector $\underline{\mathcal H}(T)$ solo depende de la variable $T$, y existe un operador
cinemático
\begin{align*}
 \hat O':\;\underline{\mathcal H}'\;\to\;\underline{\mathcal H}'.
\end{align*}
Entonces, el observable
\begin{align*}
 \hat O_T:\;\underline{\mathcal{H}}_\text{fis}\;\to\;\underline{\mathcal{H}}_\text{fis}
\end{align*}
mide la cantidad representada por $\hat O'$ ``a cierto tiempo $T$'' y es un operador
cuya acción se define mediante la siguiente secuencia de operaciones:
\begin{itemize}
 \item [(i)] Mediante un mapeo
$\underline{\hat P'}_{T}:\underline{\mathcal{H}}_\text{fis}\to\underline{\mathcal{H}}_{T}$ se
selecciona una sección de datos iniciales $\underline{\mathcal{H}}_{T}$
correspondiente al valor $T$ del tiempo interno.
\item[(ii)] Se define la transformación $\underline{\mathcal H}_{T}\to\underline{\mathcal{H}}'$ si
estos dos espacios no coinciden.
\item[(iii)] Se actúa sobre $\underline{\mathcal H}'$ con el operador cinemático $\hat O'$
correspondiente al observable que se quiere medir.
\item[(iv)] Finalmente, se deshace el mapeo\begin{align}
\underline{\hat
R}_{T}:\underline{\mathcal{H}}_\text{fis}\to\underline{\mathcal{H}}_{T}\to\underline{ \mathcal H}'
\end{align}
para buscar el elemento de $\underline{\mathcal{H}}_\text{fis}$ que
se corresponde con el resultado de (iii). 
\end{itemize}
Vemos que, en esta construcción, el espacio $\underline{\mathcal H}'$ desempeña un papel auxiliar
que permite al final referir toda la información al espacio de Hilbert físico.

Si la primera descripción da ciertamente una evolución unitaria en $T$, entonces esta
construcción también es viable. De hecho, eligiendo el mapeo $\underline{\hat P'}_{T}$ igual
al mapeo $\underline{\hat P}_{T}$ de la ec.~\eqref{mapeo-uni}, se obtiene que todos los espacios
$\underline{\mathcal H}_{T}$ pueden identificarse y la transformación del punto (ii) es
independiente de $T$. Esto asegura que el mapeo composición $\underline{\hat R}_{T}$ sea ciertamente
unitario.

En el modelo de Bianchi I bajo estudio, hemos aplicado este método a los observables cinemáticos
$\ln(\hat{v}_a)$ (donde $a=\sigma,\delta$), que actúan sobre los elementos de $\underline{\mathcal
H}_\text{grav}^{a,+}$ como un operador de multiplicación. Como tiempo interno, seguimos eligiendo
${v_\theta}$ y como espacios de datos iniciales $\underline{\mathcal{H}}_{
v_\theta}$ los definidos en la ec.~\eqref{eq:wdw-sl-def}. Por otra parte, el espacio cinemático
$\underline{\mathcal{H}}'$ es el producto
\begin{align}\label{eq:wdw-Hprime}
\underline{\mathcal{H}}' = \underline{\mathcal{H}}_{\text{grav}}^{\sigma,+}\otimes
\underline{\mathcal{H}}_{\text{grav}}^{\delta,+} = L^2((\mathbb{R}^+)^2,d v_\sigma d
v_\delta),
\end{align}
cuyos elementos expandidos en las bases dadas por las funciones $\underline{e}_{\omega_a}(v_a)$ son
de la forma
\begin{align}\label{eq:chi-wdw}
\underline{\boldsymbol{\chi}}(v_\sigma,v_\delta) =
\int_{\mathbb{R}^2}d\omega_\sigma d\omega_\delta
\tilde{\underline{\boldsymbol{\chi}}}(\omega_\sigma,\omega_\delta)
\underline{e}_{\omega_\sigma}(v_\sigma) \underline{e}_{\omega_\delta}(v_\delta).
\end{align}
Por tanto, en la representación de $\omega_a$, este espacio de Hilbert se reescribe como
\begin{align}\label{eq:wdw-Hprime-omega}
\underline{\mathcal{H}}' = \underline{\mathcal{H}}_{\text{grav}}^{\sigma,+}\otimes
\underline{\mathcal{H}}_{\text{grav}}^{\delta,+} = L^2(\mathbb{R}^2,d \omega_\sigma d
\omega_\delta).
\end{align}

Si comparamos esta expresión con la ec. \eqref{eq:wdw-sl-def}, vemos que el espacio
auxiliar $\underline{\mathcal{H}}'$ no coincide con los espacios $\underline{\mathcal
H}_{v_\theta}$. Entonces, introducimos una transformación unitaria $\underline{\mathcal
H}_{v_\theta}\to\underline{\mathcal{H}}'$. En la representación de $\omega_a$, está dada por el
mapeo
\begin{align}\label{eq:wdw-h'tr}
\tilde{\underline{\Phi}}_{v_\theta}(\omega_\sigma,\omega_\delta)\, \mapsto
&\,\tilde{\underline{\boldsymbol{\chi}}}_{v_\theta}
(\omega_\sigma,\omega_\delta)= |\omega_\sigma+\omega_\delta|^{\frac{1}{2}}
\tilde{\underline{\Phi}}_{v_\theta}(\omega_\sigma,\omega_\delta).
\end{align}
En la representación de $v_a$, esta tranformación es simplemente la aplicación
\begin{align}
 \underline{\Phi}_{v_\theta}(v_\sigma,v_\delta)\to
\underline{\boldsymbol{\chi}}_{v_\theta}(v_\sigma,v_\delta),
\end{align}
con
$\underline{\Phi}_{v_\theta}(v_\sigma,v_\delta)$ dado en la ec.~\eqref{slices-states} y
\begin{equation}\label{eq:aux-chi-wdw}
\underline{\boldsymbol{\chi}}_{v_\theta}(v_\sigma,v_\delta) =
\int_{\mathbb{R}^2}d\omega_\sigma d\omega_\delta
\tilde{\underline{\boldsymbol{\chi}}}_{v_\theta}(\omega_\sigma,\omega_\delta)
\underline{e}_{\omega_\sigma}(v_\sigma) \underline{e}_{\omega_\delta}(v_\delta).
\end{equation}

Por tanto, considerando la composición de las dos aplicaciones introducidas, obtenemos que la
relación entre los estados $\underline{\boldsymbol{\chi}}_{v_\theta}(v_\sigma,v_\delta)$
de $\underline{\mathcal{H}}'$ y los estados
$\underline{\tilde{\Phi}}(\omega_\sigma,\omega_\delta)$ de $\underline{\mathcal{H}}_\text{fis}$ es
\begin{equation}\label{eq:aux-chi-wdw-rel}
\underline{\boldsymbol{\chi}}_{v_\theta}(v_\sigma,v_\delta) =\sqrt{2\pi\bar\alpha v_\theta}
\int_{\mathbb{R}^2}d\omega_\sigma d\omega_\delta
|\omega_\sigma+\omega_\delta|^{\frac{1}{2}}
\underline{\tilde{\Phi}}(\omega_\sigma,\omega_\delta)
\underline{e}_{\omega_\theta}(v_\theta)
\underline{e}_{\omega_\sigma}(v_\sigma) \underline{e}_{\omega_\delta}(v_\delta).
\end{equation}

Ahora ya podemos actuar sobre el estado
$\underline{\boldsymbol{\chi}}_{v_\theta}(v_\sigma,v_\delta)$ con el operador cinemático
$\ln(\hat{v}_a)$ y deducir cuál es la acción que produce en el estado físico 
$\underline{\tilde{\Phi}}(\omega_\sigma,\omega_\delta)$. Teniendo en cuenta que $\ln(\hat{v}_a)$
actúa por multiplicación y que
\begin{align}
 \ln(v_a)\underline{e}_{\omega_a}(v_a)=i\bar\alpha\partial_{\omega_a}\underline{e}_{
\omega_a}(v_a),
\end{align}
como se deduce de la ec.~\eqref{eq:WDW-eig}, obtenemos
\begin{align}
\ln(\hat{v}_\sigma)\underline{\boldsymbol{\chi}}_{v_\theta}(v_\sigma,v_\delta)
&=i\bar\alpha\sqrt{2\pi\bar\alpha v_\theta}\nonumber\\&\times
\int_{\mathbb{R}^2}d\omega_\sigma d\omega_\delta
|\omega_\sigma+\omega_\delta|^{\frac{1}{2}}
\underline{\tilde{\Phi}}(\omega_\sigma,\omega_\delta)
\underline{e}_{\omega_\theta}(v_\theta)\left[\partial_{\omega_\sigma}
\underline{e}_{\omega_\sigma}(v_\sigma)\right] \underline{e}_{\omega_\delta}(v_\delta),
\end{align}
y análogamente para $\ln(\hat{v}_\delta)$.
Integrando por partes y considerando como dominio de los estados físicos el espacio de
Schwartz
$\mathcal{S}(\mathbb{R}^2)$, finalmente deducimos que el observable completo
\begin{align}
 \ln(\hat{v}_a)_{v_\theta}:\;\mathcal{S}(\mathbb{R}^2)\subset\underline{\mathcal{H}}_\text{fis}\;
\to\;\underline{\mathcal{H}}_\text{fis},
\end{align}
está definido como
\begin{align}\label{obs}
 [\ln(\hat{v}_a)_{v_\theta}\underline{\tilde{\Phi}}](\omega_\sigma,\omega_\delta)
= \frac{-i\bar\alpha} {\underline{e}_{\omega_\theta}(v_\theta)}
|\omega_\sigma+\omega_\delta|^{-\frac{1}{2}}\partial_{\omega_a}\left[
|\omega_\sigma+\omega_\delta|^{\frac{1}{2}}
\underline{\tilde{\Phi}}(\omega_\sigma,\omega_\delta)
\underline{e}_{\omega_\theta}(v_\theta)\right].
\end{align}
Este observable mide el valor que toma $\ln(v_a)$ a un ``tiempo'' fijo $v_\theta$.

\subsection{Unitariedad de la evolución}

Los observables $\ln(\hat{v}_a)_{v_\theta}$, junto con las constantes del movimiento 
$\widehat{\underline{\Omega}}_a|_{v_\theta}:={\widehat{\underline{\Omega}}_a}$
forman un conjunto completo de observables de Dirac. Los observables $\widehat{\underline{\Omega}}_a
|_{v_\theta}$ no cambian con el tiempo. Sin embargo, los observables $\ln(\hat{v}_a)_{v_\theta}$ no
coinciden a
distintos tiempos $v_\theta$ y $v_\theta^\star$, y están relacionados mediante un operador
$\underline{\widehat Q}_{v_\theta,v_\theta^\star}:\underline{\mathcal
H}_\text{fis}\rightarrow\underline{\mathcal H}_\text{fis}$, dado por
\begin{align}
[\underline{\widehat Q}_{v_\theta,v_\theta^\star}\underline{\tilde\Phi}]
(\omega_\sigma,\omega_\delta)
= \sqrt{\frac{v_\theta}{v_\theta^\star}}
\frac{\underline e_{\omega_\theta}(v_\theta)}
{\underline e_{\omega_\theta}(v_\theta^\star)}
\underline{\tilde\Phi}(\omega_\sigma,\omega_\delta).
\end{align}
La forma de las autofunciones \eqref{eq:WDW-eig} implica inmediatamente que estos operadores son
invertibles  ($\underline{\widehat Q}_{v_\theta,v_\theta^\star}^{-1}=\underline{\widehat
Q}_{v_\theta^\star,v_\theta}$) y unitarios. Por tanto, la relación entre observables a tiempos
diferentes,
\begin{align}
\ln({\hat{v}}_a)_{v_\theta^\star}
= \underline{\widehat Q}_{v_\theta,v_\theta^\star}\,\ln(\hat{ v}_a)_{v_\theta}
\,\underline{\widehat Q}_{v_\theta^\star,v_\theta},
\end{align}
es unitaria.

En definitiva, las familias uniparamétricas de observables $\ln(\hat{v}_a)_{v_\theta}$ definen en
el espacio de Hilbert físico una evolución unitaria que es local en el tiempo interno $v_\theta$.

\subsection{Análisis de los valores esperados de los observables}
\label{pred-fis}

Con el fin de comparar la dinámica predicha por las familias de observables construidas con la
dinámica clásica del sistema, hemos calculado los valores esperados de estos observables en una
clase de estados que son semiclásicos a tiempos tardíos. Estos estados son proporcionados por
estados gaussianos centrados en valores grandes $\omega_\sigma^\star$ y $\omega_\delta^\star$ de su
soporte, esto es,
\begin{align}\label{eq:wdw-gauss}
\underline{\tilde{\Phi}}(\omega_\sigma,\omega_\delta) =
\frac{K}{\sqrt{|\omega_\sigma+\omega_\delta|}}\prod_{a=\sigma}^{\delta}
e^{-\frac{(\omega_a-\omega_a^\star)^2}{2\sigma_a^2}}
e^{i\nu^a\omega_a},
\end{align}
donde $\sigma_a$ determina la anchura de la gaussiana, $\nu_a$ determina su fase, $K$ es un factor
de normalización tal que $\|\underline{\tilde{\Phi}}\|=1$ y el factor
$|\omega_\sigma+\omega_\delta|^{-1/2}$ compensa el factor no trivial en la medida del espacio de
Hilbert físico \eqref{physpaWDW} para que dicho perfil sea realmente gaussiano.

Si usamos la forma explícita de los observables \eqref{obs} y tenemos en cuenta que
\begin{align}
 \partial_{\omega_a}\underline
e_{\omega_\theta}(v_\theta)=\frac{i}{\bar\alpha}\ln(v_\theta)\left[\frac{\omega_\theta({\omega}
_\sigma,{\omega}_\delta)}{\omega_a}\right]^2\underline e_{\omega_\theta}(v_\theta),
\end{align}
como se deduce de la ec. \eqref{eq:WDW-eig}, encontramos que, para un estado general
$\underline{\tilde{\Phi}}$, se cumple
\begin{equation}\label{eq:wdw-traj-shape}
\langle \underline{\tilde{\Phi}} | \ln(\hat{v}_a)_{v_\theta}
\underline{\tilde{\Phi}} \rangle
= A_a \ln v_\theta + B_a ,
\end{equation}
donde los coeficientes constantes $A_a$ y $B_a$ son
\begin{subequations}\label{eq:traj-dir}\begin{align}
\label{eq:traj-dir-A}
A_a &= \| \omega_\theta({\omega}_\sigma,{\omega}_\delta){\omega}_a^{-1}
\underline{\tilde{\Phi}} \|^2 , \\ \label{eq:traj-dir-B}
B_a &= \bar\alpha \langle \underline{\tilde{\Phi}} |
|\omega_\sigma+\omega_\delta|^{-\frac{1}{2}} (-i\partial_{\omega_a})
|\omega_\sigma+\omega_\delta|^{\frac{1}{2}} \underline{\tilde{\Phi}} \rangle .
\end{align}\end{subequations}

Evaluemos estos coeficientes en el caso de perfiles gaussianos \eqref{eq:wdw-gauss} muy
estrechos, es decir, en el límite $\sigma_a\to 0$ para $a=\delta,\sigma$. En este límite,
\begin{align}
 A_a\simeq
\left[\frac{\omega_\theta({\omega}_\sigma^\star,{\omega}_\delta^\star)}{\omega_a^\star}\right]^2
\|\underline{\tilde{\Phi}} \|^2
=\left[\frac{\omega_\theta({\omega}_\sigma^\star,{\omega}_\delta^\star)}{\omega_a^\star}\right]^2.
\end{align}
Por otra parte, si desarrollamos el coeficiente $B_a$
\begin{align}
 B_a &= -\frac{i\bar\alpha}{2} \langle \underline{\tilde{\Phi}}|
|\omega_\sigma+\omega_\delta|^{-1}\underline{\tilde{\Phi}}\rangle
-i\bar\alpha\langle \underline{\tilde{\Phi}}|\partial_{\omega_a}\underline{\tilde{\Phi}}\rangle,
\end{align}
podemos despreciar el primer término en el límite $\sigma_a\to 0$, pues en él solo $\omega_a^\star$
da una contribución apreciable, pero su valor es muy grande, así que
$|\omega_\sigma^\star+\omega_\delta^\star|^{-1}$ es despreciable. Asimismo, en dicho límite
$\partial_{\omega_a}\underline{\tilde{\Phi}}\simeq
i\nu^a\underline{\tilde{\Phi}}$. En definitiva, obtenemos
\begin{align}\label{eq:gauss-traj}
\langle \underline{\tilde{\Phi}} | \ln(\hat{v}_a)_{v_\theta}
\underline{\tilde{\Phi}} \rangle \simeq
\bigg[\frac{\omega_\theta(\omega_\sigma^\star,\omega_\delta^\star)}
{\omega_a^\star}\bigg]^2 \ln v_\theta + \bar\alpha\nu^a.
\end{align}
Este resultado coincide con el valor clásico de $\ln[v_a(v_\theta)]$. Recordemos que la singularidad
clásica corresponde a que alguna de las cantidades $|v_i|$ tome un valor nulo. A la vista de la
ec. \eqref{eq:gauss-traj}, un valor nulo de $v_\theta$ conlleva un valor
de $v_a$ también nulo. Esto implica que, en la teoría de
WDW, las singularidades del universo vacío de Bianchi I no desaparecen
dinámicamente, en lo que respecta a las trayectorias que definen los valores esperados de los
observables en estados semiclásicos. De hecho, este fracaso en evitar la singularidad es una
propiedad general de todos los estados para los cuales los coeficientes $A_a$ y $B_a$, definidos en
la ec. \eqref{eq:traj-dir}, son finitos.

Hemos analizado también el comportamiento de las dispersiones%
\footnote{Aquí usamos la notación abreviada $\langle\hat{O}\rangle=\langle
\underline{\tilde{\Phi}}| \hat{O}\,\underline{\tilde{\Phi}}\rangle$ para cualquier operador
$\hat{O}$.}
$\langle\Delta\ln(\hat{v}_a)_{v_\theta}\rangle$.
Para ello, primero hemos buscado los valores esperados de $[\ln(\hat{v}_a)_{v_\theta}]^2$ en un
estado
general, de un modo similar a la deducción de la ec. \eqref{eq:wdw-traj-shape}. Obtenemos:
\begin{equation}\label{eq:wdw-sqr-traj}
\langle \underline{\tilde{\Phi}} | [\ln(\hat{v}_a)_{v_\theta}]^2
\underline{\tilde{\Phi}} \rangle = W_a [\ln(v_\theta)]^2 + Y_a \ln
v_\theta+X_a ,
\end{equation}
donde
\begin{subequations}\label{eq:sqr-dir}\begin{align}
\label{eq:traj-dir-W} W_a &= \|
\omega_\theta^2({\omega}_\sigma,{\omega}_\delta){\omega}_a^{-2}
\underline{\tilde{\Phi}} \|^2 , \\\label{eq:traj-dir-Y} Y_a
&= -2 i \bar\alpha \langle \underline{\tilde{\Phi}} |
|\omega_\sigma+\omega_\delta|^{-\frac{1}{2}}
\omega_\theta({\omega}_\sigma,{\omega}_\delta)
{\omega}_a^{-1} (\partial_{\omega_a})
|\omega_\sigma+\omega_\delta|^{\frac{1}{2}}
\omega_\theta({\omega}_\sigma,{\omega}_\delta)
{\omega}_a^{-1} \underline{\tilde{\Phi}} \rangle,
\\\label{eq:traj-dir-X}
X_a &= - \bar\alpha^2\langle \underline{\tilde{\Phi}} |
|\omega_\sigma+\omega_\delta|^{-\frac{1}{2}} \partial_{\omega_a}^2
|\omega_\sigma+\omega_\delta|^{\frac{1}{2}} \underline{\tilde{\Phi}} \rangle .
\end{align}\end{subequations}
Recordamos que no estamos empleando el convenio de sumación de índices repetidos y, por tanto, en la
ec.~\eqref{eq:traj-dir-Y} no hay una suma implícita sobre los índices $a$. 

A partir de la relación estándar
\begin{align}
 \langle\Delta\ln(\hat{v}_a)_{v_\theta}\rangle^2 =
\langle[\ln(\hat{v}_a)_{v_\theta}]^2\rangle -
\langle\ln(\hat{v}_a)_{v_\theta}\rangle^2
\end{align}
se encuentran fácilmente las dispersiones. En particular, observamos que, si estas
dispersiones son finitas en alguna época de la evolución, permanecen finitas durante toda ella.
Además, para estados para los que los valores esperados $B_a$, $Y_a$ y $X_a$, definidos en las
ecs.~\eqref{eq:traj-dir} y \eqref{eq:sqr-dir}, son finitos, las dispersiones relativas alcanzan
valores constantes en el límite de $v_\theta$ grande, que, a su vez, están determinados por las
dispersiones relativas de $\omega_\theta^2({\omega}_\sigma,{\omega}_\delta){\omega}_a^{-2}$, esto
es,
\begin{align}\label{eq:wdw-rel-disp}
\lim_{v_\theta\to\infty} \frac{\langle
\Delta\ln(\hat{v}_a)_{v_\theta} \rangle}{\langle
\ln(\hat{v}_a)_{v_\theta} \rangle} = \frac{\langle \Delta
[\omega_\theta^2({\omega}_\sigma,{\omega}_\delta) {\omega}_a^{-2}]
\rangle}{\langle
\omega_\theta^2({\omega}_\sigma,{\omega}_\delta){\omega}_a^{-2}
\rangle}.
\end{align}

\section{Evolución con respecto a $\beta_\theta$}
\label{sec:WDW-brep}

En la cuantización polimérica, como ya hemos comentado y luego mostraremos, la aplicación directa de
la anterior construcción de la imagen de evolución con respecto al tiempo interno $v_\theta$ no da
lugar a una evolución unitaria. No obstante, el empleo del momento conjugado de $v_\theta$ como
tiempo interno sí proporciona una buena noción de evolución unitaria. Analizaremos esta elección de
tiempo interno también en la cuantización de WDW, tanto por completitud como
por presentar el procedimiento en un contexto más sencillo, donde las dificultades
inherentes a la naturaleza más complicada del modelo de la cuantización de lazos no están presentes.

Dada la variable clásica $v_i$, relacionada con $p_i$ mediante la ec. \eqref{v-p-rel}, podemos
introducir el momento conjugado%
\footnote{De acuerdo con esta definición, la variable $\beta_i$ difiere en un factor 2
de la variable $\beta$ introducida en el Capítulo \ref{chap:introLQC}. Lo introducimos así para
seguir el criterio de la Ref. \cite{mmp2}.}
\begin{equation}\label{eq:b-def}
\beta_i=\hbar\sqrt{\Delta} c^i |p_i|^{-\frac{1}{2}},
\end{equation}
tal que
\begin{equation}\label{eq:bv-poiss}
\{\beta_i,v_j\} = 2\delta_{ij}.
\end{equation}

En la teoría de WDW, la transformación unitaria entre las representaciones de
``posiciones'' y ``momentos'' está dada por la transformada de Fourier:
\begin{align}\label{eq:v-b-trans}
[{\underline{\mathcal{F}}}\psi](\beta_i) =
\frac1{2\sqrt{\pi}}\int_{\mathbb{R}} d v_i \,\psi(v_i)
e^{-\frac{i}{2}v_i\beta_i}.
\end{align}
Es de destacar que ésta es una tranformación desde el espacio de Hilbert de las funciones de
cuadrado integrable en la representación de $v_i$ al \emph{mismo espacio de Hilbert} en la
representación de $\beta_i$, es decir,
\begin{align}
\underline{\mathcal{F}}: \underline{\mathcal{H}}_{\text{grav}}^i=
L^2(\mathbb{R},d v_i)\to\underline{\tilde{\mathcal{H}}}_{\text{grav}}^i=
L^2(\mathbb{R},d \beta_i).
\end{align}

Bajo esta transformación, los operadores cinemáticos elementales se transforman como
\begin{align}\label{eq:ops-b}
\hat{v}_i\; \to\; 2i\partial_{\beta_i}, \qquad
\partial_{v_i}\; \to\; i\hat{\beta}_i/2,
\end{align}
donde el operador $\hat{\beta}_i$ actúa por multiplicación en la nueva representación. Por tanto,
el transformado del operador $\widehat{\underline\Omega}_i$ es
\begin{equation}\label{eq:Th-trans}
\underline{\mathcal{F}}(\widehat{\underline{\Omega}}_i)=
i\bar\alpha (1+2\beta_i\partial_{\beta_i}),
\end{equation}
que coincide con el operador original
$\widehat{\underline\Omega}_i$ (salvo por un signo), ya que los dos están definidos en espacios de
Hilbert idénticos. En consecuencia, en la teoría de WDW, trabajar en la representación de
$\beta_i$ es completamente equivalente a trabajar en la representación de $v_i$.
En particular, podemos interpretar $\beta_\theta$ como el tiempo interno y cambiar la representación
solo en el subsector direccional correspondiente a la dirección~$\theta$, de modo que pasemos a
trabajar en el espacio de Hilbert cinemático
$\underline{\tilde{\mathcal{H}}}_{\text{grav}}^{\theta,+}\otimes_a
\underline{\mathcal{H}}^{a,+}_{\text{grav}}$ ($a=\delta,\sigma$), donde
$\underline{\tilde{\mathcal{H}}}_{\text{grav}}^{\theta,+}=L^2(\mathbb{R}^+,d \beta_\theta)$. Nótese
que no hay problema en restringir el estudio también en la representación de $\beta_i$ al espacio
con $\beta_i>0$, pues podemos identificar la restricción a $L^2(\mathbb{R}^\pm,dv_i)$ con la
restricción a funciones de onda pares en $L^2(\mathbb{R},dv_i)$, por ejemplo, que son mapeadas por
la transformada de Fourier a funciones de onda pares en $L^2(\mathbb{R},d\beta_i)$. Entonces, en
dicho espacio de Hilbert $\underline{\tilde{\mathcal{H}}}_{\text{grav}}^{\theta,+}\otimes_a
\underline{\mathcal{H}}^{a,+}_{\text{grav}}$ podemos definir el operador ligadura hamiltoniana 
reemplazando el operador $\widehat{\underline\Omega}_\theta$ por el operador
$\underline{\mathcal{F}}(\widehat{\underline{\Omega}}_\theta)$ en la ec.~\eqref{CWDW} y
podemos repetir exactamente la construcción de la Sección \ref{sec:WDWphys-evo}
sustituyendo $v_\theta$ por $\beta_\theta$ y
$\underline{e}_{\omega_\theta}(v_\theta)$ por $\underline{e}_{-\omega_\theta}(\beta_\theta)$.

Mientras que la representación de $\beta_i$ no introduce ninguna novedad o ventaja en la
cuantización de WDW en comparación con la representación de $v_i$, veremos en el
capítulo siguiente que hay una gran diferencia entre ambos procedimientos en la teoría de
cosmología cuántica de lazos.

\section[Límite de WDW de la cuantización de lazos]{Límite de WDW de la
cosmología cuántica de lazos}
\label{sec:wdw-limit}

\subsection{Imagen de dispersión}

La comparación de las ecs.~\eqref{physpaWDW} y \eqref{phystawdw} con las
ecs.~\eqref{eq:phys-A} y \eqref{sol-fis-A} muestra que, en este modelo, los espacios de Hilbert
físicos de la cosmología cuántica de lazos y de la teoría de WDW, así como la estructura
de las correspondondientes funciones de onda, son idénticas. La diferencia entre ambas
cuantizaciones reside en la distinta forma que poseen las autofunciones de los operadores 
$\widehat{\Omega}_i$ y $\widehat{\underline{\Omega}}_i$. Por otra parte, ya hemos visto en la
Sección \ref{3subsec:wdw}, en el modelo de FRW, que las autofunciones del operador polimérico
$\widehat{\Omega}^2$ convergen para valores grandes de $v$ a una combinación de las
autofunciones del operador análogo $\widehat{\underline{\Omega}}$ de la teoría de WDW,
que además se pueden interpretar como ondas entrantes y salientes. Recordamos, además, que estos
operadores coinciden esencialmente con los del modelo de Bianchi I presente.

Esta característica nos permite interpretar la dinámica del
universo de la cosmología cuántica de lazos como un cierto proceso de \emph{dispersión} entre
estados de la cuantización de WDW, \emph{entrantes} desde un
pasado lejano y \emph{salientes} a un futuro distante \cite{kp-posL}. Matemáticamente este
comportamiento se describe con el análogo a una matriz de {dispersión} o de \emph{colisión}
$\hat{\rho}_{s}$ actuando sobre el estado inicial entrante para dar lugar al
estado final o saliente, es decir,
\begin{subequations}\begin{align}
|\underline{\Phi}\rangle_{\rm fin} &= \hat{\rho}_{s}
|\underline{\Phi}\rangle_{\rm in},\\
\langle\underline{e}_{\omega_\sigma},\underline{e}_{\omega_\delta}|\hat{\rho}_{s}|
\underline{e}_{\omega'_\sigma},\underline{e}_{\omega'_\delta}\rangle
&= \hat{\rho}_{\theta}(\omega_\theta(\omega_\sigma,\omega_\delta),
\omega_\theta(\omega'_\sigma,\omega'_\delta))\hat{\rho}_{\sigma}(\omega_\sigma,\omega'_\sigma)
\hat{\rho}_{\delta}(\omega_\delta,\omega'_\delta),
\\
\hat{\rho}_{i}(\omega_i,\omega'_i) &:=
\langle\underline{e}_{\omega_i}|\hat{\rho}_{i}
|\underline{e}_{\omega'_i}\rangle.\label{eq:scatt-def}
\end{align}\end{subequations}

A su vez, las matrices $\hat{\rho}_{i}$ están determinadas por el límite de WDW de las
autofunciones $e^{\varepsilon_i}_{\omega_i}(v_i)$ de $\widehat{\Omega}_i$. Éstas han sido analizadas
en la Sección \ref{3sec:op-dens-lig-A}. Como ya se ha discutido en dicha sección, el soporte de
estas autofunciones se puede dividir en dos subsemirredes de paso cuatro
${}^{(4)}\!\mathcal{L}_{\tilde{\varepsilon}_i}^+$, con
$\tilde{\varepsilon}_i\in\{\varepsilon_i,\varepsilon_i+2\}$, de modo que la restricción de 
$e^{\varepsilon_i}_{\omega_i}(v_i)$ a cada subsemirred es una autofunción del operador 
$\widehat{\Omega}_i^2$ con autovalor $\omega_i^2$, es decir%
\footnote{Esta vez, por simplificar la notación, eliminaremos el superíndice $(4)$ de las
autofunciones de
$\widehat\Omega_i^2$. Recordamos que la tilde de la etiqueta $\tilde\varepsilon_i$ ya denota que
tienen soporte en redes de paso 4.},
\begin{align}\label{eq:eig-eq-restr}
\widehat{\Omega}_i^2
e^{\tilde{\varepsilon}_i}_{\omega_i}(v_i) = \omega_i^2
e^{\tilde{\varepsilon}_i}_{\omega_i}(v_i), \quad
e^{\tilde{\varepsilon}_i}_{\omega_i}(v_i) :=
e^{\varepsilon_i}_{\omega_i}(v_i)|_{v_i\in{}^{(4)}\!
\mathcal{L}_{\tilde{\varepsilon}_i}^{+}}.
\end{align}

Como la expresión de las autofunciones del modelo de FRW es la misma
que las del modelo de Bianchi I en el esquema A, los resultados descritos para el modelo de FRW
se aplican en este contexto también. En particular, podemos afirmar que las restricciones
$e^{\tilde{\varepsilon}_i}_{\omega_i}(v_i)$ tienen un límite de WDW (límite
$v_i\to\infty$) de la forma [véase la ec.~\eqref{wdw-limit-mmo}]:
\begin{align}\label{wdw-limit-bianchi}
e^{\tilde{\varepsilon}_i}_{\omega_i}(v_i)\to
\underline{e}^{\tilde{\varepsilon}_i}_{\omega_i}(v_i)=r(\omega_i)[
e^{i\phi(\omega_i)}\, \underline{e}_{\omega_i}(v_i) +
e^{-i\phi(\omega_i)}\, \underline{e}_{-\omega_i}(v_i)],
\end{align}
donde la fase $\phi(\omega_i)$ tiene el comportamiento dado en la ec.~\eqref{alpha-form}, es decir
\begin{equation}\label{alpha-form-bianchi}
\phi(\omega_i) = T(|\omega_i|) +
c_{\tilde{\varepsilon_i}} +
R_{\tilde{\varepsilon_i}}(|\omega_i|).
\end{equation}
Se ha comprobado numéricamente \cite{kp-posL}, además, que la función $T(|\omega_i|)$ es de la forma
\begin{align}\label{T-phase}
 T(|\omega_i|)=(\ln|\omega_i|+a)(|\omega_i|+b),
\end{align}
siendo $a$ y $b$ constantes.

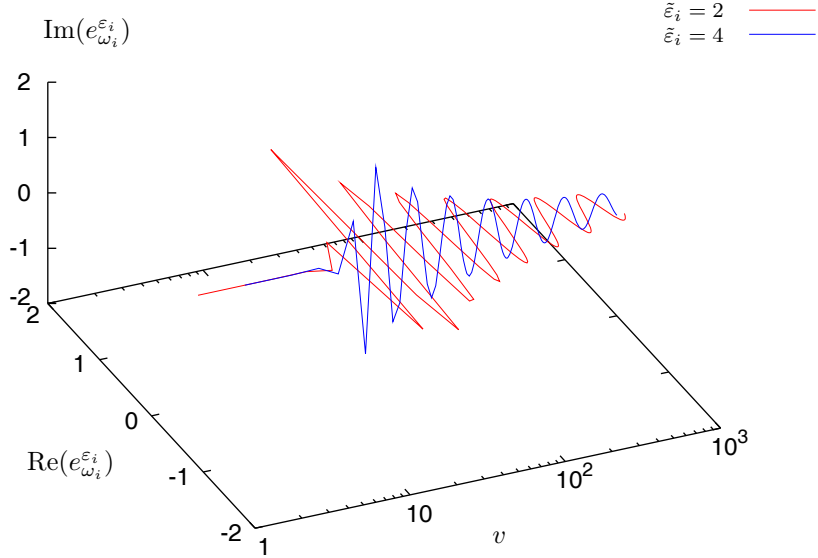
\begin{figure}[htb!]
\begin{center}
\begin{overpic}
[width=0.7\textwidth]{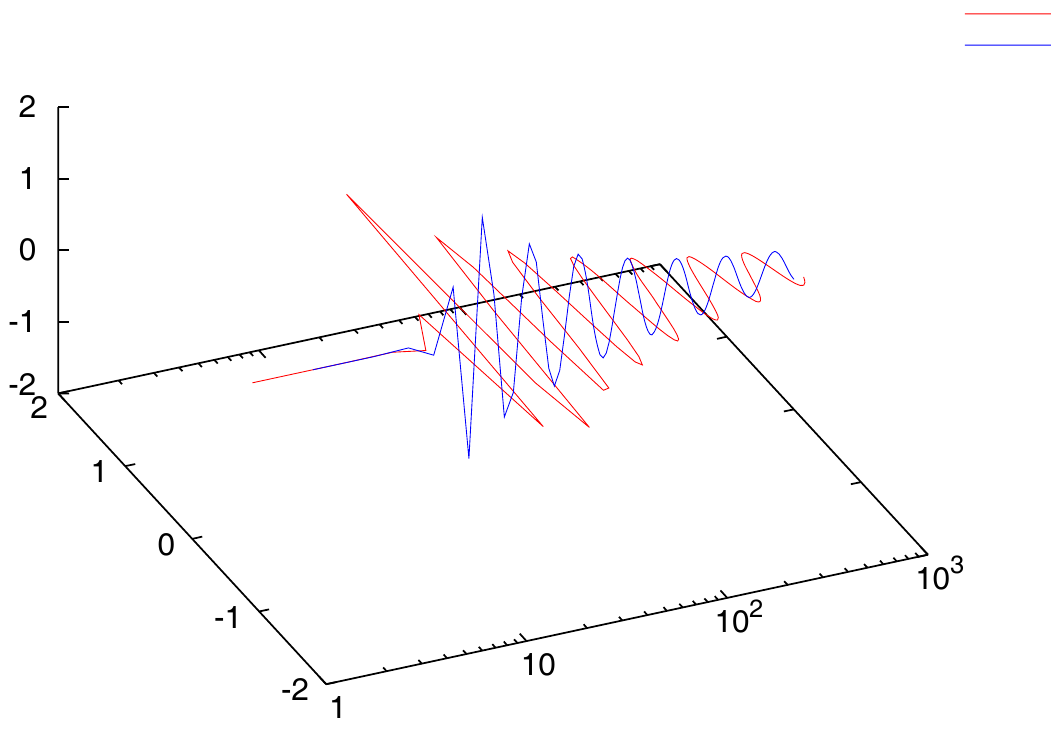}
\put(60,3){\footnotesize{$v$}}
\put(3,12){\footnotesize{$\text{Re}(e^{\varepsilon_i}_{\omega_i})$}}
\put(5,65){\footnotesize{$\text{Im}(e^{\varepsilon_i}_{\omega_i})$}}
\put(81,64.5){\scriptsize{$\tilde\varepsilon_i=4$}}
\put(81,67.5){\scriptsize{$\tilde\varepsilon_i=2$}}
\end{overpic}
\caption{Un ejemplo de autofunción del operador $\widehat{\Omega}_i$, correspondiente
al autovalor generalizado $\omega_i=100$ y al sector de superselección determinado por
$\varepsilon_i=2$. La línea roja (situada en el plano real) es la componente con soporte en
${}^{(4)}\mathcal{L}_{\tilde{\varepsilon}_i=2}^+$, mientras que la línea
azul (situada en el plano imaginario) muestra la componente con soporte
en la semirred ${}^{(4)}\mathcal{L}_{\tilde{\varepsilon}_i=4}^+$.}
\label{fig:eigenf}
\end{center}
\end{figure}

Llegados a este punto, ya podemos escribir la forma exacta de la matriz de colisión definida para
una de las subsemirredes de paso cuatro. De acuerdo con su definición \eqref{eq:scatt-def} y la
expresión \eqref{wdw-limit-bianchi}, se tiene
\begin{align}
\hat{\rho}_i(\omega_i,\omega'_i) = e^{-2i\phi(\omega_i)}
\delta(\omega_i+\omega'_i) .
\end{align}
Recordemos que, para la autofunción completa $e^{\varepsilon_i}_{\omega_i}(v_i)$ con soporte en
la semirred de paso dos $\mathcal{L}^+_{\varepsilon_i}$, cada una de sus rectricciones a las
subsemirredes de paso cuatro tiene una fase constante, y el desfase entre ambas componentes es
igual a $\pi/2$. En particular, se ha elegido la componente con soporte en
${}^{(4)}\mathcal{L}_{\tilde\varepsilon_i={\varepsilon}_i}$ de modo que es real.
En consecuencia, el límite de WDW global de la unión de ambas componentes no existe, como se
ve en la Figura \ref{fig:eigenf}.

Por tanto, no podemos escribir una expresión explícita para la matriz de colisión en la red
completa $\mathcal{L}^+_{\varepsilon_i}={}^{(4)}\mathcal{L}_{{\varepsilon}_i}^+\cup
{}^{(4)}\mathcal{L}_{{\varepsilon}_i+2}^+$. No obstante, las diferencias entre subsemirredes se
manifiestan solo mediante el desfase constante que existe entre ellas y el término 
$R_{\tilde\varepsilon_i}(|\omega_i|)$ de la ec.~\eqref{alpha-form-bianchi} que decae cuando
$\omega_i\to\infty$. Nosotros hemos considerado las propiedades de paquetes de onda asintóticamente
picados en valores grandes de $\omega_a$ y, entonces, hemos podido restringir, sin problema alguno,
el estudio a una subsemirred cualquiera.

\subsubsection{Factor $r(\omega_i)$}

Con el fin de calcular el límite de WDW de las soluciones físicas, dadas en la
ec.~\eqref{sol-fis-A}, a partir del límite de WDW \eqref{wdw-limit-bianchi} de las
autofunciones de la base del espacio de Hilbert, primero hemos tenido que calcular el valor preciso
del factor $r(\omega_i)$.

Por una parte, a partir de la ec.~\eqref{eq:estados-2}, se deduce que la
relación entre la autofunción $e^{\varepsilon_i}_{\omega_i}(v_i)$ y sus restricciones
$e^{\tilde\varepsilon_i}_{\omega_i}(v_i)$, definidas en la ec.~\eqref{eq:eig-eq-restr}, es
\begin{align}\label{eq:estados-2-bis}
 e^{\tilde\varepsilon_i}_{\omega_i}(v_i)=\frac{1}{2}\left[e^{\varepsilon_i}_{
\omega_i}(v_i)\pm e^{\varepsilon_i}_{-\omega_i}(v_i)\right)],
\end{align}
donde el signo $+$ corresponde a $\tilde{\varepsilon}_i\leq2$ y el signo $-$ a
$\tilde{\varepsilon}_i>2$. Como los estados $|e^{\varepsilon_i}_{
\omega_i}\rangle$ están ortonormalizados (a la delta de Dirac) y cada una de sus componentes
contribuye por igual a su norma, se tiene
\begin{align}\label{norm-rest}
\langle
e^{\tilde\varepsilon_i}_{\omega_i}|e^{\tilde\varepsilon_i}_{\omega'_i}\rangle=\frac{1}{2}
\delta(\omega_i-\omega'_i). 
\end{align}
Por otra parte, se demuestra que la relación entre la norma de
$|e^{\tilde\varepsilon_i}_{\omega_i}\rangle$,
\begin{align}
 \|e^{\tilde\varepsilon_i}_{\omega_i}\|^2=
\sum_{v_i\in\mathcal{L}^+_{\varepsilon_i}={}^{(4)}\mathcal{L
}_{{\tilde\varepsilon}_i}^+}\;[e^{\tilde\varepsilon_i}_{\omega_i}(v_i)]^*\;e^{\tilde\varepsilon_i}_{
\omega_i}(v_i),
\end{align}
y la norma de su límite de WDW
$|\underline{e}^{\tilde\varepsilon_i}_{\omega_i}\rangle$,
\begin{align}
 \|\underline{e}^{\tilde\varepsilon_i}_{\omega_i}\|_{\text{WDW}}^2=\int_{\mathbb{R}^+}dv_i\;
\;[\underline{e}^{\tilde\varepsilon_i}_{\omega_i}(v_i)]^*\;\underline{e}^{\tilde\varepsilon_i}_{
\omega_i}(v_i),
\end{align}
viene dada por \cite{kp-posL}
\begin{align}\label{norm-relat}
 \|e^{\tilde\varepsilon_i}_{\omega_i}\|^2=\frac1{8}\|\underline{e}^{\tilde\varepsilon_i}_{\omega_i}
\|^2_\text{WDW}.
\end{align}
Veamos cómo se demuestra este resultado:
\begin{itemize}
 \item Se escoge algún punto $v_i^\star>0$. Debido a que los coeficientes $e^{\tilde\varepsilon_i}_{
\omega_i}(v_i)$ son finitos y al grado de convergencia al límite de WDW, dado en la
ec. \eqref{eq:conv-rate}, podemos escribir
\begin{align}
 \langle e^{\tilde\varepsilon_i}_{\omega_i}| e^{\tilde\varepsilon_i}_{\omega'_i}\rangle=
O_1(\omega_i,\omega'_i)+
\sum_{v_i^\star<v_i\in{}^{(4)}\mathcal{L
}_{{\tilde\varepsilon}_i}^+}\;[\underline{e}^{\tilde\varepsilon_i}_{\omega_i}(v_i)]^*\;
\underline{e}^{\tilde\varepsilon_i}_{\omega'_i}(v_i),
\end{align}
donde $O_1(\omega_i,\omega'_i)$ es una función finita. Además, salvo por otro término finito
$O_2(\omega_i,\omega'_i)$, se puede reemplazar el anterior sumatorio por la integral
\begin{align}
O_3(\omega_i,\omega'_i)=\frac1{4}\int_{v_i>v_i^\star}dv_i\;
\;[\underline{e}^{\tilde\varepsilon_i}_{\omega_i}(v_i)]^*\;\underline{e}^{\tilde\varepsilon_i}_{
\omega'_i}(v_i),
\end{align}
de modo que $\langle e^{\tilde\varepsilon_i}_{\omega_i}|
e^{\tilde\varepsilon_i}_{\omega'_i}\rangle=O_1(\omega_i,\omega'_i)+O_2(\omega_i,
\omega'_i)+O_3(\omega_i ,\omega'_i)$.
Nótese que el factor $1/4$ da cuenta de que los coeficientes del anterior sumatorio tienen soporte
en redes de paso 4. Haciendo el cambio de variable $x=\ln(v_i)$ y teniendo en cuenta la forma de
las autofunciones $\underline{e}^{\tilde\varepsilon_i}_{\omega_i}(v_i)$, dada en la
ec.~\eqref{eq:WDW-eig}, la integral $O_3(\omega_i,\omega'_i)$ es esencialmente la integral
sobre la semirrecta real positiva de $\exp[i(\omega_i-\omega'_i)x]$. Empleando la relación
\begin{align}
 \int_{\mathbb{R}^+}dx
e^{i(\omega_i-\omega'_i)x}=\frac{N}{2}\delta(\omega_i-\omega'_i)+f(\omega_i,\omega'_i),
\end{align}
siendo $N$ un factor de normalización y $f(\omega_i,\omega'_i)$ una función conocida pero
irrelevante,
se llega a
\begin{align}
 \langle
e^{\tilde\varepsilon_i}_{\omega_i}|e^{\tilde\varepsilon_i}_{\omega'_i}\rangle=O_1(\omega_i,
\omega'_i)+O_2(\omega_i,\omega'_i)+\frac1{8}\langle\underline{e}
^{\tilde\varepsilon_i}_{\omega_i}|\underline{e}
^{\tilde\varepsilon_i}_{\omega'_i}\rangle_{\text{WDW}}+O_4(\omega_i,\omega'_i),
\end{align}
donde $O_4(\omega_i,\omega'_i)$ es otra función finita. Al tener en cuenta
que las autofunciones involucradas son ortogonales, finalmente deducimos que los términos finitos
se deben compensar, es decir, que
$\langle
e^{\tilde\varepsilon_i}_{\omega_i}|e^{\tilde\varepsilon_i}_{\omega'_i}\rangle=1/8\;\langle
\underline{e}^{\tilde\varepsilon_i}_{\omega_i}|\underline{e}^{\tilde\varepsilon_i}_{\omega_i}
\rangle_{\text{WDW}}$, como queríamos demostrar.
\end{itemize}

Además, la forma explícita del límite
$\underline{e}^{\tilde\varepsilon_i}_{\omega_i}$, definido en la ec. \eqref{wdw-limit-bianchi},
implica que
$\|\underline{e}^{\tilde\varepsilon_i}_{\omega_i}\|^2_{\text{WDW}}=2|r(\omega_i)|^2\|\underline{e}
_{\omega_i}\|^2_\text{WDW}$.
Con
este resultado y las relaciones \eqref{norm-rest} y \eqref{norm-relat}, finalmente obtenemos
\begin{equation}\label{eq:r}
  r(\omega_i)=\sqrt{2}z_i.
\end{equation}
En esta expresión, $z_i$ es una fase global que, de acuerdo con las ecs. \eqref{eq:estados-2} y
\eqref{eq:eig-eq-restr}, tomamos igual a $1$ para $\tilde{\varepsilon}_i\leq2$ e igual a
$-i\,\text{sgn}(\omega_i)$ si $\tilde{\varepsilon}_i>2$. 

\subsection[Límite de los estados físicos poliméricos]{Límite de
WDW de los estados físicos de la cuantización polimérica}
\label{scaterring-scheme}

Como hemos comentado, nosotros hemos aplicado el límite de WDW analizado en la sección
anterior en el análisis de los estados físicos. Veámoslo.

Para ello partimos de un estado general $\tilde{\Phi}(\omega_\delta,\omega_\delta)$ con
función de onda $\Phi(\vec v)$ dada en la ec. \eqref{sol-fis-A} y calculamos su límite de
WDW. Tenemos que restringir el estudio para cada dirección fiducial a las
subsemirredes de paso 4, ya que en las de paso 2 no existe un límite global.
Entonces, reemplazando en $\Phi(\vec v)$ las funciones básicas
$e^{{\varepsilon}_i}_{\omega_i}(v_i)$ por los límites
$\underline{e}^{\tilde{\varepsilon}_i}_{\omega_i}(v_i)$ de sus restricciones, encontramos el límite
de WDW,
$\underline{\Phi}(\vec{v})$, de la restricción de la función $\Phi(\vec{v})$ al sector
correspondiente. El resultado es
\begin{align}\label{eq:lim-split}
\underline{\Phi}(\vec{v})=\sum_{s_\sigma,s_\delta=\pm 1} \,\, &
\int_{\mathbb{R}^2} d\omega_\sigma d\omega_\delta\,
\tilde{\Phi}_{\vec{s}}(\omega_\sigma,\omega_\delta)
\underline{e}_{\omega_{\vec{s}}(\omega_\sigma,\omega_\delta)}(v_\theta) \,
\underline{e}_{\omega_\sigma}(v_\sigma) \, \underline{e}_{\omega_\delta}(v_\delta),
\end{align}
donde $\vec{s}=(s_\sigma,s_\delta)$,
\begin{align}
 \omega_{\vec{s}}(\omega_\sigma,\omega_\delta)=\omega_\theta(s_\sigma\omega_\sigma,
s_\delta\omega_\delta)=-\frac{s_\sigma s_\delta
\omega_\sigma\omega_\delta}{s_\sigma\omega_\sigma+s_\delta\omega_\delta}
\end{align}
y
\begin{align}
\tilde{\Phi}_{\vec{s}}(\omega_\sigma,\omega_\delta)&=
2\sqrt{2}\sum_{s_\theta=\pm 1}\tilde{\Phi}
(s_\theta s_\sigma\omega_\sigma,s_\theta s_\delta\omega_\delta)
e^{is_\theta\phi(\omega_{s_\theta\vec{s}})}
s_\theta^{({3-z_\theta^2-z_\sigma^2-z_\delta^2})/{2}}z_\theta\nonumber\\
&\times\prod_{a=\sigma}^\delta s_a^{({1-z_a^2})/{2}}
z_a e^{is_{1}s_a\phi(s_\theta s_a\omega_a)}.
\end{align}
Nótese que, para obtener esta expresión, es necesario tener en cuenta que 
\[\omega_\theta(-\omega_\sigma,-\omega_\delta)
= -\omega_\theta(\omega_\sigma,\omega_\delta)\] y que
\[r(s_j\omega_i)=s_j^{(1-z_i^2)/2}\sqrt{2}z_i\quad
(\text{con }\omega_\theta=\omega_{\vec{s}}).\] Recordamos que 
$z_i^2$ es igual a 1 para $\tilde{\varepsilon}_i=\varepsilon_i\leq2$ y $-1$ si 
$\tilde{\varepsilon}_i=\varepsilon_i+2>2$.

Si comparamos la ec. \eqref{eq:lim-split} con la ec. \eqref{phystawdw}, vemos que cada término
correspondiente a un conjunto particular de valores de $s_a$ tiene una forma muy similar a la de
un  estado de la teoría de WDW con perfil espectral 
$\underline{\tilde{\Phi}}=\tilde{\Phi}_{\vec{s}}$. La única diferencia es el reemplazo de la
función $\omega_\theta$ por la función modificada $\omega_{\vec{s}}$ en el subíndice de las
funciones básicas. En consecuencia, cada uno de estos términos puede ser considerado de forma
independiente como un estado definido por cierto análogo de la cuantización de WDW
descrita en la Sección~\ref{sec:wdw-analog}. Tal análogo hereda todas las propiedades y estructura
de la teoría de WDW original, excepto por la
transformación $\omega_\theta\to\omega_{\vec{s}}$ señalada antes.
En particular, el producto interno y la construcción de los observables 
permanecen inalterados.

Esta correspondencia nos ha permitido aplicar directamente la definición \eqref{obs} de los
observables $\ln(\hat{v}_a)_{v_\theta}$ a los términos $\tilde{\Phi}_{\vec{s}}$.
Sus valores
esperados, calculados en cada término independientemente, satisfacen un análogo de la
ec.~\eqref{eq:wdw-traj-shape} con el coeficiente direccional $A_a$ sustituido por
\begin{align}
A_{a,\vec{s}}:=s_a\|\omega_{\vec{s}}({\omega}_\sigma,{\omega}_\delta){\omega}_a^{-1}
\underline{\tilde{\Phi}} \|^2.
\end{align}
Para seleccionar los términos $\tilde{\Phi}_{\vec{s}}$ que realmente contribuyen al límite
investigado, hemos considerado que el estado está localizado, en el sentido de que
permanece picado a lo largo de las trayectorias definidas por los valores esperados de los
observables considerados. Entonces, en el régimen en el que las tres variables $v_i$ toman valores
grandes, solo contribuyen aquellos términos que corresponden a una combinación $\vec{s}$ para la
que cada
$A_{a,\vec{s}}$ es estrictamente positivo. Solo el término con $s_\sigma=s_\delta=1$ satisface este
requerimiento.

En consecuencia, el único término que contribuye apreciablemente se corresponde precisamente con el
estado de la teoría original de WDW. Por tanto, el límite de $v_i$ grande de un estado
localizado $\tilde{\Phi}$ está simplemente dado por un estado de WDW con perfil espectral
\begin{align}\label{eq:wdw-limit-phys}
\underline{\tilde{\Phi}}(\omega_\sigma,\omega_\delta)=
2\sqrt{2}\sum_{s=\pm 1}
s^{\frac{3-z_\theta^2-z_\sigma^2-z_\delta^2}{2}}
\tilde{\Phi}(s\omega_\sigma,s\omega_\delta)\;\prod_{i=\theta,\sigma,\delta} z_i \,
e^{is\phi(s\omega_i)},
\end{align}
donde $\omega_\theta$ está otra vez relacionada con $\omega_\sigma$ y
$\omega_\delta$ mediante la ec.~\eqref{eq:om-func}. 

Recordemos que la variable $v_\theta$ desempeña la función de tiempo. Por tanto, es apropiado
introducir una descomposición del estado en partes de frecuencia positiva y negativa con respecto a
ese tiempo, que se
corresponden con $\omega_\theta>0$ y $\omega_\theta<0$ respectivamente. Físicamente, estas
componentes se pueden interpretar como componentes que se mueven hacia adelante y hacia atrás en el
tiempo $v_\theta$. Se puede aplicar una correspondencia análoga en las otras direcciones
$a=\sigma,\delta$,
definiendo la descomposición en componentes entrantes y salientes, o en componentes que se contraen
y se expanden, respectivamente (como la interpretación hecha en el modelo de FRW). Entonces, el
cambio de signo de $\omega_a$ corresponde a una transformación de paridad o a una reflexión de la
variable $x_a=\ln(v_a)$ [recuérdese la expresión \eqref{eq:WDW-eig}]. En consecuencia, la
transformación $\{\omega_i\}\to\{-\omega_i\}$, la cual es de simetría de acuerdo con la discusión
anterior, es análoga a la transformación PT usual (paridad/inversión temporal). 

A partir de la ec. \eqref{eq:wdw-limit-phys} vemos inmediatamente que el límite de WDW
de un estado $\tilde{\Phi}(\omega_\sigma,\omega_\delta)$ localizado (en el sentido explicado
anteriormente)
está formado por dos componentes: la que corresponde a
$s=1$, que tiene la misma paridad y orientación temporal que el estado original
$\tilde{\Phi}$, y otra
($s=-1$) que está reflejada, es decir, que tiene paridad y orientación temporal inversas. Las
denotaremos respectivamente por $\underline{\tilde{\Phi}}_+(\omega_\sigma,\omega_\delta)$ y
$\underline{\tilde{\Phi}}_-(\omega_\sigma,\omega_\delta)$. Si el estado original $\tilde{\Phi}$
tiene una orientación temporal definida (solo un signo de $\omega_\theta$ contribuye), entonces
estas dos componentes son, respectivamente, la parte que tiene un movimiento temporal como
el del estado original y la parte reflejada temporalmente. Como ambas componentes tienen igual
norma, cualquier paquete de ondas que se mueve hacia atrás en el tiempo es completamente
reflejado en un paquete de ondas que se mueve hacia adelante en el tiempo. Este resultado, en
completa analogía con lo ya discutido para el modelo de FRW, muestra un rebote cuántico en la
dirección temporal.

\section{Discusión}

En este capítulo, hemos analizado la imagen de evolución en un escenario cosmológico
cuántico. Como hemos visto, en este contexto y, en general, en cualquier
teoría totalmente constreñida, como es la relatividad general, el concepto de evolución trasciende
el concepto que se emplea habitualmente en las teorías que introducen un tiempo
coordenado fijo. La imagen de evolución ahora no está prefijada. De hecho, es una mera
\emph{herramienta} que construimos para darle una \emph{interpretación} a la teoría en términos del
lenguaje usual al que estamos acostumbrados desde nuestro sistema de laboratorio, en el que ideamos
\emph{medidas} para calcular \emph{el valor}
que toman ciertos observables en un \emph{tiempo} concreto, en definitiva, para poder extraer
predicciones físicas de nuestra teoría.

En particular, hemos desarrollado este escenario de evolución en el ejemplo concreto del modelo de
Bianchi I en vacío cuantizado según la teoría de WDW. Debido a la representación
estándar de Schr\"odinger que se implementa en esta cuantización, es sencillo construir una imagen
de evolución unitaria y, por tanto, satisfactoria, en la que los observables que miden la misma
magnitud física, pero que están evaluados a diferentes tiempos, están relacionados unitariamente.
Esta
sencillez radica en la forma que tienen las autofunciones del operador que identificamos como el
generador de la evolución temporal. En el modelo presente, ese operador es
$\underline{\widehat{\Omega}}_\theta$ y los coeficientes a tiempo constante de sus autofunciones son
esencialmente unitarios, salvo por un prefactor constante que no depende de la frecuencia
$\omega_\theta$ y que, por tanto, se puede absorber en la normalización. Es esta unitariedad la que
luego hereda la
imagen de evolución en su conjunto.

Ya lo habíamos visto también en el contexto del modelo de FRW acoplado a un campo escalar $\phi$, en
el cual las autofunciones del generador de la evolución, $-i\partial_\phi$, son también unitarias en
cada punto de la evolución.

Una vez definida la evolución y obtenidos los observables físicos parametrizados por el
tiempo interno, se pueden hacer, como he dicho, predicciones físicas. En el modelo concreto
estudiado en este capítulo, hemos \emph{medido} la evolución en el tiempo $v_\theta$ de los
observables $\ln(v_\sigma)$ y $\ln(v_\delta)$, que están relacionados con los factores de escala
$a_i$ del espacio-tiempo de Bianchi I mediante la relación $|v_i|^{2/3}\propto|a_ja_k|$. Por tanto,
los valores nulos de alguna de estas variables $v_i$ se corresponden con los valores nulos de los
factores de escala o, equivalentemente, con la presencia de una singularidad clásica de curvatura,
de tipo \emph{big bang}. Hemos analizado las predicciones de la cuantización de WDW
en estados de interés físico, esto es, en estados que están picados sobre las trayectorias clásicas
en las regiones donde se espera que los efectos cuánticos no sean importantes. El resultado es que
la dinámica efectiva correspondiente a estos estados cuánticos de la teoría de WDW no
solo concuerda con la relatividad general
en dichas regiones semiclásicas, como es deseable, sino que coincide con ella a lo largo de toda la
evolución. Por tanto, esta cuantización presenta los mismos inconvenientes que la relatividad
general, esto es, no resuelve el problema de las singularidades.

En el caso de la cuantización polimérica, si aplicamos directamente la construcción de la evolución
desarrollada
en este capítulo, la unitariedad se pierde. Aún antes de haberla estudiado ya podemos entender por
qué. Precisamente porque los coeficientes de las autofunciones del operador análogo que
interpretamos como generador de la evolución, ${\widehat{\Omega}}_\theta$, ya no tienen norma
compleja unidad
para todo valor de $v_\theta$. En el
capítulo siguiente, veremos cómo hay que modificar la construcción para poder dar una interpretación
en una imagen de evolución bien definida.

Por otra parte, en este capítulo también hemos mostrado la relación que existe entre la cuantización
de WDW del modelo y su cuantización polimérica. En cierto sentido, podemos interpretar el
comportamiento del sistema como un proceso de dispersión. En particular, los estados físicos de la
cosmología cuántica de lazos convergen a una combinación de componentes de la teoría de
WDW, que con respecto al tiempo $v_\theta$ (o, equivalentemente, $\ln(v_\theta)$) se
pueden
interpretar como componentes de frecuencia positiva, que se mueven hacia adelante en el tiempo, y
componentes de frecuencia negativa, que se mueven hacia atrás en el tiempo. Con respecto al
``espacio'' (a $\ln(v_a)$ más precisamente), estas componentes se pueden interpretar como
componentes salientes o en expansión y entrantes o en contracción.
Además, en términos de esta interpretación hemos
identificado una simetría de paridad/inversión temporal en el sistema. Este comportamiento de los
estados físicos de la cuantización polimérica ya muestra la existencia de un rebote cuántico. Lo
veremos explícitamente en el siguiente capítulo.

\cleardoublepage

\chapter[Evolución en cosmología cuántica de lazos]{Evolución física en cosmología cuántica de
lazos}
\label{chap:evolutionLQC}

En este capítulo discutimos la dinámica cuántica del modelo de
Bianchi I en vacío cuantizado poliméricamente y según el esquema~A de la dinámica mejorada.
Por lo que sabemos, hasta la fecha, éste es el único análisis que se ha llevado a cabo en
el marco de la cosmología cuántica de lazos en el que se define la evolución con respecto a alguno
de los grados de libertad geométricos. En este contexto, la imagen de evolución, ya de por sí nada
trivial en cosmología cuántica, es aún más intrincada, como consecuencia de la naturaleza polimérica
de la geometría, lo que también la hace más interesante.

De nuevo, hemos definido observables relacionales a partir de los operadores cinemáticos $\ln(\hat
v_a)$ ($a=\sigma,\delta$). Hemos analizado tanto la construcción en la que $v_\theta$ desempeña la
función de tiempo interno, como la construcción en la que el tiempo es su
momento conjugado $\beta_\theta$. 
En el capítulo anterior, hemos mostrado que, en la cuantización de WDW, ambas
construcciones son equivalentes. Además, los observables relacionales construidos admiten una
interpretación
física precisa. No obstante, como vamos a ver, en la cuantización polimérica ambas construcciones
son muy diferentes. Por otra parte, en ambos casos se debe introducir una modificación al
procedimiento seguido en la cuantización de WDW con el fin de garantizar la unitariedad
de la evolución resultante, pero esto se consigue a costa de perder una interpretación física
nítida,
que ya solo es válida en cierta aproximación. Sin embargo, hemos podido demostrar que en los
regímenes de interés físico las predicciones obtenidas, de hecho, son muy precisas.

\section[Evolución con respecto a $\upsilon_\theta$: unitariedad]{Evolución con respecto a
$v_\theta$: unitariedad}
\label{sec:v-evo}

En esta sección, mostraremos explícitamente que, en efecto, la aplicación directa de la construcción
del anterior capítulo ahora ya no es satisfactoria. Veremos cómo, gracias a la introducción de una
modificación en dicha construcción, motivada por la relación entre los estados físicos de la teoría
de WDW y los de la cuantización polimérica, se consigue restaurar la unitariedad.

\subsection{Espacios de tiempo constante}
\label{sec:vHil}

Partimos de la forma general de la función de onda que representa a los estados físicos, dada por la
ec. \eqref{sol-fis-A}. Siguiendo exactamente el procedimiento explicado en la Sección
\ref{sec:WDWphys-evo}, las funciones de ``datos iniciales'' en cada sección
$v_\theta=\text{constante}$ se definen del modo
\begin{align}\label{eq:phi-chi}
\Phi_{v_\theta}(v_\sigma,v_\delta)\ =\ \int_{\mathbb{R}^2}
d\omega_\sigma d\omega_\delta
{|e^{\varepsilon_\theta}_{\omega_\theta}(v_\theta)|}
{\tilde{\Phi}'_{v_\theta}(\omega_\sigma,\omega_\delta)}
e^{\varepsilon_\sigma}_{\omega_\sigma}(v_\sigma)
e^{\varepsilon_\delta}_{\omega_\delta}(v_\delta) ,
\end{align}
donde los perfiles espectrales 
$\tilde{\Phi}'_{v_\theta}(\omega_\sigma,\omega_\delta)$ de
$\Phi_{v_\theta}(v_\sigma,v_\delta)$
están definidos por la transformación
\begin{align}\label{eq:chidef}\hat P'_{v_\theta}:\mathcal{H}_{\text{fis}}^{\vec\varepsilon}
\;&\to\;\mathcal{H}'_{v_\theta}=L^2(\mathbb{R}^2,
|\omega_\sigma+\omega_\delta|d\omega_\sigma d\omega_\delta),\nonumber\\
\tilde{\Phi}(\omega_\sigma,\omega_\delta)\;&\mapsto
\tilde{\Phi}'_{v_\theta}(\omega_\sigma,\omega_\delta)
= \tilde{\Phi}(\omega_\sigma,\omega_\delta)\frac{e^{\varepsilon_\theta}_{\omega_\theta}(v_\theta)}
{|e^{\varepsilon_\theta}_{\omega_\theta}(v_\theta)|}.
\end{align}
A diferencia de las autofunciones $\underline{e}_{\omega_\theta}(v_\theta)$ de la
cuantización de WDW, las autofunciones
$e^{\varepsilon_\theta}_{\omega_\theta}(v_\theta)$ tienen una fase constante en cada una de las
subsemirredes ${}^{(4)}\!\mathcal{L}_{\varepsilon_\theta}^+$ y
${}^{(4)}\!\mathcal{L}_{\varepsilon_\theta+2}^+$, con un desfase global de $\pi/2$ entre ellas. Por
tanto, el rescalado ${e^{\varepsilon_\theta}_{\omega_\theta}(v_\theta)}/
{|e^{\varepsilon_\theta}_{\omega_\theta}(v_\theta)|}$, que esencialmente extrae la fase de las
autofunciones, toma un valor constante en cada una de las subsemirredes. En consecuencia, la
aplicación 
$\hat P'_{v_\theta}$, en realidad, no introduce espacios definidos en secciones de tiempo
constante diferentes.

Si eliminamos  $|e^{\varepsilon_\theta}_{\omega_\theta}(v_\theta)|$ del mapeo, el anterior problema
desaparece. Introduzcamos por tanto la aplicación alternativa
\begin{align}\label{eq:chidef-alt}
\hat P_{v_\theta}:\mathcal{H}_{\text{fis}}^{\vec\varepsilon}
\;&\to\;\mathcal{H}_{v_\theta}=L^2(\mathbb{R}^2,
|\omega_\sigma+\omega_\delta||e^{\varepsilon_\theta}_{\omega_\theta}(v_\theta)|^{-2}d\omega_\sigma
d\omega_\delta),\nonumber\\
\tilde{\Phi}(\omega_\sigma,\omega_\delta)\;&\mapsto
\tilde{\Phi}_{v_\theta}(\omega_\sigma,\omega_\delta)
= \tilde{\Phi}(\omega_\sigma,\omega_\delta){e^{\varepsilon_\theta}_{\omega_\theta}(v_\theta)},
\end{align}
tal que
\begin{align}\label{eq:phi-chi-alt}
\Phi_{v_\theta}(v_\sigma,v_\delta)\ =\ \int_{\mathbb{R}^2}
d\omega_\sigma d\omega_\delta
{\tilde{\Phi}_{v_\theta}(\omega_\sigma,\omega_\delta)}
e^{\varepsilon_\sigma}_{\omega_\sigma}(v_\sigma)
e^{\varepsilon_\delta}_{\omega_\delta}(v_\delta) ,
\end{align}
Obviamente, ahora este mapeo depende explícitamente de $v_\theta$ y, de hecho, los espacios
$\mathcal{H}_{v_\theta}$ no coinciden para valores distintos de $v_\theta$ por la dependencia en
$|e^{\varepsilon_\theta}_{\omega_\theta}(v_\theta)|$ de la medida.
Dado que, en cada sección $v_\theta=\text{constante}$,
$e^{\varepsilon_\theta}_{\omega_\theta}(v_\theta)$ es una función de 
$\omega_\theta$ que solo se anula en un conjunto de medida nula, como se deduce de la
ec.~\eqref{eq:eigenstates-bianchi-A}, el mapeo $\hat P_{v_\theta}$ es unitario. Esta
propiedad, junto con el hecho de que tanto $e^{\varepsilon_\sigma}_{\omega_\sigma}(v_\sigma)$ como
$e^{\varepsilon_\delta}_{\omega_\delta}(v_\delta)$ forman bases de sus espacios de Hilbert
cinemáticos correspondientes $\mathcal H_{\varepsilon_a}^+$, permite determinar
$\tilde\Phi(\omega_\sigma,\omega_\delta)$ a partir de
$\Phi_{v_\theta}(v_\sigma,v_\delta)$ (salvo por un conjunto de medida nula). Como consecuencia, la
proyección en cada sección de $v_\theta$ constante contiene la misma información que la solución
física completa. 

Como en el anterior capítulo, la tranformación unitaria $\hat P_{v_\theta}$ nos permite definir un
mapeo entre espacios de datos iniciales. Cada estado del espacio de Hilbert físico
$\mathcal{H}_{\text{fis}}^{\vec\varepsilon}$ está asociado mediante $\hat P_{v_\theta}$ con una
secuencia de elementos de los espacios $\mathcal{H}_{v_\theta}$. Cada secuencia consiste en una
cadena de ``pasos en la evolución'' enumerados por $v_\theta\in\mathcal{L}_{\varepsilon_\theta}^+$.
Sin embargo, la evolución correspondiente tras la identificación de secciones temporales
diferentes no es unitaria, porque los datos iniciales que pertenecen a uno de esos
espacios $\mathcal{H}_{v_\theta}$, en general, no pertenecen a los otros, ya que no coinciden.

Por tanto, no es posible obtener una evolución unitaria en esta descripción. 
La falta de unitariedad también es heredada por la otra descripción más
elaborada, en términos de observables relacionales. Veámoslo.

\subsection{Observables relacionales}
\label{sec:pre-v-obs}

Una vez que hemos introducido los espacios de Hilbert $\mathcal{H}_{v_\theta}$, y las
transformaciones entre ellos y $\mathcal{H}_{\text{fis}}^{\vec\varepsilon}$, podemos seguir la
construcción de observables relacionales hecha en la Sección \ref{sec:WDWphys-evo}, partiendo de
los operadores cinemáticos $\ln(\hat{v}_a)$ ($a=\sigma,\delta$), que también actúan por
multiplicación en este
caso. Sin embargo, la transformación final entre $\mathcal{H}_{\text{fis}}^{\vec\varepsilon}$ y
$\mathcal{H}'$, que en la representación de $\omega_a$ viene dada por
\begin{align}
 \tilde\Phi(\omega_\sigma,\omega_\delta)\mapsto\boldsymbol{\chi}_{v_\theta}(\omega_\sigma,
\omega_\delta)=|\omega_\sigma+\omega_\delta|^{\frac1{2}}\frac{e^{\varepsilon_\theta}_{\omega_\theta
}(v_\theta)}{|e^{\varepsilon_\theta}_{
\omega_\theta}(v_\theta)|}\tilde\Phi(\omega_\sigma,\omega_\delta),
\end{align}
requiere de nuevo la introducción del rescalado ${e^{\varepsilon_\theta}_{\omega_\theta
}(v_\theta)}/{|e^{\varepsilon_\theta}_{\omega_\theta}(v_\theta)|}$, que elimina toda la información
del estado. En consecuencia, los observables construidos de esta manera no contienen
información física de interés. 

También aquí podemos intentar introducir una alternativa, adoptando un tratamiento más ingenuo, que
consiste en considerar los operadores $\ln(\hat{v}_a)$ simplemente como operadores de
multiplicación sobre los elementos de $\mathcal{H}_{v_\theta}$ en la representación de $v_a$, es
decir,
\begin{align}\label{eq:lnv-def}
[\ln(\hat{v}_a)\,\Phi_{v_\theta}](v_\sigma,v_\delta)=\
\ln(v_a)\,\Phi_{v_\theta}(v_\sigma,v_\delta).
\end{align}
Teniendo en cuenta que los elementos
$|e^{\varepsilon_a}_{\omega_a}\rangle$ forman una base de $\mathcal H_{\varepsilon_a}^+$, podemos
reescribir la acción de estos operadores en la representación de $\omega_a$ y
representarlos como operadores en el espacio de Hilbert físico, deshaciendo el mapeo
\eqref{eq:chidef-alt}. Así, obtenemos que
$\ln(\hat{v}_\sigma)_{v_\theta}:\mathcal{S}(\mathbb{R}^2)\subset\mathcal{H}_{\text{fis}}^{
\vec\varepsilon}\to\mathcal{H}_{\text{fis}}^{\vec\varepsilon}$ tiene la siguiente actuación:
\begin{align}\label{eq:dir-logv}
[&\ln(\hat{v}_\sigma)_{v_\theta}\tilde\Phi](\omega_\sigma,\omega_\delta)
=\frac{1}{e^{\varepsilon_\theta}_{\omega_\theta}(v_\theta)} \int_\mathbb{R} d\omega_\sigma'\,
\langle
e^{\varepsilon_\sigma}_{\omega_\sigma}| \ln(\hat{v}_\sigma)\,
e^{\varepsilon_\sigma}_{\omega_\sigma'} \rangle_{\mathcal
H_{\varepsilon_\sigma}^+}\,
e^{\varepsilon_\theta}_{\omega_\theta(\omega_\sigma',\omega_\delta)}(v_\theta)\,
\tilde\Phi(\omega_\sigma',\omega_\delta).
\end{align}
La acción de $\ln(\hat{v}_\delta)$ es la misma, reemplazando el subíndice $\sigma$ por
$\delta$. Esto implica que dos operadores a tiempos diferentes, por ejemplo
$\ln(\hat{v}_a)_{v_\theta}$ y $\ln(\hat{v}_a)_{v_\theta^\star}$, están relacionados mediante una
transformación definida en el espacio de Hilbert físico y dada por
\begin{align}\label{eq:dir-Q}
[\hat Q_{v_\theta,v_\theta^\star}\tilde\Phi](\omega_\sigma,\omega_\delta)
&=\frac{e^{\varepsilon_\theta}_{\omega_\theta}(v_\theta)}
{e^{\varepsilon_\theta}_{\omega_\theta}(v_\theta^\star)}
\tilde\Phi(\omega_\sigma,\omega_\delta),
\end{align}
tal que
\begin{align}\label{eq:dir-lnv-rel}
\ln(\hat{v}_a)_{v_\theta^\star}& =\hat Q_{v_\theta,v_\theta^\star}
\ln(\hat{v}_a)_{v_\theta} \hat Q_{v_\theta^\star,v_\theta}.
\end{align}

Como la amplitud $|e^{\varepsilon_\theta}_{\omega_\theta}(v_\theta)|$ cambia significativamente,
tanto cuando varía $\omega_\theta(\omega_\sigma,\omega_\delta)$, como cuando varía $v_\theta$, los
operadores $\hat Q_{v_\theta,v_\theta^\star}$ no son unitarios. Por tanto, los observables de la
familia
así construida no están unitariamente relacionados.

En definitiva, queda patente la dificultad asociada a la naturaleza polimérica del tiempo interno
a la hora de conseguir una noción de evolución unitaria no trivial. Para conseguirla
hace falta una construcción más elaborada. Nosotros la hemos obtenido empleando las propiedades
asintóticas de $e^{\varepsilon_\theta}_{\omega_\theta}(v_\theta)$ y su relación con sus análogos
$\underline{e}_{\omega_\theta}(v_\theta)$ de la cuantización de WDW, como exponemos a
continuación.

\subsection{Descripción sobre componentes rotantes}
\label{sec:v-obs}

Como ya se ha comentado, el éxito de la construcción de la Sección \ref{sec:WDWphys-evo} a la hora
de proporcionar un escenario de evolución no trivial radica en la forma que tienen las autofunciones
$\underline{e}_{\omega_\theta}(v_\theta)$ del operador interpretado como el generador de la
evolución, $\underline{\widehat\Omega}_\theta$, que esencialmente son funciones complejas rotantes
[véase la ec. \eqref{eq:WDW-eig}]. En la cuantización polimérica, las autofunciones análogas
$e^{\varepsilon_\theta}_{\omega_\theta}(v_\theta)$ oscilan en lugar de rotar. Sus componentes en
cada una de las subsemirredes ${}^{(4)}\!\mathcal{L}_{\varepsilon_\theta}^+$ y
${}^{(4)}\!\mathcal{L}_{\varepsilon_\theta+2}^+$, como se ve en la Figura \eqref{fig:eigenf},
convergen a ondas estacionarias, formadas por una combinación de autofunciones rotantes de la teoría
de WDW, una de frecuencia positiva y otra de frecuencia negativa.

Este comportamiento sugiere que, en lugar de intentar construir análogos a los observables
$\ln(\hat{v}_a)_{v_\theta}$ introducidos en la cuantización de WDW, se deberían
construir dos familias separadas $\ln(\hat{v}_a)_{v_\theta}^\pm$, cada una de ellas
en correspondencia con una u otra de las componentes que convergen a las ondas rotantes que forman 
la onda estacionaria.
Con respecto al
procedimiento especificado en la Sección \ref{5sec:obs-rel}, esto se puede ver como una elección
específica de \emph{dos} espacios de Hilbert auxiliares $\mathcal{H}'{}^+$ y
$\mathcal{H}'{}^-$, en lugar de uno.
Es decir, lo que hacemos es introducir un método que identifica y separa, en las autofunciones
de la cuantización polimérica, cada una de las componentes que, en el límite $v_\theta\to\infty$,
convergen a las autofunciones de la cuantización de WDW. 

Esta descomposición se lleva a cabo de una forma precisa introduciendo la siguiente transformación
de las autofunciones de $\widehat\Omega_\theta$, definida en sentido distribucional en cada una de
las subsemirredes ${}^{(4)}\!\mathcal{L}_{\varepsilon_\theta}^+$ y
${}^{(4)}\!\mathcal{L}_{\varepsilon_\theta+2}^+$ separadamente%
\footnote{Recordamos que $e^{\tilde{\varepsilon}_\theta}_{\omega_\theta}$, definida en la
ec.~\eqref{eq:eig-eq-restr}, denota la restricción de $e^{{\varepsilon}_\theta}_{\omega_\theta}$ a
la correspondiente subsemirred de paso 4.}:
\begin{align}\label{eq:rot-dec}
e^{\tilde{\varepsilon_\theta}}_{\omega_\theta} \ \to\
e^{\tilde{\varepsilon_\theta}s}_{\omega_\theta}\ =\
{\mathcal{F}}^{-1}\left\{\theta\left[s\left(\beta_\theta-\frac{\pi}{2}\right)\right]{\mathcal{F}}
e^{\tilde{\varepsilon_\theta}}_{\omega_\theta} \right\},
\end{align}
donde $s\in\{+,-\}$, $\tilde{\varepsilon}_\theta\in
\{\varepsilon_\theta,\varepsilon_\theta+2\}$, $\theta$ es la función de paso de Heaviside,
$\beta_\theta$ es el momento conjugado a $v_\theta$, definido en las ecs. \eqref{eq:b-def} y
\eqref{eq:bv-poiss}, y $\mathcal{F}$ es una tranformación de Fourier discreta análoga a la definida
para sistemas isótropos en la Ref. \cite{acs}:
\begin{align}\label{eq:fourier-def}
[{\mathcal{F}} f](\beta_\theta)\ =\
\sum_{v_\theta\in{}^{(4)}\!\mathcal{L}_{\tilde{\varepsilon}_\theta}^+}
f(v_\theta) v_\theta{}^{-\frac{1}{2}}
e^{-\frac{i}{2}v_\theta \beta_\theta},\quad \beta_\theta\in[0,\pi].
\end{align}
El rescalado $v_\theta{}^{-{1}/{2}}$ en esta tranformación es necesario para
llevar a cabo el análisis de la evolución en términos de $\beta _\theta$, como explicaremos
en la Sección \ref{sec:lqc-b-rep}. Por tanto, aquí hemos usado la misma tranformación de Fourier.
La tranformación \eqref{eq:rot-dec}, entonces, primero nos lleva a la representación de
$\beta_\theta$, luego separa los estados en dos componentes según su dirección de rotación y
finalmente deshace el cambio de representación, devolviéndonos a la representación de $v_\theta$. El
resultado es, como hemos dicho, la extracción en cada
subsemirred de las componentes de $e^{\tilde{\varepsilon_\theta}}_{\omega_\theta}$, etiquetadas por
$+$ y $-$, que convergen, respectivamente, en el límite de $v_\theta$ grande a las funciones
rotantes $\underline{e}_{|\omega_\theta|}$ y $\underline{e}_{-|\omega_\theta|}$ de la teoría
de WDW.

La suma de las funciones $e^{\tilde{\varepsilon}_\theta s}_{\omega_\theta}$ da la función original,
$e^{\tilde{\varepsilon_\theta}}_{\omega_\theta}=e^{\tilde{\varepsilon}_\theta+}_{\omega_\theta}
\oplus e^{\tilde{\varepsilon}_\theta-}_{\omega_\theta}$. Por tanto, podemos dividir cualquier
función de onda física $\Phi_{v_\theta}(\omega_\sigma,\omega_\delta)$, definida en la ec.
\eqref{eq:phi-chi-alt}, en componentes $\Phi_{v_\theta}^{s}$. Para ello, primero introducimos las
componentes rotantes con soporte en la red de paso dos $\mathcal L^+_{\varepsilon_\theta}$,
denominadas $e^{\varepsilon_\theta s}_{\omega_\theta}$ y obtenidas a partir de la suma directa de
$e^{(\tilde{\varepsilon}_\theta=\varepsilon_\theta)s}_{\omega_\theta}$ y
$e^{(\tilde{\varepsilon}_\theta=\varepsilon_\theta+2)s}_{\omega_\theta}$, y
simplemente reemplazamos las
autofunciones $e^{\varepsilon_\theta}_{\omega_\theta}$ en la ec. \eqref{eq:chidef-alt} por 
$e^{\varepsilon_\theta s}_{\omega_\theta}$. Sin embargo, los espacios de Hilbert
$\mathcal{H}^{s}_{v_\theta}\ni\Phi_{v_\theta}^{s}$, que son análogos a
$\mathcal{H}_{v_\theta}$ en el sentido de la definición \eqref{eq:chidef-alt}, también tienen
productos internos diferentes para distintos valores de $v_\theta$, ya que
$|e^{\varepsilon_\theta s}_{\omega_\theta}|$ depende de $v_\theta$.
Entonces, para poder establecer el mapeo entre $\mathcal{H}^{s}_{v_\theta}$ y los correspondientes 
espacios de Hilbert auxiliares $\mathcal{H}'{}^{s}$, de forma que la aplicación final 
$\hat R^{s}_{v_\theta}:\mathcal{H}_{\text{fis}}^{\vec\varepsilon}\to\mathcal{H}'{}^{s}$ sea
unitaria, es necesario ``sincronizar'' las normas, es decir, normalizar
$e^{\varepsilon_\theta s}_{\omega_\theta}$ a una fase pura. Esto es, la transformación completa
que debemos realizar es
\begin{align}\label{eq:split-eig}
e^{\varepsilon_\theta}_{\omega_\theta}(v_\theta)\ \mapsto\
{e'}^{\varepsilon_\theta s}_{\omega_\theta}(v_\theta)\ =\
|\omega_\sigma+\omega_\delta|^{\frac{1}{2}}
\frac{e^{\varepsilon_\theta s}_{\omega_\theta}(v_\theta)}{
|e^{\varepsilon_\theta s}_{\omega_\theta}(v_\theta)|},
\end{align}
de modo que definimos la proyección $\hat R^{s}_{v_\theta}$ de los estados físicos en sus
componentes rotantes de la forma
\begin{align}\label{eq:rotP-def}
\hat R^{s}_{v_\theta} : \mathcal{H}_{\text{fis}}^{\vec\varepsilon}\;&\to\;\mathcal{H}'{}^{s}=
L^2(\mathbb{R}^2,d\omega_\sigma d\omega_\delta),\nonumber \\
\tilde{\Phi}(\omega_\sigma,\omega_\delta)\;&\mapsto\;
\tilde{\boldsymbol{\chi}}^{s}_{v_\theta}(\omega_\sigma,\omega_\delta)=
\tilde{\Phi}(\omega_\sigma,\omega_\delta)
{e'}^{\varepsilon_\theta s}_{\omega_\theta}(v_\theta).
\end{align}

Con esto, podemos finalmente definir dos observables relacionales para
cada observable parcial cinemático $\ln(\hat{v}_a)$, que son
$\ln(\hat{v}_a)_{v_\theta}^{+}:\mathcal{H}_{\text{fis}}^{\vec\varepsilon}\to\mathcal{H}_{\text{fis}
}^{\vec\varepsilon}$ y
$\ln(\hat{v}_a)_{v_\theta}^{-}:\mathcal{H}_{\text{fis}}^{\vec\varepsilon}\to\mathcal{H}_{\text{fis}}
^{\vec\varepsilon}$. Su acción sobre $\mathcal{H}_{\text{fis}}^{\vec\varepsilon}$ es análoga a la de
la ec. \eqref{eq:dir-logv}, es decir,
\begin{align}\label{eq:logV-def}
[\ln(\hat{v}_\sigma)_{v_\theta}^{s}\tilde\Phi](\omega_\sigma,\omega_\delta)\
=\ \frac{1}{{e'}^{\varepsilon_\theta s}_{\omega_\theta}(v_\theta)}
\int_\mathbb{R} d\omega_\sigma'\, \langle
e^{\varepsilon_\sigma}_{\omega_\sigma}| \ln(\hat{v}_\sigma)\,
e^{\varepsilon_\sigma}_{\omega_\sigma'}
\rangle_{\mathcal{H}_{\varepsilon_\sigma}^+}\,
{e'}^{\varepsilon_\theta s}_{\omega_\theta(\omega_\sigma',\omega_\delta)}(v_\theta)\,
\tilde\Phi(\omega_\sigma',\omega_\delta),
\end{align}
y similarmente para $\ln(\hat{v}_\delta)_{v_\theta}^{s}$.

\subsection{Unitariedad de la evolución y pérdida de precisión física}
\label{perd-prec}

En cada una de las familias de observables etiquetadas por $a$ y $s$,
dos observables evaluados a tiempos diferentes $v_\theta$ y $v_\theta^\star$
están relacionados mediante los operadores $\hat Q_{v_\theta,v_\theta^\star}^{s}:
\mathcal{H}_{\text{fis}}^{\vec\varepsilon}\to
\mathcal{H}_{\text{fis}}^{\vec\varepsilon}$, definidos como
\begin{align}\label{eq:Q}
[\hat Q_{v_\theta,v_\theta^\star}^{s}\tilde\Phi](\omega_\sigma,\omega_\delta)
&=\frac{{e'}^{\varepsilon_\theta s}_{\omega_\theta}(v_\theta)}
{{e'}^{\varepsilon_\theta s}_{\omega_\theta}(v_\theta^\star)}
\tilde\Phi(\omega_\sigma,\omega_\delta),
\end{align}
de la siguiente manera
\begin{align}\label{eq:logV-rel}
\ln(\hat{v}_a)_{v_\theta^\star}^{s}\ =\hat
Q_{v_\theta,v_\theta^\star}^{s} \ln(\hat{v}_a)_{v_\theta}^{s} \hat
Q^{s}_{v_\theta^\star,v_\theta}.
\end{align}

Como, por la definición \eqref{eq:split-eig}, se cumple que
$|{e'}^{\varepsilon_\theta s}_{\omega_\theta}(v_\theta)| =
|\omega_\sigma+\omega_\delta|^{\frac{1}{2}}$ para todo
$v_\theta\in\mathcal{L}_{\varepsilon_\theta}^+$, los operadores
$\hat Q_{v_\theta,v_\theta^\star}^{s}$ son ciertamente unitarios. Por tanto,
dentro de cada familia considerada, sus observables están \emph{relacionados unitariamente}. Podemos
de nuevo extender el conjunto formado por estas familias añadiendo los operadores
$\widehat{{\Omega}}_a|_{v_\theta}:={\widehat{{\Omega}}_a}$ para conseguir un conjunto completo de
observables de Dirac.

En conclusión, los operadores definidos en la ec \eqref{eq:logV-def} proporcionan una noción
correcta de evolución unitaria. Sin embargo esta unitariedad se consigue a costa de un precio:
debido a la sicronización entre normas introducida para poder mapear los distintos espacios 
$\mathcal{H}^{s}_{v_\theta}$ en $\mathcal{H}'{}^{s}$, es decir, debido al rescalado
${e^{\varepsilon_\theta s}_{\omega_\theta}/}{|e^{\varepsilon_\theta s}_{\omega_\theta}|}$, los
observables así construidos ya no tienen una interpretación física exacta, sino aproximada, que
será más precisa cuanto mayor sea $v_\theta$, pues en el límite asintótico $v_\theta\to\infty$
las componenetes rotantes $e^{{\varepsilon}_\theta s}_{\omega_\theta}$ convergen a sus
análogos \eqref{eq:WDW-eig} de la teoría de WDW y $|e^{\varepsilon_\theta
s}_{\omega_\theta}|$ converge a una función independiente de $\omega_a$. 

Por tanto, la interpretación de que los operadores $\ln(\hat{v}_a)_{v_\theta}^{\pm}$ proporcionan
el valor de
$\ln(v_a)$ a un tiempo dado $v_\theta$ en la componente que rota hacia adelante en el tiempo (para
el signo positivo) o en la componente que rota hacia atrás (para el signo negativo) es solo
aproximada. Esta interpretación mejora a medida que $v_\theta$ aumenta; no obstante, para
$v_\theta$
del orden de $\omega_\theta$ o más pequeño (que es donde la dinámica efectiva predice un rebote en
$v_\theta$) toda la interpretación se pierde. Hemos dado un argumento ilustrativo que demuestra esta
falta de fiabilidad de la interpretación en tal régimen, argumento que mostraremos en la Sección
\ref{sec:v-results}, una
vez que hayamos explicado cómo hemos aplicado esta construcción de observables para analizar la
dinámica cuántica de estados físicos de interés.

\section[Predicciones físicas de la evolución en $\upsilon_\theta$]{Predicciones físicas de la
evolución en $v_\theta$}
\label{resultados-v}

En concreto, hemos centrado nuestro estudio en la dinámica de estados físicos que son semiclásicos a
tiempos tardíos. Éstos están proporcionados por perfiles gaussianos muy centrados en torno a valores
grandes $\omega_a^\star$ de los vectores de onda $\omega_a$, 
\begin{align}\label{eq:gauss-state}
\tilde{\Phi}(\omega_\sigma,\omega_\delta) =
\frac{K}{\sqrt{|\omega_\sigma+\omega_\delta|}}\prod_{a=\sigma}^{\delta}
e^{-\frac{(\omega_a-\omega_a^\star)^2}{2\sigma_a^2}}
e^{i\nu^a\omega_a},
\end{align}
como los analizados en el Capítulo \ref{chap:5-evol} [véase la ec. \eqref{eq:wdw-gauss}].

Este estudio requiere un análisis numérico, tanto a la hora de medir los valores
esperados de los observables en dicho tipo de estados como para obtener sus dispersiones, ya que
los productos internos que aparecen, a diferencia de los computados en el caso de
la cuantización estándar, no se pueden calcular de forma analítica. El análisis numérico no puede
llevarse a cabo para todo el rango infinito de valores del tiempo interno $v_\theta$ y éste debe
ser truncado. No obstante, sabemos que, en el límite $v_\theta\to\infty$, la cosmología cuántica de
lazos converge a la teoría de WDW de una forma muy concreta y conocida. Ya hemos visto en
la Sección \ref{sec:wdw-limit}
la relación precisa que existe entre ambas cuantizaciones del modelo. Hemos
podido usar esta relación para explorar el comportamiento de la dinámica de la
cuantización polimérica en el régimen de valores indefinidamente grandes de $v_\theta$, región que
el cálculo numérico no permite analizar.

\subsection{Comportamiento semiclásico de los observables}
\label{sec:wdw-limit-disp1}

Recordemos que, en la Sección \ref{sec:wdw-limit}, hemos demostrado que un estado físico
$\tilde\Phi(\omega_\sigma,\omega_\delta)$ de la cuantización polimérica que esté localizado, en el
sentido de que se mantiene picado a lo largo de las trayectorias definidas por los valores
esperados de los observables considerados, converge, en el límite en el que las tres
variables $v_i$ toman valores grandes, a un estado físico de la cuantización de WDW con
perfil espectral dado en la ec.~\eqref{eq:wdw-limit-phys}. Éste está formado por dos componentes,
una con la misma paridad y orientación temporal que las del estado
$\tilde\Phi(\omega_\sigma,\omega_\delta)$, denominada
$\underline{\tilde{\Phi}}_+(\omega_\sigma,\omega_\delta)$, y otra con paridad y orientación
temporal inversas, denominada $\underline{\tilde{\Phi}}_-(\omega_\sigma,\omega_\delta)$.

A la vista de este resultado, para valores de $v_\theta$ grandes podemos hacer una correspondencia
entre el valor esperado de los
observables $\ln(\hat{v}_a)_{v_\theta}^{\pm}$ de la cuantización polimérica
[ec.~\eqref{eq:logV-def}] en el estado $\tilde\Phi$ y el valor esperado
del observable $\ln(\hat{v}_a)_{v_\theta}$ de la cuantización estándar [ec.~\eqref{obs}] en
los estados $\underline{\tilde{\Phi}}_\pm$. Entonces, gracias a esta correspondencia, podemos
calcular 
\begin{align}
\langle
\underline{\tilde{\Phi}}_\pm|\ln(\hat{v}_a)_{v_\theta}\underline{\tilde{\Phi}}_\pm\rangle=:
\langle \ln(\hat{v}_a)_{v_\theta} \rangle_{\pm}.
\end{align}
para $v_\theta>>1$, usando los resultados de la Sección \ref{pred-fis}. 

Veamos cómo son las trayectorias de los valores esperados $\langle \ln(\hat{v}_a)_{v_\theta}
\rangle_{+}$ y $\langle \ln(\hat{v}_a)_{v_\theta} \rangle_{-}$. A partir de la ec.
\eqref{eq:wdw-limit-phys} observamos que los perfiles espectrales de $\underline{\tilde{\Phi}}_+$ y
$\underline{\tilde{\Phi}}_-$ están relacionados entre sí simplemente por una rotación, por la
reflexión en $\omega_a$ y por un posible cambio de signo. Por tanto, los coeficientes direccionales
$A_a$ respectivos, definidos en la ec. \eqref{eq:traj-dir-A}, son los mismos. La única diferencia
está en el parámetro $B_a$.
Como consecuencia, las trayectorias están simplemente desplazadas una con
respecto a la otra.
En el caso particular de que el estado físico polimérico tenga un perfil gaussiano
\eqref{eq:gauss-state} muy picado (esto es, en el límite $\sigma_a\to\infty$), podemos hacer una
estimación cuantitativa de dicho desplazamiento.
Para tales estados, la fase de rotación $\phi(\omega_i)$ está bien aproximada por su expansión a
primer orden:
\begin{equation}\begin{split}
\phi(\omega_i) &\approx
D(\omega^\star_i)(\omega_i-\omega^\star_i)
+E(\omega_i^\star),
\\
D(\omega^\star_i) &=
\text{sgn}{(\omega^\star_i)}(1+a+\ln|\omega^\star_i|),
\end{split}\end{equation}
donde $a$ es la constante de la ec. \eqref{T-phase}. Gracias a esta aproximación, se pueden repetir
casi directamente los cálculos de la Sección \ref{pred-fis} para encontrar la expresión
equivalente a la ec. \eqref{eq:gauss-traj}. Las trayectorias resultantes son
\begin{align}\label{eq:limit-trajs}
\langle \ln(\hat{v}_a)_{v_\theta} \rangle_{\pm}=
\bigg[\frac{\omega_\theta(\omega_\sigma^\star,\omega_\delta^\star)}
{\omega^\star_a}\bigg]^2
[\ln v_\theta - \alpha D(\omega^\star_\theta)]+ \alpha[D(\omega^\star_a)\pm\nu^a].
\end{align}
Genéricamente, estas trayectorias son disjuntas, aunque coinciden en el caso~$\nu^a=0$.

\subsubsection{Dispersiones: comportamiento semiclásico}
\label{sec:wdw-limit-disp}

En la discusión anterior, hemos tomado el límite $\sigma_a\to 0$ ($a=\sigma,\delta$)
y considerado la aproximación lineal de la variación de la fase de rotación $\phi(\omega_i)$.
Estas aproximaciones eliminan toda la información sobre el comportamiento de las dispersiones. Sin
embargo, en este escenario, en el que estamos viendo (en un límite asintótico) la dinámica de la
cosmología cuántica de lazos como la reflexión del paquete de ondas de la teoría de WDW
$\underline{\tilde\Phi}_-$ (que se mueve hacia atrás en el tiempo y se contrae) en el
paquete $\underline{\tilde\Phi}_+$ (el cual se mueve hacia adelante en el tiempo y se expande), es
importante ver cuánto crece la dispersión de $\underline{\tilde\Phi}_+$ en comparación con la de 
$\underline{\tilde\Phi}_-$, o viceversa. A este respecto, se han encontrado cotas muy restrictivas
en el posible crecimiento de la dispersión mediante el empleo de desigualdades
triangulares exactas, que involucran la dispersión en $\ln|\omega_a|$ y en
$\ln(\hat{v}_a)_{v_\theta}$ \cite{kp-posL}.

El análisis análogo en la cuantización de WDW
muestra que las dispersiones \emph{relativas} que alcanzan un valor constante no nulo para
$v_\theta$ grande son las de los observables $\ln(\hat{v}_a)_{v_\theta}$ [ec.
\eqref{eq:wdw-rel-disp}]. Por ello, aquí nos hemos centrado únicamente en buscar relaciones que
satisfagan estas cantidades relativas. Veámoslo.

Comencemos recordando que la función de onda $\Phi(\vec v)=\Phi_{v_\theta}(v_\sigma,v_\delta)$ tiene
soporte en redes cúbicas de paso dos, dadas por el producto de las tres semirredes
$\mathcal{L}_{\varepsilon_i}^+$, cada una de las cuales, a su vez, es la unión de dos subsemirredes
de
paso cuatro ${}^{(4)}\mathcal{L}_{\tilde{\varepsilon}_i}$ (para
$\tilde{\varepsilon_i} \in
\{\varepsilon_i,\varepsilon_i+2\}$). Entonces, dividimos el soporte de $\Phi(\vec v)$ en ocho
sectores, que se corresponden con las ocho posibles combinaciones de productos entre las
subsemirredes de paso cuatro y consideramos las restricciones de
$\Phi(\vec v)$ a cada uno de ellos. En cada uno de estos sectores aplicamos el escenario de
{dispersión} definido en la Sección \ref{scaterring-scheme} [ec. \eqref{eq:wdw-limit-phys}] y
resumido dos párrafos más arriba. En este escenario, el paquete de ondas $\underline{\tilde\Phi}_-$
se transforma en $\underline{\tilde\Phi}_+$ mediante la rotación unitaria
\begin{equation}\label{eq:rotation}
  U = \prod_{i=\theta,\sigma,\delta} e^{2i\phi(\omega_i)},
\end{equation}
junto con la reflexión en el signo de $\omega_a$ y un posible cambio de signo global.

Es razonable restringir nuestras consideraciones a estados en los que cada una de las componentes 
$\underline{\tilde\Phi}_-$, correspondientes a los sectores introducidos antes, sea tal que los
coeficientes $A_a$, $B_a$, $W_a$, $Y_a$, y $X_a$, definidos en las ecs.~\eqref{eq:traj-dir} y
\eqref{eq:sqr-dir}, resulten ser finitos. Esto garantiza que cada una de esas componentes tenga
asociada una
trayectoria \eqref{eq:wdw-traj-shape} bien definida y una dispersión finita para cada valor de 
$v_\theta$. Además, también garantiza que se satisfaga la relación \eqref{eq:wdw-rel-disp},
evaluada en
$\underline{\tilde\Phi}_-$:
\begin{align}\label{eq:wdw-rel-disp-bis}
\lim_{v_\theta\to\infty} \frac{\langle
\Delta\ln(\hat{v}_a)_{v_\theta} \rangle_-}{\langle
\ln(\hat{v}_a)_{v_\theta} \rangle_-} = \frac{\langle \Delta
[\omega_\theta^2({\omega}_\sigma,{\omega}_\delta) {\omega}_a^{-2}]
\rangle}{\langle
\omega_\theta^2({\omega}_\sigma,{\omega}_\delta){\omega}_a^{-2}
\rangle}.
\end{align}
Ni la rotación $U$, ni la reflexión en el signo de $\omega_a$, ni el posible cambio
de signo global de la función de onda cambian el valor esperado del operador (de multiplicación) 
$\omega_\theta^2(\omega_\sigma,\omega_\delta)\omega_a^{-2}$. Por tanto, si los coeficientes 
$A_a$, $B_a$, $W_a$, $Y_a$, y $X_a$ son finitos también para $\underline{\tilde\Phi}_+$, la
ec.~\eqref{eq:wdw-rel-disp-bis} implica inmediatamente que se verifica
\begin{equation}\label{eq:lim-disp-eq}
\lim_{v_\theta\to\infty}\frac{\langle\Delta\ln
(\hat{v})_a\rangle_+}{\langle\ln (\hat{v})_a\rangle_+}
= \lim_{v_\theta\to\infty}\frac{\langle\Delta\ln
(\hat{v})_a\rangle_-}{\langle\ln (\hat{v})_a\rangle_-}.
\end{equation}

De hecho, aplicando un método similar al propuesto en la Ref.
\cite{kp-posL}, hemos probado que la finitud de $A_a$, $B_a$, $W_a$, $Y_a$, y $X_a$ para
estados $\underline{\tilde\Phi}_-$ que cumplan ciertas condiciones (no muy restrictivas) implica que
estos coeficientes también son finitos para $\underline{\tilde\Phi}_+$ y que, por tanto, se cumple
la relación \eqref{eq:lim-disp-eq}. En realidad, los únicos coeficientes que requieren análisis son
$B_a$, $Y_a$, y $X_a$, pues $A_a$ y $W_a$ son iguales para ambas componentes (véase su definición).
En concreto, dados los operadores de multiplicación
\begin{subequations}\label{operators-dem}\begin{align}
w_a &:= \omega_\theta(\omega_\sigma,\omega_\delta)
\omega_a^{-1} , \\
\Upsilon^{(n)}_a &:= w_a^n
\ln|\omega_\theta(\omega_\sigma,\omega_\delta)| , \\
\Sigma^{(n)}_a &:= w_a^n \ln|\omega_a|,
\end{align}\end{subequations}
las condiciones que se deben verificar para que dicho resultado sea cierto son que, en el estado 
$\underline{\tilde\Phi}_-$, las dispersiones y valores esperados de $\ln|\omega_a|$,
$\Upsilon_a^{(2)}$, y $w_a^2$ sean finitos, así como los valores esperados de $\Sigma^{(2)}_a$.
Los detalles de esta demostración se exponen en la Sección \ref{dem1} del Apéndice \ref{appB}.

Más aún, la Sección \ref{dem2} de dicho apéndice recoge la prueba de que el resultado
\eqref{eq:lim-disp-eq} sobre las dispersiones relativas se puede extender al caso en el
que se considera el conjunto total de los ocho posibles sectores en los que los estados físicos de
la cuantización de lazos admiten un límite de WDW. De este modo se llega a la
siguiente conclusión:

\begin{itemize}
 \item 
Consideremos un estado físico descrito por una función de onda $\Phi(\vec v)$ con soporte en el
producto de
las semirredes $\mathcal{L}_{\varepsilon_\theta}^+ \times
\mathcal{L}_{\varepsilon_\sigma}^+ \times \mathcal{L}_{\varepsilon_\delta}^+$. Supongamos que la
restricción de $\Phi(\vec v)$ al producto de las subsemirredes
${}^{(4)}\mathcal{L}_{\tilde{\varepsilon}_\theta}^+ \times
{}^{(4)}\mathcal{L}_{\tilde{\varepsilon}_\sigma}^+ \times
{}^{(4)}\mathcal{L}_{\tilde{\varepsilon}_\delta}^+$ posee un límite de WDW tal que, sobre
la componente $\underline{\tilde\Phi}_\pm$ que se mueve hacia adelante/atrás en el tiempo, los
operadores $\ln(\hat{v}_a)_{v_\theta}$, $\ln|\omega_a|$, $\Upsilon^{(2)}_a$ y $w_a^2$
tienen valores esperados y dispersiones finitos. Supongamos también que, en este estado, los
operadores
$\Sigma^{(2)}_a$ tienen valor esperado finito. Entonces:
\begin{itemize}
\item[(i)] los operadores $\ln(\hat{v}_a)_{v_\theta}$, $\ln|\omega_a|$, $\Upsilon^{(2)}_a$ y $w_a^2$
también tienen valores esperados y dispersiones finitos sobre la otra componente correspondiente
$\underline{\tilde\Phi}_\mp$ que se mueve hacia atrás/adelante en el tiempo,
\item[(ii)] los operadores $\Sigma^{(2)}_a$ también tienen valores esperados finitos sobre esta
componente,
\item[(iii)] la relación \eqref{eq:lim-disp-eq} se satisface para el conjunto total de los límites
de
WDW que corresponden a los ocho sectores definidos por las restricciones a
las diferentes subsemirredes, construidos para reflejar las propiedades relevantes del estado
completo de la cuantización polimérica.
\end{itemize}
\end{itemize}

En conclusión, el comportamiento semiclásico para $v_\theta$ grande se conserva.

Es de destacar que, aunque en última instancia hayamos particularizado el análisis
de la dinámica a estados físicos con perfil gaussiano \eqref{eq:gauss-state}, hemos demostrado este
resultado de conservación del carácter semiclásico para estados mucho más generales, a saber,
estados
que sean localizados (en el sentido ya explicado) y que cumplan las condiciones expuestas arriba
(veáse también la discusión hecha al final del Apéndice~\ref{appB}).

\subsection{Análisis numérico}
\label{sec:v-num}

En esta sección, describimos el método numérico que hemos empleado para analizar la dinámica del
modelo
en la región no asintótica, esto es, para valores finitos de $v_\theta$. Como ya se ha dicho, hemos
particularizado el estudio a estados que son semiclásicos a tiempos tardíos y, más precisamente,
que tienen un perfil gaussiano \eqref{eq:gauss-state} con soporte muy picado en torno a valores
grandes $(\omega_\sigma^\star,\omega_\delta^\star)$. La trayectoria a tiempos tardíos de dicho tipo
de estados, como ya hemos discutido en la sección anterior, se rige por la
expresión~\eqref{eq:limit-trajs} y, por tanto, queda caracterizada por
$(\omega_\sigma^\star,\omega_\delta^\star)$ y por $(\nu^\sigma,\nu^\delta)$.

Hemos calculado las funciones de onda correspondientes a tales estados, dadas por la integral
\eqref{sol-fis-A}, usando la regla del trapecio \cite{trap-rule}, en el dominio 
$\omega_a\in[\omega_a^\star-5\sigma_a,\omega_a^\star+5 \sigma_a]$. Para ello, se ha tomado un
muestreo de $\tilde{\Phi}(\omega_\sigma,\omega_\delta)$ sobre una red uniforme definida
por la división de dicho dominio en, al menos, $2\omega_a^\star$ subintervalos en cada dirección.

Para calcular los valores esperados de $\ln(v_a)_{v_\theta}^{s}$, hemos usado las expresiones de los
elementos $\tilde{\boldsymbol{\chi}}_{v_\theta}^{s}(\omega_\sigma,\omega_\delta)$
de $\mathcal{H}'{}^{s}$ como funciones de $v_a$, es decir,
\begin{equation}\label{eq:aux-chi}
{\boldsymbol{\chi}}_{v_\theta}^{s}(v_\sigma,v_\delta) =
\int_{\mathbb{R}^2}d\omega_\sigma d\omega_\delta
\tilde{\boldsymbol{\chi}}_{v_\theta}^{s}(\omega_\sigma,\omega_\delta)
e^{\varepsilon_\sigma}_{\omega_\sigma}(v_\sigma)
e^{\varepsilon_\delta}_{\omega_\delta}(v_\delta).
\end{equation}

Como los elementos $e^{\varepsilon_a}_{\omega_a}(v_a)$ forman bases ortonormales en sus espacios de
Hilbert
cinemáticos respectivos, $\mathcal{H}_{\varepsilon_a}^+$, el producto interno de
$\mathcal{H}'{}^{s}$, que proviene del producto interno físico \eqref{eq:phys-A}, tiene una
expresión muy simple en la representación de $v_a$:
\begin{equation}\label{eq:aux-ip}
\langle{\boldsymbol{\chi}}_{v_\theta}^{s}|
{\boldsymbol{\chi}}_{v_\theta}^{\prime s}\rangle =
\sum_{\mathcal{L}_{\varepsilon_\sigma}^+\times\mathcal{L}_{\varepsilon_\delta}^+}
[{\boldsymbol{\chi}}_{v_\theta}^{s}(v_\sigma,v_\delta)]^*
{\boldsymbol{\chi}}_{v_\theta}^{\prime s}(v_\sigma,v_\delta).
\end{equation}
Además, sobre los estados ${\boldsymbol{\chi}}_{v_\theta}^{s}(v_\sigma,v_\delta)$, los observables
$\ln(\hat{v}_a)_{v_\theta}^{s}$ actúan simplemente por multiplicación:
\begin{equation}\label{eq:aux-v-act}
[\ln(\hat{v}_a)_{v_\theta}^{s}
{\boldsymbol{\chi}}_{v_\theta}^{s}](v_\sigma,v_\delta) =
\ln({v}_a){\boldsymbol{\chi}}_{v_\theta}^{s}(v_\sigma,v_\delta).
\end{equation}
Por tanto, sus valores esperados en el estado físico $\tilde\Phi(\omega_\sigma,\omega_\delta)$
son \begin{equation}\label{eq:aux-v-exp}
\langle\tilde\Phi|\ln(\hat{v}_a)_{v_\theta}^{s}\tilde\Phi\rangle =
\|{\boldsymbol{\chi}}_{v_\theta}^{s}\|^{-2}
\sum_{\mathcal{L}_{\varepsilon_\sigma}^+\times\mathcal{L}_{\varepsilon_\delta}^+}
\ln(v_a) |{\boldsymbol{\chi}}_{v_\theta}^{s}(v_\sigma,v_\delta)|^2 ,
\end{equation}
donde el perfil espectral $\tilde{\boldsymbol{\chi}}_{v_\theta}^{s}(\omega_\sigma,\omega_\delta)$
de ${\boldsymbol{\chi}}_{v_\theta}^{s}(v_\sigma,v_\delta)$ está relacionado con 
$\tilde\Phi(\omega_\sigma,\omega_\delta)$ mediante la ec. \eqref{eq:rotP-def}.

Las dispersiones de los observables $\ln(\hat{v}_a)_{v_\theta}^{s}$ están dadas por la fórmula
estándar
\begin{equation}\label{eq:disp-def}
\langle\Delta\ln(\hat{v}_a)_{v_\theta}^{s}\rangle^2 =
\langle(\ln(\hat{v}_a)_{v_\theta}^{s})^2\rangle -
\langle\ln(\hat{v}_a)_{v_\theta}^{s}\rangle^2,
\end{equation}
donde los valores esperados de $[\ln(\hat{v}_a)_{v_\theta}^{s}]^2$ se evalúan siguiendo el mismo
método que para
$\langle\tilde\Phi|\ln(\hat{v}_a)_{v_\theta}^{s}\tilde\Phi\rangle $.

En las simulaciones numéricas que hemos llevado a cabo para calcular estos valores esperados,
primero hemos evaluado las componentes de ${e}^{\varepsilon_\theta s}_{\omega_\theta}$, dadas en la
ec.~\eqref{eq:rot-dec}, mediante el uso del algoritmo denominado ``transformación de Fourier
rápida'' (FFT) \cite{FFT}. Luego hemos calculado los coeficientes normalizados
${e'}^{\varepsilon_\theta s}_{\omega_\theta}$ de acuerdo con su definición
\eqref{eq:split-eig}. Después de calcular los perfiles auxiliares
$\tilde{\boldsymbol{\chi}}^{s}_{v_\theta}(\omega_\sigma,\omega_\delta)$
definidos por la ec. \eqref{eq:rotP-def}, hemos evaluado sus funciones de onda
${\boldsymbol{\chi}}_{v_\theta}^{s}(v_\sigma,v_\delta)$, de acuerdo con la integral 
\eqref{eq:aux-chi}, del mismo modo que el explicado para el cálculo de $\Phi(\vec{v})$.
Seguidamente, hemos obtenido los valores esperados usando la ec.
\eqref{eq:aux-v-exp}. Hemos restringido los dominios de $v_a$ a
$\mathcal{L}_{\varepsilon_a}^+\cap[0,4\omega_a^\star]$. A su vez, hemos tomado valores de 
$v_\theta$ pertenecientes al conjunto $\mathcal{L}_{\varepsilon_\theta}^+\cap[0,\,2\cdot 10^5]$.
Finalmente, hemos realizado simulaciones para varios valores de $\vec{\varepsilon}$ y para valores
de $\omega_a^\star$ contenidos en el intervalo $[2\ldotp\!5\cdot 10^2,10^3]$.

\subsection{Resultados}
\label{sec:v-results}

\begin{figure}
\begin{center}
$(a)$
\\
\begin{overpic}
[width=0.7\textwidth]{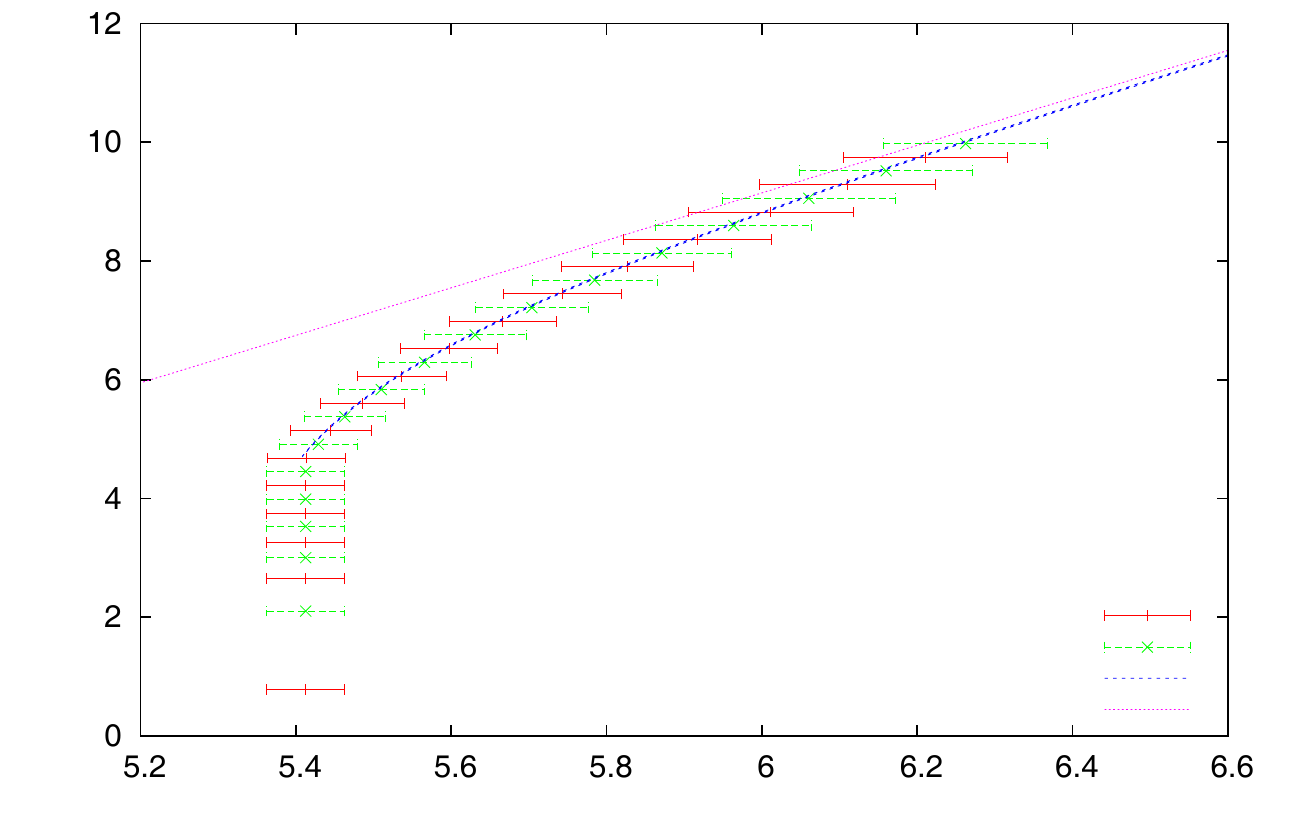}
\put(50,-1){\footnotesize{$\ln(v_\sigma)$}}
\put(-2,33.5){\footnotesize{$\ln(v_\theta)$}}
\put(76,8){\tiny{cl\'asico}}
\put(76,10.5){\tiny{efectivo}}
\put(76,13.5){\tiny{$\ln(v_\sigma)^-_{v_\theta}$}}
\put(76,16.5){\tiny{$\ln(v_\sigma)^+_{v_\theta}$}}
\end{overpic}
\vspace*{2mm}
\\
$(b)$
\\
\begin{overpic}
[width=0.7\textwidth]{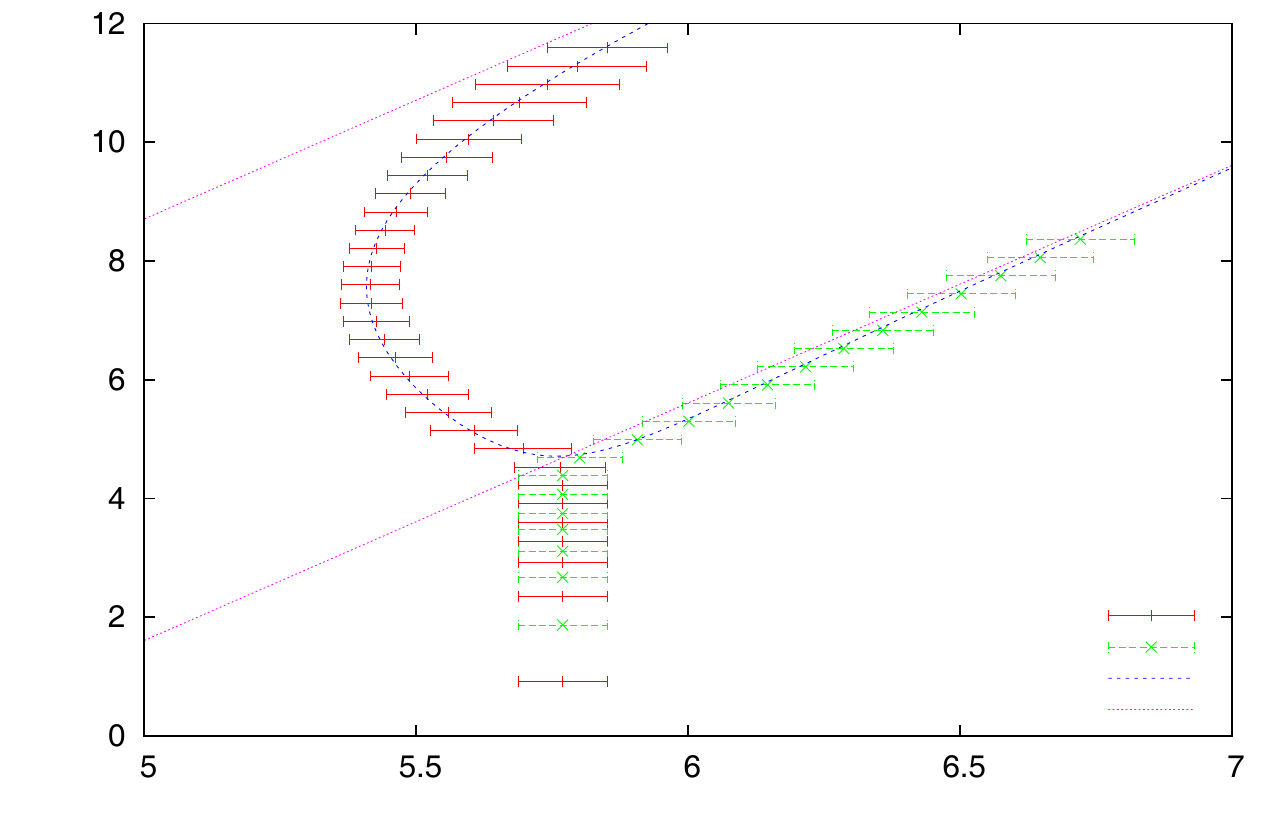}
\put(50,-1){\footnotesize{$\ln(v_\sigma)$}}
\put(-2,33.5){\footnotesize{$\ln(v_\theta)$}}
\put(76,8){\tiny{cl\'asico}}
\put(76,10.5){\tiny{efectivo}}
\put(76,13.5){\tiny{$\ln(v_\sigma)^-_{v_\theta}$}}
\put(76,16.5){\tiny{$\ln(v_\sigma)^+_{v_\theta}$}}
\end{overpic}
\caption{Para estados gaussianos [con perfil \eqref{eq:gauss-state}], los valores esperados y
dispersiones de los observables
$\ln(\hat{v}_a)_{v_\theta}^{+}$ y $\ln(\hat{v}_a)_{v_\theta}^{-}$, que se corresponden
respectivamente con épocas en las que el
paquete de onda se mueve hacia adelante en el tiempo (barras de error verdes) y en las que se mueve
hacia atrás en el tiempo (barras de error rojas), son comparados con las trayectorias clásicas
(líneas rosas) y las trayectorias clásicas efectivas (línea azul punteada).
Los estados están picados en $\omega^\star_\sigma=\omega^\star_\delta=10^3$, con dispersiones
relativas determinadas por
$\Delta\Omega_\sigma/\Omega_\sigma=\Delta\Omega_\delta/\Omega_\delta=0\ldotp\!05$.
Las fases son, respectivamente, $\nu^\sigma=\nu^\delta=0$ para el caso $(a)$, y
$\nu^\sigma=\nu^\delta=0\ldotp\!1$ para el $(b)$.
Los valores esperados siguen la trayectoria clásica efectiva hasta el punto del rebote en
$v_\theta$, en el que la dinámica ``se congela''. Por otra parte, lejos del rebote las trayectorias
cuánticas tienden a las clásicas. En particular, en $(a)$, las trayectorias antes y después del
rebote coinciden en el plano $v_\theta - v_\sigma$, ya que hemos tomado $\nu^\sigma$ y
$\nu^\delta$ nulas.}
\label{fig:v-trajecto}
\end{center}
\end{figure}
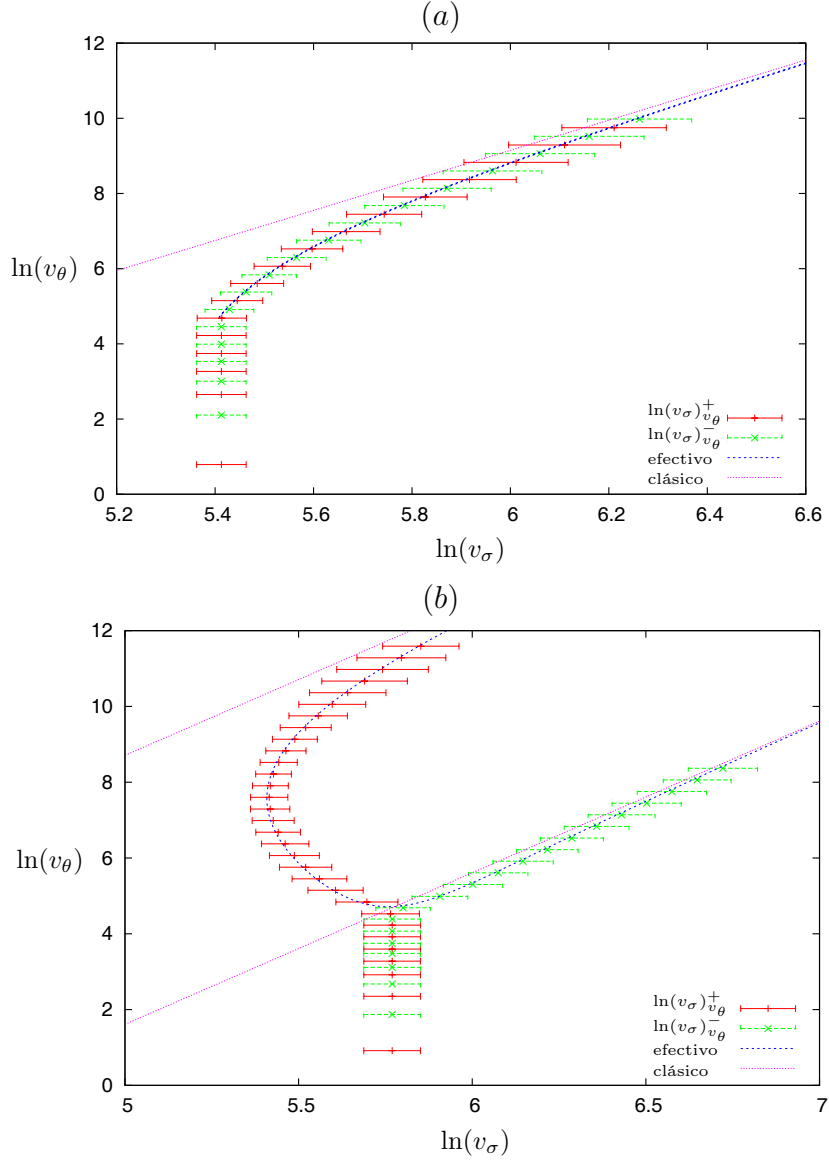

La Figura \ref{fig:v-trajecto} muestra dos ejemplos de los resultados de
nuestro estudio numérico.
En todos los casos, el análisis revela que:
\begin{itemize}
\item En lo que concierne a los valores esperados \eqref{eq:aux-v-exp}, los estados se mantienen
\emph{muy picados} para \emph{todos} los valores de $v_\theta$. Esto se aplica a \emph{ambas}
componentes: la que se mueve hacia adelante en el tiempo (+) y la que se mueve hacia atrás (-).
\item Para $v_\theta\gg\omega^\star_a$, los valores esperados 
\emph{siguen las trayectorias clásicas}, mientras que,
cuando el universo se aproxima a la singularidad clásica, los efectos de la discretización
de la geometría inducen \emph{fuerzas repulsivas} que crean \emph{rebotes}, tanto en $v_\sigma$
como en $v_\delta$. Estos rebotes se dan en los valores predichos por la dinámica efectiva, de la
que se presenta un resumen en la Sección \ref{effective} del Apéndice \ref{appB}.
\end{itemize}

Estos resultados demuestran la robustez del escenario de \emph{gran rebote} de la cosmología
cuántica de lazos, extendiendo su validez a un sistema anisótropo: las cosmologías vacías de Bianchi
I (en
el esquema A de la dinámica mejorada). Además, confirman la capacidad de la dinámica clásica
efectiva de predecir resultados correctos con errores mucho más pequeños que la dispersión de la
función de onda cuántica.

Sin embargo, existe un aspecto de los resultados que necesita un comentario especial. Como podemos
ver en la Figura \ref{fig:v-trajecto}, mientras que para valores grandes de $v_\theta$ los valores
esperados de los observables $\ln(\hat{v}_a)_{v_\theta}^{s}$ tienden a las trayectorias clásicas
(con mayor diferencia para menor valor de $v_\theta$), para $v_\theta\lesssim
0\ldotp\!2\,\omega_\theta(\omega_\sigma^\star,\omega_\delta^\star)$ los valores esperados de ambas
familias de observables (+ y -) ``se congelan'' en la misma trayectoria de $v_\sigma$ y $v_\delta$
constantes. Sin embargo, este comportamiento no se corresponde físicamente con el de la magnitud
$\ln(v_a)$ a un valor dado de $v_\theta$. El observable relacional
$\ln(\hat{v}_a)_{v_\theta}^{s}$ (para cada $a$) no representa correctamente dicha magnitud en
esa región, como hemos discutido en la Sección \ref{perd-prec}.

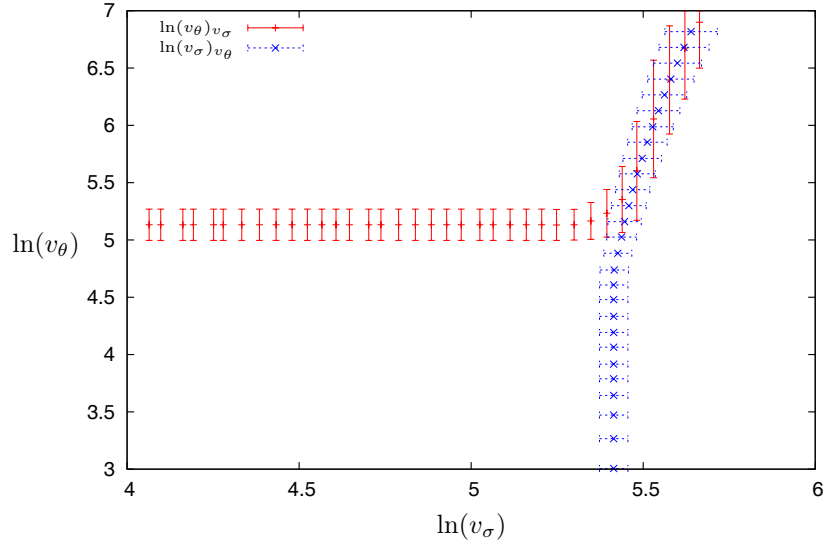
\begin{figure}[htb!]
\begin{center}
\begin{overpic}
[width=0.7\textwidth]{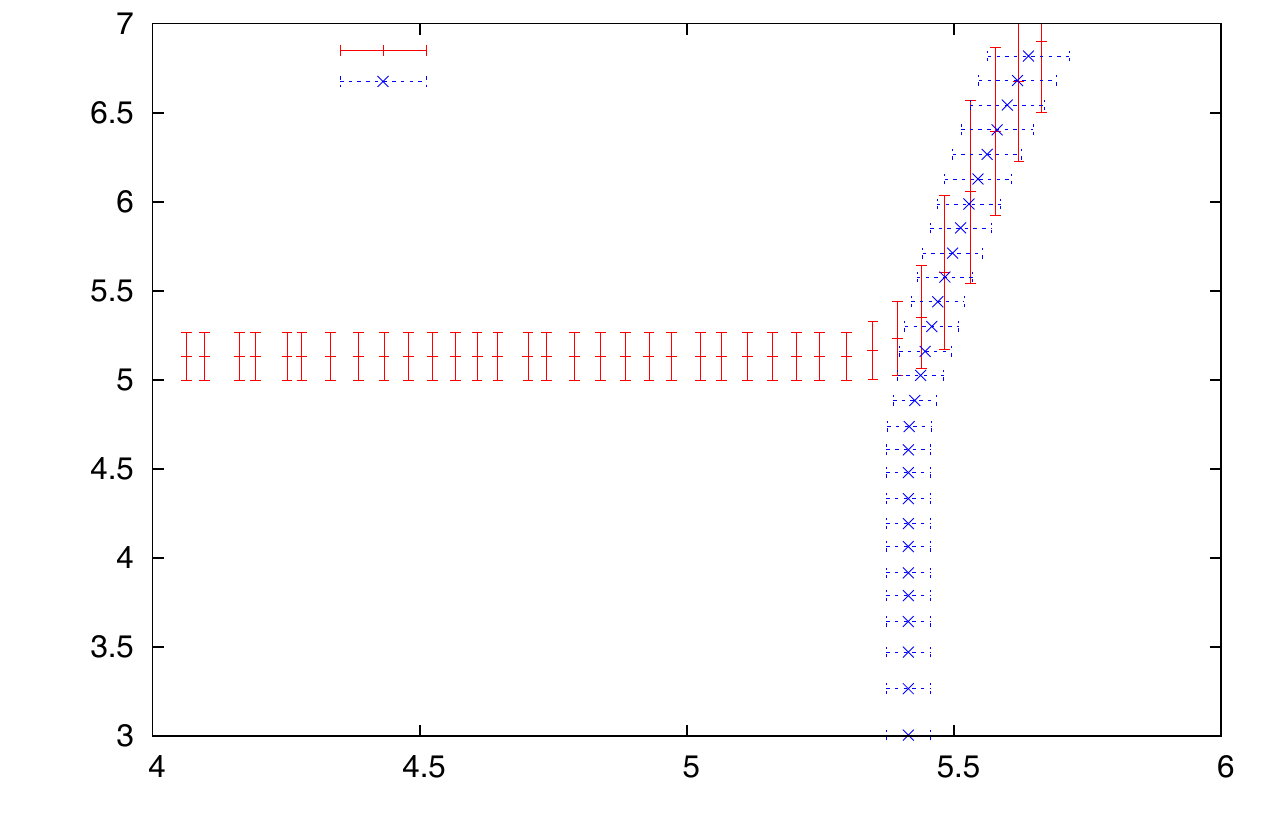}
\put(50,-1){\footnotesize{$\ln(v_\sigma)$}}
\put(-2,33.5){\footnotesize{$\ln(v_\theta)$}}
\put(16,58){\tiny{$\ln(v_\sigma)_{v_\theta}$}}
\put(16,60.5){\tiny{$\ln(v_\theta)_{v_\sigma}$}}
\end{overpic}
\caption{Comparación de los valores esperados de las dos familias de observables
$\ln(\hat{v}_\sigma)_{v_\theta}^{+}$ (azul) y $\ln(\hat{v}_\theta)_{v_\sigma}^{+}$ (roja), que se
corresponden con diferentes elecciones del tiempo interno, calculados sobre el estado considerado
en la Figura \ref{fig:v-trajecto}$(a)$.}
\label{fig:v1-v2-comp}
\end{center}
\end{figure}

Para ilustrar este hecho, hemos considerado una ligera modificación del procedimiento de
construcción de observables. Hasta ahora, hemos usado $v_\theta$ como el tiempo interno. Sin
embargo, debido a la simetría del sistema, cualquiera de las variables $v_i$ puede desempeñar esta
función. Por tanto, para un estado físico dado que es semiclásico a tiempos tardíos, podemos
considerar, por ejemplo, los siguientes observables relacionales: $\ln(\hat{v}_a)_{v_\theta}^{s}$
y $\ln(\hat{v}_{a'})_{v_\sigma}^{s}$, donde $a'=\theta,\delta$, y comparar los correspondientes
valores esperados. La comparación se muestra en la Figura~\ref{fig:v1-v2-comp}.

Fuera de las regiones donde la dinámica efectiva predice rebotes, las trayectorias formadas por las
dos familias consideradas coinciden. Sin embargo, una vez que el valor del tiempo interno
correspondiente a cada familia cae por debajo del punto donde ocurre el rebote efectivo, observamos
discrepancias significativas entre las trayectorias resultantes. De hecho, mientras que una de las
familias muestra un rebote, la otra predice ``la congelación'' de la trayectoria, y viceversa.
Esto implica que en tales regiones las trayectorias \emph{no pueden interpretarse}
como las correspondientes al valor de $\ln(v_i)$ para un tiempo interno dado, ni siquiera en sentido
aproximado. Por tanto, la aplicabilidad de los observables asociados a secciones de $v_i$
constante, con $v_i$ como tiempo interno, es limitada desde un punto de vista físico.

Para obtener una descripción más fiable, con una interpretación física válida en toda la evolución,
hemos empleado la construcción de otra familia de observables, esta vez usando $\beta_\theta$ como
el tiempo interno. Presentamos esta construcción a continuación.

\section{Evolución unitaria con respecto a $\beta_\theta$}
\label{sec:lqc-b-rep}

Una de las principales razones de por qué, en cosmología cuántica de lazos, la construcción de
observables parciales definidos en secciones de $v_\theta$ constante, análoga a la desarrollada para
la cuantización de WDW, no proporciona una evolución unitaria es el hecho de que la
variable elegida como tiempo interno no posee todas las propiedades que debería. De hecho, el
análisis de la dinámica efectiva asociada (véase la Sección \ref{effective}) y de las propiedades de
las autofunciones $e_{\omega_i}^{\varepsilon_i}(v_i)$ muestra que la variable $v_\theta$ no es
monótona
en el tiempo coordenado. Esto nos ha forzado a definir dos familias de operadores correspondientes
a contribuciones en las que el universo está, respectivamente, en contracción y en expansión con
respecto a 
$v_\theta$. Tal separación no es necesaria si identificamos como tiempo interno una variable que
realmente cambie monótonamente con el tiempo coordenado. De nuevo, la dinámica efectiva muestra que
el momento cánonicamente conjugado de $v_\theta$, $\beta_\theta$, posee dicha propiedad. Por tanto,
esto indica que $\beta_\theta$ es un candidato prometedor para desempeñar la función apropiada de
tiempo interno en la teoría cuántica genuina. Veámoslo.

\subsection{Cambio a la representación de $\beta_\theta$}
\label{sec:b-slices}

Recordemos que los momentos conjugados $\beta_i$ de $v_i$ se definen de acuerdo con las ecs.
\eqref{eq:b-def} y \eqref{eq:bv-poiss}. En nuestra cuantización polimérica, la tranformación entre
las representaciones de ``posiciones'' y de ``momentos'' viene dada por la transformación de Fourier
\begin{equation}\label{eq:lqc-v-b-trans}
[{\mathcal{F}}\psi](\beta_i) =
\sum_{\mathcal{L}^+_{\varepsilon_i}} \psi(v_i)
v_i^{-\frac{1}{2}} e^{-\frac{i}{2}v_i\beta_i}.
\end{equation}
Como hemos comentado en la Sección \ref{sec:v-obs}, introducimos un rescalado adicional en
comparación con la transformación \eqref{eq:v-b-trans}, dado por el factor $v_i^{-1/2}$.
Dos razones principales lo motivan. Por una parte este rescalado garantiza una convergencia
adecuada a la hora de computar esta transformación numéricamente y, por otra parte, también
garantiza un comportamiento más conveniente de las autofunciones transformadas, como veremos en la
Sección \ref{sec:b-num}. 

Bajo esta transformación los operadores cinemáticos que componen el operador ligadura escalar y los
observables se convierten en
\begin{align}\label{eq:lqc-ops-b}
\hat{v_i}  &\to  2i\partial_{\beta_i},  \\
 \hat{v}_i^{1/2}\left(\hat{\mathcal
N}_{2\bar\mu_i} - \hat{\mathcal
N}_{-2\bar\mu_i}\right) \hat{v}_i^{-1/2} &\to
2i\sin(\hat{\beta}_i),
\end{align}
donde $\hat{\beta}_i$ actúa en la nueva representación por multiplicación.

Como $v_i\in\mathcal{L}_{\varepsilon_i}^+$, el dominio de $\beta_i$ se puede tomar como el círculo
unidad: $\beta_i\in S^1$. Para adaptar mejor la transformación a nuestro sistema,
notamos lo siguiente:
\begin{itemize}
\item[(i)] Las restricciones de las autofunciones de 
$\widehat\Omega_i$ (con soporte en las semirredes de paso dos
$\mathcal{L}_{\varepsilon_i}^+$) a las subsemirredes 
${}^{(4)}\mathcal{L}_{\varepsilon_i}^+$ y
${}^{(4)}\mathcal{L}_{\varepsilon_i+2}^+$ son autofunciones de
$\widehat\Omega_i^2$, que (definido en un dominio apropiado) difiere del operador de evolución
del modelo isótropo de la Ref. \cite{acs} en un operador compacto.
\label{it:split}
\item[(ii)]
Ya que cualquier sección $v_\theta={\rm constante}$ contiene toda la información de los estados
físicos, la restricción del soporte del estado físico a una de las subsemirredes de 
$v_\theta$ determina unívocamente la otra. 
\label{it:subsemi}
\item[(i)] En el modelo iśotropo plano en el contexto de la cosmología cuántica de lazos
simplificada \cite{acs}, una transformación similar a \eqref{eq:lqc-v-b-trans} permite seleccionar
el
dominio del correspondiente momento $\beta$ como $[0,\pi)$. Más aún, el operador de ese modelo,
análogo a nuestro operador $\widehat\Omega^2_i$, es proporcional a $[\sin(\beta)\partial_\beta]^2$.
Esta propiedad, a su vez, permite escribir la ligadura hamiltoniana como una ecuación de
Klein-Gordon. 
\label{it:KG}
\end{itemize}

Esto indica que es conveniente aplicar la transformación \eqref{eq:lqc-v-b-trans} a las funciones
con soporte en ${}^{(4)}\mathcal{L}_{\varepsilon_i}^+$ y
${}^{(4)}\mathcal{L}_{\varepsilon_i+2}^+$ independientemente.
El dominio de $\beta_i$, entonces, consiste en dos copias en las que $\beta_i$ tiene periodicidad 
$\pi$. En consecuencia, podemos definir la transformación $\mathcal{F}$ de la forma
\begin{equation}
[{\mathcal{F}}_1\psi](\beta_i) = \sum_{v_i\in
{}^{(4)}\mathcal{L}_{\varepsilon_i}^+} \psi(v_i)
v_i{}^{-\frac{1}{2}} e^{-\frac{i}{2}v_i\beta_i}
\end{equation}
y, análogamente la transformación ${\mathcal{F}}_2$, escogiendo 
la suma sobre ${}^{(4)}\mathcal{L}_{\varepsilon_i+2}^+$ en lugar de
${}^{(4)}\mathcal{L}_{\varepsilon_i}^+$.

Usando esta transformación sobre las autofunciones de $\widehat\Omega_\theta$, hemos definido
una representación de los estados físicos ``mixta'', dada por
\begin{align}
\Phi(\beta_\theta,v_\sigma,v_\delta) &=
\int_{\mathbb{R}^2} d\omega_\sigma d\omega_\delta
\tilde{\Phi}(\omega_\sigma,\omega_\delta)
\tilde{e}^{\varepsilon_\theta}_{\omega_\theta}(\beta_\theta)
e^{\varepsilon_\sigma}_{\omega_\sigma}(v_\sigma)
e^{\varepsilon_\delta}_{\omega_\delta}(v_\delta),
\end{align}
donde $\tilde{e}^{\varepsilon_\theta}_{\omega_\theta}(\beta_\theta)$ es
\begin{equation}\label{eq:e-brep-def}
\tilde{e}^{\varepsilon_\theta}_{\omega_\theta}(\beta_\theta) =
\begin{cases}
[{\mathcal{F}}_1 e^{\varepsilon_\theta}_{\omega_\theta}](\beta_\theta),
&  \beta_\theta \in [0,\pi), \\
[{\mathcal{F}}_2
e^{\varepsilon_\theta}_{\omega_\theta}](\beta_\theta-\pi),  &  \beta_\theta \in
[\pi,2\pi).
\end{cases}
\end{equation}

Nosotros hemos restringido nuestras consideraciones a, por ejemplo, el dominio de $\beta_\theta$
que corresponde al primero de estos casos: $\beta_\theta \in [0,\pi)$. En este dominio, hemos
aplicado el análogo de la construcción de observables presentada en las Secciones
\ref{sec:wdw-analog} y \ref{sec:v-evo}.
Para ello, hemos introducido los análogos de los espacios $\mathcal{H}_{v_\theta}$ [dados en la
ec.~\eqref{eq:chidef-alt}], que ahora están definidos en secciones de $\beta_\theta$ constante,
y la transformación de los estados físicos en elementos de estos espacios:
\begin{align}
\hat P_{\beta_\theta}:\mathcal{H}_{\text{fis}}^{\vec\varepsilon}\;\
&\to\;\mathcal{H}_{\beta_\theta}=L^2(\mathbb{R}^2, |\omega_\sigma+\omega_\delta|
|\tilde{e}^{\varepsilon_\theta}_{\omega_\theta}(\beta_\theta)|^{-2}
d\omega_\sigma d\omega_\delta), \nonumber\\
\tilde{\Phi}(\omega_\sigma,\omega_\delta)\;&\mapsto\;
\tilde{\Phi}_{\beta_\theta}(\omega_\sigma,\omega_\delta)
=\ \tilde{e}^{\varepsilon_\theta}_{\omega_\theta}(\beta_\theta)
\tilde{\Phi}(\omega_\sigma,\omega_\delta).
\end{align}
Cada función $\tilde{\Phi}_{\beta_\theta}(\omega_\sigma,\omega_\delta)$, dada en una sección
cualquiera
$\beta_\theta=\text{constante}$, determina unívocamente el estado físico, ya que las autofunciones
$\tilde{e}^{\varepsilon_\theta}_{\omega_\theta}(\beta_\theta)$ no se anulan, salvo quizás en un
conjunto de medida nula. Entonces, el estado $\tilde{\Phi}_{\beta_\theta}$ puede
entenderse como el conjunto de datos iniciales en la sección correspondiente. Por tanto, al igual
que en la descipción para $v_\theta$, se puede desarrollar la noción de evolución en $\beta_\theta$
usando la correspondiente aplicación entre espacios de datos iniciales identificables entre sí. 

\subsection{Observables relacionales}
\label{sec:b-obse}

A partir de los espacios $\mathcal{H}_{\beta_\theta}$, hemos construido los tres observables
relacionales $\ln(v_i)_{\beta_\theta}$. Para $i=a=\sigma,\delta$, la construcción es
enteramente análoga a la de la Sección~\ref{sec:v-obs}, es decir, se introduce la transformación
\begin{align}\label{eq:b-norm}
\tilde{e}{}^{\varepsilon_\theta}_{\omega_\theta}(\beta_\theta)\;\to\;\tilde{e}'{}^{
\varepsilon_\theta } _ { \omega_\theta}(\beta_\theta) =
|\omega_\sigma+\omega_\delta|^{\frac{1}{2}}
\frac{\tilde{e}^{\varepsilon_\theta}_{\omega_\theta}(\beta_\theta)}
{|\tilde{e}^{\varepsilon_\theta}_{\omega_\theta}(\beta_\theta)|},
\end{align}
de modo que la aplicación análoga a la \eqref{eq:rotP-def} entre el espacio de Hilbert físico y un
espacio de Hilbert cinemático auxiliar común para todo $\beta_\theta$ está dada por 
\begin{align}\label{mapeo-b-obs}
\hat R_{\beta_\theta}:\qquad\mathcal{H}_{\text{fis}}^{\vec\varepsilon}\; &\to\;
\mathcal{H}'_{\beta_\theta}=
L^2(\mathbb{R}^2,d\omega_\sigma d\omega_\delta)\ , \nonumber\\
\tilde{\Phi}(\omega_\sigma,\omega_\delta)\; &\mapsto\;
\tilde{{\boldsymbol{\chi}}}_{\beta_\theta}(\omega_\sigma,\omega_\delta)
= \tilde{\Phi}(\omega_\sigma,\omega_\delta)
\tilde{e}'{}^{\varepsilon_\theta}_{\omega_\theta}(\beta_\theta).
\end{align}

La gran novedad es que ahora no es necesario dividir las autofunciones 
$\tilde{e}^{\varepsilon_\theta}_{\omega_\theta}(\beta_\theta)$ en componentes rotantes porque éstas
ya se comportan como tal, excepto posiblemente para valores pequeños de $\omega_\theta$. Esta
propiedad tiene una explicación cualitativa. En efecto, para $\widehat\Omega_\theta^2$ definido en
${}^{(4)}\mathcal H^+_{\varepsilon_\theta}\oplus{}^{(4)}\mathcal
H^-_{\varepsilon_\theta}$, el operador ${\mathcal{F}}_\theta(\widehat\Omega_\theta^2)$ es, salvo
por una corrección compacta, igual al operador $-[12\pi \gamma
l_{\text{Pl}}^2\sin(\beta_\theta)\partial_{\beta_\theta}]^2$ (analizado en la Ref. \cite{acs}),
cuyas
autofunciones generalizadas están formadas
por componentes básicas de la forma 
$N(k)e^{\pm ikx(\beta_\theta)}$, con $x(\beta_\theta) = \ln[\tan(\beta_\theta/2)]$
y $N(k)$ un factor de normalización. Más aún, una vez que restringimos el estudio a $v_\theta>0$,
las autofunciones se reducen a funciones puramente rotantes (con contribución nula de
una de las dos fases $\pm ikx(\beta_\theta)$). Estas autofunciones difieren de las del operador 
${\mathcal{F}}_\theta(\widehat\Omega_\theta^2)$ presente, simplemente, en pequeñas correcciones cuya
norma (cinemática) decrece a medida que $\omega_\theta$ aumenta.
En definitiva, esta vez el rescalado
${\tilde{e}^{\varepsilon_\theta}_{\omega_\theta}(\beta_\theta)}/
{|\tilde{e}^{\varepsilon_\theta}_{\omega_\theta}(\beta_\theta)|}$ es prácticamente una fase pura
para todo el rango de valores de $\beta_\theta$ y, como veremos, la imprecisión de la
interpretación física de los observables, asociada al hecho de que las autofunciones no tienen un
comportamiento rotante exacto, esta vez, es muy pequeña.

Nuevamente, considerando que el operador cinemático $\ln(\hat{v}_a)_{\beta_\theta}$
($a=\sigma,\delta$) actúa por multiplicación sobre los elementos 
\begin{equation}\label{eq:aux-chi-b}
{\boldsymbol{\chi}}_{\beta_\theta}(v_\sigma,v_\delta) =
\int_{\mathbb{R}^2} d\omega_\sigma d\omega_\delta
\tilde{{\boldsymbol{\chi}}}_{\beta_\theta}(\omega_\sigma,\omega_\delta)
e^{\varepsilon_\sigma}_{\omega_\sigma}(v_\sigma)
e^{\varepsilon_\delta}_{\omega_\delta}(v_\delta),
\end{equation}
y deshaciendo el mapeo \eqref{mapeo-b-obs}, encontramos la acción de los observables relacionales
$\ln(\hat{v}_\sigma)_{\beta_\theta}:\mathcal{H}_{\text{fis}}^{\vec\varepsilon}\to\mathcal{H}_{\text{
fis}}^{\vec\varepsilon}$,
\begin{align}\label{eq:lnv-b-def}
&[\ln(\hat{v}_\sigma)_{\beta_\theta}\tilde\Phi](\omega_\sigma,\omega_\delta) =
\frac{1}{\tilde{e}'{}^{\varepsilon_\theta}_{\omega_\theta}(\beta_\theta)}
\int_\mathbb{R} d\omega'_\sigma \langle
e^{\varepsilon_\sigma}_{\omega_\sigma}|\ln(\hat{v}_\sigma)
e^{\varepsilon_\sigma}_{\omega'_\sigma}
\rangle_{\mathcal{H}_{\varepsilon_\sigma}^+}\,
\tilde{e}'{}^{\varepsilon_\theta}_{\omega_\theta(\omega'_\sigma,\omega_\delta)}
(\beta_\theta)\tilde\Phi(\omega'_\sigma,\omega_\delta),
\end{align}
y similarmente para $a=\delta$. Los observables de estas familias están unitariamente relacionados,
exactamente por la misma razón que la explicada para $\ln(\hat{v}_a)_{v_\theta}^\pm$ en la
Sección~\ref{perd-prec}, y el conjunto que forman puede extenderse de igual modo con los observables
$\widehat\Omega_a|_{\beta_\theta}=\widehat\Omega_a$ para formar un conjunto completo de
observables de Dirac.

En este caso, además, con el fin de poder comparar los resultados de esta construcción con los
de la Sección \ref{sec:v-results}, también hemos construido el observable relacional 
$\ln(\hat{v}_\theta)_{\beta_\theta}$. Para dar una definición precisa, primero se necesita expresar
el operador $\ln(\hat{v}_\theta)_{\beta_\theta}$ en la representación de $\beta_\theta$.
Nótese que la función $\ln(v_\theta)$ se puede desarrollar en torno a un punto $v_\theta^o>0$ de la
forma
\begin{equation}
\ln(v_\theta) = \ln(v_\theta^o) - \sum_{n=1}^{\infty} \frac{1}{n
(v_\theta^o)^n}(v_\theta^o-v_\theta)^n.
\end{equation}
Promoviendo los términos $v_\theta^n$ a operadores y aplicando la relación \eqref{eq:lqc-ops-b},
hemos representado el operador $\ln(\hat{v}_\theta)$ como
\begin{equation}\label{eq:lnv-brep}
\ln(\hat{v}_\theta) = \ln(v_\theta^o) - \sum_{n=1}^{\infty}
\frac{1}{n (v_\theta^o)^n}(v_\theta^o-2i\partial_{\beta_\theta})^n,
\end{equation}
definido sobre elementos
${\boldsymbol{\chi}}_{\beta_\theta}(v_\sigma,v_\delta)$ del espacio de Hilbert cinemático.
Con el cambio a la representación de $\omega_a$, dado por la ec. \eqref{eq:aux-chi-b}, la acción
sobre $\tilde{\boldsymbol{\chi}}_{\beta_\theta}(\omega_\sigma,\omega_\delta)$ de dicho observable
tiene la expresión
\begin{equation}
[\ln(\hat{v}_\theta)\tilde{{\boldsymbol{\chi}}}_{\beta_\theta}]
(\omega_\sigma,\omega_\delta) = \tilde{\Phi}(\omega_\sigma,\omega_\delta)
[\ln(\hat{v}_\theta)
\tilde{e}'{}^{\varepsilon_\theta}_{\omega_\theta}](\beta_\theta)
\end{equation}
y, por tanto, sobre el espacio de Hilbert físico, el operador final
$\ln(\hat{v}_\theta)_{\beta_\theta}:\mathcal{H}_{\text{fis}}^{\vec\varepsilon}
\to\mathcal{H}_{\text{fis}}^{\vec\varepsilon}$ es
\begin{align}\label{eq:lnv-action}
[\ln(\hat{v}_\theta)_{\beta_\theta}\tilde{\Phi}](\omega_\sigma,\omega_\delta)
=\frac{[\ln({\hat v_\theta})
\tilde{e}'{}^{\varepsilon_\theta}_{\omega_\theta}](\beta_\theta)}%
{\tilde{e}'{}^{\varepsilon_\theta}_{\omega_\theta}(\beta_\theta)}
\tilde{\Phi}(\omega_\sigma,\omega_\delta),
\end{align}
donde $\ln({\hat v_\theta})$ está dado en la ec. \eqref{eq:lnv-brep}.

Se puede comprobar que, dentro de esta familia de observables, dos operadores diferentes,
correspondientes a tiempos ${\beta_\theta}$ diferentes, no están relacionados mediante un operador
unitario $\hat Q_{\beta_\theta,\beta_\theta^\star}: \mathcal{H}_{\text{fis}}^{\vec\varepsilon} \to
\mathcal{H}_{\text{fis}}^{\vec\varepsilon}$. Por tanto, estos operadores relacionales pueden
desempeñar solo un papel auxiliar. Como hemos dicho, se han construido simplemente por paralelismo
con
el análisis anterior, con el fin de poder obtener gráficas análogas a las de la Figura
\ref{fig:v-trajecto}, en la que se presentan los resultados en el plano $\ln(v_a)-\ln(v_\theta)$.

\section[Predicciones físicas de la evolución en $\beta_\theta$]{Predicciones físicas de la
evolución en $\beta_\theta$}
\label{resultados-b}

Hemos empleado las familias de observables anteriores para analizar la dinámica de
estados físicos que son semiclásicos a tiempos tardíos. A continuación discutimos brevemente los
métodos numéricos usados en nuestros cálculos.

\subsection{Análisis numérico}
\label{sec:b-num}

Como en la Sección \ref{resultados-v}, hemos centrado nuestro estudio en los estados
gaussianos
\eqref{eq:gauss-state}. La mayoría de los cálculos son exactamente los mismos
que los de la Sección~\ref{sec:v-num} para los observables asociados a las secciones de $v_\theta$
constante.
En particular, para calcular los valores esperados y las dispersiones de $\ln(\hat{v}_a)$ hemos
usado
las expresiones análogas a las de las ecs.~\eqref{eq:aux-chi}-\eqref{eq:disp-def}, reemplazando
${\boldsymbol{\chi}}_{v_\theta}^s$ por ${\boldsymbol{\chi}}_{\beta_\theta}$.
Las autofunciones $\tilde{e}'{}^{\varepsilon_\theta}_{\omega_\theta}$, necesarias para determinar
$\tilde{\boldsymbol{\chi}}_{\beta_\theta}$, han sido evaluadas mediante el algoritmo de FFT.
En comparación con la Sección \ref{sec:v-num}, la única diferencia relevante aparece en el cálculo
de $\langle\ln(\hat{v}_\theta)_{\beta_\theta}\rangle$ y de
$\langle\Delta\ln(\hat{v}_\theta)_{\beta_\theta}\rangle$, que ha sido realizado como se explica a
continuación.

En un primer paso, hemos calculado la acción de $\ln(\hat{v}_\theta)_{\beta_\theta}$ en
$\tilde{\Phi}(\omega_\sigma,\omega_\delta)$ de acuerdo con la ec.~\eqref{eq:lnv-action}.
Para encontrar los valores de
$\ln(\hat{v}_\theta)\tilde{e}'{}^{\varepsilon_\theta}_{\omega_\theta}$, hemos actuado con 
$\ln(\hat{v}_\theta)$ en la representación de $v_\theta$ y transformado el resultado a la
representación de $\beta_\theta$, según la transformación~\eqref{eq:lqc-v-b-trans} y aplicando el
método de FFT.

En este punto, sin embargo, es necesario comentar una sutileza técnica. Para calcular este
resultado según lo especificado en la construcción de dicho observable, es necesario normalizar la
autofunción de acuerdo con la ec. \eqref{eq:b-norm}, antes de
actuar sobre ella con $\ln(\hat{v}_\theta)$. Como la normalización debe hacerse en la representación
de $\beta_\theta$, pero la acción del operador se conoce en la representación de $v_\theta$, se
tendrían que llevar a cabo una secuencia de operaciones: una transformada de Fourier, una
normalización y una transformada de Fourier inversa. Desafortunadamente, el algoritmo de FFT que
hemos usado asume que tanto la función que es transformada como el resultado tienen soporte en
puntos que están uniformemente distribuidos en el círculo. Dado que en nuestro caso el soporte de
las
autofunciones está en una subsemirred ${}^{(4)}\mathcal{L}_{\tilde{\varepsilon}_\theta}^+$ completa,
ésta tiene que ser restringida a un conjunto de la forma
${}^{(4)}\mathcal{L}_{\tilde{\varepsilon}_\theta}^+\cap [0,v_{\rm max}]$ (donde $v_{\rm max}$ tiene
un valor grande), lo que elimina la cola de las autofunciones, la cual decae (salvo por
factores logarítmicos) como $v_\theta^{-1}$. Esta restricción daría lugar a errores numéricos que se
manifestarían cerca de $\beta_\theta=0$ y $\beta_\theta=\pi$. A su vez, esto generaría errores de
normalización que, además, se verían amplificados por la comentada secuencia de operaciones (FFT
$\to$ normalización $\to$
FFT) y los resultados obtenidos en el régimen de $v_\theta$ grande no serían fiables.

Para evitar este problema, hemos implementado un método ligeramente diferente. En lugar de
normalizar primero haciendo 
$\tilde{e}{}^{\varepsilon_\theta}_{\omega_\theta}\to
\tilde{e}'{}^{\varepsilon_\theta}_{\omega_\theta}$, nosotros
hemos actuado primero con $\ln(\hat{v}_\theta)$ en las autofunciones sin normalizar y hemos
normalizado después. Esto es, hemos hecho la transformación
\begin{equation}\label{eq:num-act}
\tilde{e}^{\varepsilon_\theta}_{\omega_\theta}(\beta_\theta) \mapsto
\tilde{f}^{\varepsilon_\theta}_{\omega_\theta}(\beta_\theta) :=
\frac{[\ln(\hat{v}_\theta)
\tilde{e}^{\varepsilon_\theta}_{\omega_\theta}](\beta_\theta)}
{|\tilde{e}^{\varepsilon_\theta}_{\omega_\theta}(\beta_\theta)|}.
\end{equation}
Este método introduce un error debido a la diferencia entre la función normalizada y la no
normalizada. No obstante, podemos argumentar que el error es despreciable para estados semiclásicos
picados en torno a valores grandes $\omega_a^\star$, teniendo en cuenta que en ese régimen
$\tilde{e}'{}^{\varepsilon_\theta}_{\omega_\theta}(\beta_\theta)$ puede ser aproximado por
\begin{equation}\label{eq:b-conv}\begin{split}
\tilde{e}'{}^{\varepsilon_\theta}_{\omega_\theta}(\beta_\theta) &\sim
|\omega_\sigma+\omega_\delta|^{\frac{1}{2}}
\left[ e^{i\omega_\theta x(\beta_\theta)} + O(\omega_\theta^{-2}) \right] ,
\\
x(\beta) &=
\ln{\left[\tan{\left(\frac{\beta}{2}\right)}\right]}.
\end{split}
\end{equation}

Una vez se han obtenido las funciones 
$\tilde{f}^{\varepsilon_\theta}_{\omega_\theta}$, hemos evaluado los perfiles
\begin{align}\label{eq:num-lnv1}
[\ln(\hat{v}_\theta){\boldsymbol{\chi}}_{\beta_\theta}](v_\sigma,v_\delta) &=
\int_{\mathbb{R}^2}d\omega_\sigma d\omega_\delta
\tilde{\Phi}(\omega_\sigma,\omega_\delta)
\tilde{f}^{\varepsilon_\theta}_{\omega_\theta}(\beta_\theta)
e^{\varepsilon_\sigma}_{\omega_\sigma}(v_\sigma)
e^{\varepsilon_\delta}_{\omega_\delta}(v_\delta).
\end{align}
En la práctica, hemos integrado el miembro derecho mediante la regla del trapecio \cite{trap-rule},
aplicada sobre
los intervalos $[\omega_a^\star-5\sigma_a^\star,\omega_a^\star+5\sigma_a^\star]$ .

Finalmente, hemos usado los perfiles así calculados para computar los valores esperados
\begin{align}\label{eq:lnv1-exp}
\langle\Phi|\ln(\hat{v}_\theta)_{\beta_\theta}\Phi\rangle &=
\|{\boldsymbol{\chi}}_{\beta_\theta}\|^{-2}
\sum_{\bar{\mathcal{L}}^{2}}
\bar{\boldsymbol{\chi}}_{\beta_\theta}(v_\sigma,v_\delta)
[\ln(\hat{v}_\theta){\boldsymbol{\chi}}_{\beta_\theta}](v_\sigma,v_\delta),
\end{align}
donde, debido a las limitaciones técnicas, hemos restringido la suma al soporte\linebreak
$\bar{\mathcal{L}}^{2}:=(\mathcal{L}_{\varepsilon_\sigma}^+
\cap[0,4\omega_\sigma^\star])\times
(\mathcal{L}_{\varepsilon_\delta}^+\cap[0,4\omega_\delta^\star])$.

Las dispersiones de estos observables se encuentran usando las relaciones usuales
(similares a la relación \eqref{eq:disp-def}). Hemos obtenido los valores esperados 
$\langle\Phi|\ln(\hat{v}_\theta)^2_{\beta_\theta}\Phi\rangle$ con un algoritmo que es completamente
paralelo al usado para
$\langle\Phi|\ln(\hat{v}_\theta)_{\beta_\theta}\Phi\rangle$.

En nuestras simulaciones numéricas, el número de puntos seleccionados para llevar a cabo la FFT de
las autofunciones es igual a $2^{17}$, distribuidos uniformemente en el conjunto
$\beta_\theta\in[0,\pi)$. Hemos realizado las simulaciones para el mismo rango de parámetros que
los de la Sección \ref{sec:v-num}, es decir, para $\omega_a^\star$ perteneciente al intervalo 
$[2\ldotp\!5\cdot 10^2,10^3]$ y para dispersiones relativas en $\omega_a$ entre $0\ldotp\!05$ y
$0\ldotp\!1$.
Pasamos a discutir los resultados obtenidos con estas simulaciones.

\subsection{Resultados}
\label{sec:b-res}

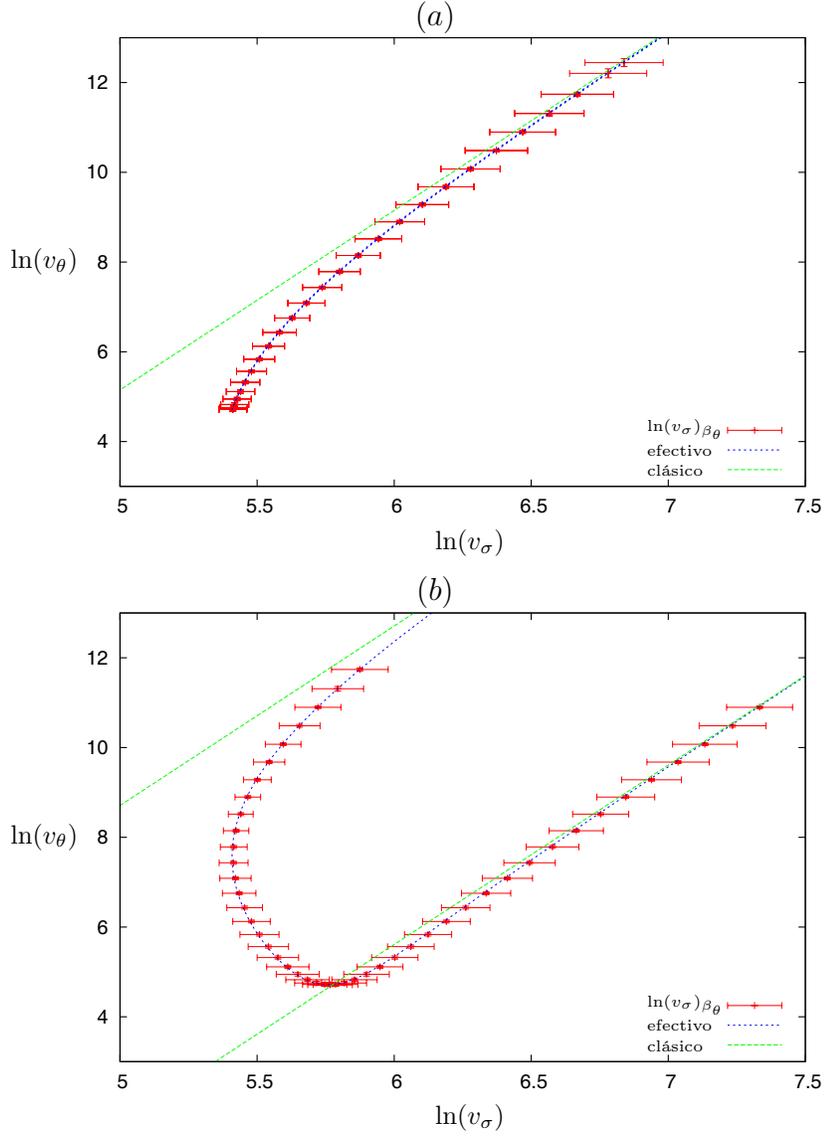
\begin{figure}[htb!]
\begin{center}
$(a)$
\\
\begin{overpic}
[width=0.7\textwidth]{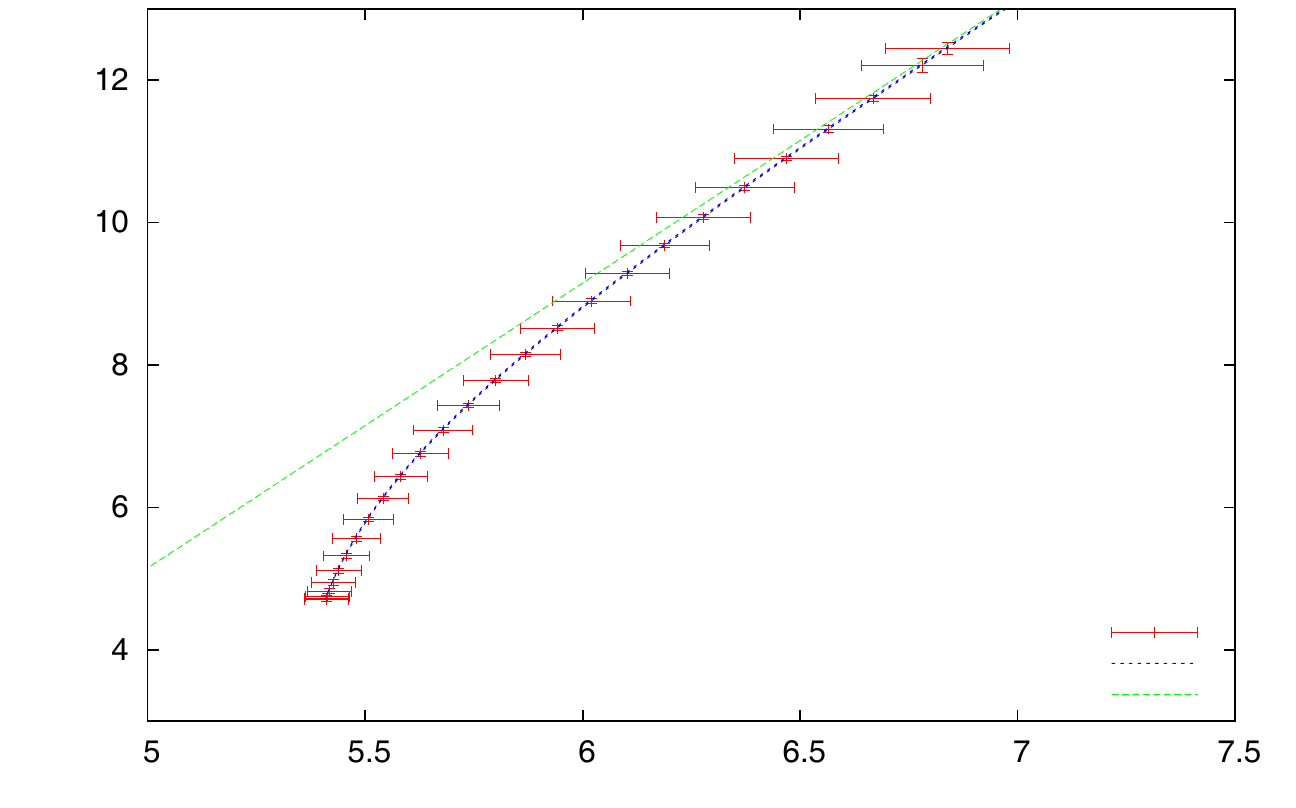}
\put(50,-1){\footnotesize{$\ln(v_\sigma)$}}
\put(-2,33.5){\footnotesize{$\ln(v_\theta)$}}
\put(76,8){\tiny{cl\'asico}}
\put(76,10.5){\tiny{efectivo}}
\put(76,13.5){\tiny{$\ln(v_\sigma)_{\beta_\theta}$}}
\end{overpic}
\\
\vspace*{2mm}
$(b)$
\\
\begin{overpic}
[width=0.7\textwidth]{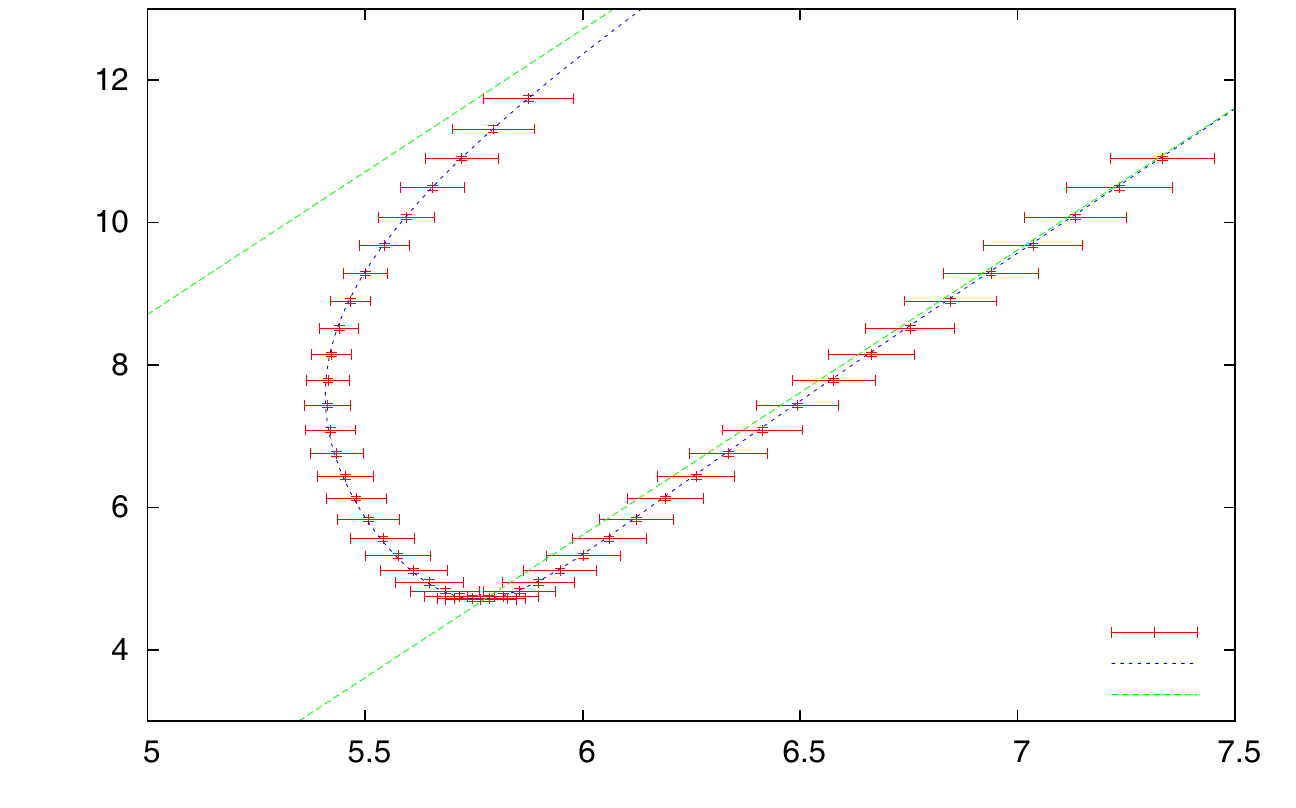}
\put(50,-1){\footnotesize{$\ln(v_\sigma)$}}
\put(-2,33.5){\footnotesize{$\ln(v_\theta)$}}
\put(76,8){\tiny{cl\'asico}}
\put(76,10.5){\tiny{efectivo}}
\put(76,13.5){\tiny{$\ln(v_\sigma)_{\beta_\theta}$}}
\end{overpic}
\caption{Para estados gaussianos [con perfil \eqref{eq:gauss-state}], los valores esperados y
las dispersiones de los observables $\ln(\hat{v}_i)_{\beta_\theta}$, presentados en el plano
$v_a-v_\theta$, son comparados con las trayectorias clásicas
(líneas verdes) y las trayectorias clásicas efectivas (línea azul punteada).
Los estados están picados en $\omega^\star_\sigma=\omega^\star_\delta=10^3$ con dispersiones
relativas determinadas por
$\Delta\Omega_\sigma/\Omega_\sigma=\Delta\Omega_\delta/\Omega_\delta=0\ldotp\!05$.
Las fases son $\nu^\sigma=\nu^\delta=0$ para el caso $(a)$ y
$\nu^\sigma=\nu^\delta=0\ldotp\!1$ para el $(b)$.
Los valores esperados siguen la trayectoria clásica efectiva a lo largo de toda la evolución.
En particular, en $(a)$ las trayectorias antes y después del
rebote coinciden, ya que hemos tomado las fases $\nu^\sigma$ y $\nu^\delta$ nulas.}
\label{fig:b-traject}
\end{center}
\end{figure}

La Figura~\ref{fig:b-traject} muestra un ejemplo representativo del análisis numérico realizado.
Las conclusiones generales que obtenemos de este estudio se resumen en los siguientes puntos:
\begin{itemize}
\item Los estados que están muy picados para algún valor de $\beta_\theta$, permanecen picados
durante toda la evolución.
\item Para valores de $\beta_\theta$ pequeños, los valores esperados de las tres familias de
observables $\ln(\hat{v}_i)_{\beta_\theta}$ siguen las trayectorias clásicas que se corresponden
con un universo en el que las tres variables $v_i$ disminuyen. A medida que $\beta_\theta$ aumenta
y los valores de $v_i$ disminuyen haciéndose del orden de $\omega_i^\star$, observamos una
desviación con respecto a la dinámica clásica, que resulta en rebotes independientes en \emph{las
tres direcciones fiduciales}. Después de cada rebote (en la dirección $i$), el valor de $v_i$
empieza a crecer y su dinámica alcanza rápidamente la del universo clásico.
\item A través de toda la evolución, las trayectorias genuinas cuánticas \emph{coinciden con gran
precisión} (con menos de un $10\%$ de dispersión) con las predichas por la dinámica efectiva
presentada en la Sección \ref{effective} del Apéndice \ref{appB}.
\end{itemize}

Estos resultados proporcionan una confirmación independiente y adicional de las conclusiones
presentadas en la Sección \ref{sec:v-results}. No obstante, el presente análisis constituye una
mejora significativa con respecto al de la Sección \ref{sec:v-evo}. En efecto, si bien es cierto que
todavía los observables usados aquí aún poseen una interpretación física que está bien definida solo
en un sentido
asintótico, la precisión de esta interpretación es buena durante toda la evolución. A diferencia de
la situación encontrada para $\ln(\hat{v}_a)_{v_\theta}^s$, en la que los observables perdían la
interpretación física en algunas épocas de la evolución, el nivel de fiabilidad ahora depende solo
de los valores $\omega^\star_a$ en los que se pica el estado físico gaussiano. Más aún, esta
fiabilidad mejora a medida que consideramos universos con carácter macroscópico más acentuado (mayor
$\omega_a^\star$).

\section{Discusión}

En este capítulo, hemos estudiado la introducción de un concepto de evolución satisfactorio
en cosmología cuántica de lazos. Nuestro objetivo principal ha sido la construcción de una imagen de
evolución unitaria bien definida, en la que la función de reloj es desempeñada por alguno de los
grados de libertad de la geometría y, por tanto, no se hace referencia a ningún campo material.
Como ``banco de pruebas'' para nuestro estudio, hemos elegido el ejemplo específico del modelo de
Bianchi I toroidal, en vacío y cuantizado de acuerdo con el esquema A (esquema factorizable) de la
dinámica mejorada.
En concreto, hemos propuesto dos construcciones diferentes de observables relacionales. En ambos 
casos, hemos construido los observables a partir de los operadores cinemáticos $\ln(\hat{v}_a)$ con
\linebreak
$a=\sigma,\delta$, y como tiempo interno que parametriza estos observables parciales hemos elegido
para cada construcción, respectivamente,  $(i)$ el parámetro afín $v_\theta$ correspondiente al
coeficiente $p_\theta$ de la tríada densitizada y $(ii)$ su momento conjugado $\beta_\theta$.

En las construcciones análogas llevadas a cabo en la cuantización de Wheler-DeWitt del modelo,
presentadas en el capítulo anterior, es fácil obtener una evolución unitaria, y los
observables obtenidos tienen una interpretación física nítida gracias a que las funciones básicas
del
espacio de Hilbert cinemático son, esencialmente, funciones rotantes. Sin embargo, en el caso
presente de la cuantización polimérica la situación es más complicada. Mientras que en el caso
$(ii)$ una construcción casi idéntica a la del modelo de WDW es satisfactoria, la
construcción $(i)$ ha supuesto un desafío mayor. En este caso, la aplicación directa del
procedimiento desarrollado en el modelo cuántico estándar da lugar a observables relacionales que, o
bien no proporcionan una familia de observables parciales unitariamente relacionados, o bien no
contienen ninguna información del sistema físicamente interesante.

Para superar este problema, hemos recurrido a nuestro conocimiento del límite de WDW de
los estados físicos de la cuantización de lazos, lo que ha motivado realizar una separación en dos
componentes, las cuales convergen por separado a funciones rotantes en dicho límite ($v_\theta\to
\infty$). La construcción de observables aplicada a cada una de estas
componentes nos ha permitido, finalmente, conseguir la unitariedad requerida en la evolución.

En contraste con la construcción en la cuantización de WDW, y debido a que ahora las
funciones básicas que caracterizan la construcción no son exactamente objetos rotantes, hemos visto
que la interpretación física de que los operadores 
construidos, tanto en el caso $(i)$ como en el $(ii)$, miden el valor de $\ln(v_a)$ a un tiempo
dado, $v_\theta$ o $\beta_\theta$, solo es
aproximadamente correcta. En particular, para el caso $(i)$ la interpretación no es nada fiable, ni
siquiera en sentido aproximado, en cierta época de la evolución, que corresponde a
valores de $v_\theta$ por debajo de una cota que depende de los parámetros que caracterizan el
estado en el que se mide el observable. Por otro lado sí es bastante precisa cuando $v_\theta$
supera dicho
valor. En el caso $(ii)$, sin embargo, mientras que la interpretación todavía es aproximada, es
bastante precisa durante \emph{toda} la evolución y es muy buena para los estados físicamente más
interesantes.

Hemos empleado los observables relacionales de ambas construcciones para analizar la dinámica de
estados con perfil espectral gaussiano muy picado en valores grandes de los vectores de ondas
$\omega_a^\star$. Hemos demostrado que dichos estados se mantienen picados a lo largo de la
evolución, siguiendo trayectorias clásicas cuando el valor de
$v_\theta/\omega_\theta(\omega_\sigma^\star,\omega_\delta^\star)$ es grande, mientras que, cuando el
valor de este cociente decrece, la dinámica cuántica se desvía de la clásica dando lugar a la
aparición de \emph{rebotes cuánticos}. Es decir, al igual que ocurre en el modelo isótropo,
universos semiclásicos con coeficientes $p_i$ de la tríada densitizada en expansión se conectan con
universos semiclásicos con coeficientes $p_i$ en contracción mediante una secuencia de
rebotes que son independientes para cada dirección.

Cabe señalar que la complejidad de los problemas numéricos que hemos tenido que resolver en
este modelo no nos ha permitido probar tantos estados semiclásicos como los que se analizaron en el
estudio del modelo isótropo \cite{aps3} y, de hecho, solo nos hemos restringido a estados
gaussianos.
Tampoco hemos podido comprobar realmente la evolución del sistema en tiempos tardíos. No obstante,
gracias a la relación entre esta cuantización y la cuantización de WDW, hemos podido
hacer un análisis mucho más general que el numérico para examinar dichos regímenes y comprobar que,
para estados bastante genéricos (no solo gaussianos), el comportamiento semiclásico se mantiene para
las dos
ramas de la evolución, antes y después de los rebotes cuánticos. 

Para concluir, conviene destacar que nuestro estudio es pionero en el desarrollo de una nueva
metodología que puede ser aplicada en cualquier situación en la que se quiera analizar la evolución
en ausencia de grados de libertad que estén cuantizados de modo estándar. Por tanto, puede ser
aplicado en una gran variedad de sistemas cuantizados poliméricamente.
En particular, se podría aplicar a la cuantización del modelo de Bianchi I según el
esquema B (esquema no factorizable) de la dinámica mejorada, discutida en el
Capítulo~\ref{chap:bianchi}, donde los análogos
de las construcciones $(i)$ y $(ii)$ presentadas aquí se construirían tomando como tiempos internos
el parámetro $v$ proporcional al volumen y su momento canónicamente conjugado, respectivamente. Por
supuesto, tal construcción requeriría conocer las propiedades espectrales del operador ligadura en
dicho esquema y la forma de sus autofunciones. Por otra parte, la cuantización del modelo de
Bianchi I según el esquema B presentaría un
inconveniente con respecto a la descripción expuesta aquí de la evolución para el
esquema A. En efecto, con la evolución aquí analizada hemos podido 
reconocer y demostrar la pérdida de una capacidad de predicción correcta de los observables
$\ln(\hat v_a)_{v_\theta}$ en cierta época de la evolución. Esto ha sido posible gracias a que
hemos podido intercambiar las
funciones de tiempo interno y de variable dinámica entre las variables $v_\theta$ y $v_a$. Esta
posibilidad no sería factible en la descripción correspondiente al modelo en el esquema B que,
no obstante, sufriría del mismo problema, ya que la variable $v$
también representaría un tiempo polimérico con dos épocas de evolución asociadas (expansión y
contracción).

\cleardoublepage

\part{Cosmología cuántica híbrida}
\label{part3}

\chapter{Cuantización híbrida del modelo de Gowdy}
\label{chap:gowdy-hyb}

Como acabamos de ver con algunos modelos homogéneos, la
singularidad cosmológica clásica es eludida en cosmología cuántica de lazos. No obstante, nos
podríamos cuestionar si la resolución cuántica de las singularidades es un mecanismo general de
dicha teoría o si, por el contrario, es simplemente un producto de la gran simetría que presentan
los modelos estudiados. Para responder a esta pregunta parece inevitable extender esta cuantización
a escenarios inhomogéneos. Más importante aún, es
imprescindible dar este paso si se quiere desarrollar una teoría realista de cosmología cuántica,
ya que, como indican los resultados obtenidos en cosmología moderna, en los primeros instantes del
universo las inhomogeneidades desempeñaron un papel fundamental en la formación de las estructuras
cósmicas que observamos hoy en día.

Con el objetivo de progresar en esta dirección,
nosotros hemos considerado la cuantización de uno de los modelos cosmológicos inhomogéneos más
sencillos: el modelo de Gowdy con secciones espaciales tres-toroidales y polarización lineal
\cite{gowdy1,gowdy2}. A pesar de que este sistema no posee ninguna isometría de tipo tiempo y además
contiene un número infinito de grados de libertad, existe esencialmente una \emph{única}
cuantización de Fock satisfactoria del mismo \cite{men2,men3}, como hemos comentado en el Capítulo
\ref{chap:introGowdy}. Por otra parte, las soluciones clásicas de este modelo de Gowdy representan,
genéricamente, espacio-tiempos con una singularidad inicial de curvatura \cite{mon1,mon2,ise}.
Además, sus soluciones homogéneas describen espacio-tiempos de Bianchi I con
la topología del tres-toro, modelo que ha sido ampliamente analizado, en particular en esta
tesis, y en el que la singularidad queda resuelta como consecuencia de la cuantización
polimérica. Por tanto, es natural preguntarse cómo afecta a la resolución de la singularidad
cosmológica la inclusión de inhomogeneidades en este modelo particular. 

A la vista de las características del modelo de Gowdy considerado, resumidas en el Capítulo
\ref{chap:introGowdy}, la posibilidad más sencilla para
investigar esta cuestión es llevar a cabo una cuantización híbrida, que combina la cuantización
polimérica de los grados de libertad que parametrizan las soluciones homogéneas con la cuantización
de Fock para las inhomogeneidades. Esta idea constituye la base de nuestro trabajo, cuyos aspectos
principales se recogen en tres de nuestras publicaciones \cite{gow-let,gow-ijmp,gow-B}. En la
literatura existen también otros estudios de la cuantización de este modelo usando el formalismo
de Ashtekar. No obstante, o bien no se emplean técnicas poliméricas en su cuantización
\cite{guille-gow}, o bien, pese a aplicar técnicas de cuantización de lazos a todos los grados de
libertad, solo se representa la estructura cinemática del modelo \cite{date1,date2}. También se
han llevado a cabo otros análises en los que se investiga el papel desempeñado por
las inhomogeneidades en el marco de la cosmología cuántica de lazos, como el de la Ref.~\cite{rv}
en el que se adopta un truncamiento de la gravedad cuántica de lazos y se emplea una aproximación
de tipo Born-Oppenheimer o el de la Ref. \cite{bhks} en el que se propone una teoría lineal de
perturbaciones que implementa de modo efectivo las correcciones cuánticas.

Nuestro procedimiento híbrido asume que existe un régimen en el que los fenómenos más
importantes emergentes de la discretización de la geometría son aquéllos que afectan al
subsistema homogéneo,
mientras que tales efectos son pequeños y pueden ser ignorados para las inhomogeneidades, incluso
si éstas todavía presentan un marcado comportamiento cuántico estándar (de tipo Fock). Aunque
nuestra
cuantización no proviene de una reducción de la gravedad cuántica de lazos, retiene características
interesantes asociadas a la naturaleza polimérica y discreta de la geometría, declarando un cierto
tipo de jerarquía perturbativa en su relevancia sobre diferentes subsectores del sistema
cosmológico.

De hecho, nuestra cuantización híbrida nos ha permitido no solo investigar si la singularidad se
resuelve debido a los efectos cuánticos de la geometría, sino también explorar si la aplicación de
la cuantización polimérica exclusivamente a los grados de libertad homogéneos
(modos cero en nuestra descripción) es suficiente para curar la singularidad inicial. Como veremos,
la respuesta es afirmativa en nuestro modelo. La simplicidad de este resultado refuerza el interés
de nuestro procedimiento híbrido, ya que la misma estrategia se puede aplicar a la cuantización de
cosmologías más generales. 

Asimismo, la cuantización híbrida del modelo de Gowdy proporciona un escenario especialmente
conveniente para llevar a cabo el análisis de otros aspectos importantes en gravedad y cosmología
cuánticas, tales como la recuperación de la teoría cuántica de campos estándar. En el modelo de
Gowdy, las inhomogeneidades se pueden ver como ondas gravitacionales (convenientemente escaladas)
que
se propagan sobre un fondo curvo y homogéneo, es decir, sobre el espacio-tiempo de Bianchi I. En una
cuantización estándar, las inhomogeneidades se pueden describir en términos de un espacio de Fock.
Entonces, nos podemos preguntar cuál es el estatus de esta descripción de Fock desde el punto de
vista de la cosmología cuántica de lazos, donde se adopta una cuantización polimérica inequivalente
a la estándar. Veremos que, con nuestro procedimiento híbrido, de hecho, recuperamos la descripción
de Fock estándar, pese a que hemos cuantizado el fondo empleando técnicas poliméricas.

\section{Descripción clásica híbrida del modelo}

Nuestro punto de partida lo constituye la descripción clásica del modelo en la parametrización
$\{(\tau,P_\tau),(\xi,P_\xi),(\bar\gamma,P_{\bar\gamma})\}$ de la Ref. \cite{men1b}, dada en las
ecs.
\eqref{eq:gowdy-metric-ccm}-\eqref{gowdy-class}.

\subsection{Fijación parcial de gauge sin deparametrización}

A diferencia del procedimiento seguido en la Ref. \cite{men1b}, donde la ligadura hamiltoniana se
fija completamente, nosotros no hemos deparametrizado completamente el sistema, con el objetivo de
tener una ligadura hamiltoniana remanente que hemos impuesto cuánticamente, como se hace en
cosmología cuántica de lazos homogénea.

Así, como ya hicimos en el Capítulo \ref{chap:introGowdy}, imponemos que el generador
$P_{\bar\gamma}$ de las transformaciones conformes de la métrica inducida en el conjunto de las
órbitas de isometría sea una función homogénea:
\begin{align}\label{eq:g1-mmg}
&g_1\equiv P_{\bar\gamma}-\frac{P_{\bar\gamma_{_0}}}{\sqrt{2\pi}}=0,
\end{align}
pero también imponemos la homogeneidad del área de dichas órbitas $\tau$:
\begin{align}
 g'_2\equiv \tau-\frac{\tau_{0}}{\sqrt{2\pi}}=0.
\end{align}
Nótese la diferencia entre nuestra condición $g'_2$ y la condición $g_2$, dada en la ec.
\eqref{eq:g2-ccm}, que fija completamente el área de las órbitas $\tau$ (que es una función
global del tiempo). Nosotros hemos dejado sin fijar su modo cero u homogéneo.

Las condiciones análogas a las expresiones \eqref{eq:gauge-property1} y \eqref{eq:gauge-property2}
para que nuestra fijación de gauge esté bien definida exigen que $P_{\bar\gamma_{_0}}$ y
$\tau_{_0}$ no se anulen.
Además, la condición adicional de compatibilidad con la dinámica implica que, tanto la componente 
$N^\theta$ del vector desplazamiento, como el lapso densitizado ${N_{_{_{\!\!\!\!\!\!\sim}}\;}}$ son
funciones homogéneas, esto es, $N^\theta=N^\theta(t)$ y
${N_{_{_{\!\!\!\!\!\!\sim}}\;}}={N_{_{_{\!\!\!\!\!\!\sim}}\;}}(t)$. Por tanto, hemos fijado toda la
libertad gauge asociada a los modos inhomogéneos (no cero) tanto de la ligadura de momentos en la
dirección inhomogénea $\theta$, $\mathcal C_\theta$, como de la ligadura escalar densitizada,
$\tilde{\mathcal C}$.

En consecuencia, hemos fijado todos los grados de libertad gauge excepto dos (en el espacio de
configuración). El espacio de fases reducido ahora está parametrizado por los grados de libertad de
tipo ``partícula puntual'' $(\tau_{_0},P_{\tau_{_0}})$ y $(\bar\gamma_{_0},P_{\bar\gamma_{_0}})$ y
por los campos $(\xi,P_\xi)$. Dos ligaduras
globales (homogéneas) permanecen constriñendo el sistema: el generador de traslaciones en el
círculo $C_\theta$ (como en la Ref. \cite{men1b}) y el generador de reparametrizaciones temporales
$\tilde C$. En términos de las variables $(\tau_{_0},P_{\tau_{_0}})$,
$(\bar\gamma_{_0},P_{\bar\gamma_{_0}})$ y $(\xi,P_\xi)$, la acción reducida del sistema es
\begin{align}
S&=\int_{t_i}^{t_f}
dt\Bigg[P_{\tau_{_0}}\dot{\tau}_{_0}\!+\!P_{\bar\gamma_{_0}}\dot{\bar\gamma}_{_0}
\!+\!\oint d\theta
P_\xi\dot{\xi}\!-\!({N_{_{_{\!\!\!\!\!\!\sim}}\;}}\tilde C\!+\!N^\theta C_\theta)\Bigg],\\
C_\theta&=\oint d\theta P_\xi\xi^\prime=0,\\
\tilde C&=\frac1{\sqrt{2\pi}}\frac{4G}{\pi}\Bigg\{-\tau_{_0}
P_{\tau_{_0}}
P_{\bar\gamma_{_0}}+\frac{P_{\bar\gamma_{_0}}^2}{8\tau_{_0}}\oint d\theta
\xi^2+\frac{\tau_{_0}}{2}\oint d\theta
\Bigg[P_\xi^2+\left(\frac{\pi}{4G}\right)^2(\xi^\prime)^2\Bigg]\Bigg\}=0,
\end{align}
obtenida a partir de la acción \eqref{gowdy-class} mediante la integración sobre $\theta$
y teniendo en cuenta la homogeneidad de $\tau$, $P_{\bar\gamma}$,
$N^\theta$ y ${N_{_{_{\!\!\!\!\!\!\sim}}\;}}$.

\subsection{Sector homogéneo en variables de Ashtekar-Barbero}

El sector homogéneo del espacio de fases está constituido por las variables de configuración
$(\xi_{_{0}},\tau_{_{0}},\bar\gamma_{_{0}})$ y por sus momentos conjugados
$(P_{\xi_{_0}},P_{\tau_{_0}},P_{\bar\gamma_{_0}})$. Con el fin de cuantizar poliméricamente este
sector, de acuerdo con el procedimiento seguido en cosmología cuántica de lazos, primero hemos
necesitado expresarlo en términos de variables de Ashtekar. Como ya hemos comentado, la subfamilia
de
soluciones homogéneas del modelo de Gowdy, parametrizada por este sector, se corresponde con el
modelo de Bianchi~I en vacío y con secciones espaciales cuya topología es la del tres-toro. Su
descripción en términos de la formulación de Ashtekar ha sido explicada en la Sección
\ref{4sec:clas}.
En esta formulación, el espacio de fases de dicho modelo se describe con las variables 
$\{(c^i,p_i),i=\theta,\sigma,\delta\}$, tales que $\{c^i,p_j\}=8\pi G\gamma\delta^i_j$.

Por tanto, hemos buscado la transformación canónica que relaciona las variables 
$\{(\xi_{_0},P_{\xi_{_0}}),(\tau_{_{0}},P_{\tau_{_0}}),(\bar\gamma_{_{0}},P_{\bar\gamma_{_0}})\}$
con las variables $\{(c^i,p_i),i=\theta,\sigma,\delta\}$. Para ello, hemos comparado la tres-métrica
de Bianchi I en el formalismo de Ashtekar:
\begin{align}
q_{ii}=\frac{|p_\theta p_\sigma p_\delta|}{(2\pi p_i)^2},
\end{align}
con su expresión en
términos del primer conjunto de variables, que se obtiene ignorando las inhomogeneidades en la
métrica de Gowdy \eqref{eq:gowdy-metric-ccm}, esto es,
\begin{align}
q_{\theta\theta}&=\frac{4G}{\pi}\exp\bigg[\frac1{\sqrt{2\pi}}\bigg(\bar\gamma_{_0}
-(2\pi)^{1/4}\frac{\xi_{_0}}{\sqrt{\tau_{_0}}}-\frac{\xi_{_0}^2}{4\tau_{_0}}\bigg)\bigg],\nonumber\\
q_{\sigma\sigma}&=\frac{\pi}{4G}\frac{\tau_{_0}^2}{2\pi}
\exp\left(-\frac1{(2\pi)^{1/4}}\frac{\xi_{_0}}{\sqrt{\tau_{_0}}}\right),\\
q_{\delta\delta}&=\frac{4G}{\pi}\exp
\left(\frac1{(2\pi)^{1/4}}\frac{\xi_{_0}}{\sqrt{\tau_{_0}}}\right).\nonumber
\end{align}

La comparación entre estas dos expresiones de la tres-métrica espacial proporciona una
transformación de contacto que puede completarse inmediatamente a una transformación canónica. El
resultado es
\begin{align}
\xi_{_0}&=\frac1{\sqrt{2\pi}}|p_\theta|^{1/2}\ln\left(\frac1{16\pi
G}\left|\frac{p_\theta p_\sigma}{p_\delta}\right|\right),\nonumber\\
\tau_{_{0}}&=\frac{\sqrt{2\pi}}{4\pi^2}|p_\theta|,\nonumber\\
\bar\gamma_{_{0}}&=\sqrt{2\pi}\left\{2\ln\left(\frac{|p_\sigma|}{16\pi
G}\right)+\frac1{4}\left[\ln\left(\frac1{16\pi
G}\left|\frac{p_\theta p_\sigma}{p_\delta}\right|\right)\right]^2\right\},\nonumber\\
P_{\xi_{_0}}&=\frac1{8\pi
G\gamma}\sqrt{2\pi}\frac1{|p_\theta|^{1/2}}\Bigg[c^\delta
p_\delta+\frac1{4}(c^\sigma p_\sigma+c^\delta
p_\delta)\ln\left(\frac1{16\pi
G}\left|\frac{p_\theta p_\sigma}{p_\delta}\right|\right)\Bigg],\label{tcanH}\\
P_{\tau_{_0}}&=\frac1{8\pi
G\gamma}\frac{4\pi^2}{\sqrt{2\pi}}\frac1{|p_\theta|}\left\{-c^\theta
p_\theta-c^\delta p_\delta-\frac1{8}(c^\sigma p_\sigma+c^\delta
p_\delta)\left[\ln\left(\frac1{16\pi
G}\left|\frac{p_\theta
p_\sigma}{p_\delta}\right|\right)\right]^2\right.\nonumber\\&\left.-\frac{c^\delta
p_\delta}{2}\ln\left(\frac1{16\pi G}\left|\frac{p_\theta
p_\sigma}{p_\delta}\right|\right)\right\},\nonumber\\
P_{\bar\gamma_{_0}}&=-\frac1{8\pi G\gamma}\frac1{2\sqrt{2\pi}}(c^\sigma
p_\sigma+c^\delta p_\delta)\nonumber.
\end{align}

\subsection{Métrica del modelo de Gowdy en las nuevas variables}

El resto de variables del modelo de Gowdy,
$\{(\xi_{_m},P_{\xi_{_m}}),m\in\mathbb{Z}-\{0\}\}$, forman el sector inhomogéneo.
En este sector, hemos implementado la cuantización de Fock realizada en la Ref. \cite{men1b} y
explicada en la Sección \ref{2sec:fock}. Por ello, hemos introducido las variables de tipo
aniquilación y creación $(a_m,a_m^*)$ que se asocian de forma natural
a un campo libre escalar sin masa, definidas en la ec. \eqref{eq:variables-fock}.

Para expresar la métrica \eqref{eq:gowdy-metric-ccm} en términos de las nuevas variables 
 $(c^i,p_i)$ y $(a_m,a_m^*)$, notamos primero que, al igual que en el Capítulo
\ref{chap:introGowdy}, $\bar\gamma$ está determinada por $\xi$, $P_\xi$ y
$P_{\bar\gamma_{_0}}\neq0$ de acuerdo con la expressión \eqref{eq:gamma}.
Teniendo esto en cuenta, la homogeneidad de $\tau$, la expresión de los campos en
modos de Fourier [usando el convenio \eqref{eq:Fourier-modes}] y las transformaciones canónicas a
las variables nuevas, obtenemos la expresión de la tres-métrica del modelo polarizado de Gowdy $T^3$
en esta parametrización. El resultado es:
\begin{align}
q_{\theta\theta}&=\frac1{4\pi^2}\left|\frac{p_\sigma
p_\delta}{p_\theta}\right|\exp\left\{\frac{2\pi}{\sqrt{|p_\theta|}}\frac{c^\delta p_\delta-c^\sigma
p_\sigma}{c^\sigma p_\sigma+c^\delta p_\delta}\tilde\xi(\theta)-\frac{\pi^2}{|p_\theta|}[
\tilde\xi(\theta)]^2 -\frac{8\pi G\gamma}{c^\sigma
p_\sigma+c^\delta p_\delta} \zeta(\theta)\right\},\nonumber\\
q_{\sigma\sigma}&=\frac1{4\pi^2}\left|\frac{p_\theta
p_\delta}{p_\sigma}\right|\exp\left\{-\frac{2\pi}{\sqrt{|p_\theta|}}\tilde\xi(\theta)\right\},\label
{newmetric} \nonumber\\q_{\delta\delta}&=\frac1{4\pi^2}\left|\frac{p_\theta
p_\sigma}{p_\delta}\right|\exp\left\{\frac{2\pi}{\sqrt{|p_\theta|}}\tilde\xi(\theta)\right\},
\end{align}
donde
\begin{align}
\tilde\xi(\theta)&=\frac1{\pi}\sum_{m\neq0}\sqrt{\frac{G}{|m|}}(a_m+a_{-m}^*)e^{im\theta},
\nonumber\\
\zeta(\theta)&=i\sum_{m\neq0}\sum_{\tilde m\neq0}\text{sgn}(m+\tilde m)\frac{\sqrt{|m+\tilde
m||\tilde m|}}{|m|}(a_{-\tilde {m}}-a^*_{\tilde m}\big)\big(a_{m+\tilde m}+a^*_{-(m+\tilde
m)}\big)e^{im\theta}.\nonumber
\end{align}
Por otra parte, debido a la homogeneidad de la función  $N^\theta$, ésta puede ser reabsorbida
mediante la siguiente redefinición de la coordenada $\theta$:
\begin{align}
d\theta+dtN^\theta(t)\rightarrow d\theta.\nonumber
\end{align}
Por tanto, la métrica espacio-temporal es [véase la ec. \eqref{eq:gowdy-metric-ccm}]
\begin{align}
ds^2=-q_{\theta\theta}\left(\frac{|p_\theta|}{4\pi^2}\right)^2
{N_{_{_{\!\!\!\!\!\!\sim}}\;}}^2dt^2+q_{\theta\theta}d\theta^2+q_{\sigma\sigma}d\sigma^2
+q_{\delta\delta}d\delta^2.\nonumber
\end{align}
Nótese que, si ignoramos las inhomogeneidades, es decir, si hacemos
$\tilde\xi=0=\zeta$ en la ec.~\eqref{newmetric}, recuperamos la métrica de Bianchi I, como era de
esperar.

\subsection{Ligaduras en las nuevas variables}

Haciendo el cambio a las nuevas variables, ahora la acción es (salvo por un término de frontera
irrelevante):
\begin{subequations}\begin{align}\label{accion}
S&=\int_{t_i}^{t_f} dt\Bigg[\frac1{8\pi G\gamma}(p_\theta\dot
c^\theta+p_\sigma\dot c^\sigma+p_\delta\dot
c^\delta)+i\sum_{m\neq0}a_m^*\dot a_m-\left(N^\theta
C_\theta+\frac1{16\pi
G}\frac{N_0}{(2\pi)^3}C_{\text{G}}
\right)\Bigg],\\
C_\theta&=\sum_{m=1}^\infty
m(a_m^*a_m-a_{-m}^*a_{-m})=0,\label{ct}\\
C_{\text{G}}&=C_\text{BI}+C_\xi=0,\quad C_\xi=\frac{G}{V}\bigg[\frac{(c^\sigma p_\sigma+
 c^\delta p_\delta)^2}{\gamma^2|p_\theta|}
H_\text{int}^\xi+32\pi^2|p_\theta|
H_0^\xi\bigg].\label{Cclas}
\end{align}\end{subequations}
En esta expresión,
\begin{align}
\label{HIclas} H_\text{int}^\xi
 &=\sum_{m\neq
0}\frac{1}{2|m|}\left[2a^*_ma_m+
a_ma_{-m}+a^*_ma^*_{-m}\right], \qquad H_0^\xi&=\sum_{m\neq
0}|m|a^*_ma_m,
\end{align}
$C_\text{BI}$ es la ligadura de Bianchi I [ec. \eqref{eq:ligBianchi}] y $V=\sqrt{|p_\theta p_\sigma
p_\delta|}$ es el volumen \emph{homogéneo} físico del universo de Bianchi I asociado al universo de
Gowdy. Es importante señalar que, gracias a la homogeneidad del lapso \emph{densitizado}, hemos
cambiado su densitización haciendo $N_0\equiv V{N_{_{_{\!\!\!\!\!\!\sim}}\;}}$, de modo que la
ligadura escalar $C_{\text{G}}$ tiene la densitización habitual de la ligadura escalar en gravedad
cuántica de lazos, y habitual también en nuestros trabajos en cosmología cuántica de lazos (véase la
discusión hecha al final de la Sección~\ref{3sec:dens}).

En la ligadura hamiltoniana las inhomogeneidades están presentes
en el término $H_0^\xi$, que se corresponde con el hamiltoniano de un campo libre escalar sin masa,
y en el término $H_\text{int}^\xi$, que representa un término de interacción cuadrático en el campo.
Nótese que las inhomogeneidades están acopladas al sector homogéneo de un modo nada trivial lo
que hace que la cuantización híbrida de este sistema
no sea sencilla en absoluto.

\section[Representación cuántica del generador de rotaciones]{Representación cuántica del generador
de\\rotaciones}

\subsection{Cuantización de Fock del sector inhomogéneo}

Como hemos dicho, para representar este sector en la teoría cuántica, hemos empleado la cuantización
de
Fock de la Ref. \cite{men1b}, descrita en la Sección \ref{2sec:fock}. Es decir, hemos promovido las
variables  $a_m$ y $a^*_m$ a operadores de aniquilación y creación, $\hat a_m$ y
$\hat a^\dagger_m$, sobre el espacio de Fock $\mathcal{F}$, tal que $[\hat a_m,\hat
a^\dagger_{\tilde m}]=\delta_{m\tilde m}$.
Estos operadores están densamente definidos en el subespacio del espacio de Fock cuyos elementos son
combinaciones lineales finitas de los estados de $n$-partículas
\begin{align}
 |\mathfrak n \rangle:=|...,n_{-2},n_{-1},n_1,n_2,...\rangle,
\end{align}
tales que $\sum_mn_m<\infty$, siendo $n_m\in\mathbb{N}$ el número de ocupación o número de
partículas del modo $m$. Denominaremos a este espacio $\mathcal S$. Nótese que los estados de
$n$-partículas proporcionan una base ortonormal del espacio de Fock, ortonormalizados con respecto
al producto interno $\langle\mathfrak n^\prime|\mathfrak n\rangle=\delta_{\mathfrak
n^\prime\mathfrak n}$.

La acción de $\hat a_m$ y $\hat a^\dagger_m$ sobre estos estados es
\begin{align}
 \hat a_m |...,n_m,...\rangle=\sqrt{n_m}|...,n_m-1,...\rangle;\quad 
\hat a_m^\dagger|...,n_m,...\rangle=\sqrt{n_m+1}|...,n_m+1,...\rangle.
\end{align}

\subsection{Ligadura de traslaciones en el círculo}

El generador de traslaciones en el círculo $C_\theta$ no afecta al sector homogéneo y es
exactamente la misma ligadura que la representada en la Sección \ref{2sec:fock}, cuyo operador
asociado es, por tanto,
\begin{align}
\widehat{C}_\theta=\hbar\sum_{m>0}^\infty m\hat{X}_m,\quad\hat{X}_m=\hat{a}^{\dagger}_m
\hat{a}_m-\hat{a}^{\dagger}_{-m} \hat{a}_{-m}.
\end{align}
Los estados de $n$-partículas aniquilados por este operador son los que satisfacen la condición
\begin{equation}\label{Fphys}
\sum_{m>0}^\infty m X_m=0,\quad X_m=n_m-n_{-m},
\end{equation}
y forman base de un subespacio propio del espacio de Fock, que denominamos $\mathcal F_f$.

\section{Operador ligadura hamiltoniana}

Los estados físicos deben ser también aniquilados por el operador que representa a la ligadura
escalar clásica $C_\text{G}$, dada en la ec. \eqref{Cclas}, la cual involucra tanto al sector
homogéneo como al inhomogéneo. 

Acabamos de ver la representación cuántica de los operadores básicos $\hat a_m$ y $\hat
a^\dagger_m$ que actúan de modo no trivial sobre el sector inhomogéneo del espacio de Hilbert
cinématico~$\mathcal F$. Por otra parte, en la construcción del sector homogéneo del espacio de
Hilbert cinemático y de los operadores definidos sobre él, hemos empleado los dos esquemas de
cuantización polimérica $A$ y $B$ del modelo de
Bianchi I, explicados en el Capítulo \ref{chap:bianchi}. Por tanto, también en el modelo de Gowdy
presente hemos propuesto dos cuantizaciones, resultantes de imponer el operador ligadura escalar
correspondiente a cada esquema.

De nuevo, como en el estudio de los modelos anteriores, hemos aplicado nuestro procedimiento de
densitización (Secciones \ref{3sec:dens} y \ref{4sec:dens}), el cual nos permite dar una descripción
equivalente y más conveniente en la que los estados físicos son aniquilados por la versión
densitizada del operador ligadura hamiltoniana.

\subsection{Densitización}
\label{7sec:dens}

Denominamos subespacio de estados con \emph{volumen homógeno nulo} al núcleo del operador volumen
$\hat V=\otimes_i\widehat{\sqrt{|p_i|}}$ que, como hemos dicho, representa el volumen físico de la
celda coordenada tres-toroidal del espacio-tiempo de Bianchi I asociado a la cosmología de Gowdy.
Sea  $\widehat C_{\text{G}}=\widehat C_{\text{BI}}+\widehat C_{\xi}$ el operador ligadura escalar
del modelo de Gowdy. Para construir $\widehat C_{\xi}$, el operador que representa al término
$C_\xi$
definido en la ec. \eqref{Cclas}, hemos aplicado el mismo procedimiento de simetrización que en la
construcción del operador ligadura del modelo de Bianchi I, $\widehat C_{\text{BI}}$. Recordamos
que,
en nuestro método, en particular, repartimos las potencias del volumen a la derecha y a la izquierda
de los factores en los que aparecen. Por tanto, el operador $\widehat C_{\xi}$ es de la forma
\begin{align}
 \widehat C_{\xi}=\widehat{\left[\frac1{V}\right]}^{\frac1{2}}
\widehat{\mathcal C}_\xi\widehat{\left[\frac{1}{V}\right]}^{\frac1{2}},
\end{align}
donde $\widehat{\mathcal C}_\xi$ es cierto operador simétrico, cuya expresión se obtiene de
promover a operador el término entre corchetes de $C_\xi$ en la ec. \eqref{Cclas}, aplicando
nuestras reglas de simetrización. En cada esquema, se obtiene un operador diferente, que
analizaremos más adelante.

Así construido, el correspondiente operador $\widehat C_{\xi}$ hereda en ambos esquemas las
propiedades de $\widehat C_{\text{BI}}$. En particular, también aniquila el subespacio de estados
con {volumen homógeno nulo} y deja invariante su complemento ortogonal. Con la notación de la
Sección \ref{4sec:dens}, este complemento ortogonal es $\widetilde{\mathcal
H}_\text{grav}\otimes\mathcal F$. Por tanto, el operador ligadura escalar del modelo de Gowdy,
$\widehat C_{\text{G}}$, desacopla los estados con {volumen homogéneo nulo} y podemos restringir el
estudio a $\widetilde{\mathcal H}_\text{grav}\otimes\mathcal F$.

Recordemos que los estados no triviales aniquilados por el operador ligadura, en general, no son
normalizables en
el espacio de Hilbert cinemático, sino que típicamente pertenecen al espacio dual algebraico de un
subespacio denso de este espacio de Hilbert, por ejemplo, al dual del producto tensorial de
$\widetilde{\text{Cil}}_\text{S}$, dado en la ec. \eqref{dual}, y de un subespacio denso de
$\mathcal F$ apropiado. La densitización, entonces, se consigue aplicando la biyección
\eqref{eq:mapeo-dens} a los elementos $(\psi|$ de dicho espacio dual. Los estados
transformados son pues soluciones de la ligadura escalar densitizada $(\psi|\widehat
{\mathcal
C}_\text{G}^\dagger=0$, con
\begin{align}\label{dens}
\widehat{{\cal C}}_{\text{G}}&=
\widehat{\left[\frac1{V}\right]}^{-\frac1{2}}
\widehat{C}_{\text{G}}\widehat{\left[\frac{1}{V}
\right]}^{-\frac1{2}}=\widehat{\mathcal C}_{\text{BI}}+
\widehat{\mathcal C}_\xi.
\end{align}

Conviene destacar, una vez más, que el volumen implicado en este proceso de densitización es el
volumen del universo de Bianchi I en lugar del de Gowdy.

\subsection{Representación cuántica de los términos del sector \\inhomogéneo}

Antes de explicar la representación de la ligadura escalar en cada esquema, veamos cómo se
representan el hamiltoniano libre $H_0^\xi$ y el término
de interacción $H_\text{int}^\xi$, definidos en la ec. \eqref{HIclas}, que son comunes a ambos
esquemas.

Escogiendo el orden normal, sus análogos cuánticos son
\begin{align}\label{H0}
\widehat{H}_0^\xi&=\sum_{m>0}^\infty m\hat N_m,\quad\hat N_m=
\hat{a}^{\dagger}_m \hat{a}_m+\hat{a}^{\dagger}_{-m} \hat{a}_{-m},
\\
\label{HI} \widehat{H}_\text{int}^\xi
 &=\sum_{m>0}^\infty\frac{\hat N_m+\hat Y_m}{m},\quad\hat Y_m=\hat{a}_m
\hat{a}_{-m}+\hat{a}^{\dagger}_m\hat{a}^{\dagger}_{-m}.
\end{align}

El sector inhomogéneo del espacio de Hilbert cinemático, es decir, el espacio de Fock~$\mathcal F$,
se puede escribir como una suma directa de subespacios de Fock dinámicamente invariantes. De hecho,
el operador $\hat Y_m$, que es el único operador en el hamiltoniano que no actúa diagonalmente
sobre los estados  $|\mathfrak n\rangle$ de la base de $\mathcal F$, aniquila y crea pares de
partículas en modos con igual número de onda, esto es, en modos con vectores de onda etiquetados con
$m$ y $-m$. En consecuencia, los números cuánticos $X_m$, definidos en la ec.~\eqref{Fphys}, se
conservan bajo la acción del operador ligadura hamiltoniana $\widehat{\mathcal C}_\text{G}$.
Por tanto, es conveniente reetiquetar los estados de $n$-partículas con los números cuánticos 
$X_m$, para todos los enteros $m$ estrictamente positivos, junto con, por ejemplo, los autovalores
$N_m=n_m+n_{-m}$ de los operadores $\hat N_m$, definidos en la ec. \eqref{H0}. De este modo,
reescribimos los estados de la base del espacio de Fock de la forma
\begin{align}
|X_1, X_2,...;N_1,N_2,...\rangle:=|\mathfrak X;\mathfrak N\rangle.
\end{align}
Nótese que los números $X_m$ pueden tomar cualquier valor entero, mientras que\linebreak
$N_m\in\{|X_m|+2k, k\in\mathbb{N}\}$.

En estos estados, la acción de los operadores relevantes es
\begin{align}\label{actX}
&\hat X_m|\mathfrak X;\mathfrak N\rangle=X_m|\mathfrak X;\mathfrak
N\rangle,\\
\label{actN} &\hat N_m|\mathfrak X;\mathfrak N\rangle=N_m|\mathfrak
X;\mathfrak N\rangle,\\\label{actF} &\hat Y_m|\mathfrak X;\mathfrak
N\rangle=\frac{\sqrt{N_m^2-X_m^2}}{2}|
\mathfrak X;...,N_m-2,...\rangle+\frac{\sqrt{(N_m+2)^2-X_m^2}}{2}|\mathfrak
X;...,N_m+2,...\rangle.
\end{align}
Por tanto, la secuencia $\mathfrak X=\{X_1,X_2,...\}$ no se ve afectada,
como habíamos señalado antes.

Denominando $\mathcal S_{\mathfrak X}$ el subespacio de $\mathcal S$ generado por los estados de 
$n$-partículas con secuencia $\mathfrak X$ fija, y $\mathcal F_{\mathfrak X}$ su respectiva
compleción, podemos escribir entonces
\begin{align}
\mathcal F&=\oplus_{\mathfrak X}\mathcal F_{\mathfrak X}.
\end{align}
En lo que sigue, por simplificar la notación, los estados de $n$-partículas serán denominados
$|\mathfrak N\rangle$, obviando así su dependencia en la secuencia fija $\mathfrak X$.

Obviamente el operador $\widehat{H}_0^\xi$ con dominio $\mathcal S_{\mathfrak
X}$ está bien definido en $\mathcal F_{\mathfrak X}$,
ya que actúa diagonalmente sobre los estados de $n$-partículas y, por tanto, deja
$\mathcal S_{\mathfrak X}$ invariante. Por otra parte, el término de interacción
$\widehat{H}_\text{int}^\xi$ crea infinitos pares de partículas y, en consecuencia, no deja
invariante el
dominio $\mathcal S_{\mathfrak X}$, que solo contiene estados con un número finito de ellas.
De las ecs. \eqref{HI}, \eqref{actN} y \eqref{actF} se deduce que
\begin{align}\label{norm}
&||\widehat{H}_\text{int}^\xi|\mathfrak N\rangle||^2=
\bigg(\sum_{m>0}^\infty\frac{N_m}{|m|}\bigg)^2+\sum_{m>0}^\infty\frac{N_m^2-X_m^2+2N_m}{2m^2}+\sum_{
m>0}^\infty
\frac1{m^2}.
\end{align}
Dado que, en los estados de $n$-partículas, solo un conjunto finito de números de ocupación son
diferentes de cero, de entre las tres sumas de arriba, únicamente la última contiene un número
infinito
de
términos no nulos. Esta suma converge y, por tanto, la norma de
$\widehat{H}_\text{int}^\xi|\mathfrak N\rangle$ es finita. En
conclusión, $\widehat{H}_\text{int}^\xi|\mathfrak N\rangle \in
\mathcal F_{\mathfrak X}$ para todo $|\mathfrak N\rangle\in\mathcal
S_{\mathfrak X}$, por lo que el operador $\widehat{H}_\text{int}^\xi$, con dominio 
$\mathcal S_{\mathfrak X}$, también está bien definido.

\subsection{Operador ligadura densitizada}

Con la representación a nuestra disposición, tanto de los términos inhomogéneos, como del operador
ligadura escalar del modelo de Bianchi I, es inmediato obtener por completo el operador ligadura
escalar del modelo de Gowdy,
\[\widehat{\mathcal{C}}_{\text{G}}^s=\widehat{\mathcal{C}}_{\text{BI}}^\text{s}+\widehat{
\mathcal{C}}_{\xi}^\text{s}, \quad s=\text{A,B}.\] Solo faltan por representar los términos de
$\tilde C_\xi\equiv VC_\xi$ [ec.\eqref{Cclas}] que afectan al sector homogéneo. 

\subsubsection{Esquema A}

La comparación del operador
$\widehat{\mathcal{C}}_{\text{BI}}^\text{A}$, definido en la ec. \eqref{eq:omega-op}, con su
análogo clásico $\tilde C_{\text{BI}}\equiv VC_{\text{BI}}$ [véase ec. \eqref{eq:ligBianchi}]
muestra que el operador $\widehat\Omega_i$, definido en la ec. \eqref{omega-op}, es el análogo
cuántico de la magnitud clásica%
\footnote{Recordamos que no hay suma implícita en índices repetidos.} $(-c^ip_i)$. Aplicando esta
prescripción en la representación de
$(c^\sigma p_\sigma+c^\delta p_\delta)^2$, finalmente obtenemos que
$\widehat{\mathcal{C}}_{\xi}^\text{A}$ es el operador simétrico
\begin{align}\label{CGA}
&\widehat{\mathcal{C}}_{\xi}^\text{A}
=l_{\text{Pl}}^2\bigg
\{\frac{(\widehat{\Omega}_\sigma
+\widehat{\Omega}_\delta)^2}{\gamma^2}
\widehat{\bigg[\frac{1}{\sqrt{|p_\theta|}}\bigg]}^2
\!\widehat{H}_\text{int}^\xi+32\pi^2\widehat{|p_\theta|}
\widehat{H}_0^\xi\bigg\},
\end{align}
donde $\widehat{H}_0^\xi$ y $\widehat{H}_\text{int}^\xi$ están dados, respectivamente, en las ecs.
\eqref{H0} y \eqref{HI}, $\widehat{|p_\theta|}$
se obtiene a partir de la ec. \eqref{repA}, y $\widehat{[1/\sqrt{|p_\theta|}]}$
está definido en la ec. \eqref{invA}.

\subsubsection{Esquema B}

En esta ocasión la comparación de $\tilde C_{\text{BI}}$ con su análogo
cuántico $\widehat{\mathcal{C}}_{\text{BI}}^\text{B}$, definido en la ec. \eqref{densCB}, muestra
que el análogo cuántico de $-2(c^jp_jc^kp_k)/\gamma^2$ es el operador $\widehat{\mathcal
C}^{(i)}$ dado en la ec. \eqref{Ci}. A la vista de esta prescripción, en el esquema B, nosotros
hemos promovido $[(c_\sigma p_\sigma+c_\delta p_\delta)/\gamma]^2$ al operador simétrico
\begin{align}\label{G}
\widehat{G}=\frac1{4\gamma^2\Delta}\widehat{\sqrt{V}}[\hat F_\sigma\hat{V}
\hat F_\sigma+\hat F_\delta\hat{V}\hat F_\delta]\widehat{\sqrt{V}}-
\widehat{\mathcal C}^{(\theta)},
\end{align}
de modo que
\begin{align}\label{CGB}
&\widehat{\mathcal{C}}_{\xi}^\text{B}
=l_{\text{Pl}}^2\bigg\{
\widehat{\bigg[\frac{1}{|p_\theta|^{\frac1{4}}}\bigg]}^2 \widehat{G}
\widehat{\bigg[\frac{1}{|p_\theta|^{\frac1{4}}}\bigg]}^2
\widehat{H}_\text{int}^\xi+32\pi^2\widehat{|p_\theta|}
\widehat{H}_0^\xi\bigg\},
\end{align}
donde $\widehat{H}_0^\xi$ y $\widehat{H}_\text{int}^\xi$ están otra vez dados por las ecs.
\eqref{H0} y \eqref{HI}, $\widehat{[1/|p_\theta|^{\frac1{4}}]}$
se encuentra en la ec. \eqref{invB} y $\widehat{|p_\theta|}$ se obtiene a partir de
la ec. \eqref{p-B}.

\subsection{Superselección}
\label{sec:super}

Gracias a la prescripción de simetrización empleada, los operadores
$\widehat{\mathcal{C}}_{\text{G}}^s$ ($s=$A,B) dejan invariantes, por construcción, los
sectores de superselección del modelo de Bianchi I respectivo, los cuales, entonces, también
proporcionan sectores de superselección en el modelo de Gowdy bajo estudio.

\subsubsection{Esquema A}
En el esquema A, dichos sectores de superselección son%
\footnote{Asumimos que los observables físicos son capaces de distinguir entre sí todos los modos
inhomogéneos y por ello no superseleccionan $\mathcal F_{\mathfrak
X}$.}
\begin{align}
\mathcal
H_{\vec\varepsilon}^+\otimes \mathcal F,\qquad{\rm con}\quad \mathcal
H_{\vec\varepsilon}^+=\otimes_i\mathcal H_{\varepsilon_i}^+,
\end{align}
y el operador ligadura escalar $\widehat{\mathcal{C}}_{\text{G}}^\text{A}$ está bien definido en
cualquiera de los subespacios $\mathcal H_{\vec\varepsilon}^+\otimes\mathcal F_{\mathfrak
X}$, con dominio denso $\otimes_i\text{Cil}_{\varepsilon_i}^+\otimes
\mathcal S_{\mathfrak X}$ [recuérdense la definición \eqref{cyl-epsilon}].
Por tanto, como ocurría en el modelo de Bianchi I en este esquema A, podemos restringir el estudio
a un espacio de Hilbert cinemático que es separable y en el que los números cuánticos que
representan los grados de libertad homogéneos son estrictamente positivos, con valores mínimos 
$\varepsilon_i\in(0,2]$ y distribuidos en redes cúbicas de paso 2.

\subsubsection{Esquema B}
Los sectores de superselección en la cuantización correspondiente al esquema B
están dados por los espacios de Hilbert
\begin{align}
\mathcal H_{\tilde\varepsilon,\lambda_\sigma^{\star},\lambda_\delta^{\star}}
\otimes\mathcal F
\end{align}
y, análogamente, el operador ligadura hamiltoniana
$\widehat{\mathcal{C}}_{\text{G}}^\text{B}$ tiene una restricción bien definida en cualquiera de
los subespacios $\mathcal H_{\tilde\varepsilon,\lambda_\sigma^{\star},
\lambda_\delta^{\star}}\otimes\mathcal F_\mathfrak{X}$, siendo el dominio denso correspondiente
el espacio
$\text{Cil}^+_{\tilde\varepsilon,\lambda_\sigma^{\star},
\lambda_\delta^{\star}}\otimes\mathcal{S}_\mathfrak{X}$ [recuérdese la definición
\eqref{cyl-lambda}]. En consecuencia, también en el modelo de Gowdy en el esquema B, el
análisis se restringe a un espacio de Hilbert cinemático separable y en el que los números cuánticos
que representan los grados de libertad homogéneos son estrictamente positivos. No obstante, en esta
ocasión, solo el número cuántico $v=\tilde\varepsilon+4k$ tiene un ínfimo
$\tilde\varepsilon\in(0,4]$ no nulo y está distribuido en redes de paso constante.

\section{Estructura física en el esquema A}
\label{sec:physA}

\subsection{Imposición de la ligadura hamiltoniana}

En esta sección analizamos qué estructura tienen las soluciones de la ligadura escalar cuántica
\begin{align}\label{lig-cua-A}
\big(\psi\big|\widehat{\mathcal{C}}_{\text{G}}^\text{A}{}^\dagger&=0.
\end{align}

Recordemos que, en el modelo de Bianchi I en el esquema A, hemos podido resolver sin demasiadas
complicaciones la ligadura análoga
gracias a que los operadores $\widehat\Omega_i$ son observables de Dirac, sabemos caracterizar
completamente sus propiedades espectrales y conocemos la resolución espectral de la identidad
\eqref{eq:spectral-ident-2} en el espacio de Hilbert cinemático correspondiente, $\mathcal
H_{\varepsilon_i}^+$, en términos de sus autofunciones generalizadas
$e^{\varepsilon_i}_{\omega_i}(v_i)$, cuya forma explícita está dada en la
ec.~\eqref{eq:eigenstates-bianchi-A}. En el modelo de Gowdy presente, los operadores
$\widehat\Omega_\sigma$ y $\widehat\Omega_\delta$ siguen siendo observables de Dirac, pues conmutan
con el término inhomogéneo $\widehat{\mathcal{C}}_{\xi}^\text{A}$ [ec.~\eqref{CGA}], mientras que
$\widehat\Omega_\theta$ ya no lo es.

Teniendo esto en cuenta, en el modelo de Gowdy, conviene desarrollar cualquier elemento genérico
$(\psi|$
en la forma
\begin{align}
(\psi|&=\!\sum_{v_\theta\in\mathcal L_{\varepsilon_\theta}^+}\!\int_{\mathbb{R}^2}d\omega_\sigma\,
d\omega_\delta\,\langle
v_\theta|\otimes\langle e^{\varepsilon_\sigma}_{\omega_\sigma}|\otimes\langle e^{\varepsilon_\delta}
_{\omega_\delta}|\otimes(\psi_{\omega_\sigma,\omega_\delta}(v_\theta)|,
\end{align}
donde $(\psi_{\omega_\sigma, \omega_\delta}(v_\theta)|$ debe de pertenecer en principio al espacio
dual de algún
subespacio denso del espacio de Fock que describe las inhomogeneidades.

Sustituyendo la anterior expresión formal en la ec. \eqref{lig-cua-A} y teniendo en cuenta la
actuación de los operadores que aparecen en la expresión de
$\widehat{\mathcal{C}}_{\text{G}}^\text{A}$, obtenemos la siguiente relación de recurrencia para
todo $k\in\mathbb{N}$:
\begin{align}\label{sol}
\big(\psi_{\omega_\sigma,
\omega_\delta}(\varepsilon_\theta+2k+2)\big|&=\big(\psi_{\omega_\sigma,
\omega_\delta}(\varepsilon_\theta+2k)\big|
\widehat{H}_{\omega_\sigma,\omega_\delta}^{\xi}(\varepsilon_\theta+2k)\nonumber\\&+
\big(\psi_{\omega_\sigma,
\omega_\delta}(\varepsilon_\theta+2k-2)\big|F(\varepsilon_\theta+2k),
\end{align}
donde 
\begin{align}\label{H}
\widehat{H}_{\omega_\sigma,\omega_\delta}^{\xi}(v_\theta)&=
\frac{i}{6\pi(\omega_\sigma+\omega_\delta)f_+(v_\theta)}
\bigg[2\omega_\sigma\omega_\delta-\frac{(\omega_\sigma+\omega_\delta)^2}{9\gamma}b^2(v_\theta)
\widehat H_{\text{int}}^\xi\nonumber\\
&-32\pi^2 (6\pi\gamma l_{\text{Pl}}^2\sqrt{\Delta})^{2/3}|v_\theta|^{2/3}\widehat
H_0^\xi\bigg].
\end{align}
En estas expresiones, $F(v)$ es la función definida en la ec. \eqref{def-F-G} y $b(v)$ es la
función definida en la ec.~\eqref{eq:triad-operator}.

La ec. \eqref{sol} relaciona la proyección del estado $(\psi|$
en tres secciones diferentes correspondientes a tres valores consecutivos del parámetro interno
$v_\theta$ y valores fijos de $\omega_\sigma$ y $\omega_\delta$, que son los autovalores de
$\widehat\Omega_\sigma$ y $\widehat\Omega_\delta$. Esta relación de recurrencia es, obviamente, una
consecuencia de la acción \eqref{eq:omega-op} del operador $\widehat\Omega_\theta$.

Por otra parte, para $k=0$, gracias a la  propiedad $F(\varepsilon_\theta)=0$, la relación anterior
se convierte en una relación de consistencia entre las proyecciones en las dos primeras secciones de
$v_\theta$ constante,
\begin{align}\label{sol2}
&\big(\psi_{\omega_\sigma,
\omega_\delta}(\varepsilon_\theta+2)\big|=\big(\psi_{\omega_\sigma,
\omega_\delta}(\varepsilon_\theta)\big|
\widehat{H}_{\omega_\sigma,\omega_\delta}^{\xi}(\varepsilon_\theta).
\end{align}
Como resultado, la solución formal de la ec. \eqref{lig-cua-A} está completamente determinada por
los datos en la sección inicial $v_ \theta=\varepsilon_\theta$, como ocurría en el modelo de Bianchi
I.

Los coeficientes de dicha solución tienen la expresión siguiente para todo $k\in\mathbb{N}^+$:
\begin{align}\label{sol3}
\big(\psi_{\omega_\sigma,
\omega_\delta}(\varepsilon_\theta+2k)\big|=\big(\psi_{\omega_\sigma,
\omega_\delta}(\varepsilon_\theta)\big|\sum_{O(0\to k)}\Big[
\prod_{\{r_p\}}F(\varepsilon_\theta+2r_p+2)\Big] \mathcal P
\Big[\prod_{\{s_q\}}
\widehat{H}_{\omega_\sigma,\omega_\delta}^{\xi}(\varepsilon_\theta+2s_q)\Big],
\end{align}
donde $O(0\to k)$, $\{r_p\}$ y $\{s_q\}$ se definen como ya hemos explicado bajo la ec.
\eqref{S}, y el símbolo $\mathcal P$ denota el operador de ordenación del camino, desde el valor más
pequeño del conjunto $\{s_q\}$ a la izquierda hasta el valor más grande a la derecha. Esto es, en
el producto de operadores $\widehat{H}_{\omega_\sigma,\omega_\delta}^{\xi}(v_\theta)$, éstos
están ordenados de acuerdo a un valor creciente de $v_\theta$.

\subsection{Análisis de las soluciones}

La expresión resultante de las soluciones generales $(\psi|$ de la ligadura hamiltoniana es solo
formal, en el sentido de que estas soluciones no pertenecen al dual del dominio de definición de
$\widehat{\mathcal C}_\text{G}^\text{A}$. En última instancia, las razones hay que buscarlas en el
hecho de que el dominio elegido no es invariante bajo la acción del operador ligadura.
En efecto, los coeficientes $(\psi_{\omega_\sigma,
\omega_\delta}(v_\theta)|\mathfrak N\rangle$ de la solución resultan no estar bien definidos cuando 
$v_\theta\geq\varepsilon_\theta+4$. La causa de estas divergencias reside en la actuación
reiterada del operador $\widehat{H}_{\omega_\sigma,\omega_\delta}^{\xi}(v_\theta)$. No es
difícil ver que el espacio $\mathcal S_\mathfrak X$ no está en el dominio del operador 
$\widehat{H}_0^\xi\widehat{H}_\text{int}^\xi$, porque contiene el término
$\sum_m\hat{N}_m\hat{Y}_m$. La acción de este término sobre un estado genérico de $\mathcal
S_\mathfrak X$ no da lugar a un estado normalizable del espacio de Fock $\mathcal F_\mathfrak X$,
ni siquiera en el sentido generalizado convencional, ya que, en particular, dicho operador
conlleva la creación de un número infinito de pares de partículas y, además, una suma sobre el
número de todas estas partículas. Por tanto, aquellos coeficientes en cuya definición aparezca
dos veces o más el operador $\widehat{H}_{\omega_\sigma,\omega_\delta}^{\xi}(v_\theta)$ divergen.

Este problema desaparecería si se eligiera un dominio que fuese invariante bajo la acción del
operador ligadura. De hecho, la determinación de un dominio invariante y tal que el operador de
ligadura escalar fuera esencialmente autoadjunto permitiría recurrir al procedimiento de promedio
sobre grupos \cite{gave1a,gave1b,gave1c,gave1d,gave2} para construir el espacio de Hilbert físico.
Sin embargo, la determinación de tal dominio es una tarea extremadamente
complicada, dada la complejidad de nuestro modelo, y no tenemos a mano ninguna elección
satisfactoria. 
Así pues, el hecho de que las soluciones de la ligadura no pertenezcan al espacio dual del dominio
en el que
hemos definido el operador ligadura indica que la estructura cinemática de nuestra cuantización no
está bien adaptada a la física y que la relación entre ellas no es inmediata a priori. No obstante,
todavía podemos completar el programa de cuantización. Para ello, simplemente necesitamos dar
sentido a las soluciones y proporcionarles una estructura de Hilbert. 

\subsection{Espacio de Hilbert físico}

Como ya hemos dicho, las soluciones de la
ligadura hamiltoniana densitizada están completamente determinadas, al menos formalmente, por un
conjunto de datos iniciales $(\psi_{\omega_\sigma,\omega_\delta}(\varepsilon_\theta)|$ y, entonces,
podemos identificar estas soluciones con los correspondientes datos. De este modo, para construir
el espacio de Hilbert físico, simplemente debemos dotar al espacio vectorial de datos iniciales
con un
producto interno. Como hemos comentado en el análisis de modelos anteriores, éste se puede fijar
(salvo equivalencia unitaria) mediante la elección de un conjunto (super) completo de observables
clásicos que satisfagan un álgebra cerrada de corchetes de Poisson y el requisito de que los
operadores que los representan, que actúan sobre el espacio de datos iniciales, cumplan las
denominadas condiciones de realidad, esto es, que las relaciones de adjunción entre ellos sean la
contrapartida directa de las relaciones clásicas de conjugación compleja.
Así, podemos elegir observables que
formen un conjunto completo en el modelo de Bianchi I (que tienen acción trivial sobre el sector
inhomogéneo), por ejemplo,
\begin{align}
 \big\{ \widehat\Omega_a, \quad -i|\omega_\sigma+\omega_\delta|^{-1/2}\partial_{\omega_a}
|\omega_\sigma+\omega_\delta|^{1/2},\quad a=\sigma,\delta\big\},
\end{align}
---o los analizados en el Capítulo \ref{chap:evolutionLQC}---
junto con un conjunto completo adecuado que actúe sobre los modos inhomogéneos. Éstos pueden ser
el conjunto de operadores que representan los coeficientes reales de Fourier, en serie de senos y
cosenos,
de los modos no cero del campo $\xi(\theta)$ y de su momento $P_\xi(\theta)$. Salvo por constantes
reales irrelevantes, estos operadores son
\begin{align}\label{realF}
\big\{(\hat a_m+\hat a_m^\dagger)\pm(\hat a_{-m}+\hat a_{-m}^\dagger),\quad
i[(\hat a_m-\hat a_m^\dagger)\pm(\hat a_{-m}-\hat a_{-m}^\dagger)];\quad m\in\mathbb{N}^+\big\}.
\end{align}
De hecho, son autoadjuntos en el espacio de Fock $\mathcal F$. Por tanto, deducimos que el
espacio de Hilbert de datos iniciales seleccionado por nuestros requisitos es, salvo
equivalencia, el espacio
\begin{align}
 L^2(\mathbb{R}^2,|\omega_\sigma+\omega_\delta|d\omega_\sigma d\omega_\delta) \otimes\mathcal F.
\end{align}

Finalmente, para obtener el verdadero espacio de Hilbert físico del modelo, todavía tenemos que
imponer la simetría adicional de traslaciones en el círculo, implementada por la condición
\eqref{Fphys}. Teniendo esto en cuenta, llegamos a la conclusión definitiva de que los datos
iniciales $(\psi_{\omega_\sigma,\omega_\delta}(\varepsilon_\theta)|$ deben pertenecer al espacio de
Hilbert
\begin{align}\label{Hinitial}
\mathcal H^\text{A}_{\text{fis}}=L^2(\mathbb{R}^2,|\omega_\sigma+\omega_\delta|d\omega_\sigma
d\omega_\delta)\otimes\mathcal F_f,
\end{align}
que, por tanto, identificamos como el espacio de Hilbert físico.

De forma alternativa, se puede argumentar que éste es el espacio de Hilbert físico siguiendo una
línea de razonamiento diferente, basada en la idea de que se puede regularizar la teoría en
términos de un corte o \emph{cutoff} en el número de onda $m>0$ y analizar el límite de valores
arbitrarios de este corte. Nótese que esta reducción del espacio de fases está permitida porque, en
el modelo de Gowdy, los modos del campo con diferentes números de onda no están mezclados
dinámicamente.

\subsubsection{Modelo regularizado}

Sea $\mathcal S_{M}$ el subespacio de $\mathcal S$ generado por los estados de 
$n$-partículas que satisfacen la condición $N_m=0$ para todo
$m>M>0$ y sea $\mathcal S_{M}^\bot$ su complemento ortogonal.
Además, sean $\widehat P_{M}$ y
$\widehat P_{M}^\bot$ los proyectores en los correspondientes subespacios ortogonales.
Ahora estamos interesados en buscar el espacio de Hilbert físico de la truncación del modelo de
Gowdy en la que todos los modos del campo con número de onda mayor que el del corte $M$ sean nulos.
Obviamente, este sistema truncado está gobernado por las ligaduras
\begin{align}\label{proj}
&\big(\psi_{\omega_\sigma, \omega_\delta}(v_\theta)|\widehat
P_{M}^\bot=0,\\\label{projCG} &\big(\psi_{\omega_\sigma,
\omega_\delta}(v_\theta)|\widehat P_{M}\,\widehat{\mathcal
C}_\text{G}^\text{A}{}^\dagger\,\widehat P_{M}=0,\\\label{projCtheta}
&\big(\psi_{\omega_\sigma, \omega_\delta}(v_\theta)|\widehat
P_{M}\,\widehat{\mathcal C}_\theta{}^\dagger\,\widehat P_{M}=0.
\end{align}

Las soluciones simultáneas de las dos primeras ecuaciones tienen la misma forma que las soluciones
de la ligadura hamiltoniana del modelo de Gowdy completo, pero contienen exclusivamente números
de onda $m$ iguales o menores que $M$, así que ahora poseen solo un número finito de términos. 
En consecuencia, las divergencias creadas por la producción infinita de partículas desaparecen, y
los
coeficientes $(\psi_{\omega_\sigma,\omega_\delta}(v_\theta)|\mathfrak N\rangle$ de las soluciones a
las ecs. \eqref{proj} y \eqref{projCG} son ahora finitos para todo $v_\theta$. Por tanto, las
soluciones (para un valor fijo de $v_\theta$, $\omega_\sigma$ y $\omega_\delta$) pertenecen al
espacio dual $\mathcal S_M^*$ del espacio generado por los estados de $n$-partículas con el corte.
De modo similar a lo que hemos hecho en el modelo completo, podemos proporcionar una estructura de
Hilbert a los datos en
la sección $v_\theta=\varepsilon_\theta$ (o en cualquier otra). Para
ello, construimos el mismo conjunto completo de observables con la diferencia de que, ahora, en el
sector inhomogéneo no tenemos un número infinito de ellos, sino solo $4M$ [véase la ec.
\eqref{realF}]. Los operadores resultantes son autoadjuntos en
$L^2(\mathbb{R}^2,|\omega_\sigma+\omega_\delta|d\omega_\sigma
d\omega_\delta) \otimes\mathcal F_M$, donde $\mathcal F_M$ es el espacio de Fock obtenido por
compleción de $\mathcal S_M$. Teniendo en cuenta la ligadura remanente \eqref{projCtheta}, que
implementa la simetría de traslaciones en $S^1$, los estados físicamente admisibles tienen
finalmente la estructura de
espacio de Hilbert
\begin{align}\label{HphysM}
L^2(\mathbb{R}^2,|\omega_\sigma+\omega_\delta|d\omega_\sigma d\omega_\delta) \otimes(\mathcal
F_f){}_M,
\end{align}
donde $(\mathcal F_f){}_M$ es el subespacio de $\mathcal F_M$ generado por los estados 
de $n$-partículas que verifican la condición $\sum_{m>0}^M m X_m=0$.

A medida que el corte $M$ aumenta, la teoría truncada tiende a la del modelo de Gowdy
completo.
Esto indica que, en el límite $M\rightarrow\infty$, deberíamos recuperar el espacio de Fock
completo: $(\mathcal F_f){}_M\rightarrow\mathcal F_f$. Este hecho apoya nuestra afirmación
inicial de que el espacio de Hilbert físico del modelo no truncado es de hecho el espacio
\eqref{Hinitial}.

La introducción del corte en el número de onda también hace factible el estudio numérico del efecto
que las inhomogeneidades tienen sobre el comportamiento dinámico del sistema y, en particular,
de los cambios que ocurren en la dinámica del fondo de Bianchi I cuando las inhomogeneidades
crecen. De hecho, este tipo de estudio ya se ha llevado a cabo y está incluido en el análisis de
la Ref. \cite{eff} (aunque los resultados allí obtenidos se extienden al modelo con infinitos
modos). En este trabajo, el
hamiltoniano clásico ha sido modificado de una manera efectiva para considerar las correcciones que
emergen de la teoría cuántica, en paralelismo con el procedimiento explicado en la Sección
\ref{effective}, y las soluciones clásicas del mismo se han analizado tanto, en parte, por medios
analíticos, como en general numéricamente.

\section{Estructura física en el esquema B}
\label{sec:physB}

\subsection{Imposición de la ligadura hamiltoniana}

A diferencia de la situación en el esquema A, ahora el sector homogéneo del espacio de Hilbert
cinemático, $\mathcal H_{\varepsilon,\lambda_\sigma^{\star},\lambda_\delta^{\star}}$, no está
factorizado en tres subsectores direccionales independientes y ninguno de los operadores 
$\widehat{\sqrt{V}}\hat{F}_i\widehat{\sqrt{V}}$ que aparecen en la expressión del operador ligadura
escalar $\widehat{\mathcal{C}}_{\text{G}}^\text{B}$, que son los análogos a los operadores 
$\widehat{\Omega}_i$ del esquema A, son constantes del movimiento. Por tanto, en este caso y al
igual que ocurría ya en el modelo de Bianchi I correspondiente, no podemos simplificar la acción del
operador ligadura diagonalizándolo en los subsectores direccionales etiquetados con $\sigma$ y con
$\delta$, como hemos podido hacer con el esquema A.

Por tanto, para resolver la ligadura hamiltoniana
\begin{align}\label{lig-cua-B}
\big(\psi\big|\widehat{\mathcal{C}}_{\text{G}}^\text{B}{}^\dagger&=0.
\end{align}
desarrollamos un elemento general $(\psi|$ usando la base de estados
$|v,\omega_{\tilde\varepsilon}\lambda_\sigma^\star,
\bar\omega_{\tilde\varepsilon}
\lambda_\delta^\star\rangle$ para el sector homogéneo.
Esto es,
\begin{align}
(\psi|&=\sum_{v\in\mathcal
L_{{\tilde\varepsilon}}}\sum_{\omega_{\tilde\varepsilon}\in\mathcal
W_{\tilde\varepsilon}}\sum_{\bar\omega_{\tilde\varepsilon}\in
\mathcal W_{\tilde\varepsilon}}\langle
v,\omega_{\tilde\varepsilon}\lambda_\sigma^\star,\bar\omega_{\tilde\varepsilon}
\lambda_\delta^\star|\otimes(
\psi(v,\omega_{\tilde\varepsilon}\lambda_\sigma^\star,
\bar\omega_{\tilde\varepsilon}\lambda_\delta^\star)|,
\end{align}
donde, recordemos, $\mathcal W_{\tilde{\tilde\varepsilon}}$ es el conjunto \eqref{W-set}.
En esta expresión, \[(\psi(v,\lambda_\sigma,\lambda_\delta)|=
(\psi(v,\omega_{{\tilde\varepsilon}}\lambda_\sigma^{\star},
\bar\omega_{{\tilde\varepsilon}}\lambda_\delta^{\star})|\] es la proyección de $(\psi|$ sobre el
estado
$|v,\lambda_\sigma,\lambda_\delta\rangle=
|v,\omega_{\tilde\varepsilon}\lambda_\sigma^\star,\bar\omega_{\tilde\varepsilon}
\lambda_\delta^\star\rangle$ del sector homogéneo
y, en principio, debe de pertenecer al espacio dual de algún dominio denso apropiado del espacio de
Fock $\mathcal F$ que
describe las inhomogeneidades.
Sustituyendo el desarrrollo anterior en la ec.~\eqref{lig-cua-B}  y teniendo en cuenta la
actuación de los operadores que aparecen en la expresión de
$\widehat{\mathcal{C}}_{\text{G}}^\text{B}$, obtenemos que las proyecciones
$(\psi(v,\lambda_\sigma,\lambda_\delta)|$
satisfacen una serie de relaciones que pueden verse como ecuaciones en diferencias en $v$ que,
genéricamente, relacionan datos en la sección $v+4$ con datos en las secciones anteriores, $v$ y
$v-4$. Para simplificar la notación de la expresión resultante, conviene introducir las proyecciones
del elemento $(\psi|$ sobre las combinaciones de estados ya definidas para el modelo de Bianchi I,
dadas en las ecs. \eqref{eq:combi}. Esto es, definimos
\begin{align}
(\psi_\pm(v\pm4,\lambda_\sigma,\lambda_\delta)|
=(\psi|v\pm4,\lambda_\sigma,\lambda_\delta\rangle_\pm,\qquad
(\psi_{0^\pm}(v,\lambda_\sigma,\lambda_\delta)|
=(\psi|v,\lambda_\sigma,\lambda_\delta\rangle_{0^\pm}.
\end{align}
Asimismo, conviene introducir las combinaciones de estados
\begin{align}
|v\pm4,\lambda_\sigma,\lambda_\delta\rangle_\pm'&=\bigg|v\pm4,\lambda_\sigma,\frac{v\pm4}
{v}\lambda_\delta\bigg\rangle+\bigg|v\pm4,\frac{v\pm4}
{v\pm2}\lambda_\sigma,\frac{v\pm2}
{v}\lambda_\delta\bigg\rangle\nonumber\\
&+\bigg|v\pm4,\frac{v\pm4}{v}\lambda_\sigma,\lambda_\delta
\bigg\rangle+\bigg|v\pm4,\frac{v\pm2}
{v}\lambda_\sigma,\frac{v\pm4}
{v\pm2}\lambda_\delta\bigg\rangle,
\end{align}
\begin{align}\label{state4}
|v,\lambda_\sigma,\lambda_\delta\rangle_{0^\pm}'=
2\bigg|v,\lambda_\sigma,
\lambda_\delta\bigg\rangle+
\bigg|v,\frac{v\pm2}{v}\lambda_\sigma,\frac{v}
{v\pm2}\lambda_\delta\bigg\rangle+
\bigg|v,\frac{v}{v\pm2}\lambda_\sigma,\frac{v\pm2}
{v}\lambda_\delta\bigg\rangle,
\end{align}
y definir las proyecciones de $(\psi|$ sobre ellos:
\begin{align}
(\psi'_\pm(v\pm4,\lambda_\sigma,\lambda_\delta)|
=(\psi|v\pm4,\lambda_\sigma,\lambda_\delta\rangle_\pm',\qquad
(\psi'_{0^\pm}(v,\lambda_\sigma,\lambda_\delta)|
=(\psi|v,\lambda_\sigma,\lambda_\delta\rangle_{0^\pm}'.
\end{align}
Con esta notación, las soluciones de la ligadura satisfacen la relación explícita
\begin{align}\label{solutionB}
(\psi_+&(v+4,\lambda_\sigma,\lambda_\delta)|-\eta
[b_\theta^\star(v,\lambda_\sigma,\lambda_\delta)b
_\theta^\star(v+4,\lambda_\sigma,\lambda_\delta)]^2\frac{
v+4}{v}(\psi{}_+'(v+4,\lambda_\sigma, \lambda_\delta)
|\widehat{H}_\text{int}^\xi\nonumber\\
&=-\frac{1}{\eta}\frac{32v^2}{\lambda_\sigma^2
\lambda_\delta^2x_+(v)}(\psi(v,
\lambda_\sigma,\lambda_\delta)|\widehat{H}_0^\xi
+\frac{x^-_0(v)}{x_+(v)}(\psi_{0^-}
(v,\lambda_\sigma,\lambda_\delta)|+
\frac{x^+_0(v)}{x_+(v)}(\psi_{0^+}
(v,\lambda_\sigma,\lambda_\delta)|
\nonumber\\&-\frac{x_-(v)}{x_+(v)}
(\psi_-(v-4,\lambda_\sigma,\lambda_\delta)|
+\eta[b_\theta^\star(v,\lambda_\sigma,
\lambda_\delta)]^4\bigg\{
\left[\frac{b_\theta^\star(v-4,\lambda_\sigma,\lambda_\delta)}{b_\theta^\star(v,\lambda_\sigma,
\lambda_\delta)}\right]^2
\frac{v-4}{v}\frac{x_-(v)}{x_+(v)}\nonumber\\
&\times(\psi_-'(v-4,\lambda_\sigma,
\lambda_\delta)|-
\bigg[\frac{x_{0}^{-}(v)}{x_+(v)}(\psi'_{0^-
}(v,\lambda_\sigma,\lambda_\delta)|+
\frac{x_{0}^{+}(v)}{x_+(v)}(\psi'_{0^+
}(v,\lambda_\sigma,\lambda_\delta)|\bigg]\bigg\}
\widehat{H}_\text{int}^\xi,
\end{align}
donde
\begin{align}\label{eta}
\eta=\left(\frac{l_{\text{Pl}}}
{4\pi\gamma\sqrt{\Delta}}\right)^{2/3},
\end{align}
$b_\theta^\star(v,\lambda_\sigma,\lambda_\delta)$ es la función \eqref{btheta} y los coeficientes
$x_\pm(v)$ y $x^\pm_0(v)$ son los definidos en la ec. \eqref{coeficientes}.

\subsection{Análisis de las soluciones}

Al igual que en el modelo de Gowdy en el esquema A y en el modelo de Bianchi I en el esquema B,
podemos investigar si la solución está completamente determinada (al menos formalmente) por los
datos de la sección inicial $v={\tilde\varepsilon}$. Nosotros hemos argumentado que, en efecto, dado
un
conjunto de datos iniciales
$(\psi({\tilde\varepsilon},\omega_{\tilde\varepsilon}\lambda_\sigma^\star,
\bar\omega_{\tilde\varepsilon}\lambda_\delta^\star)|$
pertenecientes a $\mathcal{S}^*_\mathfrak{X}$ para todo
$\omega_{{\tilde\varepsilon}},\bar\omega_{\tilde\varepsilon}\in\mathcal W_{{\tilde\varepsilon}}$, es
posible determinar
cada término $(\psi(v,\lambda_\sigma,\lambda_\delta)|$ de la solución, para cada sección 
$v>\tilde\varepsilon$ en ${\cal L}_{\tilde\varepsilon}^+$.

La presencia del término de interacción en el miembro izquierdo de la ec. \eqref{solutionB}
dificulta una demostración directa de la afirmación anterior. Sin embargo, es posible obtener
dicho resultado en términos de un análisis asintótico de las soluciones. Es de destacar que, en
nuestra teoría, aparece un parámetro sin dimensiones, $\eta$, definido en la ec.
\eqref{eta}, y podemos recurrir a él para adoptar de modo natural un procedimiento asintótico sin
la necesidad de introducir a mano ningún parámetro externo. Nótese que el salto de
área $\Delta$ es proporcional a $\gamma l_{\text{Pl}}^2$; por tanto, $\eta$ es proporcional al
inverso del parámetro de Immirzi $\gamma$. Entonces, en el límite $\eta\rightarrow 0$,
desarrollamos las soluciones en una serie asintótica, de la forma:
\begin{align}\label{exp}
&(\psi(\tilde\varepsilon+4k,\lambda_\sigma,\lambda_\delta)|
=\sum_{n\in\mathbb{N}}\eta^{n-k}\left({}
^n\psi(\tilde\varepsilon+4k, \lambda_\sigma ,
\lambda_\delta)\right|,\quad \forall k\in\mathbb{N}^+.
\end{align}
Las combinaciones lineales introducidas en la ec.  \eqref{solutionB}, como, por ejemplo,\linebreak
$(\psi_\pm(v\pm4,\lambda_\sigma,\lambda_\delta)|$, heredan desarrollos similares. Denominaremos sus
términos usando una notación obvia. Así, en este ejemplo, denominaremos
$({}^n\psi_\pm(v\pm4,\lambda_\sigma,\lambda_\delta)|$ a los términos correspondientes.

Sustituyendo el desarrollo \eqref{exp} en la ligadura \eqref{solutionB} y considerando
potencias de $\eta$ orden a orden, obtenemos una expresión para cada término 
$\left({}^n\psi_+(v+4,\lambda_\sigma,\lambda_\delta)\right|$ (para $v$ genérico) en función de los
datos en $v$ y $v-4$. El resultado explícito es el siguiente (para simplificar la notación,
ignoramos la dependencia de los estados y de la función $b_\theta^\star$ en las variables
$\lambda_\sigma$ y $\lambda_\delta$):
\begin{itemize}
 \item Término dominante:
\begin{align}\label{0}
 \big({}^0\psi_+(v+4)\big|&=-\frac{32v^2}{
\lambda_\sigma^2\lambda_\delta^2
x_+(v)}\big({}^0\psi(v)\big|\widehat{H}_0^\xi,
\end{align}
\item corrección a primer orden:
\begin{align}\label{1}
\big({}^1\psi_+(v+4)\big|&=-\frac{32v^2}{\lambda_\sigma^2
\lambda_\delta^2
x_+(v)}\big({}^1\psi(v)\big|\widehat{H}_0^\xi
+\frac {x^-_{0}(v)}{x_+(v)}
\big({}^0\psi_{0^-}(v)\big|+\frac{x^+_{0}(v)}{x_+(v)}
\big({}^0\psi_{0^+}(v)\big|\nonumber\\&
+[b_\theta^\star(v)b_\theta^\star(v+4)]^2\frac{
v+4}{v}\big({}^0\psi'_+(v+4)\big|\widehat{H}_\text{int}^\xi,
\end{align}
\item corrección a segundo orden:
\begin{align}\label{2}
\big({}^2\psi_+(v+4)\big|&=-\frac{32v^2}{\lambda_\sigma^2
\lambda_\delta^2
x_+(v)}\big({}^2\psi(v)\big|\widehat{H}_0^\xi
+\frac{x^-_{0}(v)}{x_+(v)}\big({}^1\psi_{0^-}(v)\big|+
\frac{x^+_{0}(v)}{x_+(v)}\big({}^1\psi_{0^+}
(v)\big|\nonumber\\&-\frac{x_-(v)}{x_+(v)}
\big({}^0\psi_-(v-4)\big|-[b_\theta^\star(v)]^4
\bigg\{\bigg[\frac{x^-_0(v)}{x_+(v)}\big({}^0\psi'_{0^-}(v)\big|\nonumber\\&
+\frac{x^+_0(v)}{x_+(v)}\big({}^0\psi'_{0^+}(v)\big|\bigg]
-\left[\frac{b_\theta^\star(v+4)}{b_\theta^\star(v)}\right]^2\frac{v+4}{v}\big({}^1\psi'_+(v+4)
\big|\bigg]\bigg\}\widehat{H}_\text{int}^\xi,
\end{align}
\item corrección a orden $n$-ésimo ($n\geq3$):
\begin{align}\label{n}
\big({}^n\psi_+(v+4)\big|&=-\frac{32v^2}{\lambda_\sigma^2
\lambda_\delta^2x_+(v)}\big({}^n\psi(v)\big|\widehat{H}_0^\xi
+\frac{x^-_{0}(v)}{x_+(v)}\big({}^{n-1}\psi_{0^-}(v)\big|+
\frac{x^+_{0}(v)}{x_+(v)}\big({}^{n-1}
\psi_ {
0^+}(v)\big|\nonumber\\&-\frac{x_-(v)}{x_+(v)}\big({}^{n-2}
\psi_-(v-4)\big|-[b_\theta^\star(v)]^4\bigg\{\bigg[\frac{
x^-_0(v)}{x_+(v)}\big({}^{n-2}\psi'_{0^-}(v)\big|\nonumber\\&+\frac{
x^+_0(v)}{x_+(v)}\big({}^{n-2}\psi'_{0^+}(v)\big|\bigg]
-\left[\frac{b_\theta^\star(v+4)}{b_\theta^\star(v)}\right]^2\frac{v+4}{v}\big({}^{n-1}
\psi'_+(v+4)\big|\nonumber\\&
-\left[\frac{b_\theta^\star(v-4)}{b_\theta^\star(v)}\right]^2\frac{v-4}{v}\big({}^{n-3}\psi'_-(v-4)
\big|\bigg\}\widehat{H}_\text{int}^\xi.
\end{align}
\end{itemize}

Las expresiones anteriores se simplifican considerablemente en los casos $v=\tilde\varepsilon$
y $v=\tilde\varepsilon+4$. Esto se debe al hecho de que, por una parte, los datos en la primera
sección no están dados en principio por una serie asintótica, esto es, 
$(\psi(\tilde\varepsilon)|=({}^0\psi(\tilde\varepsilon)|$ y, en consecuencia, 
$({}^n\psi(\tilde\varepsilon)|=0$ para todo $\tilde\varepsilon\in(0,4]$ y $n>0$. Por otra parte, 
$x_-(\tilde\varepsilon)=0$ para todo $\tilde\varepsilon\in(0,4]$, mientras que 
$x^-_{0}(\tilde\varepsilon)=0$ si  $\tilde\varepsilon\leq2$ [véase la ec. \eqref{coeficientes}].

A la vista de estas ecuaciones, vemos que, para $v>4$, el conocimiento de la solución en las
secciones $v-4$ y $v$, junto con los términos de orden asintótico $n-1$ en la sección $v+4$,
determinan en esta última sección los términos de la solución para orden $n$ a través de las
siguientes combinaciones lineales:
\begin{align}
({}^n\psi_+(v+4,\lambda_\sigma,\lambda_\delta)|
&=\bigg({}^n\psi\bigg(v+4,\lambda_\sigma,\frac{v+4}
{v+2}\lambda_\delta\bigg)\bigg|+\bigg({}^n\psi
\bigg(v+4,\lambda_\sigma, \frac{v+2 }
{v}\lambda_\delta\bigg)\bigg|\nonumber\\&+\bigg({}^n\psi
\bigg(v+4,\frac{v+4 }
{v+2}\lambda_\sigma,\frac{v+2}
{v}\lambda_\delta\bigg)\bigg|+\bigg({}^n\psi
\bigg(v+4,\frac{v+4}
{v+2}\lambda_\sigma,\lambda_\delta\bigg)\bigg|
\nonumber\\&+\bigg({}^n\psi\bigg(v+4, \frac{v+2}{v}
\lambda_\sigma,\frac{v+4}{v+2}\lambda_\delta\bigg)\bigg|
+\bigg({}^n\psi\bigg(v+4,\frac{
v+2}{v}\lambda_\sigma,\lambda_\delta\bigg)\bigg|.
\end{align}
Por otra parte, para $0<v<4$, los datos en la sección $v-4$ son espurios y no se requiere su
conocimiento para determinar los términos de la solución en la sección $v+4$.

Nótese que, de hecho, esta estructura es similar a la estructura de las soluciones de la ligadura
hamiltoniana del modelo de Bianchi I en el esquema B, en tanto en cuanto la solución en la
sección $v+4$, ec. \eqref{eq:combinations-v4}, está determinada en términos del mismo tipo de
combinaciones lineales. Por tanto, según hemos aducido en la Sección \ref{phys-b}, para todo
$v>0$, el conjunto de combinaciones lineales
\begin{align*}
\{\left({}^n\psi_+(v+4,\omega_{{\tilde\varepsilon}}\lambda_\sigma^{\star},
\bar\omega_{\tilde\varepsilon}\lambda_\delta^{\star})\right|;\;
\omega_{{\tilde\varepsilon}},\bar\omega_{{\tilde\varepsilon}}\in\mathcal W_{{\tilde\varepsilon}},\}
\end{align*}
determina el conjunto de términos individuales
\begin{align*}
\{\left({}^n\psi(v+4,\omega_{{\tilde\varepsilon}}\lambda_\sigma^{\star},
\bar\omega_{\tilde\varepsilon}\lambda_\delta^{\star})\right|;\;
\omega_{{\tilde\varepsilon}},\bar\omega_{\tilde\varepsilon}\in\mathcal W_{{\tilde\varepsilon}}\}
\end{align*}
a través de una aplicación biyectiva. En consecuencia, comenzando con los datos iniciales, podemos
obtener, paso por paso, los términos de la solución ---hasta el orden asintótico deseado--- en
todas las secciones de $v$ constante.

En conclusión, los datos iniciales $(\psi({\tilde\varepsilon},\lambda_\sigma,\lambda_\delta)|$
(donde $\lambda_\sigma$ y $\lambda_\delta$ toman todos los posibles valores en sus sectores de
superselección correspondientes) determinan completamente la solución, como queríamos probar.
Las soluciones construidas de esta manera son formales, como ocurría en el esquema A, en el
sentido de que los objetos $({}^n\psi(v+4,\lambda_\sigma,\lambda_\delta)|$ no pertenecen en general
al espacio dual del dominio escogido para definir el operador ligadura hamiltoniana, debido a la
presencia del operador $\widehat{H}_\text{int}^\xi$ en sus expresiones.

\subsection{Espacio de Hilbert físico}

Para proporcionar una estructura de espacio de Hilbert a las soluciones, procedemos exactamente de
la misma
manera que en el caso A. Una vez que hemos justificado que el conjunto de datos iniciales
\begin{align*}
\{\left(\psi({\tilde\varepsilon},\omega_{{\tilde\varepsilon}}\lambda_\sigma^{\star},
\bar\omega_{\tilde\varepsilon}\lambda_\delta^{\star})\right|;\;
\omega_{{\tilde\varepsilon}},\bar\omega_{\tilde\varepsilon}\in\mathcal W_{{\tilde\varepsilon}}\}
\end{align*}
caracteriza la solución, podemos identificar las soluciones con sus correspondientes datos
iniciales, y el espacio de Hilbert físico con el espacio de Hilbert de tales datos iniciales.

Las condiciones de realidad sobre un conjunto completo de observables que actúen sobre estos datos
iniciales determina unívocamente, una vez más, el producto interno que proporciona la estructura
de Hilbert. Tales observables están dados, por ejemplo, por el conjunto (super) completo de
observables del modelo de Bianchi I en vacío, introducido en la ec.~\eqref{conjunto-completo}, y por
los observables escogidos en el esquema A para las inhomogeneidades, dados en la ec.~\eqref{realF}.
Imponiendo la simetría traslacional sobre $S^1$ remanente
en el espacio de Hilbert resultante, finalmente obtenemos una estructura similar a la encontrada en
el caso
A para el espacio de Hilbert físico, a saber, el producto tensorial del subespacio de Fock 
$\mathcal F_f$ por el espacio de Hilbert físico del modelo de Bianchi I en vacío, pero ahora en el
esquema B:
\begin{align}\label{HphysB}
\mathcal H^\text{B}_{\text{fis}}=\mathcal H_{\lambda_\sigma^\star,\lambda_\delta^\star}
\otimes\mathcal F_f.
\end{align}

\section{Discusión}
\label{sec:results}

Como hemos comentado al inicio de este capítulo, una de las motivaciones de este trabajo
ha sido investigar la plausibilidad de recuperar la teoría cuántica de campos estándar en
el contexto de la cuantización de lazos. En particular, queríamos mostrar que, en el espacio de los
estados físicos del modelo de Gowdy, se obtiene una descripcion de tipo Fock de las
inhomogeneidades sobre un fondo de tipo Bianchi I cuantizado poliméricamente, a partir de una
cuantización híbrida implementada en el marco cinemático. De hecho, hemos probado que éste es el
caso, ya que el espacio de Hilbert físico obtenido en ambos esquemas tiene la estructura de un
producto tensorial del espacio de Hilbert físico del modelo de Bianchi I por un espacio de Fock que
resulta ser equivalente al espacio obtenido en la cuantización de Fock estándar  \cite{men1a,
men1b}.
Este resultado respalda la validez de la cuantización híbrida, porque ésta debería dar lugar a la
cuantización estándar del sistema en el límite en el que los efectos que emergen de la
discretización de la geometría se vuelvan despreciables. Merece la pena resaltar que este resultado
no es trivial, dado que la cuantización híbrida se introduce en el escenario cinemático, y la
relación entre las estructuras cinemática y física no se puede anticipar antes de completar la
cuantización,
aún menos si tenemos en cuenta la complejidad del modelo, que se describe mediante grados de
libertad locales.

Por otra parte, este análisis también nos ha permitido investigar el destino de la singularidad
cosmológica del modelo. En efecto, las soluciones clásicas del modelo de Gowdy $T^3$ polarizado
linealmente presentan genéricamente una singularidad cosmológica. En la Ref. \cite{mon2}, por
ejemplo, se calculó un invariante de curvatura y se demostró que diverge a tiempo inicial casi en
todo punto. En términos de las variables que hemos empleado, la singularidad cosmológica
corresponde a los valores nulos de las componentes $p_i$ de la tríada densitizada. De hecho, como
podemos ver en la ec. \eqref{newmetric}, la métrica está mal definida si alguno de los coeficientes
$p_i$ es cero. En nuestra teoría cuántica, la cuantización polimérica llevada a cabo en el sector
homogéneo tiene éxito a la hora de eliminar la singularidad. Más específicamente, hemos sido
capaces de
eliminar el núcleo de todos los operadores $\hat p_i$ y, como consecuencia, cuánticamente ya no
existe ningún análogo de la singularidad cosmológica clásica. Esta resolución de la singularidad
se obtiene cinemáticamente y, por tanto, es independiente de la dinámica. Por supuesto, persiste en
el espacio de Hilbert de los estados físicos, ya que éstos no tienen proyección en los
autoespacios de autovalor nulo de los operadores $\hat p_i$. Más áun, solo tienen soporte en un
sector con orientaciones de las componentes de la tríada constantes y, entonces, no cruzan la
singularidad hacia otras
ramas del universo que correspondan a orientaciones diferentes.

Por otra parte, como complemento a esta resolución cinemática de la singularidad cosmológica, cabe
resaltar que, al menos en el esquema A y en el marco de la descripción efectiva
correspondiente a nuestra cuantización híbrida, las simulaciones numéricas realizadas para el
modelo de Gowdy  muestran la presencia de un rebote que reemplaza la singularidad y que emerge
de las correcciones  que la geometría cuántica induce en la relatividad general \cite{eff}.
Actualmente se están desarrollando cálculos numéricos similares para el esquema B \cite{mpw}, con
el fin de validar este resultado más fuerte que el cinemático acerca de la resolución de la
singularidad.

Para finalizar, quisiéramos enfatizar el hecho de que los métodos cuánticos de tipo estándar (no
poliméricos) no logran
resolver la singularidad cosmológica. Por una parte, en la cuantización de Fock del sistema
deparametrizado, que ha sido ampliamente discutida en la literatura
\cite{misner,berger1,berger2,berger3,pierri,ccq,torre} y que se ha llevado a cabo satisfactoriamente
en las
Refs.~\cite{men1a,men1b,men2,men3,men4}, un parámetro temporal clásico está presente explícitamente
en la descripción cuántica, y el invariante de curvatura calculado en la Ref. \cite{mon2} depende de
su inverso, de tal modo que dicho invariante todavía diverge en el instante inicial. Por otra parte,
si no se deparametriza el sistema, siguiendo nuestra reducción de gauge, y se cuantiza el sector
homogéneo de una manera estándar, como en el procedimiento de WDW, el autovalor cero estaría
incluido en el espectro continuo del operador tríada, en lugar de en el espectro discreto, y no
existiría un subespacio propio asociado a este autovalor que pudiera ser desacoplado y eliminado de
la teoría.

\cleardoublepage

\part*{Conclusiones}


{\renewcommand{\thechapter}{}\renewcommand{\chaptername}{}
\addtocounter{chapter}{0}
\chapter*{Conclusiones}\markboth{\sl CONCLUSIONES}{\sl CONCLUSIONES}}
\addcontentsline{toc}{part}{Conclusiones}

La motivación principal de esta tesis es extender la aplicabilidad de la cosmología
cuántica de lazos
a situaciones con complejidad suficiente como para permitir la extracción de predicciones físicas
realistas. Con el fin de progresar en esta dirección, hemos analizado la cuantización de diversos
modelos con un orden jerárquico de complejidad, proporcionado por la inclusión de anisotropías e
inhomogeneidades.

\section*{Resultados específicos}

\begin{list}{\labelitemi}{\leftmargin=1em}
 \item Hemos propuesto una nueva prescripción de simetrización y de densitización para el operador
ligadura hamiltoniana, tal que en todos los modelos estudiados:
\begin{itemize}
\item[(i)] El operador resultante desacopla los estados de volumen (homogéneo) nulo. Gracias a
ello, el problema de la singularidad cosmológica inicial queda resuelto cuánticamente incluso sin
restringirse a estados físicos.
\item[(ii)] Dicho operador deja invariantes espacios de Hilbert simples, que son generados
por autoestados de las componentes (homogéneas) de la tríada densitizada, cada una de ellas con una
orientación fija. Como consecuencia, los estados que contienen la información acerca de la
discretización de la geometría verifican una descripción de tipo {ausencia de frontera}, ya
que surgen en una sola sección de volumen (homogéneo) mínimo no nulo. Dichos espacios de
Hilbert proporcionan sectores de superselección que a su vez son simples y tienen propiedades
físicas óptimas.
\end{itemize}
\item En el modelo de FRW acoplado a un campo escalar sin masa, hemos demostrado que los estados
que encierran la información sobre la geometría cuántica convergen a una onda estacionaria
\emph{exacta} en el límite de volumen grande. Este comportamiento, junto con la descripción de tipo
{ausencia de frontera}, muestra la existencia de un rebote cuántico genérico
que reemplaza dinámicamente la singularidad clásica. Por tanto, en este contexto, hemos
demostrado que el mecanismo de resolución de singularidades
mediante un rebote cuántico es una consecuencia directa de los efectos discretos de la geometría que
subyacen en cosmología cuántica de lazos y, por consiguiente, que es una propiedad fundamental de la
teoría.
\item En el modelo de Bianchi I en vacío y en el denominado esquema A para la dinámica mejorada
(esquema factorizable),
el estudio se reduce esencialmente a considerar tres copias del sector gravitacional del modelo de
FRW. Como
consecuencia, los estados que codifican aquí la información sobre la geometría cuántica poseen el
mismo tipo de comportamiento que los análogos del modelo de FRW: descripción de tipo {ausencia
de frontera} y límite de WDW de tipo onda estacionaria. Por tanto, una vez más, existen
rebotes cuánticos genéricos que curan las singularidades.
\item En el modelo de Bianchi I en vacío y en el esquema B (no factorizable), hemos demostrado que
no solo el volumen
toma valores en conjuntos discretos, sino también las variables que representan las anisotropías.
Éstas, sin embargo, no poseen realmente un mínimo, aunque sí un ínfimo nulo. Más aún, en nuestro
análisis,
restringido al sector en el que los autovalores de los operadores
que representan todas las componentes de la tríada densitizada son estrictamente positivos, el
rango de valores de las anisotropías cubre densamente la semirrecta real positiva.
\item Para este modelo, hemos argumentado que los datos iniciales de la sección de volumen mínimo
(no nulo) determinan completamente las soluciones de la ligadura hamiltoniana, y hemos
caracterizado el espacio de Hilbert físico como el espacio de Hilbert de los datos iniciales, cuya
estructura, a su vez, ha sido determinada imponiendo
condiciones de realidad en un conjunto (super) completo de observables físicos.
\item Usando el modelo de Bianchi I en vacío como ejemplo, hemos sido pioneros en analizar y
desarrollar metodológicamente la imagen de evolución física en cosmología cuántica de lazos usando
como tiempo interno alguno de los grados de libertad de la geometría. En particular, hemos llevado a
cabo dos construcciones distintas de observables relacionales o completos, en las que la variable
del espacio de fases que desempeña la función de tiempo interno es, respectivamente: $(i)$ un
parámetro afin asociado a uno de los coeficientes de la tríada densitizada, $(ii)$ su momento
canónicamente conjugado.
\item En ambas construcciones, hemos llevado a cabo un análisis numérico de los valores esperados y
de las dispersiones de los observables, evaluados en estados semiclásicos con perfil gaussiano, que
revela las siguientes predicciones:
\begin{itemize}
 \item[(a)] Los estados gaussianos permanecen picados a lo largo de toda la evolución, en el sentido
de que las dispersiones de los observables se mantienen acotadas.
\item[(b)] En los regímenes de volumen grande, las trayectorias definidas por los valores esperados
de los observables coinciden con las predichas por la relatividad general.
\item[(c)] A medida que las trayectorias se acercan a la ubicación de la singularidad, la dinámica
cuántica se desvía con respecto a la clásica y aparecen rebotes cuánticos que evitan la
singularidad. 
\item[(d)] Además, el límite de WDW de tipo onda estacionaria de las autofunciones de la
cosmología cuántica de lazos nos ha permitido analizar la región asintótica que el cálculo
numérico no permite explorar. Hemos demostrado que el comportamiento semiclásico de estados físicos
muy generales (no solo correspondientes a perfiles gaussianos) se conserva a través del rebote
cuántico.
\end{itemize}
Por el contrario, el mismo tipo de análisis aplicado a la cuantización de WDW del modelo
muestra que esta teoría no cura las singularidades.
\item En el contexto particular del modelo de Gowdy $T^3$, hemos propuesto un procedimiento híbrido,
que combina las cuantizaciones de lazos y de Fock, para tratar la cuantización de cosmologías
inhomogéneas. La cuantización de lazos del sector homogéneo es suficiente para curar la singularidad
cosmológica inicial incluso en el espacio de Hilbert cinemático.
\item Hemos obtenido un operador ligadura hamiltoniana bien definido en un dominio
denso y que conserva los sectores de superselección obtenidos en el modelo de Bianchi I, que, por
tanto, proporcionan en este sistema el sector homogéneo de espacios de Hilbert separables que
también están superseleccionados.
\item Como en modelos anteriores, la ligadura hamiltoniana da lugar a una ecuación en
diferencias en un parámetro interno discreto que tiene un valor mínimo estrictamente positivo.
Hemos demostrado que las soluciones de la ligadura están unívocamente determinadas por los datos
proporcionados en la sección inicial de tal parámetro. Identificando datos iniciales con soluciones,
hemos podido caracterizar el espacio de Hilbert físico como el espacio de Hilbert de los datos
iniciales, cuyo producto interno queda fijado al imponer condiciones de realidad en un conjunto
adecuado de observables.
\item El espacio de Hilbert físico obtenido es el producto tensorial del espacio de Hilbert físico
del modelo de Bianchi I (el correspondiente a cada esquema, A o B) por un espacio de Fock
unitariamente equivalente al obtenido en la cuantización estándar del modelo deparametrizado.
Por tanto, recuperamos la teoría cuántica de campos estándar para las inhomogeneidades, que pueden
enterderse como grados de libertad que se propagan sobre un fondo de tipo Bianchi I.
\item En definitiva, hemos sido capaces de mejorar las técnicas de la cosmología cuántica de lazos
ya antes desarrolladas en contextos homogéneos y de completar la cuantización de cosmologías
inhomogéneas por primera vez dentro del marco de la teoría de lazos.
\end{list}

\cleardoublepage


{\renewcommand{\thechapter}{}\renewcommand{\chaptername}{}
\addtocounter{chapter}{0}
\chapter*{Conclusions}\markboth{\sl CONCLUSIONS}{\sl CONCLUSIONS}}
\addcontentsline{toc}{part}{Conclusions}

The main motivation of this thesis is to extend the application of Loop Quantum Cosmology
(LQC) to situations with sufficient complexity as to allow one to extract realistic physical
predictions. In order to progress in this direction,  we have studied several models with increasing
complexity, provided by the inclusion of anisotropies and inhomogeneities. 

\section*{Specific outcomes}

\begin{list}{\labelitemi}{\leftmargin=1em}
 \item We have proposed a new symmetrization and densitization prescription for the
Hamiltonian constraint operator such that for all the studied models:
\begin{itemize}
\item[(i)] This operator decouples the zero (homogeneous) volume states. Thanks to this, the
initial
cosmological singularity becomes quantum-mechanically resolved, even without restricting the study
to physical states.
\item[(ii)] The operator leaves simple separable Hilbert spaces invariant, which are spanned by
eigenstates of the (homogeneous) densitized triad components, each of these components with a fixed
orientation. As a consequence, the states that encode the information about the discreteness of the
geometry satisfy a {no-boundary} description, arising in a single section of non-zero minimum
(homogeneous) volume.
Those spaces provide sectors of superselection which in turn are simple and have optimal
physical
properties. 
\end{itemize}
\item In the flat FRW model coupled to a massless scalar field, we have shown that the states that
encode the information about the quantum geometry converge to an \emph{exact} standing wave in the
large volume regime. This behavior, together with the {no-boundary} description, indicates the
presence of a generic quantum bounce that dynamically replaces the classical singularity.
Hence, within this context, we have shown that the mechanism to cure the
singularities by means of a quantum bounce is a direct consequence of the discrete effects of the
geometry underlying LQC and,
thus, it is a fundamental feature of the theory.
\item In the vacuum Bianchi I model within the so-called scheme A for the improved dynamics
(factorizable scheme), the
study essentially reduces to the consideration of three copies of the gravitational sector of the
FRW model. As a
consequence, the states that encode here the information about the quantum geometry possess the same
kind of behavior as their analogs for the FRW model: {no-boundary} description and standing
wave
WDW limit. Therefore, once again generic quantum bounces cure the singularities.
\item In the vacuum Bianchi I model within scheme B (non-factorizable), we have shown that not only
the
volume takes values in discrete sets, but so do the variables that represent the anisotropies. These
variables, nonetheless, do not really possess a minimum, although they have a vanishing infimum.
Moreover,
in
our analysis, restricted to the sector in which all the operators
that represent the nontrivial densitized triad coefficients are positive, the range of values of
the anisotropies densely cover the positive real line. 
\item For this model, we have argued that the initial data on the section of minimum non-zero volume
completely determine the solutions to the Hamiltonian constraint, and we have characterized the
physical Hilbert space as the Hilbert space of initial data, whose structure in turn has been
determined by imposing reality conditions on a(n over) complete set of physical observables.
\item Using the vacuum Bianchi I model as an example, we have pioneered the proposal and
methodological development of a formalism of physical evolution in LQC using as
internal time one of the geometry degrees of freedom. In particular, we have carried out two
different constructions of relational observables, in which the variable that plays the role of
internal time is, respectively: $(i)$ an affine parameter associated with one of the densitized
triad coefficients, $(ii)$ its canonically conjugate momentum.
\item In both constructions, we have performed a numerical analysis of the expectation values and
dispersions of the observables, measured on semiclassical states provided by Gaussian profiles.
Such analyses reveal the following predictions:
\begin{itemize}
 \item[(a)] The Gaussian states remain sharply peaked along the whole evolution, in the sense that
the dispersions of the observables remain bounded.
\item[(b)] In the large volume regime, the trajectories defined by the expectation values of the
observables agree with those of General Relativity.
\item[(c)] As these trajectories approach the singularity location, the quantum dynamics differs
from the classical dynamics and there appear quantum bounces that avoid the singularity.
\item[(d)] In addition, the stationary-wave type WDW limit of the LQC eigenfunctions
has allowed us to study the asymptotic region that the numerical analysis does not permit to
explore. We have shown that the semiclassical behavior of quite general physical states (more
general than Gaussian ones) is preserved through the quantum bounce.
\end{itemize}
In contrast, the same analysis applied to the WDW quantization of the model shows that
this theory does not cure the singularities.
\item In the particular context of the Gowdy $T^3$ model, we have proposed a hybrid loop/Fock
approach to deal with the quantization of inhomogeneous cosmologies. The loop quantization of the
homogeneous sector suffices to cure the classical initial singularity, even in the kinematical
Hilbert
space. 
\item We have obtained a Hamiltonian constraint operator that is well defined in a dense domain
and that preserves the superselection sectors of the Bianchi I model, which therefore provide in
this system the homogeneous sector of separable Hilbert spaces that are also superselected.
\item As in previous models, the Hamiltonian constraint provides a difference equation with
respect to an internal parameter which has a strictly positive minimum. We have shown that the
solutions of the constraint are completely determined by the data on the initial section of that
parameter. Identifying initial data with solutions, we have been able to characterize the physical
Hilbert space as the Hilbert space of initial data, whose inner product is fixed by
imposing reality conditions on a given suitable set of observables.
\item The attained physical Hilbert space is the tensor product of the Bianchi I physical Hilbert
space (corresponding to each scheme A or B) times a Fock space which is unitarily equivalent to the
space obtained in the standard quantization of the deparametrized model. Therefore, we recover
standard quantum field theory for the inhomogeneities, which can be regarded as degrees of
freedom which propagate in a Bianchi I background.
\item In short, we have been able to improve the LQC techniques that had already been developed in
homogeneous contexts and complete the quantization of inhomogeneous cosmologies for the first
time within the framework of the loop theory.
\end{list}

\cleardoublepage

\appendix
\part*{Apéndices}
\addcontentsline{toc}{part}{Apéndices} 

\chapter{Teoría de operadores}
\label{appA}

Este apéndice contiene un resumen de los conceptos y teoremas de la teoría de operadores que se han
empleado en esta tesis, con el fin de facilitar su entendimiento sin necesidad de recurrir a la
bibliografía especializada en el tema. Está basado esencialmente en las Refs.
\cite{functional1,functional2,kato,functional3}, e incluso contiene párrafos directamente
reproducidos de alguna de estas referencias. Remitimos a ellas si se quieren ver las demostraciones
de los teoremas y proposiciones enunciados en este apéndice.

\section{Definiciones básicas}

Sea $\mathcal H$ un espacio de Hilbert, con producto interno $\langle\,\cdot\,|\,\cdot\,\rangle$.

\begin{definition}[Dominio denso y recorrido:] Sea $A: \mathcal D(A)\subset\mathcal H\to\mathcal H$
un operador lineal%
\footnote{A lo largo de este apéndice, no distinguiremos los operadores con un símbolo
 de ``gorro'' a menos que
pueda existir confusión.}
definido en el subespacio
$\mathcal D(A)$. Decimos que $\mathcal D(A)$ es el \emph{dominio} del operador $A$ y que\linebreak
$\mathcal R(A)=A\mathcal D(A)\subset\mathcal H$ es su imagen o \emph{recorrido}. El dominio
$\mathcal D(A)$ es \emph{denso} en $\mathcal H$ si su cierre o compleción de Cauchy
$\overline{\mathcal D(A)}$ con respecto a la norma de $\mathcal H$ verifica $\overline{\mathcal
D(A)}=\mathcal H$.
\end{definition}

\begin{definition}[Operador acotado:] El operador $A$ se dice \emph{acotado} si $||A\phi||\leq
M||\phi||<\infty$ para todo $\phi\in\mathcal H$, en cuyo caso, el operador está definido en todo
$\mathcal H$.
\end{definition}

\begin{definition}[Gráfico de un operador:] El operador $A$ queda caracterizado por su
\emph{gráfico},
definido como 
\begin{align}
 \Gamma(A)=\{(\phi,\psi)\in\mathcal H\times\mathcal H;\;\phi\in\mathcal D(A),\;\psi=A\phi\}.
\end{align}
\end{definition}

\begin{definition}[Operador cerrado, cerrable y cierre de un operador:] Se dice que $A$ es
\emph{cerrado} si $\Gamma(A)$ es un subespacio cerrado de $\mathcal H\times\mathcal H$. Dados dos
operadores en $\mathcal H$, $A_1$ y $A_2$, se dice que $A_2$ es \emph{extensión} de $A_1$ y se
escribe $A_1\subset A_2$ si $\Gamma(A_1)\subset\Gamma(A_2)$. Equivalentemente, $A_1\subset A_2$ si
$\mathcal D(A_1)\subset\mathcal D(A_2)$ y $A_1\phi=A_2\phi$ para todo $\phi\in\mathcal D(A_1)$. Un
operador se dice \emph{cerrable} si tiene una extensión cerrada. Todo operador $A$ cerrable tiene
una extensión cerrada mínima llamada su \emph{cierre} y denotada por $\bar{A}$. Se cumple que
$\Gamma(\bar A)=\overline{\Gamma(A)}$. Sin embargo, $\mathcal D(\bar A)$ no tiene por
qué coincidir con $\overline{\mathcal D(A)}$.
\end{definition}

\begin{definition}[Operador adjunto:] Si $A$ está densamente definido, es decir, $\overline{\mathcal
D(A)}=\mathcal H$, existe el operador \emph{adjunto} $A^\dagger$ en $\mathcal H$
definido mediante la expresión
\begin{align}\label{adjunto}
 \langle\phi|A\psi\rangle=\langle A^\dagger\phi|\psi\rangle,
\end{align}
para todo $\psi\in\mathcal D(A)$ y para todo $\phi\in\mathcal D(A^\dagger)\subset\mathcal H$. Nótese
que, si
$\mathcal D(A^\dagger)$ es denso en $\mathcal H$, entonces podemos definir
$A^{\dagger\dagger}=(A^\dagger)^\dagger$.
\end{definition}

\begin{theorem}
Sea $A$ un operador densamente definido en $\mathcal H$. Entonces $A^\dagger$ es cerrado, y
$A$ es cerrable si y solo si  $\mathcal D(A^\dagger)$ es denso en $\mathcal H$, en cuyo caso
$\bar{A}=A^{\dagger\dagger}$ y $(\bar A)^\dagger=A^\dagger$.
\end{theorem}

\subsection{Resolvente y espectro de un operador}

Dado un operador $A$ y $\lambda\in\mathbb{C}$, se dice que $\lambda$ pertenece al conjunto
\emph{resolvente} $\rho(A)$ si $\mathcal R(\lambda\mathbb{I}-A)$ es denso en $\mathcal H$, existe
$(\lambda\mathbb{I}-A)^{-1}$ y es acotado ($\mathbb{I}$ denota el operador identidad y, a veces, se
omite su escritura). El conjunto $\rho(A)$ es abierto. Su complemento
$\mathbb{C}\setminus\rho(A)=\sigma(A)$ es cerrado y se denomina \emph{espectro} de A. A su vez, los
puntos
del espectro se pueden clasificar del siguiente modo:
\begin{itemize}
 \item Espectro continuo, $\sigma_c(A)$: $\lambda\in\sigma_c(A)$ si $\mathcal
R(\lambda\mathbb{I}-A)$ es denso en $\mathcal H$, existe $(\lambda\mathbb{I}-A)^{-1}$ y no es
acotado.
\item Espectro residual, $\sigma_r(A)$: $\lambda\in\sigma_r(A)$ si $\mathcal
R(\lambda\mathbb{I}-A)$ no es denso y existe $(\lambda\mathbb{I}-A)^{-1}$.
\item Espectro puntual, $\sigma_p(A)$: $\lambda\in\sigma_p(A)$ si no existe
$(\lambda\mathbb{I}-A)^{-1}$. Por tanto, la ecuación, que denominaremos \emph{de autovalores},
$A\phi=\lambda\phi$ tiene solución no trivial $\phi\in\mathcal H$. $\lambda$ se denomina
\emph{autovalor o valor propio} de $A$ y $\phi$ \emph{autovector}, \emph{autoestado} o \emph{vector
propio}.
\end{itemize}
$\sigma_c,\sigma_r,\sigma_p$ son disjuntos dos a dos y
$\sigma(A)=\sigma_c(A)\cup\sigma_r(A)\cup\sigma_p(A)$.

\begin{proposition}
 Si $A$ es cerrable, entonces $\sigma(A)=\sigma(\bar A)$
\end{proposition}

\begin{proposition}
 Si $A$ es acotado y densamente definido, entonces su espectro es no-vacío.
\end{proposition}

\subsection{Operadores simétricos, autoadjuntos y esencialmente autoadjuntos}

\begin{definition}[Operador simétrico:] Un operador $A$ densamente definido en $\mathcal H$ se dice
que es \emph{simétrico} si $A\subset A^\dagger$, esto es, si $\mathcal
D(A)\subset\mathcal D(A^\dagger)\subset\mathcal H$ y $A\phi=A^\dagger\phi$ para todo
$\phi\in\mathcal D(A)$.
Equivalentemente, $A$ es simétrico si $\langle\phi|A\psi\rangle=\langle
A\phi|\psi\rangle$ para todo $\phi,\psi\in\mathcal D(A)$.
\end{definition}

\begin{proposition}
Un operador simétrico $A$ es cerrable. Se tiene 
$A\subset\bar A= A^{\dagger\dagger}\subset A^\dagger$.
\end{proposition}

\begin{definition}[Operador autoadjunto:] $A$ es \emph{autoadjunto} si $A=A^\dagger$, esto es, si
$A$ es simétrico y $\mathcal D(A)=\mathcal D(A^\dagger)$
\end{definition}

\begin{proposition}
Si $A$ es autoadjunto, se tiene $A=\bar A= A^{\dagger\dagger}=A^\dagger$.
\end{proposition}

\begin{definition}[Operador esencialmente autoadjunto:] $A$ es \emph{esencialmente autoadjunto} si
su cierre $\bar A$ es autoadjunto. Entonces, a $\bar A$ se lo denomina la extensión autoadjunta
de $A$.
\end{definition}

\begin{theorem}\label{th:esa}
Si $A$ es un operador
simétrico definido en $\mathcal H$ y $\rho$ es cualquier número no real,
entonces $A$ es esencialmente autoadjunto si y solo si la llamada {ecuación de
índices de defecto} de $A$, $A^\dagger\phi=\rho\phi$, no tiene solución en $\mathcal H$.
\end{theorem}

\begin{proposition}
Para todo operador simétrico, $\sigma_c$ y $\sigma_p$ yacen en el eje real. $\sigma_r$ es también
real si y solo si el operador, además, es esencialmente autoadjunto. Finalmente, $\sigma_r$ es vacío
si el operador es autoadjunto.
\end{proposition}

\section{Teorema espectral para operadores autoadjuntos}
\label{teorema-espectral}

Sea un operador $A$ autoadjunto y $B$ un conjunto boreliano en
$\mathbb{R}$, entonces existe una única medida espectral regular $E^A(B)$ asociada al operador $A$
tal que $E^A(\sigma(A))=1$. 

A su vez, esta medida espectral tiene asociada una única familia espectral $\{E^A_\lambda\}$ de
proyectores ortogonales definidos por $E^A_\lambda=E^A(-\infty,\lambda]$.

La familia espectral del operador autoadjunto $A$ definido en el espacio de Hilbert $\mathcal H$
define la \emph{resolución espectral de la identidad} en $\mathcal H$:
\begin{align}
 \mathbb{I}=\int_\mathbb{R} dE^A_\lambda,
\end{align}
y la \emph{descomposición espectral} del operador $A$
\begin{align}
 A=\int_\mathbb{R} \lambda dE^A_\lambda.
\end{align}
Esta integral debe ser entendida en el sentido:
\begin{align}
 \langle\psi_1|A\psi_2\rangle=\int_\mathbb{R} \lambda d\langle\psi_1|E^A_\lambda\psi_2\rangle,
\end{align}
para todo $\psi_1\in\mathcal H$ y para todo $\psi_2\in\mathcal
D(A)=\{\psi;\;\int_\mathbb{R}\lambda^2d||E^A_\lambda\psi||^2<\infty\}$.

Para toda función $f$ definida sobre $\mathbb{R}$, con valores finitos reales o complejos, y Borel
(basta, por ejemplo, que sea continua) entonces se define el operador $f(A)$ como
\begin{align}
f(A)&=\int_\mathbb{R} f(\lambda) dE^A_\lambda.
\end{align}

En definitiva, en el lenguaje de esta tesis, $\widehat{f(A)}=f(\widehat A)$. Asimismo, en general, 
por descomposición espectral asociada a un operador
esencialmente autoadjunto nos referimos a la descomposición
espectral de su extensión autoadjunta.

\subsection{El espectro de un operador autoadjunto}

Sea $\mu(\,\cdot\,)$ la medida espectral regular $E^A(\,\cdot\,)$ dada por el teorema espectral. Se
define su
parte \emph{puramente puntual} $\mu_{p.p}$ como
$\mu_{p.p}(B)=\sum_{\lambda\in\sigma(A)}\mu(\{\lambda\})\chi_B(\lambda)$, donde $\chi_B(\lambda)$
es la función característica de B: vale 1 si $\lambda\in B$ y se anula si $\lambda\notin B$.
Entonces, se obtiene la descomposición $\mu=\mu_{p.p}+\mu_c$, siendo $\mu_c$ \emph{continua}, esto
es, carece de puntos puros: $\mu_c(\{\lambda\})=0$ para todo $\lambda$. A su vez, $\mu_c$ puede
descomponerse en una parte \emph{absolutamente continua} $\mu_{a.c}$ (equivalente a la medida de
Lebesgue)%
\footnote{Recordamos que dos medidas se dicen equivalentes si comparten los mismos conjuntos de
medida nula.} y otra singularmente continua $\mu_{s.c}$ (es continua y existe $B$ nulo Lebesgue
tal que $\mu_{s.c}(\mathbb{R}-B)=0$). En resumen $\mu=\mu_{p.p}+\mu_{a.c}+\mu_{s.c}$ y,
correspondientemente,
\begin{align}
 \mathcal H=\mathcal H_{p.p}\oplus\mathcal H_{a.c}\oplus\mathcal H_{s.c}.
\end{align}
Estos subespacios son estables bajo $A$ y tales que las restricciones del operador $A$ a cada uno
de ellos,
\begin{align}
 A_{p.p}=A\upharpoonright\mathcal H_{p.p},\qquad A_{a.c}=A\upharpoonright\mathcal H_{a.c},
\qquad A_{s.c}=A\upharpoonright\mathcal H_{s.c},
\end{align}
son autoadjuntos con medidas $\mu_{p.p}$, $\mu_{a.c}$ y $\mu_{s.c}$ asociadas por el teorema
espectral.

Se definen los espectros \emph{puramente puntual}, $\sigma_{p.p}(A)=\sigma(A_{p.p})$,
\emph{absolutamente continuo}, $\sigma_{a.c}(A)=\sigma(A_{a.c})$, y \emph{singularmente continuo},
$\sigma_{s.c}(A)=\sigma(A_{s.c})$. Obsérvese que $\sigma_{p.p}=\bar{\sigma}_p$ y que
$\sigma=\sigma_{p.p}\cup\sigma_{a.c}\cup\sigma_{s.c}$,
aunque estos conjuntos no son necesariamente
disjuntos. Para el operador $A_{p.p}$, existe una base ortonormal de vectores propios. Por abuso del
lenguaje, si $A=A_{a.c}$, diremos que existe una base ortonormal generalizada o normalizable en
sentido generalizado (normalizable a la delta de Dirac) y a los puntos de $\sigma_{a.c}$ también los
llamaremos autovalores (en sentido generalizado).

Otra descompisición útil del espectro de $A$ es la siguiente: 
\begin{align}
 \sigma(A)=\sigma_{\text{disc}}(A)\cup\sigma_{\text{es}}(A),
\end{align}
donde el espectro \emph{discreto} $\sigma_{\text{disc}}$ está formado por los puntos del espectro
puntual que son de multiplicidad finita y aislados en $\sigma(A)$, y el espectro esencial
$\sigma_{\text{es}}(A)$ es su complemento en $\sigma(A)$ y es cerrado.

\section{Perturbación de operadores autoadjuntos}
\label{perturbaciones}

\begin{definition}[Operador pequeño:]
 Sea $A$ esencialmente autoadjunto y $V$ simétrico. Diremos que $V$ es $A$-\emph{pequeño} si
$\mathcal D(A)\subset\mathcal D(V)$, y existen $a<1$ y $b>0$ tales que
\begin{align}
 ||V\psi||\leq a||A\psi||+b||\psi||,\qquad\text{para todo }\psi\in\mathcal D(A).
\end{align}
\end{definition}

\begin{definition}[Operador compacto:] Un operador acotado $A$ se dice compacto (o completamente
continuo) si la imagen de cualquir sucesión acotada contenida en $\mathcal H$ contiene alguna
sucesión parcial convergente en $\mathcal H$.
\end{definition}

\begin{definition}[Operador de la clase de traza:] Un operador acotado $A$ pertenece a la
\emph{clase de traza} si para alguna base ortonormal $\{|v\rangle\}$ de $\mathcal H$ se verifica
\begin{align}
 \sum_{v}\langle v|(A^\dagger A)^{1/2}v\rangle<\infty.
\end{align}
Los operadores de la clase de traza, en particular, son compactos.
\end{definition}

\begin{theorem}\emph{(Teorema de Kato-Rellich)}\label{kato-rellich} Si $V$ es $A$-pequeño, entonces
$A+V$ es
esencialmente autoadjunto en $\mathcal D(A)$.
\end{theorem}

\begin{theorem} \label{weyl}Sea $A$ un operador autoadjunto y $V$ un operador compacto y
autoadjunto; entonces se verifica $\sigma_{\text{es}}(A+V)=\sigma_{\text{es}}(A)$.
\end{theorem}

\begin{theorem}\emph{(Teorema de Kato-Birman)}\label{kato-birman} Sean $A_1$ y $A_2$ dos operadores
autoadjuntos tales
que $A_1=A_2+V$, siendo $V$ de la clase de traza; entonces las partes absolutamente continuas de
$A_1$ y $A_2$ son unitariamente equivalentes.
\end{theorem}

\section{Teoría espectral de operadores normales}
\label{normales}

\begin{definition}[Operador normal:] Un operador acotado $A$ se dice normal si $[A,A^\dagger]=0$
\end{definition}

\begin{proposition}
 Sea $A$ un operador normal; entonces siempre se puede descomponer de la forma $A=N+iM$ con $N$ y
$M$
operadores autoadjuntos que conmutan entre sí.
\end{proposition}

Gracias a esta propiedad se puede definir la descomposición espectral de $A$ en
términos de las descomposiciones espectrales de los operadores autoadjuntos $N$ y $M$. En efecto,
sean $E^N(\,\cdot\,)$ y $E^M(\,\cdot\,)$ sus medidas espectrales respectivas.
A partir de ellas, se obtiene de forma única una medida espectral regular asociada a $A$, dada
por $E^A(\,\cdot\,,\,\cdot\,)=E^N(\,\cdot\,)E^M(\,\cdot\,)$, que define un cálculo
funcional conjunto de $A$ y $A^\dagger$: 
\begin{align}\label{des-nor}
 f(A,A^\dagger)=\int_{\sigma(N)}\int_{\sigma(M)}f(z,z^*)dE^{A}(x,y),
\end{align}
con $z=x+iy$, $z^*=x-iy$ y $f$ una función acotada Borel.

\cleardoublepage

\chapter[Consideraciones sobre la dinámica cuántica]{Otras consideraciones sobre la
dinámica del modelo de Bianchi I en cosmología cuántica de lazos}
\label{appB}

\section[Conservación del comportamiento semiclásico]{Conservación del comportamiento semiclásico en
el límite de WDW}

\subsection[En cada sector con límite de WDW bien definido]{En cada sector con límite
de WDW bien definido}
\label{dem1}

Esta sección contiene la demostración del siguiente resultado:

\begin{proof}[Proposición]
Sea la función de onda $\underline{\tilde\Phi}_-(\vec v)$ de la teoría de WDW, con
soporte en el producto de tres subsemirredes de paso cuatro (una por cada $v_i$) y cuyo perfil
espectral $\underline{\tilde\Phi}_-(\omega_\sigma,\omega_\delta)$ es el correspondiente a la
componente con $s=-1$ de la ec. \eqref{eq:wdw-limit-phys}, y sea $\underline{\tilde\Phi}_+(\vec v)$
la función definida de forma análoga y correspondiente a $s=1$. 

Si en el estado $\underline{\tilde\Phi}_-(\omega_\sigma,\omega_\delta)$ son finitos:
\begin{itemize}
 \item[(a)] los coeficientes $B_a$, $Y_a$, y $X_a$, definidos en las ecs.~\eqref{eq:traj-dir} y
\eqref{eq:sqr-dir},
\item[(b)] los valores esperados de los operadores de multiplicación $\ln|\omega_a|$,
$\Upsilon_a^{(2)}$, $w_a^2$ y $\Sigma^{(2)}_a$, definidos en las ecs.~\eqref{operators-dem},
\item[(c)] las dispersiones de los operadores $\ln|\omega_a|$, $\Upsilon_a^{(2)}$ y $w_a^2$,
\end{itemize}
entonces los coeficientes $B_a$, $Y_a$, y $X_a$ evaluados en
$\underline{\tilde\Phi}_+(\omega_\sigma,\omega_\delta)$ también son finitos.
\end{proof}

\begin{proof}[Demostración]
 Primero, recordemos que
\begin{align}\label{eq:coeff-reorg}
B_a = \langle \mathcal{D}_a \rangle, \qquad\quad Y_a =
\langle \mathcal{D}'_a \rangle,\qquad\quad
X_a-B_a^2 = \langle \Delta \mathcal{D}_a \rangle^2 =:
\sigma_{\mathcal{D}_a}^2,
\end{align}
donde
\begin{subequations}\label{eq:difXB}\begin{align}
\mathcal{D}_a &=
-i\alpha|\omega_\sigma+\omega_\delta|^{-\frac{1}{2}}
(\partial_{\omega_a})|\omega_\sigma+\omega_\delta|^{\frac{1}{2}} ,
\\ \mathcal{D}'_a &=
2\omega_\theta(\omega_\sigma,\omega_\delta)\omega_a^{-1} \,
\mathcal{D}_a \,
\omega_\theta(\omega_\sigma,\omega_\delta)\omega_a^{-1}.
\end{align}\end{subequations}
A partir de la relación entre $\underline{\tilde\Phi}_-$ y
$\underline{\tilde\Phi}_+$, buscamos las relaciones entre sus correspondientes valores esperados
\eqref{eq:coeff-reorg}. Obtenemos:
\begin{subequations}\label{eq:exp-disp-trans}\begin{align}
\langle \mathcal{D}_a \rangle_+ &= - \langle
\mathcal{D}_a \rangle_- -2\alpha\sum_{i=1}^3 \langle
[\partial_{\omega_a}
\phi(-\omega_i)] \rangle_- ,\label{eq:rel-exp}\\
\langle \mathcal{D}'_a \rangle_+ &= - \langle
\mathcal{D}'_a \rangle_- -4\alpha\sum_{i=1}^3 \langle
\frac{\omega_\theta^2}{\omega_a^{2}} [\partial_{\omega_a}
\phi(-\omega_i)] \rangle_-,
\label{eq:rel-exp2}\\
\sigma_{\mathcal{D}_a+} &\leq \sigma_{\mathcal{D}_a-}
+2\alpha\sum_{i=1}^3
\langle\Delta[\partial_{\omega_a}\phi(-\omega_i)]\rangle_-.
\label{eq:rel-disp}
\end{align}\end{subequations}
De hecho, es posible estimar los términos en los que aparece $\partial_{\omega_a}\phi(-\omega_i)$.
Para ello, hemos tenido en cuenta que la función $\phi(\omega)$ posee las siguientes propiedades
\cite{kp-posL}:
\begin{subequations}\begin{align}
&|\partial_{\omega}\phi (\omega)| \leq C_\theta
\left|\ln|\omega|\right| + C_0 , \label{eq:phi-1}\\
&|\omega\partial_{\omega}^2\phi (\omega)|
\leq C_\sigma \label{eq:phi-2} ,
\end{align}\end{subequations}
donde $C_0$, $C_\theta$ y $C_\sigma$ son constantes finitas positivas, que, sin embargo, pueden
depender del
valor de la etiqueta $\tilde{\varepsilon}_i$ de la subsemirred y, en particular, pueden no tener
una cota global (en todo el intervalo $\tilde{\varepsilon}_i\in(0,4]$).

Estas desigualdades pueden usarse para relacionar los términos de las ecs.
\eqref{eq:exp-disp-trans} con los valores esperados y dispersiones de los operadores 
$\ln|\omega_i|$. En el caso de la relación \eqref{eq:rel-exp}, el último término tiene una cota de
la forma [recuérdense las definiciones~\eqref{operators-dem}]:
\begin{align}\label{eq:bound-exp}
\left| \sum_{i=1}^3 \langle
[\partial_{\omega_a}\phi(-\omega_i)] \rangle_- \right|
\leq C_\theta \left[ \langle | \ln{|\omega_a|}
|\rangle_- + \langle |\Upsilon^{(2)}_a| \rangle_-
\right] + C_0(1+\langle w_a^2 \rangle) ,
\end{align}
mientras que para el término de la ec. \eqref{eq:rel-exp2} se cumple 
\begin{align}\label{eq:bound-exp2}
\left| \sum_{i=1}^3 \langle
\frac{\omega_\theta^2}{\omega_a^{2}}
[\partial_{\omega_a}\phi(-\omega_i)] \rangle_- \right|
\leq C_\theta \left[ \langle| \Sigma^{(2)}_a
|\rangle_- + \langle|\Upsilon^{(4)}_a|\rangle_- \right]
+ C_0\left[\langle w_a^2\rangle+\langle
w_a^4\rangle\right].
\end{align}
De modo similar, la suma de la ec. \eqref{eq:rel-disp} satisface
\begin{align}\label{eq:bound-disp}
\sum_{i=1}^3 \langle\Delta[\partial_{\omega_a}
\phi(-\omega_i)]\rangle_- \leq C_\sigma\left[
\langle\Delta\ln|\omega_a|\rangle_- + \langle\Delta
\Upsilon^{(2)}_a\rangle_- \right].
\end{align}

Las condiciones (a), (b) y (c) que verifica el estado $\underline{\tilde\Phi}_-$ considerado
implican inmediatamente que los miembros derechos de las ecs.
\eqref{eq:bound-exp}-\eqref{eq:bound-disp} son finitos. Este resultado, junto con el hecho de que 
$C_0$ ,$C_\theta$ y $C_\sigma$ son finitas, permite comprobar directamente con las ecs.
\eqref{eq:exp-disp-trans}
que, si los coeficientes $B_a$, $X_a$ y $Y_a$ son finitos para $\underline{\tilde\Phi}_-$, entonces
también lo son los correspondientes a $\underline{\tilde\Phi}_+$, como queríamos demostrar.
\end{proof}

Nótese que, en toda esta discusión, los papeles que $\underline{\tilde\Phi}_+$ y
$\underline{\tilde\Phi}_-$ desempeñan pueden ser intercambiados, de modo que se pueden imponer
condiciones del tipo explicado arriba sobre $\underline{\tilde\Phi}_+$ y entonces asegurar el
buen comportamiento de las dispersiones relativas correspondientes a $\underline{\tilde\Phi}_-$.

\subsection[En el conjunto de sectores]{En el conjunto de los ocho sectores}
\label{dem2}

En esta sección analizamos el comportamiento de las dispersiones en el régimen de WDW
teniendo en cuenta los ocho sectores en los que las funciones de onda de la cosmología cuántica de
lazos tienen un límite bien definido, en lugar de considerar cada uno de esos sectores por separado.
Queremos definir el conjunto de estos sectores de tal modo que las propiedades de los valores
esperados y dispersiones de los observables de la cosmología cuántica de lazos queden
reflejados en características de sus análogos en este conjunto.

Para ello, comenzamos notando las siguientes propiedades de los operadores
$\ln(\hat{v}_a)_{v_\theta}$:
\begin{enumerate}[(a)]
  \item Como cada operador está definido por un valor concreto de $v_\theta$, que pertenece a una
de las subsemirredes de paso cuatro, ${}^{(4)}\mathcal{L}_{\varepsilon_\theta}$
ó ${}^{(4)}\mathcal{L}_{\varepsilon_\theta+2}$, queremos considerar estados con soporte en cada una
de ellas por separado. En consecuencia, podemos considerar sin problema dos límites independientes,
en correspondencia con la división del soporte en $v_\theta$.
\item Para un $v_\theta$ dado, la acción del operador $\ln(\hat{v}_a)_{v_\theta}$ se puede
representar como la acción de la suma de las respectivas restricciones a cada uno de los cuatro
sectores obtenidos al dividir en subsemirredes el soporte en las direcciones $a=\sigma,\delta$.
\end{enumerate}
Aplicando el método de cálculo explicado en la Sección \ref{sec:v-num} a los valores esperados y
dispersiones, se ve inmediatamente que
\begin{itemize}
  \item[(i)] los valores esperados de $\ln(\hat{v}_a)_{v_\theta}$ son promedios aritméticos de los
valores esperados de las restricciones, y que
  \item[(ii)] las dispersiones están acotadas por un promedio aritmético de las dispersiones de cada
uno de los cuatro sectores que contribuyen, al que hay que añadir términos de la forma
    $\sqrt{\langle \hat{D}_{m}^2 \rangle}$, donde el operador $\hat{D}_{m}$
    mide la diferencia entre el valor esperado total y el valor esperado en el sector considerado,
etiquetado por $m$, donde $m=1,...4$.
\end{itemize}

Por otra parte, hemos probado ya que, para cada uno de los ocho sectores, la dispersión
relativa de $\ln(\hat{v}_a)_{v_\theta}$ ($a=2,3$) es la misma en las dos componentes que se mueven
hacia adelante o hacia atrás en el tiempo (etiquetadas con + y -), asumiendo que una de tales
componentes satisface ciertas condiciones. Para extender este resultado al conjunto de sectores,
hay que tener en cuenta las posibles diferencias entre los valores esperados de
$\ln(\hat{v}_a)_{v_\theta}$ mencionadas arriba en el punto (ii).
Para verificar que estas diferencias están acotadas, observamos que la relación entre las
restricciones de las autofunciones a cada sector [ec. \eqref{eq:estados-2-bis}], junto con la
relación
entre la normalización de estas restricciones y la de sus límites de WDW, implican que
los coeficientes $A_a$ y $W_a$ son iguales para todos los sectores. De la invariancia de estos
coeficientes bajo la
tranformación $\underline{\tilde\Phi}_-\to\underline{\tilde\Phi}_+$ y las consideraciones hechas en
la Sección \ref{sec:wdw-limit-disp}, se deduce que las diferencias analizadas tienen, de hecho, un
límite finito bien definido cuando $v_\theta\to\infty$.

Nótese también que, como en el caso de la discusión para los sectores por separado, los 
argumentos anteriores pueden repetirse considerando la transformación
$\underline{\tilde\Phi}_+\to\underline{\tilde\Phi}_-$, opuesta a la estudiada aquí.

Finalmente, cabe señalar que el requerimiento de que los operadores $\Upsilon^{(2)}_a$ tengan
valores esperados y dispersiones finitos en el estado estudiado, aunque represente una
restricción en el espacio de estados físicos, se puede considerar bastante razonable desde un punto
de vista físico. De hecho, dado que la elección de $v_\theta$ como tiempo interno y, por tanto, la
interpretación de $\omega_\theta$ como una ``frecuencia temporal'', es arbitraria, se pueden
cambiar los papeles desempeñados por $\omega_\theta$ y $\omega_\sigma$, por ejemplo, y considerar
la transformación siguiente del estado $\tilde{\Phi}(\omega_\sigma,\omega_\delta)$:
\begin{equation}\label{eq:trans1}
\tilde{\Phi}\to\tilde{\Phi}' : \
\tilde{\Phi}'(\omega_\sigma,\omega_\delta)=\tilde{\Phi}
(\omega_\theta(\omega_\sigma,\omega_\delta),\omega_\delta).
\end{equation}
Por otra parte la transformación
\begin{equation}\label{eq:trans2}
\tilde{\Phi}(\omega_\sigma,\omega_\delta) \mapsto
\frac{\omega_\theta^2(\omega_\sigma,\omega_\delta)}{\omega_\sigma^2}
\tilde{\Phi}(\omega_\theta(\omega_\sigma,\omega_\delta),\omega_\delta)
\end{equation}
se corresponde justamente con el intercambio de la variable $v_\theta$ por $v_\sigma$.
Obviamente, se pueden definir transformaciones similares también para el intercambio 
$v_\theta \leftrightarrow v_\delta$. Combinando ambos intercambios, se puede reescribir el cuadrado
de $\Upsilon^{(2)}_a$ como el cuadrado de $\ln|\omega_\sigma|$ actuando en el estado representado
por 
$\tilde{\Phi}'$. Por tanto, como este cuadrado es la mayor contribución a la dispersión, se puede
relacionar $\langle\Delta\Upsilon^{(2)}_a\rangle$ con la dispersión del operador $\ln|\omega_a|$ en
$\tilde{\Phi}'$, al menos de forma heurística. Asimismo, argumentos similares se pueden aplicar a
$\Sigma^{(2)}_a$.

\section[Dinámica clásica efectiva]{Dinámica clásica efectiva: modelo de Bianchi I en el esquema A}
\label{effective}

La dinámica cuántica asociada a la cosmología cuántica de lazos resulta estar muy bien reproducida
por cierta dinámica clásica efectiva, construida simplemente reemplazando en el operador ligadura
hamiltoniana las holonomías y los flujos por sus valores esperados \cite{victor,sv-eff}. Se ha
demostrado en muchas situaciones que esta construcción, en principio bastante ingenua, reproduce con
mucha precisión la evolución cuántica genuina, con discrepancias mucho menores que las dispersiones
cuánticas (véase, por ejemplo, las Refs. \cite{aps3,luc2}). 

En esta sección resumimos su formulación para el caso del modelo de Bianchi I en vacío y cuantizado
según el esquema A de la dinámica mejorada. Partimos del operador ligadura
\eqref{eq:lig-dens-bian-A} y le aplicamos las sustituciones que acabamos de mencionar. El
hamiltoniano clásico \emph{efectivo} resultante tiene la forma
\begin{subequations}\label{eq:effH-def}\begin{align}
H^{\text{ef}} &= -\frac{2}{\gamma^2}
[\Omega^{\text{ef}}_\theta\Omega^{\text{ef}}_\sigma+\Omega^{\text{ef}}_\theta
\Omega^{\text{ef}}_\delta+\Omega^{\text{ef}}_\sigma\Omega^{\text{ef}}_\delta] , \\
\Omega^{\text{ef}}_i &:= 6\pi\gamma G v_i \sin(\beta_i) ,
\end{align}\end{subequations}
donde, recordemos, la variable $\beta_i$ ($i=1,2,3$) es el momento conjugado a $v_i$ [véanse las
ecs.~\eqref{eq:b-def} y \eqref{eq:bv-poiss}]. Este momento está compactificado como el círculo
unidad.
Dado el hamiltoniano \eqref{eq:effH-def}, podemos derivar fácilmente las ecuaciones de
Hamilton-~Jacobi para $\partial_{\tau}v_i$ y $\partial_{\tau}\beta_i$, donde $\tau$ es el tiempo
coordenado asociado a este hamiltoniano. Definiendo las constantes de movimiento 
$\Omega^{\text{ef}}_i = 8\pi\gamma G\mathcal{K}_i$ y $\mathcal{K} =
\mathcal{K}_\theta + \mathcal{K}_\sigma + \mathcal{K}_\delta$ (como en la Ref. \cite{chi2}) tenemos
\begin{subequations}\begin{align}
\label{veff} (\partial_{\tau} v_i)^2 &= 9 (8\pi G)^4
(\mathcal{K}-\mathcal{K}_i)^2\left[v_i^2 - \left(\frac{4\mathcal{K}_i}{3}
\right)^2\right], \\
\label{b}
\partial_{\tau}b_i &= 3(8\pi G)^2
(\mathcal{K}_i-\mathcal{K})\sin(\beta_i).
\end{align}\end{subequations}

La ec. \eqref{b} implica inmediatamente que, dentro de los intervalos $\beta_i\in[0,\pi)$ y
$\beta_i\in[\pi,2\pi)$ que en la cuantización polimérica se corresponden respectivamente con las
subsemirredes ${}^{(4)}\!\mathcal{L}_{\varepsilon_i}^+$ y
${}^{(4)}\!\mathcal{L}_{\varepsilon_i+2}^+$ (véase la Sec.
\ref{sec:lqc-b-rep}), cada una de las variables $\beta_i$ proporciona un buen candidato para ser
interpretado como un tiempo interno, por ser monótono en $\tau$. Por otra parte, la solución general
de la ec. \eqref{veff} es
\begin{align}
v_i(\tau) &= \frac{4}{3}\mathcal{K}_i \cosh[3 (8\pi G)^2
(\mathcal{K}-\mathcal{K}_i)(\tau-\tau_o)],
\end{align}
donde $\tau_o$ es una constante que representa el instante del rebote. Por tanto, $v_i$ no es
monótona en el tiempo coordenado.

\cleardoublepage


{\renewcommand{\thechapter}{}\renewcommand{\chaptername}{}
\addtocounter{chapter}{0}
\chapter*{Publicaciones}\markboth{\sl PUBLICACIONES}{\sl PUBLICACIONES}}
\addcontentsline{toc}{chapter}{Publicaciones} 

El trabajo de investigación aquí contenido ha dado lugar a cinco publicaciones y a dos trabajos
enviados a Phys. Rev. D:
\vspace{0.5cm}
\begin{itemize}
\item M. Mart\'{\i}n-Benito, L. J. Garay y G. A. Mena Marug\'{a}n, Hybrid Quantum Gowdy
Cosmology: Combining Loop and Fock Quantizations, Phys. Rev. D \textbf{78}, 083516 (2008).
\vspace{0.3cm}
 \item M. Mart\'{\i}n-Benito, G. A. Mena Marug\'{a}n y T. Paw{\l}owski, Loop Quantization
of Vacuum Bianchi I Cosmology, Phys. Rev. D {\bf 78}, 064008 (2008).
\vspace{0.3cm}
 \item G. A. Mena Marug\'{a}n y M. Mart\'{\i}n-Benito, Hybrid Quantum Cosmology: Combining
Loop and Fock Quantizations, Int. J. Mod. Phys. A {\bf24}, 2820 (2009).
\vspace{0.3cm}
 \item M. Mart\'{\i}n-Benito, G. A. Mena Marug\'{a}n y T. Paw{\l}owski, Physical
evolution in Loop Quantum Cosmology: The Example of vacuum Bianchi I, Phys. Rev. D {\bf80}, 084038
(2009).
\vspace{0.3cm}
\item M. Mart\'{\i}n-Benito, G. A. Mena Marug\'{a}n y J. Olmedo, Further Improvements in
the Understanding of Isotropic Loop Quantum Cosmology, Phys. Rev. D {\bf80}, 104015 (2009).
\vspace{0.3cm}
\item L. J. Garay, M. Mart\'{\i}n-Benito y G. A. Mena Marug\'{a}n, Inhomogeneous\linebreak Loop
Quantum
Cosmology: Hybrid Quantization of the Gowdy Model, Phys. Rev. D {\bf 82}, 044048 (2010).
\vspace{0.3cm}
\item M. Mart\'{\i}n-Benito, G. A. Mena Marug\'{a}n y E. Wilson-Ewing, Hybrid Quantization: from
Bianchi I to the Gowdy Model, Phys. Rev. D {\bf82}, 084012 (2010).
\end{itemize}

\cleardoublepage

\cleardoublepage\addcontentsline{toc}{chapter}{Bibliografía}



\end{document}